\documentclass[12pt]{article} 
\usepackage[a4paper,textwidth=490.0pt,textheight=703.1pt]{geometry}
\usepackage{eucal}
\usepackage{url}
\usepackage{booktabs}
\usepackage{epsfig} %
\usepackage{float} %
\usepackage{amssymb} %
\usepackage{amsmath} %
\usepackage{verbatim} %
\usepackage{cite} %
\usepackage{color} %
\usepackage{eufrak}
\usepackage[english]{babel}
\usepackage[latin1]{inputenc}
\usepackage[T1]{fontenc}
\usepackage{ae}
\usepackage{url}   
\usepackage{graphicx} 
\usepackage{booktabs}
\usepackage{slashed}

\newcommand{\HA}{{\rm H}}

\newcommand{\ep}{\varepsilon}

\newcommand{\Mvec}{\mbox{\rm\bf M}}

\newcommand{\beq}{\begin{equation}}
\newcommand{\eeq}{\end{equation}}
\newcommand{\bea}{\begin{eqnarray}}
\newcommand{\eea}{\end{eqnarray}}

\allowdisplaybreaks[4]

\begin{document} 
\setlength{\baselineskip}{0.515cm}

\sloppy 
\thispagestyle{empty} 
\begin{flushleft} 
DESY 22--123 \hfill 
\\ 
DO--TH 22/20
\\ 
TTP 22--057
\\ 
RISC Report Series 22--12
\\
SAGEX--22--30
\\ 
August 2022 
\end{flushleft}

\mbox{} \vspace*{\fill} \begin{center}

{\Large\bf The massless three-loop Wilson coefficients for the}

\vspace*{2mm} 
{\Large\bf deep-inelastic structure functions \boldmath $F_2, F_L, xF_3$ and $g_1$}

\vspace{3cm} 
\large 
{\large J.~Bl\"umlein$^a$, P.~Marquard$^a$, C.~Schneider$^b$ and K.~Sch\"onwald$^{c}$ }

\normalsize 

\vspace{1.cm} 
{\it $^a$~Deutsches Elektronen--Synchrotron DESY,}\\ {\it Platanenallee 6, 15738 Zeuthen, Germany}

\vspace*{2mm} 
{\it $^b$~
Johannes Kepler University Linz, Research Institute for Symbolic
Computation (RISC), Altenberger Stra\ss{}e 69, A--4040 Linz, Austria}

\vspace*{2mm} 
{\it $^c$~Institut f\"ur Theoretische Teilchenphysik,\\ Karlsruher Institut f\"ur Technologie (KIT) D--76128 
Karlsruhe, Germany}


\end{center} 
\normalsize 
\vspace{\fill} 
\begin{abstract} 
\noindent
We calculate the massless unpolarized Wilson coefficients for deeply inelastic scattering for the
structure functions $F_2(x,Q^2), F_L(x,Q^2), x F_3(x,Q^2)$ in the $\overline{\sf MS}$ scheme and the 
polarized Wilson coefficients of the structure function $g_1(x,Q^2)$ in the Larin scheme up to three--loop 
order in QCD in a fully automated way based on the method of arbitrary high Mellin moments. We work
in the Larin scheme in the case of contributing axial--vector couplings or polarized nucleons. For the 
unpolarized structure functions we compare to results given in the literature. The polarized three--loop 
Wilson coefficients are calculated for the first time. As a by--product we also obtain the quarkonic 
three--loop anomalous dimensions from the $O(1/\ep)$ terms of the unrenormalized forward Compton amplitude. 
Expansions for small and large values of the Bjorken variable $x$ are provided.
\end{abstract}

\vspace*{\fill} \noindent
\newpage

\section{Introduction} 
\label{sec:1}

\vspace*{1mm} 
\noindent
The scaling violations of the deep--inelastic structure functions $F_k(x,Q^2)$ \cite{DIS,REV,
Politzer:1974fr,Blumlein:2012bf} in the massless limit and twist 2 approximation are described by those 
of the twist 2 parton distribution functions $f_i$ and massless Wilson coefficients $\mathbb{C}_j$ in a 
perturbative expansion in the strong coupling constant $a_s(Q^2) = \alpha_s(Q^2)/(4\pi)$, where 
$Q^2 = - q^2$ denotes the space--like virtuality of the scattering process and $x = Q^2/(2 p.q)$ 
is the Bjorken variable, with $p$ the nucleon momentum. In the present paper we  
calculate the Wilson coefficients in the case of virtual photon exchange in the unpolarized and polarized 
case to three--loop order, where the latter results are new. For the structure function $xF_3$ we consider
the exchange of {both} the weak $W^\pm$ bosons. The present results are of importance for the
experimental data analysis of the deep--inelastic scattering data at HERA and future lepton--nucleon colliders,
such as the EIC \cite{Boer:2011fh} and the LHeC--project \cite{LHeCStudyGroup:2012zhm} for precision measurements in 
the unpolarized and polarized case, to extract the parton distribution functions \cite{Accardi:2016ndt}
and to measure the strong 
coupling constant at highest possible precision \cite{alphas}.

The calculation is based on  massless virtual forward Compton amplitudes for on--shell partonic states, which 
are gauge--invariant quantities. The structure functions are described by
\begin{eqnarray}
F_k(x,Q^2)  = \sum_i \mathbb{C}_{k,i}\left(x,\frac{Q^2}{\mu^2}\right) \otimes f_i(x,\mu^2)
\label{eq:SF}
\end{eqnarray}
with
\begin{eqnarray}
\mathbb{C}_{k,i}\left(x,\frac{Q^2}{\mu^2}\right) = \delta_{iq} + \sum_{l=1}^\infty a_s^l \sum_{n=0}^{l} 
\ln^n\left(\frac{Q^2}{\mu^2}\right) C_{k,i}^{(l,n)}(x), 
\label{eq:WC1}
\end{eqnarray}
where $\mu^2$ denotes the factorization scale and 
$\otimes$  the Mellin convolution
\begin{eqnarray}
\label{eq:MEL1}
A(x) \otimes B(x) = \int_0^1 dx_1 \int_0^1 dx_2  \delta(x-x_1 x_2) A(x_1) B(x_2).
\end{eqnarray}
The Mellin transform
\begin{eqnarray}
\label{eq:MEL2}
\Mvec[A(x)](N) = \int_0^1 dx x^{N-1} A(x)
\end{eqnarray}
diagonalizes the expression in Eq.~($\ref{eq:MEL1}$) into the product $\Mvec[A](N) \cdot \Mvec[B](N)$ and will therefore be often used
in the subsequent calculation. The dependence on the factorization scale diminishes for higher and higher 
orders in the coupling constant. One may also derive representations in Mellin $N$ space which are scheme--invariant
order by order, see e.g. \cite{Blumlein:2000wh,Blumlein:2004xs,SI}. Here the preferred scale of the 
deep--inelastic process is $\mu^2 = Q^2$.

The massless Wilson coefficients for the structure functions $F_2(x,Q^2), F_L(x,Q^2)$ and 
$x F_3(x,Q^2)$ 
have been calculated at one-- \cite{Furmanski:1981cw}, two-- \cite{TWOLOOP,Zijlstra:1992qd} and three--loop 
order 
\cite{Vermaseren:2005qc,Moch:2007rq,Moch:2008fj}. Furthermore, also their Mellin moments have been 
computed at 2-- \cite{Larin:1991fv} and 3--loop order \cite{THRLMOM}. In the polarized massless case the 
corresponding 1-- and 2--loop Wilson coefficients for the structure function $g_1(x,Q^2)$ have been obtained 
in \cite{Kodaira:1978sh,Zijlstra:1993sh,FORTRAN,Vogt:2008yw,Blumlein:2019zux}.

In the present paper we re--calculate the massless Wilson coefficients for the structure functions 
$F_2(x,Q^2)$ and $F_L(x,Q^2)$ as well as for the structure function $xF_3(x,Q^2) \equiv 
xF_3^{W^+}(x,Q^2) + xF_3^{W^-}(x,Q^2)$ and calculate newly those for the polarized structure function
$g_1(x,Q^2)$ in the Larin scheme \cite{Larin:1993tq,Matiounine:1998re}. The Wilson coefficients are 
calculated using the method of the forward Compton amplitude. Here the parameter $x = Q^2/(2p.q)$
is the variable on which the master integrals and the associated differential equations depend. 
{The Mellin moments of the structure functions correspond to the expansion coefficients of 
the forward Compton amplitude in the unphysical limit $\omega = 1/x \ll 1$ which is frequently 
exploited
to compute low Mellin moments. 
Therefore, the variable $\omega$ corresponds to the parameter $t$ used in Refs.~
\cite{Blumlein:2021enk,Blumlein:2021ryt,Blumlein:2022ndg} and the subsequent technical steps are quite
similar to those in the calculation of the three--loop anomalous dimensions.
}

The paper is organized as follows. In Section~\ref{sec:2} we discuss the basic formalism, including the 
renormalization, and summarize the technical details of the calculation in Section~\ref{sec:3}.
In Section~\ref{sec:4} we calculate the one-- and two--loop Wilson coefficients to the order in the 
dimensional parameter $\ep = D - 4$, needed for the present calculation at three--loop order and present
the 2--loop Wilson coefficients. In Section~\ref{sec:5} the three--loop Wilson coefficients for the structure 
functions $F_2(x,Q^2)$ and $F_L(x,Q^2)$ are given and in Section~\ref{sec:6} those for the structure function
$xF_3(x,Q^2)$. Finally, we present in Section~\ref{sec:7} the three--loop Wilson coefficients for the polarized 
structure function $g_1(x,Q^2)$ and also remark that the quarkonic anomalous dimensions being obtained as a 
by--product of the present calculation from the $O(1/\ep)$ terms of the {unrenormalized} Compton amplitude,
agree with the results in the literature. In Section~\ref{sec:8} we present the small--$x$ and large--$x$
expansions of the respective Wilson coefficients and compare to the large $N_F$ behaviour predicted in the 
literature. Here $N_F$ denotes the number of massless flavors. Section~\ref{sec:9} contains the conclusions.
To shorten the presentation we will identify the factorization scale $\mu^2 = Q^2$ and only print the Mellin 
$N$ space representation for the complete expressions. In an ancillary file to this paper we present all 
Wilson coefficients in Mellin $N$ and $z$--space, in full form, including their scale dependence.
Here $z$ denotes the parton momentum fraction in the 
nucleon, $z \in [0,1]$, and is, for twist--2, identical to the Bjorken variable $x$.
\section{Basic Formalism}
\label{sec:2}

\vspace*{1mm}
\noindent
We consider the forward Compton amplitude for neutral current photon exchange in the unpolarized 
case for the deep--inelastic structure functions $F_2(x,Q^2)$ and $F_L(x,Q^2)$, as well {as} for the 
structure functions $g_1(x,Q^2)$ {in the polarized 
case} in the twist--2 approximation to three--loop order. 
Furthermore, we study the charged current structure function $xF_3(x,Q^2)$.

The hadronic tensor has the principal structure \cite{Blumlein:1998nv,Blumlein:2012bf}
\begin{eqnarray}
\label{eq:Wmunu}
W_{\mu\nu} &=& \left( -g_{\mu\nu} + \frac{q_\mu q_\nu}{q^2}\right) F_1(x,Q^2) + \frac{\hat{p}_\mu 
\hat{p}_\nu}{p.q} F_2(x,Q^2) 
- i \ep_{\mu\nu\lambda\sigma} \frac{q^\lambda p^\sigma}{2 p.q} F_3(x,Q^2)
\nonumber\\ &&
+ i \ep_{\mu\nu\lambda\sigma} \frac{q^\lambda S^\sigma}{p.q} g_1(x,Q^2) + \cdots ~.
\end{eqnarray}
Here the ellipses denote contributions from other structure functions not considered in the present paper, $p$ 
denotes the nucleon 4--momentum and $S$ the polarization vector of the nucleon,
which can be taken in the longitudinal case in the nucleon rest frame
\begin{eqnarray}
S_L = (M,0,0,0) ~,
\end{eqnarray}
where only the energy component is non--vanishing and $M$ denotes the nucleon mass. The vector $\hat{p}$ is given by
\begin{eqnarray}
\hat{p} = p_\mu - \frac{p.q}{q^2} q_\mu ~.
\end{eqnarray}
All structure functions can be isolated by using corresponding projectors in $D = 4 + \ep$ space--time dimensions.
In the polarized case one furthermore considers the polarization asymmetry, i.e. the difference for $S = S_L$ and 
$-S_L$. In the following we will consider the structure functions $F_{2,L}(x,Q^2)$ and $g_1(x,Q^2)$ in the case of 
pure photon exchange, while $xF_3(x,Q^2)$ is measured for the arithmetic mean of $W^\pm$ boson exchange.

The associated forward Compton amplitudes are given by the Fourier transform of the time--ordered
product of current operators
\begin{eqnarray}
T_{\mu\nu} = i \int dz~{\rm e}^{iqz} \langle p|T\left(J^\dagger_\mu(z) J_\nu(0)\right)|p\rangle ~,
\end{eqnarray}
and
\begin{eqnarray}
W_{\mu\nu} =  \frac{1}{\pi} {\sf Im} T_{\mu\nu} ~.
\end{eqnarray}
We consider the deep--inelastic scattering cross section for pure photon exchange \cite{Blumlein:1996vs} 
\begin{eqnarray}
\frac{d^2\sigma^{\gamma^*}(\lambda, \pm S_L)}{dxdy}
&=& 2 \pi S  \frac{\alpha^2}{Q^4}
\Biggl \{ y^2 2xF_1(x,Q^2) + 2\left (1 - y
- \frac{x y M^2}{S} \right) F_2(x,Q^2)
\nonumber\\
&\pm& \left [
-2\lambda y \left( 2-y-\frac{2 x y M^2}{S} \right) xg_1(x,Q^2) +
8 \lambda \frac{y x^2 M^2}{S} g_2(x,Q^2)
\right] 
\Biggr\} ~,
\end{eqnarray}
with $\lambda$ the longitudinal charged lepton polarization, $M$ the nucleon mass, $y = Q^2/Sx$ a
Bjorken variable, and $S$ the cms energy squared. We will use the collinear parton model \cite{FEYNM} in the 
present calculation, where the parton momentum is given by $zp$, with $p$ the nucleon momentum and $z \in [0,1]$. 
Due to this the Wilson coefficients of the structure function $g_2(x,Q^2)$ cannot be calculated. It requires the 
use of the covariant parton model \cite{G2rel}, already at tree level.

One furthermore has
\begin{eqnarray}
F_L(x,Q^2) = F_2(x,Q^2) - 2 xF_1(x,Q^2) ~.
\end{eqnarray}
The structure functions are related to the parton densities for pure photon exchange at tree level as
\cite{Blumlein:2012bf}
\begin{eqnarray}
F_2(x,Q^2) &=& x \sum_{f=1}^{N_F} e_f^2 [q_f(x,Q^2) +\bar{q}_f(x,Q^2)] ~,
\\
g_1(x,Q^2) &=& \frac{1}{2} \sum_{f=1}^{N_F} e_f^2 [\Delta q_f(x,Q^2) +\Delta \bar{q}_f(x,Q^2)] ~,
\end{eqnarray}
for pure photon exchange, with $q_i (\bar{q}_i)$ the unpolarized and  $\Delta q_i (\Delta \bar{q}_i)$ the 
polarized parton densities and $F_L = 0$.

In the charged current case we consider the massless Wilson coefficients of the structure function $xF_3(x,Q^2)$.
One should notice, that here also strange--charm transitions occur, which imply heavy flavor corrections
\cite{CCheavy,Blumlein:2014fqa,Behring:2015roa,Behring:2016hpa}. They are of importance in quantitative analyses. 
Here we 
consider all contributing quark flavors as massless. For incoming charged leptons the scattering cross section 
reads, cf.~\cite{Blumlein:1987xk,Arbuzov:1995id,SCHMITZ},
\begin{eqnarray}
\frac{d^2\sigma^{\rm cc}}{dxdy}
&=& \frac{G_\mu^2 S}{8 \pi}  \left[\frac{M_W^2}{Q^2+M_W^2}\right]^2 \Biggl\{Y_+ W_2(x,Q^2) + Y_- W_3(x,Q^2)
- y^2 W_L(x,Q^2) \Biggr\} ~,
\end{eqnarray}
where $G_\mu$ denotes the Fermi constant and $Y_\pm = 1 \pm (1-y)^2$ and both unpolarized leptons and nucleons 
are considered. The structure functions are given by
\begin{eqnarray}
W_2(x,Q^2) &=& F_2^{{\rm cc}, Q_l}(x,Q^2) ~,
\\
W_3(x,Q^2) &=& - {\rm sign}(Q_l) xF_3^{{\rm cc}, Q_l}(x,Q^2) ~,
\end{eqnarray}
where $Q_l$ is the charge of the incoming charged lepton. At tree level these structure functions have the
following quark flavor decomposition \cite{Arbuzov:1995id}
\begin{eqnarray}
F_2^{{\rm cc}, +}(x,Q^2)  &=& 2x \sum_i \left[d_i(x,Q^2) + \bar{u}_i(x,Q^2) \right] ~,
\\
F_2^{{\rm cc}, -}(x,Q^2)  &=& 2x \sum_i \left[u_i(x,Q^2) + \bar{d}_i(x,Q^2) \right] ~,
\\
xF_3^{{\rm cc}, +}(x,Q^2) &=& 2x \sum_i \left[d_i(x,Q^2) - \bar{u}_i(x,Q^2) \right] ~,
\\
xF_3^{{\rm cc}, -}(x,Q^2) &=& 2x \sum_i \left[u_i(x,Q^2) - \bar{d}_i(x,Q^2) \right] ~,
\end{eqnarray}
e.g. for four flavors\footnote{More generally, one has to account for the Cabibbo--Kobayashi--Maskawa 
mixing, cf. e.g.~\cite{Blumlein:2014fqa},
effectively redefining the down--quark densities in the charged current case.}, 
and $W_L = F_L^{{\rm cc}, Q_l}$ denotes the longitudinal structure function with
$F_L^{{\rm cc}, \pm} = F_2^{{\rm cc}, \pm} - 2xF_1^{{\rm cc}, \pm}$.

In the present paper we consider in the charged current case the flavor non--singlet structure function 
combination $xF_3^{{\rm cc}, +}(x,Q^2) + xF_3^{{\rm cc}, -}(x,Q^2)$. At the experimental side one might 
consider the scattering off deuteron targets, obeying the $SU(2)$ isospin flavor symmetry 
\begin{eqnarray}
u(x,Q^2) =  d(x,Q^2) ~, ~~~~
\bar{u}(x,Q^2) =  \bar{d}(x,Q^2) ~,
\end{eqnarray}
after deuteron wave function corrections. Through this one obtains at tree level
\begin{eqnarray}
F_2^{{\rm cc},d,+} &=& x[u+d+2s +\bar{u} + \bar{d} + 2 \bar{c}] ~,
\\
F_2^{{\rm cc},d,-} &=& x[u+d+2c +\bar{u} + \bar{d} + 2 \bar{s}] ~,
\\
F_2^{{\rm cc},d,+} + F_2^{{\rm cc},d,-} &=&  2x \Sigma  ~,
\\
xF_3^{{\rm cc},d,+} &=& x[u+d+2s -\bar{u} - \bar{d} - 2 \bar{c}]  ~,
\\
xF_3^{{\rm cc},d,-} &=& x[u+d+2sc -\bar{u} - \bar{d} - 2 \bar{s}]  ~,
\\
xF_3^{{\rm cc},d,+} + xF_3^{{\rm cc},d,-} &=&  2x [u+d -\bar{u} -\bar{d} + s - \bar{s} +c - \bar{c}]  ~,
\nonumber\\ & \equiv & 2x [u_v + d_v],
\end{eqnarray}
and
\begin{eqnarray}
xF_3 &=&
-\frac{8\pi}{Y_- G_\mu^2 S} \Biggl[
\frac{d^2\sigma^{{\rm cc},d,+}}{dxdy}
- \frac{d^2\sigma^{{\rm cc},d,-}}{dxdy}\Biggr]
+ \frac{Y_+}{Y_-} \left(
  F_2^{{\rm cc},d,+}
- F_2^{{\rm cc},d,+}\right) - \frac{y^2}{Y_-}\left(W_L^{+,d} - W_L^{-d}\right) ~. 
\end{eqnarray}
Here the non--singlet combinations for $F_2^d$ and $W_L^d$ $\propto Y_+, y^2$ correspond to the quark flavor 
combination
\begin{eqnarray}
f^{\rm NS, +} = 2x[(s - \bar{s}) - (c - \bar{c})]  ~,
\end{eqnarray}
which vanishes under the assumption of symmetric sea--quarks of the same flavor and one obtains 
the direct projection on $xF_3^d$ by the weighted differential cross section difference.  In other 
cases one has to perform fits in Bjorken $y$ to separate the $Y_-$ from the $Y_+$ and $y^2$ contributions.

The projectors of the hadronic tensor allowing to isolate the Wilson coefficients of the different structure 
functions are given by
\begin{eqnarray}
    P_L^{\mu\nu} &=& \frac{8 x^3}{Q^2} p^\mu p^\nu
    ~, \\
    P_2^{\mu\nu} &=& - \frac{2x}{D-2} \left( g^{\mu\nu} -
\bigl(D-1\bigr) \frac{4x^2}{Q^2} p^\mu p^\nu \right)
    ~, \\
    P_3^{\mu\nu} &=& \frac{-i}{(D-2)(D-3)} \frac{4x}{Q^2}
\varepsilon^{\mu\nu\rho\sigma} p_\rho q_\sigma 
    ~, \\
    P_{g_1}^{\mu\nu} &=& \frac{i}{(D-2)(D-3)} \frac{2x}{Q^2}
\varepsilon^{\mu\nu\rho\sigma} p_\rho q_\sigma
    ~, 
\end{eqnarray}
applied to $W_{\mu\nu}$, cf. also \cite{IZ}.
The twist--2 contributions to the structure functions in the massless case are given by Eq.~(\ref{eq:SF}).

The perturbative expansion of the structure functions 
\begin{eqnarray}
F_i(x,Q^2) = \sum_{k=0}^\infty a_s^k(Q^2) F_{i,k}(x)
\end{eqnarray}
obey the following renormalization group equation (RGE) in 
the massless limit \cite{Symanzik:1970rt,Callan:1970yg}, see also \cite{Blumlein:2000wh}, 
\begin{eqnarray}
\label{eq:RGE1}
\left[\mathcal{D} + \gamma_{J_1} + \gamma_{J_2} - n_\psi \gamma_\psi - n_A \gamma_A \right]
F_i(N,Q^2) &=& 0
\end{eqnarray}
with
\begin{eqnarray}
\mathcal{D} = \mu^2 \frac{\partial}{\partial \mu^2} + \beta(a_s) \frac{\partial}{\partial a_s}  
\end{eqnarray}
in Mellin space with $a_s = a_s(\mu^2)$, $\beta(a_s)$ the $\beta$ function of Quantum 
Chromodynamics (QCD), and setting $\mu 
\equiv \mu_F = \mu_R$, where $\mu_{F,R}$ 
are the factorization and renormalization scale. $\gamma_{J_i}$ are the anomalous dimensions of 
the currents forming the forward Compton amplitude and $n_{\psi(A)}$ and $\gamma_{\psi(A)}$ are 
the numbers of external quark (gluon fields) and their anomalous dimensions. One may now split
(\ref{eq:RGE1}) for those of the renormalized Wilson coefficients and renormalized parton 
densities, 
\begin{eqnarray}
\sum_j \left[\mathcal{D} (\mu^2) \delta_{ij} + \gamma_{ij}^{\rm S,NS}  
- n_\psi \gamma_\psi - n_A \gamma_A \right] f_j(N,\mu^2) &=& 0
~, \\
\sum_j \left[\mathcal{D} (\mu^2) \delta_{ij} 
+ \gamma_{J_1} + \gamma_{J_2} - \gamma_{ij}^{\rm S,NS}  \right] 
C_i\left(N,\frac{Q^2}{\mu^2}\right) &=& 0 ~.
\end{eqnarray}
In this way the $Z$ factors of the massless Wilson coefficients and the massless parton densities
are related.\footnote{This is different in the massive case, cf.~\cite{Bierenbaum:2009mv}, and 
even in treating heavy quarks as light at asymptotic scales.}

Before we turn to the renormalized Wilson coefficients we will study the unrenormalized ones, as
those emerge in the calculation. They  have the following representation. Here and in the 
following we work in Mellin $N$ space. 
\begin{eqnarray} \label{eq:unrenWC1}
    \hat{C}_{i,q}^{{\rm NS}} &=& 1
    + \hat{a}_s 
    \biggl\{
          \frac{1}{\ep} c_{i,q}^{{\rm NS},(1,-1)}
        + c_{i,q}^{{\rm NS},(1,0)}
        + \ep c_{i,q}^{{\rm NS},(1,1)}
        + \ep^2 c_{i,q}^{{\rm NS},(1,2)}
        + {O}(\ep^3)
    \biggr\} 
    \nonumber \\ &&
    + \hat{a}_s^2
    \biggl\{
          \frac{1}{\ep^2} c_{i,q}^{{\rm NS},(2,-2)}
        + \frac{1}{\ep} c_{i,q}^{{\rm NS},(2,-1)}
        + c_{i,q}^{{\rm NS},(2,0)}
        + \ep c_{i,q}^{{\rm NS},(2,1)}
        + {O}(\ep^2)
    \biggr\}
    \nonumber \\ &&
    + \hat{a}_s^3
    \biggl\{
          \frac{1}{\ep^3} c_{i,q}^{{\rm NS},(3,-3)}
        + \frac{1}{\ep^2} c_{i,q}^{{\rm NS},(3,-2)}
        + \frac{1}{\ep} c_{i,q}^{{\rm NS},(3,-1)}
        + c_{i,q}^{{\rm NS},(3,0)}
        + {O}(\ep)
    \biggr\}
    + {O} \left(\hat{a}_s^4 \right) ~,
\end{eqnarray}
\begin{eqnarray}
    \hat{C}_{i,q}^{{\rm PS}} &=&
    \hat{a}_s^2
    \biggl\{
          \frac{1}{\ep^2} c_{i,q}^{{\rm PS},(2,-2)}
        + \frac{1}{\ep} c_{i,q}^{{\rm PS},(2,-1)}
        + c_{i,q}^{{\rm PS},(2,0)}
        + \ep c_{i,q}^{{\rm PS},(2,1)}
        + {O}(\ep^2)
    \biggr\}
    \nonumber \\ &&
    + \hat{a}_s^3
    \biggl\{
          \frac{1}{\ep^3} c_{i,q}^{{\rm PS},(3,-3)}
        + \frac{1}{\ep^2} c_{i,q}^{{\rm PS},(3,-2)}
        + \frac{1}{\ep} c_{i,q}^{{\rm PS},(3,-1)}
        + c_{i,q}^{{\rm PS},(3,0)}
        + {O}(\ep)
    \biggr\}
    + {O} \left(\hat{a}_s^4 \right) ~,
\end{eqnarray}
\begin{eqnarray}
    \hat{C}_{i,g} &=&
    \hat{a}_s 
    \biggl\{
          \frac{1}{\ep} c_{i,g}^{(1,-1)}
        + c_{i,g}^{(1,0)}
        + \ep c_{i,g}^{(1,1)}
        + \ep^2 c_{i,g}^{(1,2)}
        + {O}(\ep^3)
    \biggr\}
    \nonumber \\ &&
    + \hat{a}_s^2
    \biggl\{
          \frac{1}{\ep^2} c_{i,g}^{(2,-2)}
        + \frac{1}{\ep} c_{i,g}^{(2,-1)}
        + c_{i,g}^{(2,0)}
        + \ep c_{i,g}^{(2,1)}
        + {O}(\ep^2)
    \biggr\}
    \nonumber \\ &&
    + \hat{a}_s^3
    \biggl\{
          \frac{1}{\ep^3} c_{i,g}^{(3,-3)}
        + \frac{1}{\ep^2} c_{i,g}^{(3,-2)}
        + \frac{1}{\ep} c_{i,g}^{(3,-1)}
        + c_{i,g}^{(3,0)}
        + {O}(\ep)
    \biggr\}
    + O\left(\hat{a}_s^4 \right) ~,
\label{eq:unrenWC2}
\end{eqnarray}
with the bare coupling constant $\hat{a}_s$. These relations apply synonymously to  the structure functions
$i = F_2, F_3, g_1$. 

For those of the structure function $F_L$ one obtains
\begin{eqnarray}
    \hat{C}_{F_L,q}^{{\rm NS}} &=&
	 \hat{a}_s 
    \biggl\{
          c_{F_L,q}^{{\rm NS},(1,0)}
        + \ep c_{F_L,q}^{{\rm NS},(1,1)}
        + \ep^2 c_{F_L,q}^{{\rm NS},(1,2)}
        + {O}(\ep^3)
    \biggr\}
    \nonumber \\ &&
    + \hat{a}_s^2
    \biggl\{
          \frac{1}{\ep} c_{F_L,q}^{{\rm NS},(2,-1)}
        + c_{F_L,q}^{{\rm NS},(2,0)}
        + \ep c_{F_L,q}^{{\rm NS},(2,1)}
        + {O}(\ep^2)
    \biggr\}
    \nonumber \\ &&
    + \hat{a}_s^3
    \biggl\{
          \frac{1}{\ep^2} c_{F_L,q}^{{\rm NS},(3,-2)}
        + \frac{1}{\ep} c_{F_L,q}^{{\rm NS},(3,-1)}
        + c_{F_L,q}^{{\rm NS},(3,0)}
        + {O}(\ep)
    \biggr\}
    + {O} \left(\hat{a}_s^4 \right) ~,
\end{eqnarray}
\begin{eqnarray}
    \hat{C}_{F_L,q}^{{\rm PS}} &=&
    \hat{a}_s^2
    \biggl\{
          \frac{1}{\ep} c_{F_L,q}^{{\rm PS},(2,-1)}
        + c_{F_L,q}^{{\rm PS},(2,0)}
        + \ep c_{F_L,q}^{{\rm PS},(2,1)}
        + {O}(\ep^2)
    \biggr\}
    \nonumber \\ &&
    + \hat{a}_s^3
    \biggl\{
          \frac{1}{\ep^2} c_{F_L,q}^{{\rm PS},(3,-2)}
        + \frac{1}{\ep} c_{F_L,q}^{{\rm PS},(3,-1)}
        + c_{F_L,q}^{{\rm PS},(3,0)}
        + {O}(\ep)
    \biggr\}
    + {O} \left(\hat{a}_s^4 \right) ~,
\end{eqnarray}
\begin{eqnarray}
    \hat{C}_{F_L,g} &=&
    \hat{a}_s 
    \biggl\{
          c_{F_L,g}^{(1,0)}
        + \ep c_{F_L,g}^{(1,1)}
        + \ep^2 c_{F_L,g}^{(1,2)}
        + {O}(\ep^3)
    \biggr\}
    \nonumber \\ &&
    + \hat{a}_s.^2
    \biggl\{
          \frac{1}{\ep} c_{F_L,g}^{(2,-1)}
        + c_{F_L,g}^{(2,0)}
        + \ep c_{F_L,g}^{(2,1)}
        + {O}(\ep^2)
    \biggr\}
    \nonumber \\ &&
    + \hat{a}_s^3
    \biggl\{
          \frac{1}{\ep^2} c_{F_L,g}^{(3,-2)}
        + \frac{1}{\ep} c_{F_L,g}^{(3,-1)}
        + c_{F_L,g}^{(3,0)}
        + {O}(\ep)
    \biggr\}
    + {O} \left(\hat{a}_s^4 \right) ~.
\end{eqnarray}

The renormalization proceeds in the following way.
The coupling constant is renormalized in the $\overline{{\rm MS}}$ scheme
\begin{eqnarray}
    \hat{a}_s &=& a_s 
    \left( 
        1 
        + \frac{2 \beta_0}{\ep} a_s 
        + \left[ \frac{4 \beta_0^2}{\ep^2} + \frac{\beta_1}{\ep} \right] a_s^2
        + O(a_s^3)    
    \right) ~.
\end{eqnarray}
Here $\beta_i$ denote the expansion coefficients of the QCD $\beta$--function, \cite{BETA}.
\begin{eqnarray}
\beta_0(N_F) &=& \frac{11}{3} \textcolor{blue}{C_A} - \frac{4}{3} \textcolor{blue}{T_F N_F} ~,
\\
\beta_1(N_F) &=& \frac{34}{3} \textcolor{blue}{C_A^2} - 4 \left( \frac{5}{3} \textcolor{blue}{C_A} +
\textcolor{blue}{C_F} \right) \textcolor{blue}{T_F N_F} ~.
\end{eqnarray}
The color factors are $\textcolor{blue}{C_F} = (N_C^2-1)/(2 N_C), \textcolor{blue}{C_A} = N_C, 
\textcolor{blue}{T_F} = 1/2$ for $SU(N_C)$ and $N_C = 3$ for QCD; $\textcolor{blue}{N_F}$ denotes 
the number of massless quark flavors.
{
Later we will also need the color factor\footnote{We follow the notation of {\tt COLOR}, 
see Ref.~\cite{vanRitbergen:1998pn}.}
\begin{eqnarray}
        d_{abc} d^{abc} &=& \frac{(N_C^2-1)(N_C^2-4)}{16 N_C}
\end{eqnarray}
and the dimension of the adjoint representation $N_A = N_C^2 -1$.
} 
We are carrying out a partial renormalization of the 
Wilson coefficient of the structure function $xF_3$ for the axial vector coupling, which is 
performed by multiplying the unrenormalized Wilson coefficient by $Z_A$, \cite{Moch:2008fj}, see also
\cite{Larin:1991tj,Larin:1993tq},
\begin{eqnarray}
Z_A &=& 1
+\frac{\hat{a}_s^2}{\ep} \Biggl[
        \frac{22}{3} \textcolor{blue}{C_A C_F}
        -\frac{8}{3} \textcolor{blue}{C_F T_F N_F}
\Biggr]
+\hat{a}_s^3 \Biggl[
        \frac{1}{\ep^2} \Biggl(-\frac{484}{27} \textcolor{blue}{C_A^2 C_F}
        +\frac{352}{27} \textcolor{blue}{C_A C_F T_F N_F}
\nonumber\\ &&    
    -\frac{64}{27} \textcolor{blue}{C_F T_F^2 N_F^2} 
        \Biggr)
+        \frac{1}{\ep} \Biggl(
                \frac{3578}{81} \textcolor{blue}{C_A^2 C_F}
                -\frac{308}{9} \textcolor{blue}{C_A C_F^2}
                -\frac{1664}{81} \textcolor{blue}{C_A C_F T_F N_F} 
                +\frac{64}{9} \textcolor{blue}{C_F^2 T_F N_F}
\nonumber\\ && 
                +\frac{32}{81} \textcolor{blue}{C_F T_F^2 N_F^2} 
        \Biggr)
\Biggr] ~.
\end{eqnarray}

The non--singlet Wilson coefficients are renormalized via 
\begin{eqnarray}
    C_{i,q}^{{\rm NS}} &=& Z_{qq}^{{\rm NS}} \hat{C}_{i,q}^{{\rm NS}}
\end{eqnarray}
and the singlet Wilson coefficients via 
\begin{eqnarray}
    \begin{pmatrix}
        C_{i,q}^{{\rm S}} \\
        C_{i,g}
    \end{pmatrix}
    &=& 
    \left. Z^{{\rm S}} \right.^T \cdot 
    \begin{pmatrix}
        \hat{C}_{i,q}^{{\rm S}} \\
        \hat{C}_{i,g}
    \end{pmatrix} ~,
\end{eqnarray}
with 
\begin{eqnarray}
    Z_{qq}^{{\rm NS}} &=& 
        1 
        + a_s \frac{\gamma_{qq}^{(0),\rm NS}}{\ep}
        + a_s^2 
        \Biggl[
                \frac{1}{\ep^2} 
                \Biggl(
                        \frac{1}{2} {\gamma_{qq}^{(0),\rm NS}}^2 
                        + \beta_0 \gamma_{qq}^{(0),\rm NS}
                \Biggr) 
                + \frac{1}{2\ep} \gamma_{qq}^{(0),\rm NS} 
        \Biggr]
        + a_s^3 
        \Biggl[ 
                \frac{1}{\ep^3} 
                \Biggl( 
                        \frac{1}{6} {\gamma_{qq}^{(0),\rm NS}}^3 
                        \nonumber\\ &&
                        + \beta_0 {\gamma_{qq}^{(0),\rm NS}}^2 
                        + \frac{4}{3} \beta_0^2 \gamma_{qq}^{(0),\rm NS}
                \Biggr)
                + \frac{1}{\ep^2} 
                \Biggl( 
                        \frac{1}{2} \gamma_{qq}^{(0),\rm NS} \gamma_{qq}^{(1),\rm NS}
                        + \frac{2}{3} \beta_0 \gamma_{qq}^{(1),\rm NS} 
                        + \frac{2}{3} \beta_1 \gamma_{qq}^{(0),\rm NS} 
                \Biggr) 
                \nonumber\\ &&
                + \frac{1}{3 \ep} \gamma_{qq}^{(2),\rm NS}
        \Biggr]
        + {O}(a_s^4) ~,
    \\
        Z^{\rm S}_{ij} &=& 
        \delta_{ij}  
        + a_s \frac{ \gamma_{ij}^{(0)}}{\ep}
        + a_s^2 
        \Biggl[
                \frac{1}{\ep^2} 
                \Biggl(
                        \frac{1}{2} \gamma_{il}^{(0)} \gamma_{lj}^{(0)} 
                        + \beta_0 \gamma_{ij}^{(0)}
                \Biggr) 
                + \frac{1}{2\ep}  \gamma_{ij}^{(1)} 
        \Biggr]
        \nonumber\\ && 
        + a_s^3 
        \Biggl[
                \frac{1}{\ep^3} 
                \Biggl( 
                        \frac{1}{6}  \gamma_{il}^{(0)} \gamma_{lk}^{(0)} \gamma_{kj}^{(0)} 
                        + \beta_0  \gamma_{il}^{(0)}  \gamma_{lj}^{(0)} 
                        + \frac{4}{3} \beta_0^2  \gamma_{ij}^{(0)}
                \Biggr)
        \nonumber\\ && 
                + \frac{1}{\ep^2} 
                \Biggl( 
                        \frac{1}{6}  \gamma_{il}^{(1)}  \gamma_{lj}^{(0)}
                        + \frac{1}{3}  \gamma_{il}^{(0)}  \gamma_{lj}^{(1)}
                        + \frac{2}{3} \beta_0  \gamma_{ij}^{(1)} 
                        + \frac{2}{3} \beta_1  \gamma_{ij}^{(0)} 
                \Biggr) 
                + \frac{1}{3 \ep}  \gamma_{ij}^{(2)}
        \Biggr]
        + {O}(a_s^4) ~.
\end{eqnarray}

The pole terms of the Wilson coefficients are predicted by the anomalous dimensions and the expansion
coefficients of the unrenormalized Wilson coefficients in lower orders, where in the polarized 
case the anomalous dimensions $\gamma_{ij}$ have to be replaced by $\Delta \gamma_{ij}$.
\begin{eqnarray}
    c_{i,q}^{{\rm NS},(1,-1)} &=& - \gamma_{qq}^{{\rm NS},(0)} 
    ~, \\
    c_{i,q}^{{\rm NS},(2,-2)} &=& 
        \gamma_{qq}^{{\rm NS},(0)} 
        \left(
            \beta_0 + \frac{1}{2} \gamma_{qq}^{{\rm NS},(0)} 
        \right)
    ~,\\
    c_{i,q}^{{\rm NS},(2,-1)} &=&
        - c_{i,q}^{{\rm NS},(1,0)}
        \left(
             \gamma_{qq}^{{\rm NS},(0)} 
            + 2 \beta_0
        \right)
        - \frac{1}{2} \gamma_{qq}^{{\rm NS},(1)}
        ~, \\
    c_{i,q}^{{\rm NS},(3,-3)} &=& 
        - \gamma_{qq}^{{\rm NS},(0)}
        \left(
            \frac{4}{3} \beta_0^2
            + \beta_0 \gamma_{qq}^{{\rm NS},(0)}
            + \frac{1}{6} \left. \gamma_{qq}^{{\rm NS},(0)} \right.^2 
        \right)
        ~, \\
    c_{i,q}^{{\rm NS},(3,-2)} &=& 
        4 \beta_0^2 c_{i,q}^{{\rm NS},(1,0)}
        + \frac{1}{3} \beta_1 \gamma_{qq}^{{\rm NS},(0)}
        \nonumber \\ &&
        + 3 \beta_0 c_{i,q}^{{\rm NS},(1,0)} \gamma_{qq}^{{\rm NS},(0)}
        + \frac{1}{2} c_{i,q}^{{\rm NS},(1,0)} \left. \gamma_{qq}^{{\rm NS},(0)} \right.^2
        + \frac{4}{3} \beta_0 \gamma_{qq}^{{\rm NS},(1)}
        + \frac{1}{2} \gamma_{qq}^{{\rm NS},(0)} \gamma_{qq}^{{\rm NS},(1)}
        ~, \\
    c_{i,q}^{{\rm NS},(3,-1)} &=&
        - \beta_1 c_{i,q}^{{\rm NS},(1,0)}
        - 4 \beta_0^2 c_{i,q}^{{\rm NS},(1,1)}
        - 4 \beta_0 c_{i,q}^{{\rm NS},(2,0)}
        - 3 \beta_0 c_{i,q}^{{\rm NS},(1,1)} \gamma_{qq}^{{\rm NS},(0)}
        - c_{i,q}^{{\rm NS},(2,0)} \gamma_{qq}^{{\rm NS},(0)}
        \nonumber \\ &&
        - \frac{1}{2} c_{i,q}^{{\rm NS},(1,1)} \left. \gamma_{qq}^{{\rm NS},(0)} \right.^2
        - \frac{1}{2} c_{i,q}^{{\rm NS},(1,0)} \gamma_{qq}^{{\rm NS},(1)}
        - \frac{1}{3} \gamma_{qq}^{{\rm NS},(2)} ~,
\end{eqnarray}
\begin{eqnarray}
    c_{i,q}^{{\rm PS},(2,-2)} &=& 
        \frac{1}{2} \gamma_{gq}^{(0)} \gamma_{qg}^{(0)}
        ~, \\
    c_{i,q}^{{\rm PS},(2,-1)} &=& 
        - \gamma_{gq}^{(0)} c_{i,g}^{(1,0)}
        - \frac{1}{2} \gamma_{qq}^{{\rm PS},(1)}
        ~, \\
    c_{i,q}^{{\rm PS},(3,-3)} &=& 
        - \beta_0 \gamma_{gq}^{(0)} \gamma_{qg}^{(0)}
        - \frac{1}{6} \gamma_{gg}^{(0)} \gamma_{gq}^{(0)} \gamma_{qg}^{(0)}
        - \frac{1}{3} \gamma_{qq}^{{\rm NS},(0)} \gamma_{gq}^{(0)} \gamma_{qg}^{(0)}
        ~, \\
    c_{i,q}^{{\rm PS},(3,-2)} &=& 
        3 \beta_0 \gamma_{gq}^{(0)} c_{i,g}^{(1,0)}
        + \frac{1}{2} \gamma_{gg}^{(0)} \gamma_{gq}^{(0)} c_{i,g}^{(1,0)}
        + \frac{1}{2} \gamma_{gq}^{(0)} \gamma_{qg}^{(0)} c_{i,q}^{{\rm NS},(1,0)}
        + \frac{1}{6} \gamma_{gq}^{(1)} \gamma_{qg}^{(0)}
        + \frac{1}{3} \gamma_{gq}^{(0)} \gamma_{qg}^{(1)}
        \nonumber \\ &&
        + \frac{1}{2} \gamma_{gq}^{(0)} \gamma_{qq}^{{\rm NS},(0)} c_{i,g}^{(1,0)}
        + \frac{4}{3} \beta_0 \gamma_{qq}^{{\rm PS},(1)}
        + \frac{1}{2} \gamma_{qq}^{{\rm NS},(0)} \gamma_{qq}^{{\rm PS},(1)}
        ~, \\
    c_{i,q}^{{\rm PS},(3,-1)} &=&
        - 4 \beta_0 c_{i,q}^{{\rm PS},(2,0)}
        - 3 \beta_0 \gamma_{gq}^{(0)} c_{i,g}^{(1,1)}
        - \gamma_{gq}^{(0)} c_{i,g}^{(2,0)}
        - \frac{1}{2} \gamma_{gg}^{(0)} \gamma_{gq}^{(0)} c_{i,g}^{(1,1)}
        - \frac{1}{2} \gamma_{gq}^{(1)} c_{i,g}^{(1,0)}
        \nonumber \\ &&
        - \frac{1}{2} \gamma_{gq}^{(0)} \gamma_{qg}^{(0)} c_{i,q}^{{\rm NS},(1,1)}
        - \gamma_{qq}^{{\rm NS},(0)} c_{i,q}^{{\rm PS},(2,0)}
        - \frac{1}{2} \gamma_{gq}^{(0)} \gamma_{qq}^{{\rm NS},(0)} c_{i,g}^{(1,1)}
        - \frac{1}{2} \gamma_{qq}^{{\rm PS},(1)} c_{i,q}^{{\rm NS},(1,0)}
        \nonumber \\ &&
        - \frac{1}{3} \gamma_{qq}^{{\rm PS},(2)}
        ~,
\end{eqnarray}
and
\begin{eqnarray}
    c_{i,g}^{(1,-1)} &=& - \gamma_{qg}^{(0)}
    ~, \\
    c_{i,g}^{(2,-2)} &=& 
        \beta_0 \gamma_{qg}^{(0)}
        + \frac{1}{2} \gamma_{gg}^{(0)} \gamma_{qg}^{(0)}
        + \frac{1}{2} \gamma_{qg}^{(0)} \gamma_{qq}^{{\rm NS},(0)} 
        ~, \\
    c_{i,g}^{(2,-1)} &=&
        - 2 \beta_0 c_{i,g}^{(1,0)}
        - \gamma_{gg}^{(0)} c_{i,g}^{(1,0)}
        - \gamma_{qg}^{(0)} c_{i,q}^{{\rm NS},(1,0)}
        - \frac{1}{2} \gamma_{qg}^{(1)}
        ~, \\
    c_{i,g}^{(3,-3)} &=& 
    	- \frac{4}{3} \beta_0^2 \gamma_{qg}^{(0)} 
    	- \beta_0 \gamma_{gg}^{(0)} \gamma_{qg}^{(0)} 
    	- \frac{1}{6} \left. \gamma_{gg}^{(0)} \right.^2 \gamma_{qg}^{(0)} 
    	- \frac{1}{6} \gamma_{gq}^{(0)} \left. \gamma_{qg}^{(0)} \right.^2
    	- \beta_0 \gamma_{qg}^{(0)} \gamma_{qq}^{{\rm NS},(0)} 
        \nonumber \\ &&
    	- \frac{1}{6} \gamma_{gg}^{(0)} \gamma_{qg}^{(0)} \gamma_{qq}^{{\rm NS},(0)} 
    	- \frac{1}{6} \gamma_{qg}^{(0)} \left. \gamma_{qq}^{{\rm NS},(0)} \right.^2
        ~, \\
    c_{i,g}^{(3,-2)} &=& 
    	  4 \beta_0^2 c_{i,g}^{(1,0)} 
    	+ 3 \beta_0 c_{i,g}^{(1,0)} \gamma_{gg}^{(0)} 
    	+ \frac{1}{2} c_{i,g}^{(1,0)} \left. \gamma_{gg}^{(0)} \right.^2
    	+ \frac{1}{3} \beta_1 \gamma_{qg}^{(0)}
    	+ 3 \beta_0 c_{i,q}^{{\rm NS},(1,0)} \gamma_{qg}^{(0)} 
        \nonumber \\ &&
    	+ \frac{1}{2} c_{i,q}^{{\rm NS},(1,0)} \gamma_{gg}^{(0)} \gamma_{qg}^{(0)}
    	+ \frac{1}{6} \gamma_{gg}^{(1)} \gamma_{qg}^{(0)}
    	+ \frac{1}{2} c_{i,g}^{(1,0)} \gamma_{gq}^{(0)} \gamma_{qg}^{(0)} 
    	+ \frac{4}{3} \beta_0 \gamma_{qg}^{(1)} 
        + \frac{1}{3} \gamma_{gg}^{(0)} \gamma_{qg}^{(1)}
        \nonumber \\ &&
    	+ \frac{1}{2} c_{i,q}^{{\rm NS},(1,0)} \gamma_{qg}^{(0)} \gamma_{qq}^{{\rm NS},(0)}
    	+ \frac{1}{6} \gamma_{qg}^{(1)} \gamma_{qq}^{{\rm NS},(0)}
    	+ \frac{1}{3} \gamma_{qg}^{(0)} \gamma_{qq}^{{\rm NS},(1)} 
    	+ \frac{1}{3} \gamma_{qg}^{(0)} \gamma_{qq}^{{\rm PS},(1)}
        ~, \\
    c_{i,g}^{(3,-1)} &=&
    	- \beta_1 c_{i,g}^{(1,0)} 
    	- 4 \beta_0^2 c_{i,g}^{(1,1)} 
    	- 4 \beta_0 c_{i,g}^{(2,0)} 
    	- 3 \beta_0 c_{i,g}^{(1,1)} \gamma_{gg}^{(0)} 
    	- c_{i,g}^{(2,0)} \gamma_{gg}^{(0)} 
        \nonumber \\ &&
    	- \frac{1}{2} c_{i,g}^{(1,1)} \left. \gamma_{gg}^{(0)} \right.^2
    	- \frac{1}{2} c_{i,g}^{(1,0)} \gamma_{gg}^{(1)}
    	- 3 \beta_0 c_{i,q}^{{\rm NS},(1,1)} \gamma_{qg}^{(0)} 
    	- c_{i,q}^{{\rm NS},(2,0)} \gamma_{qg}^{(0)} 
    	- c_{i,q}^{{\rm PS},(2,0)} \gamma_{qg}^{(0)} 
        \nonumber \\ &&
    	- \frac{1}{2} c_{i,q}^{{\rm NS},(1,1)} \gamma_{gg}^{(0)} \gamma_{qg}^{(0)}
    	- \frac{1}{2} c_{i,g}^{(1,1)} \gamma_{gq}^{(0)} \gamma_{qg}^{(0)}
    	- \frac{1}{2} c_{i,q}^{{\rm NS},(1,0)} \gamma_{qg}^{(1)}
    	- \frac{1}{3} \gamma_{qg}^{(2)}
\nonumber\\ &&
    	- \frac{1}{2} c_{i,q}^{{\rm NS},(1,1)} \gamma_{qg}^{(0)} \gamma_{qq}^{{\rm NS},(0)}
        ~.
\end{eqnarray}
For the longitudinal structure function $F_L$ the poles are predicted by
\begin{eqnarray}
    c_{F_L,q}^{{\rm NS},(2,-1)} &=&
        - c_{F_L,q}^{{\rm NS},(1,0)}
        \left(
            \gamma_{qq}^{{\rm NS},(0)}
            + 2 \beta_0
        \right)
    ~, \\
    c_{F_L,q}^{{\rm NS},(3,-2)} &=& 
        4 \beta_0^2 c_{F_L,q}^{{\rm NS},(1,0)}
        + 3 \beta_0 c_{F_L,q}^{{\rm NS},(1,0)} \gamma_{qq}^{{\rm NS},(0)}
        + \frac{1}{2} c_{F_L,q}^{{\rm NS},(1,0)} \left. \gamma_{qq}^{{\rm NS},(0)} \right.^2
    ~, \\
    c_{F_L,q}^{{\rm NS},(3,-1)} &=&
        - \beta_1 c_{F_L,q}^{{\rm NS},(1,0)}
        - 4 \beta_0^2 c_{F_L,q}^{{\rm NS},(1,1)}
        - 4 \beta_0 c_{F_L,q}^{{\rm NS},(2,0)}
        - 3 \beta_0 c_{F_L,q}^{{\rm NS},(1,1)} \gamma_{qq}^{{\rm NS},(0)}
        - c_{F_L,q}^{{\rm NS},(2,0)} \gamma_{qq}^{{\rm NS},(0)}
        \nonumber \\ &&
        - \frac{1}{2} c_{F_L,q}^{{\rm NS},(1,1)} \left. \gamma_{qq}^{{\rm NS},(0)} \right.^2
        - \frac{1}{2} c_{F_L,q}^{{\rm NS},(1,0)} \gamma_{qq}^{{\rm NS},(1)}
        ~,
\end{eqnarray}
\begin{eqnarray}
    c_{F_L,q}^{{\rm PS},(2,-1)} &=& 
        - \gamma_{gq}^{(0)} c_{F_L,g}^{(1,0)}
        ~, \\
    c_{F_L,q}^{{\rm PS},(3,-2)} &=& 
        3 \beta_0 \gamma_{gq}^{(0)} c_{F_L,g}^{(1,0)}
        + \frac{1}{2} \gamma_{gg}^{(0)} \gamma_{gq}^{(0)} c_{F_L,g}^{(1,0)}
        + \frac{1}{2} \gamma_{gq}^{(0)} \gamma_{qg}^{(0)} c_{F_L,q}^{{\rm NS},(1,0)}
        + \frac{1}{2} \gamma_{gq}^{(0)} \gamma_{qq}^{{\rm NS},(0)} c_{F_L,g}^{(1,0)}
        ~, \\
    c_{F_L,q}^{{\rm PS},(3,-1)} &=&
        - 4 \beta_0 c_{F_L,q}^{{\rm PS},(2,0)}
        - 3 \beta_0 \gamma_{gq}^{(0)} c_{F_L,g}^{(1,1)}
        - \gamma_{gq}^{(0)} c_{F_L,g}^{(2,0)}
        - \frac{1}{2} \gamma_{gg}^{(0)} \gamma_{gq}^{(0)} c_{F_L,g}^{(1,1)}
        - \frac{1}{2} \gamma_{gq}^{(1)} c_{F_L,g}^{(1,0)}
        \nonumber \\ &&
        - \frac{1}{2} \gamma_{gq}^{(0)} \gamma_{qg}^{(0)} c_{F_L,q}^{{\rm NS},(1,1)}
        - \gamma_{qq}^{{\rm NS},(0)} c_{F_L,q}^{{\rm PS},(2,0)}
        - \frac{1}{2} \gamma_{gq}^{(0)} \gamma_{qq}^{{\rm NS},(0)} c_{F_L,g}^{(1,1)}
        - \frac{1}{2} \gamma_{qq}^{{\rm PS},(1)} c_{F_L,q}^{{\rm NS},(1,0)}
        ~,
\end{eqnarray}
\begin{eqnarray}
    c_{F_L,g}^{(2,-1)} &=&
        - 2 \beta_0 c_{F_L,g}^{(1,0)}
        - \gamma_{gg}^{(0)} c_{F_L,g}^{(1,0)}
        - \gamma_{qg}^{(0)} c_{F_L,q}^{{\rm NS},(1,0)}
        ~, \\
    c_{F_L,g}^{(3,-2)} &=& 
    	  4 \beta_0^2 c_{F_L,g}^{(1,0)} 
    	+ 3 \beta_0 c_{F_L,g}^{(1,0)} \gamma_{gg}^{(0)} 
    	+ \frac{1}{2} c_{F_L,g}^{(1,0)} \left. \gamma_{gg}^{(0)} \right.^2
    	+ 3 \beta_0 c_{F_L,q}^{{\rm NS},(1,0)} \gamma_{qg}^{(0)} 
    	+ \frac{1}{2} c_{F_L,q}^{{\rm NS},(1,0)} \gamma_{gg}^{(0)} \gamma_{qg}^{(0)}
    	\nonumber \\ &&
    	+ \frac{1}{2} c_{F_L,g}^{(1,0)} \gamma_{gq}^{(0)} \gamma_{qg}^{(0)} 
    	+ \frac{1}{2} c_{F_L,q}^{{\rm NS},(1,0)} \gamma_{qg}^{(0)} \gamma_{qq}^{{\rm NS},(0)}
        ~, \\
    c_{F_L,g}^{(3,-1)} &=&
    	- \beta_1 c_{F_L,g}^{(1,0)} 
    	- 4 \beta_0^2 c_{F_L,g}^{(1,1)} 
    	- 4 \beta_0 c_{F_L,g}^{(2,0)} 
    	- 3 \beta_0 c_{F_L,g}^{(1,1)} \gamma_{gg}^{(0)} 
    	- c_{F_L,g}^{(2,0)} \gamma_{gg}^{(0)} 
    	- \frac{1}{2} c_{F_L,g}^{(1,1)} \left. \gamma_{gg}^{(0)} \right.^2
    	\nonumber \\ &&
    	- \frac{1}{2} c_{F_L,g}^{(1,0)} \gamma_{gg}^{(1)}
    	- 3 \beta_0 c_{F_L,q}^{{\rm NS},(1,1)} \gamma_{qg}^{(0)} 
    	- c_{F_L,q}^{{\rm NS},(2,0)} \gamma_{qg}^{(0)} 
    	- c_{F_L,q}^{{\rm PS},(2,0)} \gamma_{qg}^{(0)} 
    	- \frac{1}{2} c_{F_L,q}^{{\rm NS},(1,1)} \gamma_{gg}^{(0)} \gamma_{qg}^{(0)}
    	\nonumber \\ &&
    	- \frac{1}{2} c_{F_L,g}^{(1,1)} \gamma_{gq}^{(0)} \gamma_{qg}^{(0)}
    	- \frac{1}{2} c_{F_L,q}^{{\rm NS},(1,0)} \gamma_{qg}^{(1)}
    	- \frac{1}{2} c_{F_L,q}^{{\rm NS},(1,1)} \gamma_{qg}^{(0)} \gamma_{qq}^{{\rm NS},(0)}
        ~.
\end{eqnarray}

The renormalized Wilson coefficients for the structure functions $F_2, F_3$ and $g_1$ have the following 
structure
\begin{eqnarray}
    C_{i,q}^{{\rm NS}} &=& 
    1 
    + a_s
    \biggr\{
        \ln\left( \frac{Q^2}{\mu^2} \right)
        \bigl[
        	- \frac{1}{2} \gamma_{qq}^{{\rm NS},(0)}
        \bigr]
        + c_{i,q}^{{\rm NS},(1,0)}
    \biggr\}
    + a_s^2
    \biggl\{
        \ln^2\left( \frac{Q^2}{\mu^2} \right)
        \bigl[
        	  \frac{1}{4} \beta_0 \gamma_{qq}^{{\rm NS},(0)}
        \nonumber \\ &&
        	+ \frac{1}{8} \left. \gamma_{qq}^{{\rm NS},(0)} \right.^2
        \bigr]
        - \ln\left( \frac{Q^2}{\mu^2} \right)
        \bigl[
        	 \beta_0 c_{i,q}^{{\rm NS},(1,0)} 
        	+ \frac{1}{2} c_{i,q}^{{\rm NS},(1,0)} \gamma_{qq}^{{\rm NS},(0)}
        	+ \frac{1}{2} \gamma_{qq}^{{\rm NS},(1)}
        \bigr]
        + 2 \beta_0 c_{i,q}^{{\rm NS},(1,1)} 
    \nonumber \\ &&
        + c_{i,q}^{{\rm NS},(1,1)} \gamma_{qq}^{{\rm NS},(0)}
        + c_{i,q}^{{\rm NS},(2,0)} 
     \biggr\}
    + a_s^3
    \biggl\{
        -\ln^3\left( \frac{Q^2}{\mu^2} \right)
        \bigl[
            \frac{1}{6} \beta_0^2 \gamma_{qq}^{{\rm NS},(0)} 
        	+ \frac{1}{8} \beta_0 \left. \gamma_{qq}^{{\rm NS},(0)} \right.^2
                \nonumber \\ &&
        	+ \frac{1}{48} \left. \gamma_{qq}^{{\rm NS},(0)} \right.^3
        \bigr]
        + \ln^2\left( \frac{Q^2}{\mu^2} \right)
        \bigl[
        	\beta_0^2 c_{i,q}^{{\rm NS},(1,0)} 
        	+ \frac{1}{4} \beta_1 \gamma_{qq}^{{\rm NS},(0)}
        	+ \frac{3}{4} \beta_0 c_{i,q}^{{\rm NS},(1,0)} \gamma_{qq}^{{\rm NS},(0)}
        \nonumber \\ &&
        	+ \frac{1}{8} c_{i,q}^{{\rm NS},(1,0)} \left. \gamma_{qq}^{{\rm NS},(0)} \right.^2 
        	+ \frac{1}{2} \beta_0 \gamma_{qq}^{{\rm NS},(1)}
        	+ \frac{1}{4} \gamma_{qq}^{{\rm NS},(0)} \gamma_{qq}^{{\rm NS},(1)}
        \bigr]
        - \ln\left( \frac{Q^2}{\mu^2} \right)
        \bigl[
        	\beta_1 c_{i,q}^{{\rm NS},(1,0)}
        \nonumber \\ &&
        	+ 4 \beta_0^2 c_{i,q}^{{\rm NS},(1,1)} 
        	+ 2 \beta_0 c_{i,q}^{{\rm NS},(2,0)} 
        	+ 3 \beta_0 c_{i,q}^{{\rm NS},(1,1)} \gamma_{qq}^{{\rm NS},(0)} 
        	+ \frac{1}{2} c_{i,q}^{{\rm NS},(2,0)} \gamma_{qq}^{{\rm NS},(0)}
        	+ \frac{1}{2} c_{i,q}^{{\rm NS},(1,1)} \left. \gamma_{qq}^{{\rm NS},(0)} \right.^2 
        \nonumber \\ &&
                + \frac{1}{2} c_{i,q}^{{\rm NS},(1,0)} \gamma_{qq}^{{\rm NS},(1)}
        	+ \frac{1}{2} \gamma_{qq}^{{\rm NS},(2)}
        \bigr]
        + \beta_1 c_{i,q}^{{\rm NS},(1,1)} 
        + 4 \beta_0^2 c_{i,q}^{{\rm NS},(1,2)} 
        + 4 \beta_0 c_{i,q}^{{\rm NS},(2,1)} 
        \nonumber \\ &&
        + 3 \beta_0 c_{i,q}^{{\rm NS},(1,2)} \gamma_{qq}^{{\rm NS},(0)} 
        + c_{i,q}^{{\rm NS},(2,1)} \gamma_{qq}^{{\rm NS},(0)} 
        + \frac{1}{2} c_{i,q}^{{\rm NS},(1,2)} \left. \gamma_{qq}^{{\rm NS},(0)} \right.^2
        + \frac{1}{2} c_{i,q}^{{\rm NS},(1,1)} \gamma_{qq}^{{\rm NS},(1)}
        + c_{i,q}^{{\rm NS},(3,0)}
    \biggr\}
    \nonumber \\ &&
    + O(a_s^4) ~,
\\
    C_{i,q}^{{\rm PS}} &=&
    a_s^2
    \biggl\{
        \ln^2\left( \frac{Q^2}{\mu^2} \right)
        \bigl[
        	\frac{1}{8} \gamma_{gq}^{(0)} \gamma_{qg}^{(0)}
        \bigr]
        - \ln\left( \frac{Q^2}{\mu^2} \right)
        \bigl[
        	\frac{1}{2} (c_{i,g}^{(1,0)} \gamma_{gq}^{(0)}) 
        	+\frac{1}{2} \gamma_{qq}^{{\rm PS},(1)}
        \bigr]
        + c_{i,g}^{(1,1)} \gamma_{gq}^{(0)}
    \nonumber \\ &&
        + c_{i,q}^{{\rm PS},(2,0)} 
        \biggr\}
    + a_s^3
    \biggl\{
        -\ln^3\left( \frac{Q^2}{\mu^2} \right)
        \bigl[
        	 \frac{1}{8} (\beta_0 \gamma_{gq}^{(0)} \gamma_{qg}^{(0)}) 
        	+ \frac{1}{48} \gamma_{gg}^{(0)} \gamma_{gq}^{(0)} \gamma_{qg}^{(0)}
        	+ \frac{1}{24} \gamma_{gq}^{(0)} \gamma_{qg}^{(0)} \gamma_{qq}^{{\rm NS},(0)}
        \bigr]
        \nonumber \\ &&
        + \ln^2\left( \frac{Q^2}{\mu^2} \right)
        \bigl[
            \frac{3}{4} \beta_0 c_{i,g}^{(1,0)} \gamma_{gq}^{(0)}
        	+ \frac{1}{8} c_{i,g}^{(1,0)} \gamma_{gg}^{(0)} \gamma_{gq}^{(0)} 
        	+ \frac{1}{8} c_{i,q}^{{\rm NS},(1,0)} \gamma_{gq}^{(0)} \gamma_{qg}^{(0)} 
        	+ \frac{1}{8} \gamma_{gq}^{(1)} \gamma_{qg}^{(0)}
            \nonumber \\ &&
        	+ \frac{1}{8} \gamma_{gq}^{(0)} \gamma_{qg}^{(1)} 
        	+ \frac{1}{8} c_{i,g}^{(1,0)} \gamma_{gq}^{(0)} \gamma_{qq}^{{\rm NS},(0)}
        	+ \frac{1}{2} \beta_0 \gamma_{qq}^{{\rm PS},(1)}
        	+ \frac{1}{4} \gamma_{qq}^{{\rm NS},(0)} \gamma_{qq}^{{\rm PS},(1)}
        \bigr]
        - \ln\left( \frac{Q^2}{\mu^2} \right)
             \nonumber \\ &&
\times
        \bigl[
        	 2 \beta_0 c_{i,q}^{{\rm PS},(2,0)} 
        	+ 3 \beta_0 c_{i,g}^{(1,1)} \gamma_{gq}^{(0)} 
        	+ \frac{1}{2} c_{i,g}^{(2,0)} \gamma_{gq}^{(0)} 
        	+ \frac{1}{2} c_{i,g}^{(1,1)} \gamma_{gg}^{(0)} \gamma_{gq}^{(0)} 
        	+ \frac{1}{2} c_{i,g}^{(1,0)} \gamma_{gq}^{(1)}
            \nonumber \\ &&
        	+ \frac{1}{2} c_{i,q}^{{\rm NS},(1,1)} \gamma_{gq}^{(0)} \gamma_{qg}^{(0)} 
        	+ \frac{1}{2} c_{i,q}^{{\rm PS},(2,0)} \gamma_{qq}^{{\rm NS},(0)}
        	+ \frac{1}{2} c_{i,g}^{(1,1)} \gamma_{gq}^{(0)} \gamma_{qq}^{{\rm NS},(0)} 
        	+ \frac{1}{2} c_{i,q}^{{\rm NS},(1,0)} \gamma_{qq}^{{\rm PS},(1)}
        	+ \frac{1}{2} \gamma_{qq}^{{\rm PS},(2)}
        \bigr]
        \nonumber \\ &&
        + 4 \beta_0 c_{i,q}^{{\rm PS},(2,1)} 
        + 3 \beta_0 c_{i,g}^{(1,2)} \gamma_{gq}^{(0)} 
        + c_{i,g}^{(2,1)} \gamma_{gq}^{(0)} 
        + \frac{1}{2} c_{i,g}^{(1,2)} \gamma_{gg}^{(0)} \gamma_{gq}^{(0)}
        + \frac{1}{2} c_{i,g}^{(1,1)} \gamma_{gq}^{(1)}
        \nonumber \\ &&
        + \frac{1}{2} c_{i,q}^{{\rm NS},(1,2)} \gamma_{gq}^{(0)} \gamma_{qg}^{(0)}
        + c_{i,q}^{{\rm PS},(2,1)} \gamma_{qq}^{{\rm NS},(0)} 
        + \frac{1}{2} c_{i,g}^{(1,2)} \gamma_{gq}^{(0)} \gamma_{qq}^{{\rm NS},(0)}
        + \frac{1}{2} c_{i,q}^{{\rm NS},(1,1)} \gamma_{qq}^{{\rm PS},(1)}
        + c_{i,q}^{{\rm PS},(3,0)} 
    \biggr\}
\nonumber\\ &&
    + O(a_s^4) ~,
    \\
    C_{i,g} &=&
    a_s
    \biggr\{
        -\ln\left( \frac{Q^2}{\mu^2} \right)
        \bigl[
        	\frac{1}{2} \gamma_{qg}^{(0)}
        \bigr]
        + c_{i,g}^{(1,0)}
    \biggr\}
    + a_s^2
    \biggl\{
        \ln^2\left( \frac{Q^2}{\mu^2} \right)
        \bigl[
            \frac{1}{4} \beta_0 \gamma_{qg}^{(0)}
        	+ \frac{1}{8} \gamma_{gg}^{(0)} \gamma_{qg}^{(0)} 
            \nonumber \\ &&
        	+ \frac{1}{8} \gamma_{qg}^{(0)} \gamma_{qq}^{{\rm NS},(0)}
        \bigr]
        - \ln\left( \frac{Q^2}{\mu^2} \right)
        \bigl[
        	 \beta_0 c_{i,g}^{(1,0)}
        	+ \frac{1}{2} c_{i,g}^{(1,0)} \gamma_{gg}^{(0)}
        	+ \frac{1}{2} c_{i,q}^{{\rm NS},(1,0)} \gamma_{qg}^{(0)} 
        	+ \frac{1}{2} \gamma_{qg}^{(1)}
        \bigr]
        \nonumber \\ &&
        + 2 \beta_0 c_{i,g}^{(1,1)} 
        + c_{i,g}^{(1,1)} \gamma_{gg}^{(0)} 
        + c_{i,q}^{{\rm NS},(1,1)} \gamma_{qg}^{(0)}
        + c_{i,g}^{(2,0)} 
    \biggr\}
    + a_s^3
    \biggl\{
        -\ln^3\left( \frac{Q^2}{\mu^2} \right)
        \bigl[
        	\frac{1}{6} \beta_0^2 \gamma_{qg}^{(0)}
            \nonumber \\ &&
        	+ \frac{1}{8} \beta_0 \gamma_{gg}^{(0)} \gamma_{qg}^{(0)}
        	+ \frac{1}{48} \left. \gamma_{gg}^{(0)} \right.^2 \gamma_{qg}^{(0)} 
        	+ \frac{1}{48} \gamma_{gq}^{(0)} \left. \gamma_{qg}^{(0)} \right.^2 
        	+ \frac{1}{8} \beta_0 \gamma_{qg}^{(0)} \gamma_{qq}^{{\rm NS},(0)} 
        	+ \frac{1}{48} \gamma_{gg}^{(0)} \gamma_{qg}^{(0)} \gamma_{qq}^{{\rm NS},(0)}
            \nonumber \\ &&
        	+ \frac{1}{48} \gamma_{qg}^{(0)} \left. \gamma_{qq}^{{\rm NS},(0)} \right.^2
        \bigr]
        + \ln^2\left( \frac{Q^2}{\mu^2} \right)
        \bigl[
        	\beta_0^2 c_{i,g}^{(1,0)} 
        	+ \frac{3}{4} \beta_0 c_{i,g}^{(1,0)} \gamma_{gg}^{(0)}
        	+ \frac{1}{8} c_{i,g}^{(1,0)} \left. \gamma_{gg}^{(0)} \right.^2 
        	+ \frac{1}{4} \beta_1 \gamma_{qg}^{(0)}
            \nonumber \\ &&
        	+ \frac{3}{4} \beta_0 c_{i,q}^{{\rm NS},(1,0)} \gamma_{qg}^{(0)}
        	+ \frac{1}{8} c_{i,q}^{{\rm NS},(1,0)} \gamma_{gg}^{(0)} \gamma_{qg}^{(0)} 
        	+ \frac{1}{8} \gamma_{gg}^{(1)} \gamma_{qg}^{(0)}
        	+ \frac{1}{8} c_{i,g}^{(1,0)} \gamma_{gq}^{(0)} \gamma_{qg}^{(0)} 
        	+ \frac{1}{2} \beta_0 \gamma_{qg}^{(1)}
        	+ \frac{1}{8} \gamma_{gg}^{(0)} \gamma_{qg}^{(1)} 
            \nonumber \\ &&
        	+ \frac{1}{8} c_{i,q}^{{\rm NS},(1,0)} \gamma_{qg}^{(0)} \gamma_{qq}^{{\rm NS},(0)} 
        	+ \frac{1}{8} \gamma_{qg}^{(1)} \gamma_{qq}^{{\rm NS},(0)}
        	+ \frac{1}{8} \gamma_{qg}^{(0)} \gamma_{qq}^{{\rm NS},(1)} 
        	+ \frac{1}{8} \gamma_{qg}^{(0)} \gamma_{qq}^{{\rm PS},(1)}
        \bigr]
        \nonumber \\ &&
        - \ln\left( \frac{Q^2}{\mu^2} \right)
        \bigl[
        	 (\beta_1 c_{i,g}^{(1,0)}) 
        	+ 4 \beta_0^2 c_{i,g}^{(1,1)} 
        	+ 2 \beta_0 c_{i,g}^{(2,0)} 
        	+ 3 \beta_0 c_{i,g}^{(1,1)} \gamma_{gg}^{(0)} 
        	+ \frac{1}{2} c_{i,g}^{(2,0)} \gamma_{gg}^{(0)}
            \nonumber \\ &&
        	+ \frac{1}{2} c_{i,g}^{(1,1)} \left. \gamma_{gg}^{(0)} \right.^2
        	+ \frac{1}{2} c_{i,g}^{(1,0)} \gamma_{gg}^{(1)}
        	+ 3 \beta_0 c_{i,q}^{{\rm NS},(1,1)} \gamma_{qg}^{(0)} 
        	+ \frac{1}{2} c_{i,q}^{{\rm NS},(2,0)} \gamma_{qg}^{(0)} 
        	+ \frac{1}{2} c_{i,q}^{{\rm PS},(2,0)} \gamma_{qg}^{(0)} 
            \nonumber \\ &&
        	+ \frac{1}{2} c_{i,q}^{{\rm NS},(1,1)} \gamma_{gg}^{(0)} \gamma_{qg}^{(0)}
        	+ \frac{1}{2} c_{i,g}^{(1,1)} \gamma_{gq}^{(0)} \gamma_{qg}^{(0)}
        	+ \frac{1}{2} c_{i,q}^{{\rm NS},(1,0)} \gamma_{qg}^{(1)}
        	+ \frac{1}{2} c_{i,q}^{{\rm NS},(1,1)} \gamma_{qg}^{(0)} \gamma_{qq}^{{\rm NS},(0)}
        	+ \frac{1}{2} \gamma_{qg}^{(2)}
        \bigr]
        \nonumber \\ &&
        + \beta_1 c_{i,g}^{(1,1)} 
        + 4 \beta_0^2 c_{i,g}^{(1,2)} 
        + 4 \beta_0 c_{i,g}^{(2,1)} 
        + 3 \beta_0 c_{i,g}^{(1,2)} \gamma_{gg}^{(0)} 
        + c_{i,g}^{(2,1)} \gamma_{gg}^{(0)} 
        + \frac{1}{2} c_{i,g}^{(1,2)} \left. \gamma_{gg}^{(0)} \right.^2
        \nonumber \\ &&
        + \frac{1}{2} c_{i,g}^{(1,1)} \gamma_{gg}^{(1)}
        + 3 \beta_0 c_{i,q}^{{\rm NS},(1,2)} \gamma_{qg}^{(0)} 
        + c_{i,q}^{{\rm NS},(2,1)} \gamma_{qg}^{(0)} 
        + c_{i,q}^{{\rm PS},(2,1)} \gamma_{qg}^{(0)} 
        + \frac{1}{2} c_{i,q}^{{\rm NS},(1,2)} \gamma_{gg}^{(0)} \gamma_{qg}^{(0)} 
        \nonumber \\ &&
        + \frac{1}{2} c_{i,g}^{(1,2)} \gamma_{gq}^{(0)} \gamma_{qg}^{(0)} 
        + \frac{1}{2} c_{i,q}^{{\rm NS},(1,1)} \gamma_{qg}^{(1)}
        + \frac{1}{2} c_{i,q}^{{\rm NS},(1,2)} \gamma_{qg}^{(0)} \gamma_{qq}^{{\rm NS},(0)}
        + c_{i,g}^{(3,0)} 
    \biggr\}
    + O(a_s^4) ~.
\end{eqnarray}
Accordingly, one obtains for the structure function $F_L$,
\begin{eqnarray}
    C_{F_L,q}^{{\rm NS}} &=& 
    a_s
    \biggr\{
        c_{F_L,q}^{{\rm NS},(1,0)}
    \biggr\}
    + a_s^2
    \biggl\{
        - \ln\left( \frac{Q^2}{\mu^2} \right)
        \bigl[
        	 \beta_0 c_{F_L,q}^{{\rm NS},(1,0)} 
        	+ \frac{1}{2} c_{F_L,q}^{{\rm NS},(1,0)} \gamma_{qq}^{{\rm NS},(0)}
        \bigr]
        + 2 \beta_0 c_{F_L,q}^{{\rm NS},(1,1)} 
        \nonumber \\ &&
        + c_{F_L,q}^{{\rm NS},(1,1)} \gamma_{qq}^{{\rm NS},(0)}
        + c_{F_L,q}^{{\rm NS},(2,0)} 
    \biggr\}
    + a_s^3
    \biggl\{
        \ln^2\left( \frac{Q^2}{\mu^2} \right)
        \bigl[
        	\beta_0^2 c_{F_L,q}^{{\rm NS},(1,0)} 
        	+ \frac{3}{4} \beta_0 c_{F_L,q}^{{\rm NS},(1,0)} \gamma_{qq}^{{\rm NS},(0)}
        	\nonumber \\ &&
        	+ \frac{1}{8} c_{F_L,q}^{{\rm NS},(1,0)} \left. \gamma_{qq}^{{\rm NS},(0)} \right.^2 
        \bigr]
        - \ln\left( \frac{Q^2}{\mu^2} \right)
        \bigl[
        	\beta_1 c_{F_L,q}^{{\rm NS},(1,0)}
        	+ 4 \beta_0^2 c_{F_L,q}^{{\rm NS},(1,1)} 
        	+ 2 \beta_0 c_{F_L,q}^{{\rm NS},(2,0)} 
        \nonumber \\ &&
        	+ 3 \beta_0 c_{F_L,q}^{{\rm NS},(1,1)} \gamma_{qq}^{{\rm NS},(0)} 
        	+ \frac{1}{2} c_{F_L,q}^{{\rm NS},(2,0)} \gamma_{qq}^{{\rm NS},(0)}
        	+ \frac{1}{2} c_{F_L,q}^{{\rm NS},(1,1)} \left. \gamma_{qq}^{{\rm NS},(0)} \right.^2 
        	+ \frac{1}{2} c_{F_L,q}^{{\rm NS},(1,0)} \gamma_{qq}^{{\rm NS},(1)}
        \bigr]
        \nonumber \\ &&
        + \beta_1 c_{F_L,q}^{{\rm NS},(1,1)} 
        + 4 \beta_0^2 c_{F_L,q}^{{\rm NS},(1,2)} 
        + 4 \beta_0 c_{F_L,q}^{{\rm NS},(2,1)} 
        + 3 \beta_0 c_{F_L,q}^{{\rm NS},(1,2)} \gamma_{qq}^{{\rm NS},(0)} 
        + c_{F_L,q}^{{\rm NS},(2,1)} \gamma_{qq}^{{\rm NS},(0)} 
        \nonumber \\ &&
        + \frac{1}{2} c_{F_L,q}^{{\rm NS},(1,2)} \left. \gamma_{qq}^{{\rm NS},(0)} \right.^2
        + \frac{1}{2} c_{F_L,q}^{{\rm NS},(1,1)} \gamma_{qq}^{{\rm NS},(1)}
        + c_{F_L,q}^{\rm NS,(3,0)}
    \biggr\}
    + {O}(a_s^4) ~,
    \\
    C_{F_L,q}^{{\rm PS}} &=&
    a_s^2
    \biggl\{
        - \ln\left( \frac{Q^2}{\mu^2} \right)
        \bigl[
        	\frac{1}{2} (c_{F_L,g}^{(1,0)} \gamma_{gq}^{(0)}) 
        \bigr]
        + c_{F_L,g}^{(1,1)} \gamma_{gq}^{(0)}
        + c_{F_L,q}^{{\rm PS},(2,0)} 
    \biggr\}
    \nonumber \\ &&
    + a_s^3
    \biggl\{
        \ln^2\left( \frac{Q^2}{\mu^2} \right)
        \bigl[
            \frac{3}{4} \beta_0 c_{F_L,g}^{(1,0)} \gamma_{gq}^{(0)}
        	+ \frac{1}{8} c_{F_L,g}^{(1,0)} \gamma_{gg}^{(0)} \gamma_{gq}^{(0)} 
        	+ \frac{1}{8} c_{F_L,q}^{{\rm NS},(1,0)} \gamma_{gq}^{(0)} \gamma_{qg}^{(0)} 
        	\nonumber \\ &&
        	+ \frac{1}{8} c_{F_L,g}^{(1,0)} \gamma_{gq}^{(0)} \gamma_{qq}^{{\rm NS},(0)}
        \bigr]
        - \ln\left( \frac{Q^2}{\mu^2} \right)
        \bigl[
        	 2 \beta_0 c_{F_L,q}^{{\rm PS},(2,0)} 
        	+ 3 \beta_0 c_{F_L,g}^{(1,1)} \gamma_{gq}^{(0)} 
        	+ \frac{1}{2} c_{F_L,g}^{(2,0)} \gamma_{gq}^{(0)} 
        	\nonumber \\ &&
        	+ \frac{1}{2} c_{F_L,g}^{(1,1)} \gamma_{gg}^{(0)} \gamma_{gq}^{(0)} 
        	+ \frac{1}{2} c_{F_L,g}^{(1,0)} \gamma_{gq}^{(1)}
        	+ \frac{1}{2} c_{F_L,q}^{{\rm NS},(1,1)} \gamma_{gq}^{(0)} \gamma_{qg}^{(0)} 
        	+ \frac{1}{2} c_{F_L,q}^{{\rm PS},(2,0)} \gamma_{qq}^{{\rm NS},(0)}
        	\nonumber \\ &&
        	+ \frac{1}{2} c_{F_L,g}^{(1,1)} \gamma_{gq}^{(0)} \gamma_{qq}^{{\rm NS},(0)} 
        	+ \frac{1}{2} c_{F_L,q}^{{\rm NS},(1,0)} \gamma_{qq}^{{\rm PS},(1)}
        \bigr]
        + 4 \beta_0 c_{F_L,q}^{{\rm PS},(2,1)} 
        + 3 \beta_0 c_{F_L,g}^{(1,2)} \gamma_{gq}^{(0)} 
        \nonumber \\ &&
        + c_{F_L,g}^{(2,1)} \gamma_{gq}^{(0)} 
        + \frac{1}{2} c_{F_L,g}^{(1,2)} \gamma_{gg}^{(0)} \gamma_{gq}^{(0)}
        + \frac{1}{2} c_{F_L,g}^{(1,1)} \gamma_{gq}^{(1)}
        + \frac{1}{2} c_{F_L,q}^{{\rm NS},(1,2)} \gamma_{gq}^{(0)} \gamma_{qg}^{(0)}
        \nonumber \\ &&
        + c_{F_L,q}^{{\rm PS},(2,1)} \gamma_{qq}^{{\rm NS},(0)} 
        + \frac{1}{2} c_{F_L,g}^{(1,2)} \gamma_{gq}^{(0)} \gamma_{qq}^{{\rm NS},(0)}
        + \frac{1}{2} c_{F_L,q}^{{\rm NS},(1,1)} \gamma_{qq}^{{\rm PS},(1)}
        + c_{q,(3,0)}^{{\rm PS},{(L)}}
    \biggr\}
    + {O}(a_s^4) ~,
    \\
    C_{F_L,g} &=&
    a_s
    \biggr\{
         c_{F_L,g}^{(1,0)}
    \biggr\}
    + a_s^2
    \biggl\{
        - \ln\left( \frac{Q^2}{\mu^2} \right)
        \bigl[
        	 \beta_0 c_{F_L,g}^{(1,0)}
        	+ \frac{1}{2} c_{F_L,g}^{(1,0)} \gamma_{gg}^{(0)}
        	+ \frac{1}{2} c_{F_L,q}^{{\rm NS},(1,0)} \gamma_{qg}^{(0)} 
        \bigr]
        \nonumber \\ &&
        + 2 \beta_0 c_{F_L,g}^{(1,1)} 
        + c_{F_L,g}^{(1,1)} \gamma_{gg}^{(0)} 
        + c_{F_L,q}^{{\rm NS},(1,1)} \gamma_{qg}^{(0)}
        + c_{F_L,g}^{(2,0)} 
    \biggr\}
    + a_s^3
    \biggl\{
         \ln^2\left( \frac{Q^2}{\mu^2} \right)
        \bigl[
        	\beta_0^2 c_{F_L,g}^{(1,0)} 
        	\nonumber \\ &&
        	+ \frac{3}{4} \beta_0 c_{F_L,g}^{(1,0)} \gamma_{gg}^{(0)}
        	+ \frac{1}{8} c_{F_L,g}^{(1,0)} \left. \gamma_{gg}^{(0)} \right.^2 
        	+ \frac{3}{4} \beta_0 c_{F_L,q}^{{\rm NS},(1,0)} \gamma_{qg}^{(0)}
        	+ \frac{1}{8} c_{F_L,q}^{{\rm NS},(1,0)} \gamma_{gg}^{(0)} \gamma_{qg}^{(0)} 
        	\nonumber \\ &&
        	+ \frac{1}{8} c_{F_L,g}^{(1,0)} \gamma_{gq}^{(0)} \gamma_{qg}^{(0)} 
        	+ \frac{1}{8} c_{F_L,q}^{{\rm NS},(1,0)} \gamma_{qg}^{(0)} \gamma_{qq}^{{\rm NS},(0)} 
        \bigr]
        - \ln\left( \frac{Q^2}{\mu^2} \right)
        \bigl[
        	  \beta_1 c_{F_L,g}^{(1,0)}
        	+ 4 \beta_0^2 c_{F_L,g}^{(1,1)} 
        	\nonumber \\ &&
        	+ 2 \beta_0 c_{F_L,g}^{(2,0)} 
        	+ 3 \beta_0 c_{F_L,g}^{(1,1)} \gamma_{gg}^{(0)} 
        	+ \frac{1}{2} c_{F_L,g}^{(2,0)} \gamma_{gg}^{(0)}
        	+ \frac{1}{2} c_{F_L,g}^{(1,1)} \left. \gamma_{gg}^{(0)} \right.^2
        	+ \frac{1}{2} c_{F_L,g}^{(1,0)} \gamma_{gg}^{(1)}
        	\nonumber \\ &&
        	+ 3 \beta_0 c_{F_L,q}^{{\rm NS},(1,1)} \gamma_{qg}^{(0)} 
        	+ \frac{1}{2} c_{F_L,q}^{{\rm NS},(2,0)} \gamma_{qg}^{(0)} 
        	+ \frac{1}{2} c_{F_L,q}^{{\rm PS},(2,0)} \gamma_{qg}^{(0)} 
        	+ \frac{1}{2} c_{F_L,q}^{{\rm NS},(1,1)} \gamma_{gg}^{(0)} \gamma_{qg}^{(0)}
        	\nonumber \\ &&
        	+ \frac{1}{2} c_{F_L,g}^{(1,1)} \gamma_{gq}^{(0)} \gamma_{qg}^{(0)}
        	+ \frac{1}{2} c_{F_L,q}^{{\rm NS},(1,0)} \gamma_{qg}^{(1)}
        	+ \frac{1}{2} c_{F_L,q}^{{\rm NS},(1,1)} \gamma_{qg}^{(0)} \gamma_{qq}^{{\rm NS},(0)}
        \bigr]
        + \beta_1 c_{F_L,g}^{(1,1)} 
        + 4 \beta_0^2 c_{F_L,g}^{(1,2)} 
        \nonumber \\ &&
        + 4 \beta_0 c_{F_L,g}^{(2,1)} 
        + 3 \beta_0 c_{F_L,g}^{(1,2)} \gamma_{gg}^{(0)} 
        + c_{F_L,g}^{(2,1)} \gamma_{gg}^{(0)} 
        + \frac{1}{2} c_{F_L,g}^{(1,2)} \left. \gamma_{gg}^{(0)} \right.^2
        + \frac{1}{2} c_{F_L,g}^{(1,1)} \gamma_{gg}^{(1)}
        \nonumber \\ &&
        + 3 \beta_0 c_{F_L,q}^{{\rm NS},(1,2)} \gamma_{qg}^{(0)} 
        + c_{F_L,q}^{{\rm NS},(2,1)} \gamma_{qg}^{(0)} 
        + c_{F_L,q}^{{\rm PS},(2,1)} \gamma_{qg}^{(0)} 
        + \frac{1}{2} c_{F_L,q}^{{\rm NS},(1,2)} \gamma_{gg}^{(0)} \gamma_{qg}^{(0)} 
        \nonumber \\ &&
        + \frac{1}{2} c_{F_L,g}^{(1,2)} \gamma_{gq}^{(0)} \gamma_{qg}^{(0)} 
        + \frac{1}{2} c_{F_L,q}^{{\rm NS},(1,1)} \gamma_{qg}^{(1)}
        + \frac{1}{2} c_{F_L,q}^{{\rm NS},(1,2)} \gamma_{qg}^{(0)} \gamma_{qq}^{{\rm NS},(0)}
        + c_{F_L,g}^{(3,0)} 
    \biggr\}
    + {O}(a_s^4) ~.
\end{eqnarray}

The Wilson coefficients for the different deep--inelastic structure functions are split
w.r.t. the parton contents they are mapping to, the observables and the specific 
couplings of the gauge bosons exchanged, see Ref.~\cite{Larin:1996wd}. 

For pure photon exchange there are two contributions in the flavor non--singlet case. One 
corresponds to a through-flowing external fermion line to which the photons are coupling to, 
which is of relative weight ${\sf w}_1$ = 1. The second case corresponds to {a photon coupling
to one through-flowing line and another to a closed internal fermion line, with relative weight}
\begin{eqnarray}
{\sf w_2} = 
\frac{{\rm tr}(\hat{Q}_f \lambda_\alpha)}
     {{\rm tr}(\hat{Q}_f^2 \lambda_\alpha)} 
\frac{1}{N_F} \sum_{f=1}^{N_F} e_f,~~~~ 
\frac{{\rm tr}(\hat{Q}_f \lambda_\alpha)}
     {{\rm tr}(\hat{Q}_f^2 \lambda_\alpha)} = 3 ~,
\end{eqnarray}
for $SU(N_F)$ and $\lambda_a$ the generalized Pauli--Gell--Mann matrices. $e_f$ denotes the 
fermion charge and $\hat{Q}_f = 
{\rm diag}(2/3,-1/3,-1/3,2/3,-1/3)$ the quark charge matrix of the electromagnetic 
current. 

The flavor non--singlet structure functions for photon exchange read
\cite{Zijlstra:1992qd}
\begin{eqnarray}
F_i^{\rm NS,+}(x,Q^2) = \sum_{f=1}^{N_F} e_f^2 \Biggl[C_{i,q}^{\rm NS}
\left(x,\frac{Q^2}{\mu^2}\right) 
+ {\sf w_2} C_{i,q}^{d_{abc}}\left(x,\frac{Q^2}{\mu^2}\right) \Biggr]
\otimes f_{q,f}^{\rm NS,+}\left(x,\mu^2\right) ~, 
\end{eqnarray}
with
\begin{eqnarray}
{\sf w}_2 = \frac{{\rm tr}(\hat{Q}_f)}{N_F} \frac{{\rm tr}(\hat{Q}_f \lambda_\alpha)}
     {{\rm tr}(\hat{Q}_f^2 \lambda_\alpha)}~,
~~~{\sf w}_2(N_F=3) = 0 ~,~~
  {\sf w}_2(N_F=4) = \frac{1}{2} ~,~~
  {\sf w}_2(N_F=5) = \frac{1}{5} ~.
\end{eqnarray}
The non--singlet distribution function is given by
\begin{eqnarray}
f_{q,f}^{\rm NS,+}\left(x,\mu^2\right) = f_{q,f}(x,\mu^2) + \bar{f}_{q,f}(x,\mu^2) - \frac{1}{N_F} 
\Sigma(x,\mu^2) ~,
\end{eqnarray}
with the quark singlet distribution
\begin{eqnarray}
\Sigma(x,\mu^2) = \sum_{f=1}^{N_F} f_{q,f}(x,\mu^2) + \bar{f}_{q,f}(x,\mu^2) ~. 
\end{eqnarray}

In the flavor singlet case the weight factor ${\sf w}_3$ appears
\begin{eqnarray}
{\sf w}_3 = \frac{1}{N_F} \frac{\left(\sum_{f=1}^{N_F} e_f\right)^2}{\sum_{f=1}^{N_F} e_f^2} ~, 
~~~{\sf w}_3(N_F=3) = 0 ~,~~
   {\sf w}_3(N_F=4) = \frac{1}{10} ~,~~
   {\sf w}_3(N_F=5) = \frac{1}{55} ~,
\end{eqnarray}
because of the overall normalization to the sum of the quark charge squares for all diagrams in which the 
electromagnetic current couples to two different fermion lines.

The flavor singlet structure functions for photon exchange  are given by \cite{Zijlstra:1992qd}
\begin{eqnarray}
F_i^{\rm S,+}(x,Q^2) = \left(\frac{1}{N_F} \sum_{f=1}^{N_F} e_f^2\right) \Biggl[
C_{i,q}^{\rm S}
\left(x,\frac{Q^2}{\mu^2}\right) \otimes \Sigma(x,\mu^2)
+
C_{i,g}
\left(x,\frac{Q^2}{\mu^2}\right) \otimes G(x,\mu^2)\Biggr] ~,
\end{eqnarray}
with 
{
\begin{eqnarray}
C_{i,q}^{\rm S} &=& C_{i,q}^{\rm NS} + {\sf w}_3 C_{i,q}^{d_{abc}} + C_{i,q}^{\rm PS} ~,
\\
C_{i,g} &=& C_{i,g}^{a} +  {\sf w}_3 C_{i,g}^{d_{abc}} ~,
\end{eqnarray}
}

\noindent
and $G(x,\mu^2)$ denotes the gluon distribution and $C_{i,q}^{\rm S,1}$ the non--weighted contributions
and $C_{i,q}^{\rm S,2}$  the one corresponding to the charge weight factor ${\sf w}_3$. 
Synonymous relations apply to the structure function $g_1(x,Q^2)$ by replacing the unpolarized 
quantities by the corresponding polarized ones.

Usually the deep--inelastic structure functions are represented referring to $N_F = 3$ massless 
Wilson coefficients and massive Wilson coefficients \cite{Bierenbaum:2009mv,HEAVY,
Ablinger:2019etw,
Ablinger:2019gpu,
Ablinger:2020snj,
Behring:2015zaa,
Behring:2015roa,Behring:2016hpa,Blumlein:2021xlc} due to charm and bottom quark corrections.
The scaling violations of the heavy flavor Wilson coefficients are very different compared to those in 
the massless case.

The renormalization of the forward Compton amplitude is performed by renormalizing the strong coupling constant
and removing the collinear singularities into the running of the parton distribution functions. In the case of the
non--singlet structure function $xF_3(x,Q^2)$ the original calculation is performed in the Larin scheme
and we finally switch to the $\overline{\sf MS}$ scheme by a finite renormalization. In the case of the structure
function $g_1(x,Q^2)$ this last step requires the knowledge of the polarized four--loop anomalous dimensions in the 
singlet case. Because of this we present the corresponding Wilson coefficients in the Larin scheme and switch to the 
$\overline{\sf MS}$ in addition only for the non--singlet Wilson coefficient, since here the respective Ward 
identity is known in explicit form. Thereby, we also derive the $Z$-factor {$Z_5^{\rm NS}(N,a_s)$} from the
ratio of off--shell massless non--singlet operator matrix elements to three--loop order.
\section{Details of the calculation}
\label{sec:3}

\vspace*{1mm} 
\noindent 
The Feynman diagrams for the different massless Wilson coefficients are generated by {\tt QGRAF} 
\cite{Nogueira:1991ex} using the forward Compton amplitude. 

We maintain all contributions up to the first 
order in the $R_\xi$ gauge parameter $\xi$, which is canceling already for the unrenormalized result. 
For the Compton amplitude the corresponding crossing relations have to be observed \cite{Politzer:1974fr,Blumlein:1996vs}. 
We decompose the Wilson coefficients into their flavor non--singlet $\mathbb{C}_{i,q}^{\rm NS}$, 
{pure} singlet 
$\mathbb{C}_{i,q}^{\rm PS}$ and gluonic $\mathbb{C}_{i,g}$ parts. At one--loop order the contributions up to $O(\ep^2)$ 
and two--loop order up to $O(\ep)$ are required to extract the three--loop Wilson coefficients. We perform the 
calculation keeping the gauge parameter $\xi$ to first order as a test for gauge invariance and show that the corresponding 
terms disappear. The Dirac and Lorentz algebra is performed by {\tt FORM} \cite{FORM} and the color algebra is performed 
by using {\tt Color} \cite{vanRitbergen:1998pn}. The crossing relations \cite{Politzer:1974fr,Blumlein:1996vs} 
imply that for the Wilson coefficients contributing to 
the structure functions $F_2(x,Q^2)$ and $F_L(x,Q^2)$ only the even moments contribute and to $x F_3(x,Q^2)$
and $g_1(x,Q^2)$ only the odd moments. 
The corresponding generating variable $\omega = 1/x$ appears therefore dominantly quadratically, 
except of 
a factor $\omega$ in the case of $x F_3(x,Q^2)$ and $g_1(x,Q^2)$.

The irreducible three--loop diagrams are reduced to 293 master integrals using the code {\tt Crusher} 
\cite{CRUSHER} by applying the integration--by--parts relations \cite{IBP,Chetyrkin:1981qh}. 
The master integrals are mapped to $\omega^2$ dependent structures and a sufficient number of 
their Mellin moments
is calculated and inserted into the forward Compton amplitude. The required numbers of moments are summarized
in Table~\ref{TAB1}.
{For the gluonic and non-singlet structures we do not map to a manifestly $\omega^2$ dependent 
structure a priori.}

The number of moments in Table~\ref{TAB1} in the lower loop orders includes also the terms needed to 
renormalize the Wilson coefficients at three--loop order. As has been observed before in 
Ref.~\cite{Blumlein:2021enk} there is a small number of master--integral relations which are difficult
to prove for vanishing powers in $\omega$. We have verified them explicitly for their moments up to a much 
higher 
number than requested and to the respective contributing power in $\ep$, cf.~Table~\ref{TAB1}. We determine 
recurrence 
relations for the corresponding color and $\zeta$--value projections \cite{Blumlein:2009cf}
of the Wilson coefficients using the
method of arbitrary large moments \cite{Blumlein:2017dxp} implemented within the package
\texttt{SolveCoupledSystem}~\cite{Blumlein:2019hfc}. 

This is done by using the  method of guessing 
\cite{GUESS,Blumlein:2009tj} and its implementation in {\tt Sage} \cite{SAGE,GSAGE}. For the calculation of 
the necessary initial values for the difference equations we use the results given in \cite{Chetyrkin:1981qh,INIT}. 
In the three--loop case the calculation is based on {5000} even 
or odd moments. We have explicitly checked, that the other moments vanish, which is implied by 
the amplitude crossing relations. 
The difference equations are solved by using 
methods from difference field theory \cite{DRING} implemented in the package {\tt Sigma} \cite{SIG1,SIG2} utilizing 
functions from {\tt HarmonicSums} \cite{HARMSU,Blumlein:2009ta,Vermaseren:1998uu,Blumlein:1998if,Remiddi:1999ew,
Ablinger:2011te,Ablinger:2013cf,Ablinger:2014bra,Blumlein:2003gb}, to obtain the three--loop 
Wilson coefficients. 
The largest difference equation for the individual color, gauge parameter, and $\zeta_k$--factors contributing 
in the present case {has order {\sf o} = {25} and degree  {\sf d} = {778} and needed 4300 moments.} 
In parallel, we calculated the Wilson coefficients by using the differential equations for the master 
integrals directly in the variable $\omega$, using the method presented in Ref.~\cite{Ablinger:2018zwz}. 
The depth of 
the initial conditions in the dimensional parameter $\ep$ is the same as in the method described 
previously. The corresponding systems were decoupled using the formalisms of Ref.~\cite{ZUE} 
implemented 
in the package {\tt ORESYS} \cite{ORESYS} and we further proceeded to find the $N$-space solution 
by using algorithms contained in the package {\tt HarmonicSums}.

Comparing to 
the reconstruction of the anomalous dimensions and unpolarized Wilson coefficients out of their moments 
performed in Ref.~\cite{Blumlein:2009tj} in 2008 the largest difference equation had order {\sf o} = 35 and 
degree {\sf d} = 938 
requiring 5114 moments. Here, however, even and odd moments had been used. 
The overall computation time using the automated chain of codes described amounted to about one year on 
{\tt Intel(R) Xeon(R) CPU E5-2643 v4} processors, using also parallelizations.

{Working in the variable $\omega^2$ rather than $\omega$ we 
obtain the results of the recurrences 
first for only the even or odd moments expressed in terms of cyclotomic 
harmonic sums at argument $2N$.
For a systematic study of this class of nested sums see 
\cite{Ablinger:2011te}.  These objects can be algorithmically reduced 
to simple harmonic sums at argument $N$ using 
{\tt HarmonicSums}.
}

Therefore, all Wilson coefficients can be expressed by harmonic sums \cite{Vermaseren:1998uu,Blumlein:1998if}
\begin{eqnarray}
S_{b,\vec{a}}(N) &=& \sum_{k=1}^N \frac{({\rm sign}(b))^k}{k^{|b|}} S_{\vec{a}}(k),~~~S_\emptyset 
= 
1,~~~b_, a_i \in \mathbb{Z} \backslash \{0\}, N \in \mathbb{N}  \backslash \{0\}. 
\end{eqnarray}
If it is clear from the context we will write $S_{\vec{a}}$
instead of $S_{\vec{a}}(N)$.

Their Mellin inversion to momentum fraction $z$--space 
\begin{eqnarray}
C(N) =  \int_0^1 dz z^{N-1} \bar{C}(z) 
\end{eqnarray}
can be performed using routines of the packages {\tt HarmonicSums} and is expressed in terms of harmonic 
polylogarithms 
\cite{Remiddi:1999ew} given by
\begin{eqnarray}
\HA_{b,\vec{a}}(z) &=& \int_0^z dx f_b(x) \HA_{\vec{a}}(x),~~~\HA_\emptyset = 1,~~~b, a_i \in 
\{-1,0,1\},
\end{eqnarray}
with  the  alphabet of letters
\begin{eqnarray}
\mathfrak{A}_{\rm H} = \Biggl\{f_{0}(z) = \frac{1}{z},~~~f_{-1}(z) = \frac{1}{1+z},~~f_{1}(z) = 
\frac{1}{1-z}\Biggr\}.
\end{eqnarray}
In $z$--space one distinguishes three contributions to the individual Wilson coefficients 
because of their different treatment in Mellin convolutions,
\begin{eqnarray}
\bar{C}(z) = \bar{C}^\delta(z)  + \bar{C}^{\rm plu}(z) + \bar{C}^{\rm reg}(z),
\end{eqnarray}
where $\bar{C}^\delta(z) = c_0 \delta(1-z)$, $\bar{C}^{\rm reg}(z)$ is a regular function in $z \in [0,1]$ and 
$\bar{C}^{\rm plu}(z)$ denotes the remaining genuine $+$-distribution, the Mellin transformation of 
which is given by
\begin{eqnarray}
C^{\rm plu}(N) = \int_0^1 dz (z^{N-1} - 1) \bar{C}^{\rm plu}(z).
\end{eqnarray}
We will use this representation later on.

\renewcommand{\arraystretch}{1.15}
\setcounter{table}{0}
\begin{table}[H]
        \centering
        \begin{tabular}{|l|r|r|r|}
                \hline\hline
                Wilson coefficient   & 1 loop & 2 loop & 3 loop
                \\ \hline\hline
                $F_1^{\rm NS}$ & 126 & 1219 & 4300    \\
                \hline
                $F_1^{\rm PS}$ &   0 &  374 & 1708  \\
                \hline
                $F_1^{g}$      &  104 & 960 & 3534
                \\ \hline\hline
                $F_L^{\rm NS}$ &   48 & 560 & 2387  \\
                \hline
                $F_L^{\rm PS}$ &   0 &  175 & 774 \\
                \hline
                $F_L^{g}$      &  54 & 434  & 2046 \\
                \hline\hline
                $x F_3^{\rm NS}$ & 126 & 1219 & 4171  \\
                 \hline\hline
                $g_1^{\rm NS}$ & 126 & 1219 & {4171}  \\
\hline             
                $g_1^{\rm PS}$ &   0 &  175 & 1458 \\
\hline     
                $g_1^{g}$      &  84 & 1166 & 2998  \\
                 \hline\hline
        \end{tabular}
\caption{\sf The necessary maximal number of non--vanishing even (resp. odd) Mellin moments for 
$F_1, F_L, (xF_3, g_1)$ to determine the Wilson coefficients.}
\label{TAB1}
\end{table}
\renewcommand{\arraystretch}{1.}

In the Mellin $N$ space representation it can technically occur that there are factors of up to $1/(N-2)^2$
contributing and structures of $1/(N-1)$, in the cases that they 
{are physically not allowed}. However, these are all tractable 
poles, which can be shown by expanding at $N=2$ and $N=1$ using {\tt HarmonicSums}.
No Kronecker symbols have to be introduced for this case and the usual analytic continuation, described in
Ref.~\cite{Blumlein:2009ta}, can be applied to the $N$--space expressions directly.

The Wilson coefficients can be represented in $N$ space by harmonic sums weighted by 
rational functions 
in $N$. Here the degree of the numerator needs not to be smaller than that of the denominator, which 
needs special care in performing the inverse Mellin transform to $z$--space. After applying the algebraic 
relations between harmonic sums, c.f. e.g.  \cite{Blumlein:2003gb}, one obtains the following set of 
60 harmonic sums
\begin{eqnarray}
&&\{
S_1;
S_2,
S_{-2};
S_3,
S_{-3},
S_{2,1},
S_{-2,1};
S_4,
S_{-4},
S_{-2,2},
S_{3,1},
S_{-3,1},
S_{2,1,1},
S_{-2,1,1};
S_5,
S_{-5},
S_{-2,3},
S_{2,3},
\nonumber\\ &&
S_{2,-3},
S_{-2,-3},
S_{2,2,1},
S_{-2,1,-2},
S_{-2,2,1}
S_{4,1},
S_{-4,1},
S_{2,1,-2},
S_{3,1,1},
S_{-3,1,1},
S_{2,1,1,1},
S_{-2,1,1,1};
S_6,
S_{-6},
\nonumber\\ && 
S_{-3,3},
S_{4,2},
S_{4,-2},
S_{-4,2},
S_{-4,-2},
S_{5,1},
S_{-5,1},
S_{-2,2,-2},
S_{-2,2,2},
S_{2,-3,1},
S_{-2,3,1},
S_{-3,1,-2},
S_{-3,-2,1},
\nonumber\\ &&
S_{-3,2,1},
S_{-4,1,1},
S_{2,3,1},
S_{3,1,-2},
S_{3,2,1},
S_{4,1,1},
S_{-2,-2,1,1},
S_{-2,1,1,2},
S_{-2,2,1,1},
S_{2,-2,1,1},
S_{2,2,1,1},
S_{3,1,1,1},
\nonumber\\ &&
S_{-3,1,1,1},
S_{2,1,1,1,1},
S_{-2,1,1,1,1}
\}.
\end{eqnarray}
One may, furthermore, apply also the structural relations \cite{Blumlein:2009ta,Blumlein:2009fz}
and  obtains the following 31 harmonic sums. 
\begin{eqnarray}
&& \{
S_1,
S_{-1,1},
S_{-2,1},
S_{-3,1},
S_{-4,1},
S_{-5,1},
S_{2,1},
S_{4,1},
S_{-1,1,1},
S_{2,1,1},
S_{1,2,-1},
S_{2,1,-1},
S_{-2,1,-2},
S_{2,1,-2},
\nonumber\\ &&
S_{-3,1,1},
S_{3,1,1},
S_{-2,2,-2},
S_{-3,-2,1},
S_{3,1,-2},
S_{-4,1,1},
S_{4,1,1},
S_{-2,1,1,1},
S_{2,1,1,1},
S_{-3,1,1,1},
S_{-2,-2,1,1},
\nonumber\\ &&
S_{-2,1,1,2},
S_{-2,2,1,1},
S_{2,-2,1,1},
S_{2,2,1,1},
S_{-2,1,1,1,1},
S_{2,1,1,1,1}\}
\end{eqnarray}
spanning all quantities. In $z$ space, the number of harmonic polylogarithms is, usually, higher if compared
to the objects needed in Mellin $N$ space. Here 68 harmonic polylogarithms of up to weight {\sf w = 5}, weighted 
by one more letter, contribute
\begin{eqnarray}
&& 
\Bigl\{\HA_{-1},\HA_0,\HA_1,\HA_{0,-1},\HA_{0,1},\HA_{0,-1,-1},\HA_{0,-1,1},\HA_{0,0,-1},\HA_{0,0,1},
\HA_{0,1,-1},\HA_{0,1,1},\HA_{0,-1,-1,-1},\HA_{0,-1,-1,1},
\nonumber\\ &&
\HA_{0,-1,0,1},\HA_{0,-1,1,-1},\HA_{0,-1,1,1},
\HA_{0,0,-1,-1},\HA_{0,0,-1,1},\HA_{0,0,0,-1},\HA_{0,0,0,1},\HA_{0,0,1,-1},\HA_{0,0,1,1},\HA_{0,1,-1,-1},
\nonumber\\ &&
\HA_{0,1,-1,1},\HA_{0,1,1,-1},\HA_{0,1,1,1},
\HA_{0,-1,-1,-1,-1},\HA_{0,-1,-1,-1,1},\HA_{0,-1,-1,0,1},\HA_{0,-1,-1,1,-1},\HA_{0,-1,-1,1,1},
\nonumber\\ &&
\HA_{0,-1,0,-1,-1},\HA_{0,-1,0,-1,1},
\HA_{0,-1,0,1,-1},\HA_{0,-1,0,1,1},\HA_{0,-1,1,-1,-1},\HA_{0,-1,1,-1,1},\HA_{0,-1,1,0,1},\HA_{0,-1,1,1,-1},
\nonumber\\ &&
\HA_{0,-1,1,1,1},\HA_{0,0,-1,-1,-1},
\HA_{0,0,-1,-1,1},\HA_{0,0,-1,0,-1},\HA_{0,0,-1,0,1},\HA_{0,0,-1,1,-1},\HA_{0,0,-1,1,1},\HA_{0,0,0,-1,-1},
\nonumber\\ &&
\HA_{0,0,0,-1,1},\HA_{0,0,0,0,-1},
\HA_{0,0,0,0,1},\HA_{0,0,0,1,-1},\HA_{0,0,0,1,1},\HA_{0,0,1,-1,-1},\HA_{0,0,1,-1,1},\HA_{0,0,1,0,-1},
\HA_{0,0,1,0,1},
\nonumber\\ && 
\HA_{0,0,1,1,-1},\HA_{0,0,1,1,1},\HA_{0,1,-1,-1,-1},
\HA_{0,1,-1,-1,1},\HA_{0,1,-1,1,-1},\HA_{0,1,-1,1,1},\HA_{0,1,0,1,-1},\HA_{0,1,0,1,1},
\nonumber\\ &&
\HA_{0,1,1,-1,-1},
\HA_{0,1,1,-1,1},\HA_{0,1,1,1,-1},\HA_{0,1,1,1,1}\Bigr\}
\end{eqnarray}
after algebraic reduction.
\section{The one-- and two--loop Wilson coefficients}
\label{sec:4}

\vspace*{1mm}
\noindent
In the following we present the results for the massless Wilson coefficients contributing to the structure 
functions $F_2, F_L, xF_3$ and $g_1$ at one- and two--loop  order, together with the higher expansion 
coefficients in $\ep$ of the corresponding forward Compton amplitudes in the unrenormalized case, see 
Section~\ref{sec:2}. These contributions are needed for the calculation of the three--loop Wilson 
coefficients. We will work in Mellin $N$ space and also present the renormalized one- and two--loop 
Wilson coefficients.
\subsection{One-Loop Order}
\label{sec:41}

\vspace*{1mm}
\noindent
At one loop--order one obtains the following expansion coefficients, 
cf.~(\ref{eq:unrenWC1}--\ref{eq:unrenWC2}), 
\begin{eqnarray}
c_{F_2,q}^{(1,0)} &=&
\textcolor{blue}{C_F} \Biggl\{
        \frac{2+5 N-2 N^2-9 N^3}{N^2 (1+N)}
        +\frac{-2+3 N+3 N^2}{N (1+N)} S_1
        +2 S_1^2
        -2 S_2
\Biggr\} ~,
\\
c_{F_2,q}^{(1,1)} &=&
\textcolor{blue}{C_F} \Biggl\{
        \frac{P_3}{2 N^3 (1+N)}
        +\Biggl(
                \frac{-2-5 N-14 N^2-7 N^3}{2 N^2 (1+N)}
                +S_2
                +\zeta_2
        \Biggr) S_1
       -\frac{1}{3} S_1^3
\nonumber\\ && 
        +\frac{\big(
                2-3 N-3 N^2\big)}{4 N (1+N)} S_1^2
        +\frac{\big(
                -2+3 N+3 N^2\big) }{4 N (1+N)} S_2
        +\frac{\big(
                -2-3 N-3 N^2\big)}{4 N (1+N)} \zeta_2
\nonumber\\ &&
        -\frac{2}{3} S_3
\Biggr\} ~,
\\
c_{F_2,q}^{(1,2)} &=& 
\textcolor{blue}{C_F} \Biggl\{
        \frac{P_5}{4 N^4 (1+N)}
        +\Biggl(
                \frac{P_2}{4 N^3 (1+N)}
                +\frac{\big(
                        2-3 N-3 N^2\big)}{8 N (1+N)} [S_2 + \zeta_2]
                +\frac{1}{3} S_3
                -\frac{7}{3} \zeta_3
        \Biggr) S_1
\nonumber\\ && 
        +\Biggl(
                \frac{2+5 N+14 N^2+7 N^3}{8 N^2 (1+N)}
                -\frac{1}{4} S_2
                -\frac{1}{4} \zeta_2
        \Biggr) S_1^2
        +\frac{\big(
                -2+3 N+3 N^2\big)}{24 N (1+N)} S_1^3
        +\frac{1}{24} S_1^4
\nonumber\\ && 
        +\Biggl(
                \frac{-2-5 N-14 N^2-7 N^3}{8 N^2 (1+N)}
                +\frac{1}{4} \zeta_2
        \Biggr) S_2
        +\frac{1}{8} S_2^2
        +\frac{\big(
                -2+3 N+3 N^2\big)}{12 N (1+N)} S_3
        -\frac{1}{4} S_4
\nonumber\\ &&         
+\frac{\big(
                -2-5 N+2 N^2+9 N^3\big)}{8 N^2 (1+N)} \zeta_2
        +
        \frac{7 \big(
                2+3 N+3 N^2\big)}{12 N (1+N)} \zeta_3
\Biggr\} ~,
\\
c_{F_2,g}^{(1,0)} &=&
\textcolor{blue}{T_F N_F} \Biggl\{
        -\frac{4 \big(
                -2-N-4 N^2+N^3\big)}{N^2 (1+N) (2+N)}
        -\frac{4 \big(
                2+N+N^2\big)}{N (1+N) (2+N)} S_1
\Biggr\} ~,
\\
c_{F_2,g}^{(1,1)} &=&
\textcolor{blue}{T_F N_F}\Biggr\{
        -\frac{2 P_1}{N^3 (1+N) (2+N)}
        +\frac{2 \big(
                -2-N-4 N^2+N^3\big)}{N^2 (1+N) (2+N)} S_1
\nonumber\\ &&
        +\frac{\big(
                2+N+N^2\big)}{N (1+N) (2+N)} 
[S_1^2 - S_2 - \zeta_2]
\Biggr\} ~, 
\\
c_{F_2,g}^{(1,2)} &=&
\textcolor{blue}{T_F N_F} \Biggl\{
        \frac{P_6}{N^4 (1+N) (2+N)}
        +\Biggl(
                \frac{P_1}{N^3 (1+N) (2+N)}
                +\frac{\big(
                        2+N+N^2\big)}{2 N (1+N) (2+N)} 
\nonumber\\ &&    
\times [S_2 + 
\zeta_2]
        \Biggr) 
S_1
     +\frac{\big(
                2+N+4 N^2-N^3\big)}{2 N^2 (1+N) (2+N)} [S_1^2 - S_2 - 
\zeta_2]
        +\frac{\big(
                -2-N-N^2\big)}{6 N (1+N) (2+N)} S_1^3
\nonumber\\ &&         
+\frac{\big(
                -2-N-N^2\big)}{3 N (1+N) (2+N)} S_3
        +\frac{7 \big(
                2+N+N^2\big)}{3 N (1+N) (2+N)} \zeta_3
\Biggr\} ~,
\\
c_{F_L,q}^{(1,0)} &=&
\textcolor{blue}{C_F} \frac{4}{1+N} ~,
\\
c_{F_L,q}^{(1,1)} &=&
\textcolor{blue}{C_F} \Biggl\{
        -\frac{2}{1+N} [1 + S_1]
\Biggr\} ~,
\\
c_{F_L,q}^{(1,2)} &=&
\textcolor{blue}{C_F} \Biggl\{
        \frac{2}{1+N}
        +\frac{1}{1+N} S_1
        +\frac{1}{2 (1+N)} [S_1^2 - S_2 - \zeta_2]
\Biggr\} ~,
\\
c_{F_L,g}^{(1,0)} &=&
\textcolor{blue}{T_F N_F}
\frac{16}{(1+N) (2+N)} ~,
\\
c_{F_L,g}^{(1,1)} &=&
\textcolor{blue}{T_F N_F} \Biggl\{
        -\frac{16}{(1+N) (2+N)}
        -\frac{8}{(1+N) (2+N)} S_1
\Biggr\} ~,
\\
c_{F_L,g}^{(1,2)} &=&
\textcolor{blue}{T_F N_F} \Biggl\{
        \frac{16}{(1+N) (2+N)}
        +\frac{8}{(1+N) (2+N)} S_1 
        +\frac{2}{(1+N) (2+N)} 
\nonumber\\  & &
[S_1^2 - S_2 - \zeta_2] 
\Biggr\} ~,
\\
c_{F_3,q}^{(1,0),\rm L} &=&
\textcolor{blue}{C_F} 
\Biggl\{
        \frac{2+3 N-2 N^2-5 N^3}{N^2 (1+N)}
        +\frac{\big(-2+3 N+3 N^2\big)}{N (1+N)} S_1
        +2 S_1^2
        -2 S_2
\Biggr\} ~,
\\
c_{F_3,q}^{(1,1),\rm L} &=&
\textcolor{blue}{C_F} 
\Biggl\{
        \frac{2+3 N+3 N^3+8 N^4}{2 N^3 (1+N)}
        +\biggl(
                \frac{-2-3 N-10 N^2-7 N^3}{2 N^2 (1+N)}
                +S_2
        \biggr) S_1
\nonumber \\ &&
        +\frac{\big(2-3 N-3 N^2\big)}{4 N (1+N)} S_1^2
        +\frac{\big(-2+3 N+3 N^2\big)}{4 N (1+N)} S_2
        -\frac{1}{3} S_1^3
        -\frac{2}{3} S_3
\nonumber \\ &&
        +\biggl(
                \frac{-2-3 N-3 N^2}{4 N (1+N)}
                +S_1
        \biggr) \zeta_2
\Biggr\} ~,
\\
c_{F_3,q}^{(1,2),\rm L} &=&
\textcolor{blue}{C_F} 
\Biggl\{
        \frac{P_4}{4 N^4 (1+N)}
        +\biggl(
                \frac{2+3 N+10 N^2+7 N^3}{8 N^2 (1+N)}
                -\frac{1}{4} S_2
        \biggr) S_1^2
\nonumber \\ &&
        +\biggl(
                \frac{-2-3 N+19 N^3+14 N^4}{4 N^3 (1+N)}
                +\frac{\big(2-3 N-3 N^2\big)}{8 N (1+N)} S_2
                +\frac{1}{3} S_3
        \biggr) S_1
\nonumber \\ &&
        +\frac{\big(-2+3 N+3 N^2\big)}{24 N (1+N)}  S_1^3
        +\frac{1}{24} S_1^4
        +\frac{\big(-2-3 N-10 N^2-7 N^3\big)}{8 N^2 (1+N)}  S_2
        +\frac{1}{8} S_2^2
\nonumber \\ &&
        +\frac{\big(-2+3 N+3 N^2\big) }{12 N (1+N)} S_3
        -\frac{1}{4} S_4
        +\biggl(
                \frac{-2-3 N+2 N^2+5 N^3}{8 N^2 (1+N)}
\nonumber \\ &&
                +\frac{\big(2-3 N-3 N^2\big) S_1}{8 N (1+N)}
                -\frac{1}{4} S_1^2
                +\frac{1}{4} S_2
        \biggr) \zeta_2
        +\biggl(
                \frac{7 \big(2+3 N+3 N^2\big)}{12 N (1+N)}
                -\frac{7}{3} S_1
        \biggr) \zeta_3
\Biggr\} ~,
\\
c_{g_1,q}^{(1,0),\rm NS,L} &=&
\textcolor{blue}{C_F} \Biggl\{
        \frac{2-5 N-6 N^2-9 N^3}{N^2 (1+N)}
        +\frac{\big(
                -2+3 N+3 N^2\big)}{N (1+N)} S_1
        +2 S_1^2
        -2 S_2
\Biggr\} ~,
\\
c_{g_1,q}^{(1,1),\rm NS,L} &=&
\textcolor{blue}{C_F} \Biggl\{
        \frac{P_8}{2 N^3 (1+N)}
        +\Biggl(
                \frac{-2+5 N-10 N^2-7 N^3}{2 N^2 (1+N)}
                +S_2
        \Biggr) S_1
        +\frac{\big(
                2-3 N-3 N^2\big)}{4 N (1+N)} S_1^2
\nonumber\\ && 
        -\frac{1}{3} S_1^3
        +\frac{\big(
                -2+3 N+3 N^2\big)}{4 N (1+N)} S_2
        -\frac{2}{3} S_3
        +\Biggl(
                \frac{-2-3 N-3 N^2}{4 N (1+N)}
                +S_1
        \Biggr) \zeta_2
\Biggr\} ~,
\\
c_{g_1,q}^{(1,2),\rm NS,L} &=&
\textcolor{blue}{C_F} \Biggl\{
        \frac{P_9}{4 N^4 (1+N)}
        +\Biggl(
                \frac{P_7}{4 N^3 (1+N)}
                +\frac{\big(
                        2-3 N-3 N^2\big)}{8 N (1+N)} S_2
                +\frac{1}{3} S_3
        \Biggr) S_1
\nonumber\\ && 
        +\Biggl(
                \frac{2-5 N+10 N^2+7 N^3}{8 N^2 (1+N)}
                -\frac{1}{4} S_2
        \Biggr) S_1^2
        +\frac{\big(
                -2+3 N+3 N^2\big)}{24 N (1+N)} S_1^3
        +\frac{1}{24} S_1^4
\nonumber\\ && 
        +\frac{\big(
                -2+5 N-10 N^2-7 N^3\big) S_2}{8 N^2 (1+N)}
        +\frac{1}{8} S_2^2
        +\frac{\big(
                -2+3 N+3 N^2\big)}{12 N (1+N)} S_3
        -\frac{1}{4} S_4
\nonumber\\ &&    
     +\Biggl(
                \frac{-2+5 N+6 N^2+9 N^3}{8 N^2 (1+N)}
                +\frac{\big(
                        2-3 N-3 N^2\big)}{8 N (1+N)} S_1
                -\frac{1}{4} S_1^2
                +\frac{1}{4} S_2
        \Biggr) \zeta_2
\nonumber\\ &&   
      +\Biggl(
                \frac{7 \big(
                        2+3 N+3 N^2\big)}{12 N (1+N)}
                -\frac{7}{3} S_1
        \Biggr) \zeta_3
\Biggr\} ~,
\\
c_{g_1,g}^{(1,0)} &=&
\textcolor{blue}{T_F N_F} (N-1)\Biggl\{
        -\frac{4 (N-1)}{N^2 (1+N)}
        -\frac{4}{N (1+N)} S_1
\Biggr\} ~,
\\
c_{g_1,g}^{(1,1)} &=&
\textcolor{blue}{T_F N_F} (N-1)\Biggl\{
        \frac{2 \big(
                1-N+2 N^2\big)}{N^3 (1+N)}
        +\frac{2 (N-1)}{N^2 (1+N)} S_1
        +\frac{1}{N (1+N)} S_1^2
\nonumber\\ &&   
        -\frac{1}{N (1+N)} S_2
        -\frac{1}{N (1+N)} \zeta_2
\Biggr\} ~,
\\
c_{g_1,g}^{(1,2)} &=& 
\textcolor{blue}{T_F N_F} (N-1) \Biggl\{
        \frac{ \big(
                1-N+2 N^2-4 N^3\big)}{N^4 (1+N)}
        +\Biggl(
                -\frac{ \big(
                        1-N+2 N^2\big)}{N^3 (1+N)}
\nonumber\\ &&             
    +\frac{1}{2 N (1+N)} S_2
        \Biggr) S_1
        -\frac{(N-1)}{2 N^2 (1+N)} S_1^2
        -\frac{1}{6 N (1+N)} S_1^3
        +\frac{(N-1)}{2 N^2 (1+N)} S_2
\nonumber\\ && 
        -\frac{1}{3 N (1+N)} S_3
        +\Biggl(
                \frac{(N-1)}{2 N^2 (1+N)}
                +\frac{1}{2 N (1+N)} S_1
        \Biggr) \zeta_2
        +\frac{7}{3 N (1+N)} \zeta_3
\Biggr\} ~,
\end{eqnarray}


\noindent
with
\begin{eqnarray}
    P_1 &=& -3 N^4+6 N^3-4 N^2-N-2,\\
    P_2 &=& 14 N^4+27 N^3+2 N^2-5 N-2,\\
    P_3 &=& 18 N^4+5 N^3-2 N^2+5 N+2,\\
    P_4 &=& -14 N^5-6 N^4-N^3+3 N+2,\\
    P_5 &=& -36 N^5-10 N^4+5 N^3-2 N^2+5 N+2,\\
    P_6 &=& -7 N^5+12 N^4-6 N^3+4 N^2+N+2,\\ 
    P_7 &=& 14 N^4+23 N^3-8 N^2+5 N-2,\\
    P_8 &=& 18 N^4+9 N^3+8 N^2-5 N+2,\\
    P_9 &=& -36 N^5-18 N^4-15 N^3+8 N^2-5 N+2.
\end{eqnarray}

\noindent
The one--loop Wilson coefficients at $\mu^2 = Q^2$ are given by 
the expansion coefficients $c_{F_k,q(g)}^{(1,0)}$, $k=2,L$. 
There are finite renormalizations for the Wilson coefficients of the structure functions 
$xF_3$ and $g_1$.
The respective Wilson coefficients are given by
\begin{eqnarray}
    \label{eq:F3NS1}
    C_{F_3,q}^{(1),\rm NS} &=& 
    \textcolor{blue}{C_F} \Biggl[
            -\frac{(2+3 N) \big(
                    -1+3 N^2\big)}{N^2 (1+N)}
            +\frac{\big(
                    -2+3 N+3 N^2\big) S_1}{N (1+N)}
            +2 S_1^2
            -2 S_2
    \Biggr]
    ~, \\
    \label{eq:g1NS1}
    C_{g_1,q}^{(1),\rm NS,L} &=&  \textcolor{blue}{C_F} \Biggl[
         \frac{2 - 5 N - 6 N^2 - 9 N^3}{N^2 (1+N)}
        +\frac{\big(-2+3 N+3 N^2\big) S_1}{N (1+N)}
        +2 S_1^2
        -2 S_2
    \Biggr]
    ~, \\
    C_{g_1,q}^{(1),\rm NS} &=& C_{g_1,q}^{(1),\rm NS,L} - z_{qq}^{(1)} = C_{g_1,q}^{(1),\rm NS,L} + \frac{8 C_F}{N(N+1)}
    = C_{F_3,q}^{(1),\rm NS}
    ~,
\end{eqnarray}

\noindent
where the superscript $L$ denotes the Wilson coefficient obtained in the 
Larin scheme.

\subsection{Two-Loop Order}
\label{sec:42}

\vspace*{1mm}
\noindent
The corresponding expansion coefficients at two--loop order read, 
cf.~(\ref{eq:unrenWC1}--\ref{eq:unrenWC2}),
 

\noindent
with the polynomials
\begin{eqnarray}
P_{10} &=& -1239 N^4-1554 N^3-259 N^2+584 N+396~,\\ 
P_{11} &=& -1079 N^4-2410 N^3-1379 N^2-192 N-108~,\\ 
P_{12} &=& -75 N^4-248 N^3-225 N^2-108 N-50~,\\ 
P_{13} &=& N^4-12 N^3-11 N^2+14 N+24~,\\ 
P_{14} &=& N^4-7 N^3-7 N^2+27 N+18~,\\ 
P_{15} &=& N^4+2 N^3-N^2-2 N-6~,\\ 
P_{16} &=& N^4+2 N^3+11 N^2+10 N+4~,\\ 
P_{17} &=& N^4+2 N^3+14 N^2+13 N+4~,\\ 
P_{18} &=& N^4+3 N^3-6 N^2-2 N-6~,\\ 
P_{19} &=& 2 N^4+6 N^3-9 N^2-15 N-8~,\\ 
P_{20} &=& 3 N^4+4 N^3-N^2-6 N-8~,\\ 
P_{21} &=& 4 N^4+N^3-7 N^2+2 N-12~,\\ 
P_{22} &=& 4 N^4+6 N^3-9 N^2-9 N-4~,\\ 
P_{23} &=& 5 N^4+31 N^3+38 N^2+24 N+11~,\\ 
P_{24} &=& 6 N^4+7 N^3+8 N^2+5 N-6~,\\ 
P_{25} &=& 6 N^4+9 N^3+4 N^2-5 N-10~,\\ 
P_{26} &=& 12 N^4+18 N^3-23 N^2-23 N-8~,\\ 
P_{27} &=& 15 N^4+23 N^3+13 N^2-9 N-22~,\\ 
P_{28} &=& 15 N^4+28 N^3-21 N^2-38 N-24~,\\ 
P_{29} &=& 26 N^4+45 N^3+11 N^2-8 N-5~,\\ 
P_{30} &=& 111 N^4+138 N^3+11 N^2-64 N-36~,\\ 
P_{31} &=& 281 N^4+430 N^3-1927 N^2-1284 N+396~,\\ 
P_{32} &=& 373 N^4+998 N^3+673 N^2+192 N+108~,\\ 
P_{33} &=& 541 N^4+782 N^3+199 N^2-42 N+36~,\\ 
P_{34} &=& -4541 N^5-17529 N^4-20631 N^3-8099 N^2-1968 N-1188~,\\ 
P_{35} &=& N^5-5 N^3-6 N^2-8 N-4~,\\ P_{36} &=& N^5-10 N^4-37 N^3-6 N^2+56 N+48~,\\ 
P_{37} &=& N^5-2 N^4-10 N^3-5 N^2-6 N-2~,\\ 
P_{38} &=& N^5-N^4-5 N^3+3 N^2+14 N+12~,\\ 
P_{39} &=& 3 N^5+10 N^4+16 N^3+11 N^2-4 N-4~,\\ 
P_{40} &=& 8 N^5-7 N^4-11 N^3+68 N^2+78 N+32~,\\ 
P_{41} &=& 13 N^5+40 N^4+45 N^3+18 N^2-12 N-8~,\\ 
P_{42} &=& 13 N^5+80 N^4+94 N^3-2 N^2-51 N-54~,\\ 
P_{43} &=& 14 N^5+35 N^4-20 N^3-65 N^2-42 N-72~,\\ 
P_{44} &=& 179 N^5+1219 N^4+1691 N^3+111 N^2-396 N-108~,\\ 
P_{45} &=& 229 N^5+821 N^4+1081 N^3+795 N^2+198 N+108~,\\ 
P_{46} &=& 461 N^5+999 N^4-2037 N^3-4507 N^2-1320 N+396~,\\ 
P_{47} &=& 4541 N^5+17151 N^4+20199 N^3+8477 N^2+2292 N+1188~,\\ 
P_{48} &=& -2311 N^6-5529 N^5+6417 N^4+19097 N^3+15042 N^2+4284 N+4536~,\\ 
P_{49} &=& -237 N^6-699 N^5-683 N^4-321 N^3-316 N^2-272 N-72~,\\ 
P_{50} &=& -37 N^6-171 N^5-231 N^4+7 N^3+388 N^2+352 N+96~,\\ 
P_{51} &=& N^6-12 N^4-13 N^3-12 N^2-8 N-4~,\\ 
P_{52} &=& N^6-2 N^4+7 N^3+10 N^2-4~,\\ 
P_{53} &=& N^6-6 N^5-22 N^4-2 N^3+5 N^2+16 N+12~,\\ 
P_{54} &=& N^6-5 N^5+N^4+61 N^3-26 N^2-128 N-96~,\\ 
P_{55} &=& N^6-3 N^5-11 N^4-5 N^3+22 N^2+36 N+8~,\\ 
P_{56} &=& N^6+N^5-18 N^4-27 N^3-17 N^2+16 N+12~,\\ 
P_{57} &=& N^6+3 N^5-2 N^4-7 N^3+3 N^2+6 N+24~,\\ 
P_{58} &=& N^6+3 N^5+13 N^4+21 N^3-2 N^2-12 N+72~,\\ 
P_{59} &=& N^6+7 N^5-7 N^4-39 N^3+14 N^2+40 N+48~,\\ 
P_{60} &=& 2 N^6-2 N^5-5 N^4+20 N^3-39 N^2-12 N-36~,\\ 
P_{61} &=& 2 N^6-2 N^5-3 N^4+26 N^3-45 N^2-34 N-48~,\\ 
P_{62} &=& 2 N^6-2 N^5+N^4+38 N^3-57 N^2-78 N-72~,\\ 
P_{63} &=& 3 N^6-6 N^5-28 N^4-11 N^3-10 N^2+4~,\\ 
P_{64} &=& 6 N^6+35 N^5+57 N^4+73 N^3+77 N^2+56 N+20~,\\ 
P_{65} &=& 7 N^6-14 N^5-64 N^4-23 N^3-18 N^2+4 N+12~,\\ 
P_{66} &=& 7 N^6-10 N^5-58 N^4-27 N^3-24 N^2+4 N+12~,\\ 
P_{67} &=& 12 N^6-9 N^5-133 N^4+295 N^3+99 N^2-524 N-492~,\\ 
P_{68} &=& 13 N^6+25 N^5+16 N^4-27 N^3-55 N^2-40 N-12~,\\ 
P_{69} &=& 15 N^6+53 N^5-7 N^4-137 N^3+40 N^2+204 N+144~,\\ 
P_{70} &=& 18 N^6-18 N^5-31 N^4+244 N^3+479 N^2+168 N-44~,\\ 
P_{71} &=& 23 N^6+57 N^5-171 N^4-357 N^3-120 N^2+80 N+24~,\\ 
P_{72} &=& 24 N^6-25 N^5-21 N^4+433 N^3-403 N^2-1012 N-516~,\\ 
P_{73} &=& 27 N^6+107 N^5+139 N^4+169 N^3+238 N^2+136 N+48~,\\ 
P_{74} &=& 31 N^6+130 N^5+187 N^4+192 N^3+176 N^2+128 N+48~,\\ 
P_{75} &=& 67 N^6+375 N^5+963 N^4+1625 N^3+1656 N^2+918 N+324~,\\ 
P_{76} &=& 105 N^6+231 N^5+159 N^4-59 N^3-172 N^2-112 N-24~,\\ 
P_{77} &=& 834 N^6+1773 N^5-2738 N^4-7729 N^3-2150 N^2+2550 N+2916~,\\ 
P_{78} &=& 2667 N^6+5469 N^5+3269 N^4+463 N^3+860 N^2+1296 N+504~,\\ 
P_{79} &=& 7081 N^6+28839 N^5+35439 N^4+13693 N^3-2580 N^2-3888 N-1512~,\\ 
P_{80} &=& 14311 N^6+85344 N^5+179550 N^4+159592 N^3+57195 N^2+9036 N+3564~,\\ 
P_{81} &=& -37 N^7-157 N^6-335 N^5-639 N^4-924 N^3-708 N^2-320 N-96~,\\ 
P_{82} &=& N^7+N^6-N^5+3 N^4+8 N^3-60 N^2-208 N-192~,\\ 
P_{83} &=& 2 N^7-25 N^5-84 N^4-135 N^3-114 N^2-68 N-24~,\\ 
P_{84} &=& 13 N^7+56 N^6+52 N^5-60 N^4-121 N^3-56 N^2+12 N+8~,\\ 
P_{85} &=& 55 N^7+219 N^6-41 N^5-1119 N^4-1418 N^3-320 N^2+184 N+48~,\\ 
P_{86} &=& 108 N^7+362 N^6-6 N^5-908 N^4-1447 N^3-2073 N^2-1604 N-924~,\\ 
P_{87} &=& 198 N^7+765 N^6+114 N^5-2865 N^4-3904 N^3-324 N^2+1984 N+1632~,\\ 
P_{88} &=& -56163 N^8-166680 N^7-175082 N^6-69468 N^5-1323 N^4+15676 N^3+27792 N^2
\nonumber\\ &&
+22608 N+6480~,\\ 
P_{89} &=& -30651 N^8-94248 N^7+83888 N^6+331530 N^5+210723 N^4+5534 N^3+17376 N^2
\nonumber\\ &&
+67032 N+33264~,\\ 
P_{90} &=& -5885 N^8-20168 N^7+51756 N^6+169186 N^5-154415 N^4-399114 N^3
\nonumber\\ &&
-107208 N^2+34344 N-14256~,\\ 
P_{91} &=& -397 N^8-2944 N^7-8680 N^6-12850 N^5-10191 N^4-5186 N^3-2760 N^2
\nonumber\\ &&
-1392 N-288~,\\ 
P_{92} &=& 55 N^8+111 N^7-227 N^6-263 N^5+1108 N^4+2112 N^3
\nonumber \\ &&
+1632 N^2+688 N+160~, 
\\
P_{93} &=& -86393 N^9-545545 N^8-734144 N^7+1327702 N^6+4117751 N^5+3541963 N^4
\nonumber\\ &&
+1033098 N^3-171576 N^2-254232 N-99792~,\\ 
P_{94} &=& -17 N^9-85 N^8-63 N^7-3 N^6-552 N^5-1424 N^4-1584 N^3-1040 N^2
\nonumber\\ &&
-480 N-128~,\\ 
P_{95} &=& N^9+6 N^8+7 N^7+25 N^6+124 N^5+285 N^4+384 N^3+304 N^2+160 N+48~,\\ 
P_{96} &=& 19 N^9+41 N^8-71 N^7-71 N^6+452 N^5+894 N^4+800 N^3+432 N^2
\nonumber\\ &&
+160 N+32~,\\ 
P_{97} &=& 2 N^{10}+8 N^9-2 N^8+8 N^7+271 N^6+571 N^5+491 N^4+117 N^3+98 N^2
\nonumber\\ &&
+156 N+72~,\\ 
P_{98} &=& 3 N^{10}+15 N^9-11 N^8-93 N^7-24 N^6+52 N^5+304 N^4+998 N^3-356 N^2
\nonumber\\ &&
-1272 N-720~,\\ 
P_{99} &=& 7 N^{10}+19 N^9-60 N^8-186 N^7+139 N^6+571 N^5-542 N^4-1940 N^3
\nonumber\\ &&
+1224 N^2
+3456 N+1728~,\\ 
P_{100} &=& 9 N^{10}+44 N^9+83 N^8+107 N^7+291 N^6+879 N^5+1579 N^4+1672 N^3
\nonumber\\ &&
+1128 N^2
+496 N+112~,\\ 
P_{101} &=& 9 N^{10}+66 N^9+147 N^8-41 N^7-787 N^6-1627 N^5-1655 N^4-904 N^3
\nonumber\\ &&
-296 N^2
-112 N-48~,\\ 
P_{102} &=& 18 N^{10}+114 N^9+109 N^8-331 N^7-718 N^6-904 N^5-623 N^4+545 N^3
\nonumber\\ &&
+1386 N^2
+988 N+264~,\\ 
P_{103} &=& 34 N^{10}+136 N^9+10 N^8-372 N^7-447 N^6-99 N^5-215 N^4-921 N^3
\nonumber\\ &&
-1210 N^2
-716 N-168~,\\ 
P_{104} &=& 36 N^{10}+248 N^9+451 N^8-523 N^7-3189 N^6-5209 N^5-4382 N^4
\nonumber\\ &&
-2104 N^3
-704 N^2-272 N-96~,\\ 
P_{105} &=& 133 N^{10}+573 N^9-1052 N^8-5646 N^7+2901 N^6+21505 N^5+18130 N^4
\nonumber\\ &&
+2320 N^3
-160 N^2+1200 N+288~,\\ 
P_{106} &=& 763 N^{10}+3027 N^9-128 N^8-9930 N^7-11309 N^6-2945 N^5-350 N^4
\nonumber\\ &&
-5224 N^3
-7296 N^2-3824 N-672~,\\ 
P_{107} &=& 2 N^{11}+17 N^{10}+34 N^9-34 N^8-190 N^7-255 N^6-102 N^5+648 N^4
\nonumber\\ &&
+1592 N^3
+1136 N^2+32 N-192~,\\ 
P_{108} &=& 1203 N^{11}+9097 N^{10}+19614 N^9-6786 N^8-83057 N^7-122227 N^6
\nonumber\\ &&
-74000 N^5
-9540 N^4+26544 N^3+34608 N^2+19072 N+3648~,\\ 
P_{109} &=& 4 N^{12}+34 N^{11}+76 N^{10}-4 N^9-257 N^8-653 N^7-1597 N^6
\nonumber\\ &&
-3349 
N^5-4822 N^4-4520 N^3-2784 N^2-1104 N-224~,\\ 
P_{110} &=& 68 N^{12}+360 N^{11}+367 N^{10}-1033 N^9-2581 N^8-2317 N^7
\nonumber\\ &&
-3258 N^6-7354 N^5
-9852 N^4-7648 N^3-3776 N^2-1184 N-192~,\\ 
P_{111} &=& 648699 N^{12}+3241902 N^{11}-1257119 N^{10}-24261820 N^9
-20205047 N^8
\nonumber\\ &&
+35553886 N^7
+60343151 N^6+28400368 N^5+5406396 N^4+1441440 N^3
\nonumber\\ &&
-4831920 N^2-7008768 N
-2566080~,\\ 
P_{112} &=& -4521 N^{14}-26467 N^{13}-9301 N^{12}+178761 N^{11}+272257 N^{10}
-131725 N^9
\nonumber\\ &&
-554287 N^8
-465045 N^7-251812 N^6-157892 N^5+36656 N^4+261568 N^3
\nonumber\\ &&
+265824 N^2+114048 N+17280~,\\ 
P_{113} &=& -92 N^{14}-560 N^{13}-355 N^{12}+3542 N^{11}+7046 N^{10}+469 N^9
-11572 N^8
\nonumber\\ &&
-16160 N^7
-14914 N^6-9595 N^5+5423 N^4+21784 N^3+22728 N^2+11088 N
\nonumber\\ &&
+2160~,\\ 
P_{114} &=& 28 N^{15}+246 N^{14}+715 N^{13}+259 N^{12}-3044 N^{11}-7282 N^{10}
-6158 N^9+2882 N^8
\nonumber\\ &&
+18427 N^7+44383 N^6+79244 N^5+99004 N^4+83536 N^3+47056 N^2
+16704 N
\nonumber\\ &&
+2880~,\\ 
P_{115} &=& 14 N^{17}+174 N^{16}+731 N^{15}+818 N^{14}-2448 N^{13}-7521 N^{12}
-3298 N^{11}+11946 N^{10}
\nonumber\\ &&
+15938 N^9-10959 N^8-71053 N^7-162458 N^6-255636 N^5
-282952 N^4
\nonumber\\ &&
-216304 N^3-111584 N^2-36288 N-5760.
\end{eqnarray}

\noindent
The expansion coefficients for the Compton amplitudes of $xF_3(x,Q^2)$ and $g_1(x,Q^2)$ are
\begin{eqnarray}
c_{F_3,q}^{(2,0),\rm L} &=& 
\textcolor{blue}{C_F} \Biggl\{
        \textcolor{blue}{T_F N_F} \Biggl[
                \frac{P_{136}}{54 N^3 (1+N)^3}
                -\frac{4 \big(
                        -6+19 N+19 N^2\big)}{9 N (1+N)} S_1^2
                +\Biggl(
                        -\frac{2 P_{122}}{27 N^2 (1+N)^2}
\nonumber\\ &&                    
     +\frac{16}{3} S_2
                        +\frac{8}{3} \zeta_2
                \Biggr) S_1
                -\frac{16}{9} S_1^3
                +\frac{4 \big(
                        -6+47 N+47 N^2\big) S_2}{9 N (1+N)}
                -\frac{104}{9} S_3
                +\frac{16}{3} S_{2,1}
\nonumber\\ &&                
 -\frac{2 \big(
                        2+3 N+3 N^2\big) \zeta_2}{3 N (1+N)}
        \Biggr]
        +\textcolor{blue}{C_A} \Biggl[
                \frac{P_{144}}{216 (N-1) N^4 (1+N)^4 (2+N)}
                +\Biggl(
                        \frac{P_{137}}{54 N^3 (1+N)^3}
\nonumber\\ &&                
         -\frac{4 \big(
                                3+11 N+11 N^2\big) S_2}{3 N (1+N)}
                        +24 S_3
                        -16 S_{2,1}
                        -16 S_{-2,1}
                        -\frac{22}{3} \zeta_2
                        -48 \zeta_3
                \Biggr) S_1
\nonumber\\ &&                
 +\Biggl(
                        \frac{-66+233 N+233 N^2}{9 N (1+N)}
                        +4 S_2
                \Biggr) S_1^2
                +\frac{\big(
                        66-481 N-1166 N^2-583 N^3\big)}{9 N (1+N)^2} S_2
\nonumber\\ && 
                +\frac{44}{9} S_1^3
                -4 S_2^2
                +\frac{2 \big(
                        -36+143 N+143 N^2\big)}{9 N (1+N)} S_3
                +\Biggl(
                        \frac{4 P_{123}}{(N-1) N (1+N)^2 (2+N)}
\nonumber\\ &&                
         -\frac{8 S_1}{N (1+N)}
                        +16 S_1^2
                        -8 S_2
                \Biggr) S_{-2}
                -8 S_4
                -12 S_{-2}^2
                +\Biggl(
                        \frac{8}{N (1+N)}
                        -8 S_1
                \Biggr) S_{-3}
\nonumber\\ &&                
 -20 S_{-4}
                -\frac{4 \big(
                        -6+11 N+11 N^2\big)}{3 N (1+N)} S_{2,1}
                -24 S_{3,1}
                +8 S_{-2,2}
                +16 S_{-3,1}
                +24 S_{2,1,1}
\nonumber\\ &&                
 +
                \frac{11 \big(
                        2+3 N+3 N^2\big)}{6 N (1+N)} \zeta_2
                +\frac{6 (1+3 N) (2+3 N)}{N (1+N)} \zeta_3
        \Biggr]
\Biggr\}
+\textcolor{blue}{C_F^2} \Biggl[
        \frac{2 S_2 P_{119}}{N^2 (1+N)^2}
\nonumber\\ &&    
     +\frac{P_{145}}{8 (N-1) N^4 (1+N)^4 (2+N)}
        +\Biggl(
                \frac{P_{130}}{2 N^3 (1+N)^3}
                -\frac{2 \big(
                        -14+9 N+9 N^2\big)}{N (1+N)} S_2
\nonumber\\ &&                
 -\frac{56}{3} S_3
                +16 S_{2,1}
                +32 S_{-2,1}
                +\frac{4 \big(
                        2+3 N+3 N^2\big)}{N (1+N)} \zeta_2
                +48 \zeta_3
        \Biggr) S_1
        +\Biggl(
                \frac{2 P_{118}}{N^2 (1+N)^2}
\nonumber\\ &&                
 -28 S_2
                -8 \zeta_2
        \Biggr) S_1^2
        +\frac{2 \big(
                -14+15 N+15 N^2\big)}{3 N (1+N)} S_1^3
        -\frac{2 \big(
                -2+33 N+33 N^2\big)}{3 N (1+N)} S_3
\nonumber\\ && 
        +\frac{14}{3} S_1^4
        +6 S_2^2
        +12 S_4
        +\Biggl(
                -\frac{8 P_{123}}{(N-1) N (1+N)^2 (2+N)}
                +\frac{16 S_1}{N (1+N)}
                -32 S_1^2
\nonumber\\ &&                
 +16 S_2
        \Biggr) S_{-2}
        +\Biggl(
                -\frac{16}{N (1+N)}
                +16 S_1
        \Biggr) S_{-3}
        +\frac{4 \big(
                -2+3 N+3 N^2\big)}{N (1+N)} S_{2,1}
\nonumber\\ && 
        +40 S_{-4}
        +24 S_{-2}^2
        +40 S_{3,1}
        -16 S_{-2,2}
        -32 S_{-3,1}
        -24 S_{2,1,1}
        -\frac{\big(
                2+3 N+3 N^2\big)^2}{2 N^2 (1+N)^2} \zeta_2
\nonumber\\ &&  
      -72 \zeta_3
\Biggr] ~,
\\
c_{F_3,q}^{(2,1),L} &=& 
\textcolor{blue}{C_F} \Biggl\{
        \textcolor{blue}{C_A} \Biggl[
                \frac{2 S_{2,1} P_{121}}{9 N^2 (1+N) (2+N)}
                +
                \frac{P_{129}}{72 N^3 (1+N)^3} \zeta_2
                -\frac{2 P_{135}}{9 (N-1) N^2 (1+N)^2 (2+N)} \zeta_3
\nonumber\\ && 
                +\frac{P_{148}}{2592 (N-1)^2 N^5 (1+N)^5 (2+N)^2}
                +\Biggl(
                        \frac{P_{143}}{648 (N-1) N^4 (1+N)^4 (2+N)}
\nonumber\\ &&                         
+\frac{S_2 P_{126}}{9 N^2 (1+N)^2 (2+N)}
                        +8 S_2^2
                        -\frac{4 \big(
                                -9+22 N+22 N^2\big) S_3}{9 N (1+N)}
                        +32 S_4
                        -\frac{8 S_{2,1}}{N (1+N)}
\nonumber\\ &&                
         -4 S_{3,1}
                        -8 S_{-2,2}
                        -16 S_{-3,1}
                        -20 S_{2,1,1}
                        +\frac{\big(
                                -66+233 N+233 N^2\big) \zeta_2}{18 N (1+N)}
                        +\frac{72}{5} \zeta_2^2
\nonumber\\ &&               
         +\frac{514}{9} \zeta_3
                \Biggr) S_1
                +\Biggl(
                        \frac{P_{127}}{108 N^3 (1+N)^3}
                        +\frac{\big(
                                9+22 N+22 N^2\big) S_2}{3 N (1+N)}
                        -14 S_3
                        +12 S_{2,1}
\nonumber\\ &&                
         +8 S_{-2,1}
                        +\frac{11}{3} \zeta_2
                        +24 \zeta_3
                \Biggr) S_1^2
                +\Biggl(
                        \frac{66-233 N-233 N^2}{27 N (1+N)}
                        -\frac{8}{3} S_2
                \Biggr) S_1^3
                -\frac{11}{9} S_1^4
\nonumber\\ &&                
 +\Biggl(
                        \frac{P_{140}}{108 N^3 (1+N)^3 (2+N)}
                        -\frac{22}{3} S_3
                        -8 S_{2,1}
                        -8 S_{-2,1}
                        -\frac{22}{3} \zeta_2
                        -48 \zeta_3
                \Biggr) S_2
\nonumber\\ &&               
 +\frac{\big(
                        -2-17 N-17 N^2\big) S_2^2}{N (1+N)}
                +\Biggl(
                        \frac{P_{128}}{27 (N-1) N^2 (1+N)^2 (2+N)}
                        +2 \zeta_2
                \Biggr) S_3
\nonumber\\ &&                
 +\frac{\big(
                        -33+59 N+59 N^2\big) S_4}{3 N (1+N)}
                -22 S_5
                +\Biggl(
                        -
                        \frac{4 P_{142}}{(N-1)^2 N (1+N)^3 (2+N)^2}
\nonumber\\ &&                
         +\Biggl(
                                -\frac{4 P_{123}}{(N-1) N (1+N)^2 (2+N)}
                                +16 S_2
                                +4 \zeta_2
                        \Biggr) S_1
                        +\frac{4 S_1^2}{N (1+N)}
                        -8 S_1^3
                        -\frac{4 S_2}{N (1+N)}
\nonumber\\ &&                
         -32 S_3
                        -8 S_{-2,1}
                        -\frac{2 \zeta_2}{N (1+N)}
                        -72 \zeta_3
                \Biggr) S_{-2}
                +\Biggl(
                        -\frac{4}{N (1+N)}
                        +20 S_1
                \Biggr) S_{-2}^2
\nonumber\\ &&                
 +\Biggl(
                        \frac{4 P_{123}}{(N-1) N (1+N)^2 (2+N)}
                        -\frac{8 S_1}{N (1+N)}
                        +12 S_1^2
                        +4 S_2
                        -16 S_{-2}
                        +2 \zeta_2
                \Biggr) S_{-3}
\nonumber\\ &&                
 +\Biggl(
                        \frac{4}{N (1+N)}
                        +12 S_1
                \Biggr) S_{-4}
                -38 S_{-5}
                +20 S_{2,3}
                -\frac{2 \big(
                        -4+5 N+5 N^2\big)}{N (1+N)} S_{3,1}
\nonumber\\ && 
                -8 S_{2,-3}
                -44 S_{4,1}
                +44 S_{-2,3}
                +8 S_{-4,1}
                +\frac{4 \big(
                        3+11 N+11 N^2\big)}{3 N (1+N)} S_{2,1,1}
\nonumber\\ &&             
    +4 S_{2,2,1}
                +40 S_{3,1,1}
                -8 S_{-2,1,-2}
                +4 S_{2,1,1,1}
                -4 S_{-2,1} \zeta_2
                -\frac{9 (1+3 N) (2+3 N) \zeta_2^2}{5 N (1+N)}
        \Biggr]
\nonumber\\ && 
        +\textcolor{blue}{T_F N_F} \Biggl[
                \frac{P_{120}}{18 N^2 (1+N)^2} \zeta_2
                +\frac{P_{141}}{648 N^4 (1+N)^4}
                +\Biggl(
                        \frac{P_{138}}{162 N^3 (1+N)^3}
\nonumber\\ &&                
         -\frac{4 \big(
                                -6+19 N+19 N^2\big) S_2}{9 N (1+N)}
                        +\frac{32}{9} S_3
                        -\frac{2 \big(
                                -6+19 N+19 N^2\big) \zeta_2}{9 N (1+N)}
                        -
                        \frac{128}{9} \zeta_3
                \Biggr) S_1
\nonumber\\ &&                
 +\Biggl(
                        \frac{P_{122}}{27 N^2 (1+N)^2}
                        -\frac{8}{3} S_2
                        -\frac{4}{3} \zeta_2
                \Biggr) S_1^2
                +\frac{4 \big(
                        -6+19 N+19 N^2\big) S_1^3}{27 N (1+N)}
                +\frac{4}{9} S_1^4
\nonumber\\ &&                
 +\Biggl(
                        \frac{P_{116}}{27 N^2 (1+N)^2}
                        +\frac{8 \zeta_2}{3}
                \Biggr) S_2
                +4 S_2^2
                +\frac{16 \big(
                        -3+41 N+41 N^2\big) S_3}{27 N (1+N)}
                -\frac{28}{3} S_4
\nonumber\\ &&                 
-\frac{112}{9} S_{2,1}
                +8 S_{3,1}
                -\frac{16}{3} S_{2,1,1}
                +\frac{32 \big(
                        2+3 N+3 N^2\big) \zeta_3}{9 N (1+N)}
        \Biggr]
\Biggr\}
\nonumber\\ && 
+\textcolor{blue}{C_F^2} \Biggl[
        -\frac{2 S_{2,1} P_{124}}{N^2 (1+N)^2 (2+N)} 
        +\frac{P_{132}}{8 N^3 (1+N)^3} \zeta_2
        +\frac{4 P_{134}}{3 (N-1) N^2 (1+N)^2 (2+N)} \zeta_3
\nonumber\\ &&    
     +\frac{P_{147}}{32 (N-1)^2 N^5 (1+N)^5 (2+N)^2}
        +\Biggl(
                \frac{P_{146}}{8 (N-1) N^4 (1+N)^4 (2+N)}
\nonumber\\ &&                
 +\Biggl(
                        \frac{P_{125}}{2 N^2 (1+N)^2 (2+N)}
                        -8 \zeta_2
                \Biggr) S_2
                -23 S_2^2
                -\frac{4 \big(
                        -2+5 N+5 N^2\big) S_3}{N (1+N)}
                -38 S_4
\nonumber\\ &&                
 +\frac{4 \big(2+3 N+3 N^2\big) S_{2,1}}{N (1+N)}
                -8 S_{3,1}
                +16 S_{-2,2}
                +32 S_{-3,1}
                +32 S_{2,1,1}
\nonumber\\ &&                
                +\frac{\big(16+12 N-17 N^2-29 N^3\big)}{2 N^2 (1+N)} \zeta_2
 -\frac{72}{5} \zeta_2^2
                -\frac{4 \big(
                        32+21 N+21 N^2\big)}{3 N (1+N)} \zeta_3
        \Biggr) S_1
\nonumber\\ &&    
     +\Biggl(
                \frac{P_{131}}{4 N^3 (1+N)^3}
                +\frac{\big(
                        -34+27 N+27 N^2\big) S_2}{2 N (1+N)}
                +12 S_3
                -16 S_{2,1}
                -16 S_{-2,1}
\nonumber\\ &&                
 +\frac{3 (N-1) (2+N) }{N (1+N)} \zeta_2
                -\frac{16}{3} \zeta_3
        \Biggr) S_1^2
        +\Biggl(
                \frac{P_{117}}{6 N^2 (1+N)^2}
                +\frac{38}{3} S_2
                +4 \zeta_2
        \Biggr) S_1^3
\nonumber\\ &&   
      +\frac{\big(
                10-13 N-13 N^2\big)}{4 N (1+N)} S_1^4
        -S_1^5
        +\Biggl(
                \frac{P_{139}}{4 N^3 (1+N)^3 (2+N)}
                +\frac{28}{3} S_3
                +16 S_{2,1}
\nonumber\\ &&                
 +16 S_{-2,1}
                +\frac{2 \big(
                        2+3 N+3 N^2\big)}{N (1+N)} \zeta_2
                +48 \zeta_3
        \Biggr) S_2
        +\frac{3 \big(
                10+27 N+27 N^2\big)}{4 N (1+N)} S_2^2
\nonumber\\ &&     
    +\Biggl(
                \frac{2 P_{133}}{3 (N-1) N^2 (1+N)^2 (2+N)}
                -4 \zeta_2
        \Biggr) S_3
        +\frac{3 (2-N) (3+N) S_4}{2 N (1+N)}
        +36 S_5
\nonumber\\ &&     
    +\Biggl(
                \frac{8 P_{142}}{(N-1)^2 N (1+N)^3 (2+N)^2}
                +\Biggl(
                        \frac{8 P_{123}}{(N-1) N (1+N)^2 (2+N)}
                        -32 S_2
\nonumber\\ &&                  
       -8 \zeta_2
                \Biggr) S_1
                -\frac{8}{N (1+N)} S_1^2
                +16 S_1^3
                +\frac{8}{N (1+N)} S_2
                +64 S_3
                +16 S_{-2,1}
\nonumber\\ && 
                +\frac{4}{N (1+N)} \zeta_2
                +144 \zeta_3
        \Biggr) S_{-2}
        +\Biggl(
                -40 S_1
                + \frac{8}{N (1+N)}
        \Biggr) S_{-2}^2
        +\Biggl(
	                \frac{16}{N (1+N)} S_1
\nonumber\\ && 
                -\frac{8 P_{123}}{(N-1) N (1+N)^2 (2+N)}
                -24 S_1^2
                -8 S_2
                +32 S_{-2}
                -4 \zeta_2
        \Biggr) S_{-3}
        -\Biggl(
                24 S_1
\nonumber\\ &&                 
+\frac{8}{N (1+N)}
        \Biggr) S_{-4}
        +76 S_{-5}
        -32 S_{2,3}
        +16 S_{2,-3}
        -\frac{8 }{N (1+N)} S_{3,1}
        +76 S_{4,1}
\nonumber\\ &&        
 -88 S_{-2,3}
        -16 S_{-4,1}
        -\frac{2 \big(
                2+15 N+15 N^2\big)}{N (1+N)} S_{2,1,1}
        -48 S_{3,1,1}
        +16 S_{-2,1,-2}
\nonumber\\ &&       
 -20 S_{2,1,1,1}
        +8 S_{-2,1} \zeta_2
        +\frac{108}{5} \zeta_2^2
\Biggr] ~,
\end{eqnarray}

\noindent
with
\begin{eqnarray}
P_{116} &=& -1007 N^4-2122 N^3-1091 N^2-120 N-108~,
\\
P_{117} &=& -83 N^4-208 N^3-149 N^2-80 N-50~,
\\
P_{118} &=& 9 N^4+27 N^3+24 N^2+18 N+11~,
\\
P_{119} &=& 22 N^4+41 N^3+13 N^2-6 N-5~,
\\
P_{120} &=& 87 N^4+138 N^3+59 N^2-40 N-36~,
\\
P_{121} &=& 229 N^4+507 N^3+188 N^2+72~,
\\
P_{122} &=& 373 N^4+854 N^3+457 N^2+120 N+108~,
\\
P_{123} &=& 2 N^5+6 N^4+5 N^3+2 N^2+7 N+2~,
\\
P_{124} &=& 27 N^5+66 N^4+33 N^3-2 N^2+24 N+16~,
\\
P_{125} &=& 75 N^5+246 N^4+285 N^3+202 N^2+198 N+108~,
\\
P_{126} &=& 179 N^5+896 N^4+1261 N^3+394 N^2-204 N-72~,
\\
P_{127} &=& -4541 N^6-14649 N^5-16023 N^4-7523 N^3-2256 N^2-324 N+432~,
\\
P_{128} &=& -2311 N^6-6393 N^5-1423 N^4+6333 N^3+2654 N^2-372 N+216~,
\\
P_{129} &=& -975 N^6-2529 N^5-2341 N^4-467 N^3+428 N^2+108 N-144~,
\\
P_{130} &=& -133 N^6-355 N^5-379 N^4-281 N^3-228 N^2-160 N-40~,
\\
P_{131} &=& -89 N^6-351 N^5-331 N^4+83 N^3+308 N^2+224 N+64~,
\\
P_{132} &=& 57 N^6+135 N^5+103 N^4-43 N^3-68 N^2-32 N+8~,
\\
P_{133} &=& 104 N^6+246 N^5+14 N^4-208 N^3-3 N^2+N-10~,
\\
P_{134} &=& 180 N^6+351 N^5-120 N^4-447 N^3-232 N^2-76 N-88~,
\\
P_{135} &=& 726 N^6+1557 N^5-320 N^4-1921 N^3-1108 N^2-122 N-108~,
\\
P_{136} &=& 1923 N^6+4725 N^5+3893 N^4+679 N^3+20 N^2+864 N+504~,
\\
P_{137} &=& 4541 N^6+14703 N^5+15807 N^4+7037 N^3+1932 N^2+324 N-432~,
\\
P_{138} &=& 7081 N^6+24375 N^5+26871 N^4+10813 N^3-60 N^2-2592 N-1512~,
\\
P_{139} &=& -345 N^7-1629 N^6-2849 N^5-2367 N^4-1142 N^3-456 N^2-208 N-32~,
\\
P_{140} &=& 13519 N^7+67649 N^6+122235 N^5+96259 N^4+31370 N^3+3540 N^2
\nonumber\\ &&
+216 N-864~,
\\
P_{141} &=& -38715 N^8-131784 N^7-168338 N^6-89028 N^5-11355 N^4+4564 N^3
\nonumber\\ &&
+11808 N^2
+16560 N+6480~,
\\
P_{142} &=& 3 N^8+15 N^7+13 N^6-41 N^5-94 N^4-78 N^3-50 N^2-36 N-20~,
\\
P_{143} &=& -86393 N^{10}-469657 N^9-848380 N^8-376714 N^7+589247 N^6+803483 N^5
\nonumber\\ &&
+303838 N^4
-78072 N^3-102168 N^2+432 N+15552~,
\\
P_{144} &=& -20163 N^{10}-89115 N^9-115804 N^8+11018 N^7+141341 N^6+95849 N^5
\nonumber\\ &&
+15002 N^4
+21272 N^3+28872 N^2-144 N-5184~,
\\
P_{145} &=& 187 N^{10}+707 N^9+468 N^8-1258 N^7-2037 N^6-289 N^5-202 N^4-1848 N^3
\nonumber\\ &&
-1856 N^2
-176 N+160~,
\\
P_{146} &=& 619 N^{10}+3003 N^9+4988 N^8+2654 N^7-2525 N^6-5553 N^5-3274 N^4
\nonumber\\ &&
+1480 N^3
+3136 N^2+1392 N+224~,
\\
P_{147} &=& -41 N^{14}+461 N^{13}+4075 N^{12}+10649 N^{11}+5585 N^{10}-28621 N^9-66255 N^8
\nonumber\\ &&
-64133 N^7-59540 N^6-74900 N^5-47952 N^4+10560 N^3+16480 N^2+384 N
\nonumber\\ &&
-1664~,
\\
P_{148} &=& 370659 N^{14}+2352657 N^{13}+4903159 N^{12}+1576653 N^{11}-7789003 N^{10}
\nonumber\\ &&
-9288369 N^9+1194501 N^8+8087991 N^7+5995676 N^6+3609228 N^5
\nonumber\\ &&
+2013520 N^4-526896 N^3
-789120 N^2+88128 N+145152.
\end{eqnarray}

\noindent
and


\noindent
with
\begin{eqnarray}
P_{149} &=& -1079 N^4-2266 N^3-1019 N^2+24 N-108~,
\\
P_{150} &=& -75 N^4-192 N^3-29 N^2+32 N-50~,
\\
P_{151} &=& N^4+2 N^3-12 N^2-13 N+4~,
\\
P_{152} &=& N^4+2 N^3-9 N^2-10 N+4~,
\\
P_{153} &=& N^4+4 N^3-13 N^2-40 N-20~,
\\
P_{154} &=& 5 N^4+19 N^3-4 N^2-6 N+11~,
\\
P_{155} &=& 13 N^4+20 N^3+14 N^2-N-6~,
\\
P_{156} &=& 26 N^4+49 N^3+33 N^2+10 N-5~,
\\
P_{157} &=& 111 N^4+186 N^3+131 N^2+8 N-36~,
\\
P_{158} &=&229 N^4+507 N^3+116 N^2-144 N+72~,
\\
P_{159} &=&373 N^4+854 N^3+313 N^2-24 N+108~,
\\
P_{160} &=&N^5+N^4-12 N^3+11 N^2+N-10~,
\\
P_{161} &=&N^5+2 N^4-17 N^3-26 N^2+16~,
\\
P_{162} &=&N^5+2 N^4+4 N^3+N^2+4~,
\\
P_{163} &=&N^5+3 N^4-3 N^3-9 N^2-8 N-8~,
\\
P_{164} &=&N^5+4 N^4+N^3-4 N^2+4 N+4~,
\\
P_{165} &=&2 N^5+6 N^4+N^3-6 N^2+11 N+10~,
\\
P_{166} &=&7 N^5+25 N^4+2 N^3-39 N^2+28 N+60~,
\\
P_{167} &=&13 N^5+43 N^4+27 N^3-39 N^2-16 N+36~,
\\
P_{168} &=&27 N^5+66 N^4+25 N^3-26 N^2+8 N+16~,
\\
P_{169} &=&67 N^5+214 N^4+133 N^3-150 N^2-26 N+108~,
\\
P_{170} &=&179 N^5+896 N^4+1333 N^3+610 N^2-60 N-72~,
\\
P_{171} &=&-4541 N^6-14649 N^5-14439 N^4-4355 N^3-672 N^2-324 N+432~,
\\
P_{172} &=&-2311 N^6-6393 N^5-991 N^4+7197 N^3+2222 N^2-1236 N+216~,
\\
P_{173} &=&-1239 N^6-3321 N^5-3661 N^4-1787 N^3-100 N^2+108 N-144~,
\\
P_{174} &=&-237 N^6-635 N^5-675 N^4-321 N^3+12 N^2-40~,
\\
P_{175} &=&-37 N^6-243 N^5-135 N^4+167 N^3+44 N^2+16 N+64~,
\\
P_{176} &=&N^6+3 N^5-5 N^4-7 N^3-8 N^2-24 N-8~,
\\
P_{177} &=&N^6+3 N^5-2 N^4-10 N^3-8 N^2+4~,
\\
P_{178} &=&6 N^6+19 N^5+5 N^4-19 N^3+9 N^2-4 N-20~,
\\
P_{179} &=&8 N^6+19 N^5-28 N^4-56 N^3-31 N^2-32 N+48~,
\\
P_{180} &=&12 N^6+27 N^5-97 N^4-281 N^3-177 N^2+64 N+20~,
\\
P_{181} &=&15 N^6+36 N^5-19 N^4-42 N^3+30 N^2-40 N-56~,
\\
P_{182} &=&18 N^6+36 N^5-31 N^4-134 N^3-357 N^2-200 N+236~,
\\
P_{183} &=&24 N^6+59 N^5-211 N^4-487 N^3-29 N^2+280 N+76~,
\\
P_{184} &=&31 N^6+90 N^5+23 N^4-84 N^3-8 N^2-16 N-48~,
\\
P_{185} &=&43 N^6+102 N^5-37 N^4-100 N^3+44 N^2-104 N-128~,
\\
P_{186} &=&105 N^6+279 N^5+375 N^4+261 N^3+124 N^2+32 N+8~,
\\
P_{187} &=&108 N^6+258 N^5+14 N^4-228 N^3-7 N^2+9 N-10~,
\\
P_{188} &=&198 N^6+405 N^5-102 N^4-501 N^3-268 N^2-76 N-88~,
\\
P_{189} &=&834 N^6+1881 N^5-104 N^4-2029 N^3-1432 N^2-338 N-108~,
\\
P_{190} &=&2667 N^6+6669 N^5+6605 N^4+2095 N^3-940 N^2+504~,
\\
P_{191} &=&4541 N^6+14703 N^5+14223 N^4+3869 N^3+348 N^2+324 N-432~,
\\
P_{192} &=&7081 N^6+25239 N^5+25431 N^4+8797 N^3+2820 N^2-1512~,
\\
P_{193} &=&-397 N^7-1873 N^6-3293 N^5-2811 N^4-1302 N^3-200 N^2+16 N-32~,
\\
P_{194} &=&-37 N^7-160 N^6-85 N^5+426 N^4+612 N^3+184 N^2+32 N+64~,
\\
P_{195} &=&-13 N^7-80 N^6-121 N^5+30 N^4+28 N^3-96 N^2+208 N+192~,
\\
P_{196} &=&14311 N^7+71609 N^6+126915 N^5+92875 N^4+23306 N^3+372 N^2
\nonumber\\ &&
+216 N-864~,
\\
P_{197} &=&-56163 N^8-192648 N^7-268610 N^6-170340 N^5-27651 N^4+4084 N^3
\nonumber\\ &&
-8928 N^2
+4464 N+6480~,
\\
P_{198} &=&2 N^8+11 N^7+10 N^6-26 N^5-2 N^4+151 N^3+262 N^2+160 N+8~,
\\
P_{199} &=&3 N^8+15 N^7+11 N^6-45 N^5-74 N^4-30 N^3-52 N^2-80 N-36~,
\\
P_{200} &=&10 N^8+34 N^7+4 N^6-64 N^5-51 N^4-48 N^3-53 N^2+12 N+28~,
\\
P_{201} &=&-86393 N^{10}-479161 N^9-842908 N^8-273610 N^7+665567 N^6
\nonumber\\ &&
+681947 N^5
+206494 N^4-29400 N^3-86616 N^2-20304 N+15552~,
\\
P_{202} &=&-30651 N^{10}-138387 N^9-203260 N^8-49510 N^7+185861 N^6+211505 N^5
\nonumber\\ &&
+73610 N^4+8504 N^3+23688 N^2+6768 N-5184~,
\\
P_{203} &=&4 N^{10}+18 N^9-18 N^8-130 N^7-3 N^6+253 N^5+131 N^4+115 N^3+146 N^2
\nonumber\\ &&
-44 N-88~,
\\
P_{204} &=&16 N^{10}+96 N^9+137 N^8-155 N^7-584 N^6-618 N^5-431 N^4-67 N^3
\nonumber\\ &&
+ 178 N^2
-20 N-88~,
\\
P_{205} &=&34 N^{10}+144 N^9+134 N^8-156 N^7-147 N^6+449 N^5+825 N^4+355 N^3
\nonumber\\ &&
-130 N^2
-28 N+56~,
\\
P_{206} &=&50 N^{10}+214 N^9+99 N^8-533 N^7-355 N^6+605 N^5+562 N^4+338 N^3
\nonumber\\ &&
+244 N^2
-216 N-240~,
\\
P_{207} &=&763 N^{10}+3395 N^9+5268 N^8+2262 N^7-3701 N^6-5729 N^5-4554 N^4
\nonumber\\ &&
-2872 N^3
-1216 N^2
+80 N+160~,
\\
P_{208} &=&1203 N^{10}+5811 N^9+9468 N^8+4366 N^7-5429 N^6-8825 N^5-3450 N^4
\nonumber\\ &&
+1256 N^3
+1152 N^2+368 N+224~,
\\
P_{209} &=&32 N^{12}+172 N^{11}+187 N^{10}-377 N^9-635 N^8+560 N^7+1634 N^6
\nonumber\\ &&
+1246 N^5
+488 N^4
-145 N^3-294 N^2+84 N+120~,
\\
P_{210} &=&-4521 N^{14}-28595 N^{13}-64213 N^{12}-42855 N^{11}+63121 N^{10}+130099 N^9
\nonumber\\ &&
+49137 N^8-79237 N^7-137428 N^6-123924 N^5-65104 N^4-448 N^3
\nonumber\\ &&
+11360 N^2
-640 N-1664~,
\\
P_{211} &=&-84 N^{14}-520 N^{13}-1019 N^{12}-186 N^{11}+1538 N^{10}+1433 N^9+2372 N^8
\nonumber\\ &&
+9376 N^7
+14990 N^6+9125 N^5+163 N^4-868 N^3+720 N^2+64 N-240~,
\\
P_{212} &=&12 N^{14}+76 N^{13}+33 N^{12}-576 N^{11}-806 N^{10}+1161 N^9+1790 N^8
\nonumber\\ &&
-2494 N^7
-5322 N^6-3207 N^5-715 N^4+664 N^3+840 N^2-304 N-368~,
\\
P_{213} &=&648699 N^{14}+4173081 N^{13}+9225583 N^{12}+4949301 N^{11}-11882659 N^{10}
\nonumber\\ &&
-20345337 N^9-5717331 N^8+11889903 N^7+12065660 N^6+4779564 N^5
\nonumber\\ &&
+2272528 N^4+613584 N^3
-713088 N^2-160704 N+145152.
\end{eqnarray}

The evanescent poles at $N = 2$ are tractable, and the rightmost singularities are
at $N = 1$ for the singlet contributions and at $N = 0$ for the non--singlet contributions,
as generally expected, cf.~\cite{SMALLX}.
\subsection{The Two--Loop Wilson Coefficients} 
\label{eq:43}

\vspace*{1mm}
\noindent
The two--loop Wilson coefficients for $\mu^2 = Q^2$ are given by
\begin{eqnarray}
C_{F_2,q}^{\rm NS,(2)} &=&
\textcolor{blue}{C_F^2} \Biggl[
        \frac{S_2 P_{218}}{2 N^2 (1+N)^2}
        +\frac{P_{233}}{8 (N-2) N^4 (1+N)^4 (3+N)}
        +\Biggl(
                \frac{P_{225}}{2 N^3 (1+N)^3}
\nonumber\\ &&                
 -\frac{2 (-2+3 N) (5+3 N) S_2}{N (1+N)}
                -24 S_3
                +16 S_{2,1}
                +32 S_{-2,1}
                +48 \zeta_3
        \Biggr) S_1
        +\Biggl(
                \frac{P_{214}}{2 N^2 (1+N)^2}
\nonumber\\ &&                
 -20 S_2
        \Biggr) S_1^2
        +\frac{2 \big(
                -2+3 N+3 N^2\big) S_1^3}{N (1+N)}
        +2 S_1^4
        +6 S_2^2
        -\frac{2 \big(
                -10+25 N+9 N^2\big) S_3}{N (1+N)}
\nonumber\\ &&         
+12 S_4
        +\Biggl(
                -\frac{8 P_{61}}{(N-2) N^2 (1+N)^2 (3+N)}
                -\frac{16 (-3+4 N) S_1}{N (1+N)}
                -32 S_1^2
                +16 S_2
        \Biggr) S_{-2}
\nonumber\\ &&         
+24 S_{-2}^2
        +\Biggl(
                -\frac{32}{1+N}
                +16 S_1
        \Biggr) S_{-3}
        +40 S_{-4}
        +\frac{4 \big(
                -2+3 N+3 N^2\big) S_{2,1}}{N (1+N)}
        +40 S_{3,1}
\nonumber\\ &&      
  +\frac{32 (-1+2 N) S_{-2,1}}{N (1+N)}
        -16 S_{-2,2}
        -32 S_{-3,1}
        -24 S_{2,1,1}
        -\frac{24 \big(
                2-N+3 N^2\big) \zeta_3}{N (1+N)}
\Biggr]
\nonumber\\ && +\textcolor{blue}{C_A C_F} \Biggl[
        \frac{P_{229}}{216 (N-2) N^3 (1+N)^3 (3+N)}
        +\Biggl(
                -\frac{2 \big(
                        6+11 N+11 N^2\big) S_2}{3 N (1+N)}
\nonumber\\ && 
+                \frac{P_{224}}{54 N^2 (1+N)^3}
                +24 S_3
                -16 S_{2,1}
                -16 S_{-2,1}
                -48 \zeta_3
        \Biggr) S_1
        +\Biggl(
                \frac{-66+367 N+367 N^2}{18 N (1+N)}
\nonumber\\ &&                
 +4 S_2
        \Biggr) S_1^2
        +
        \frac{\big(
                66-929 N-2134 N^2-1067 N^3\big) S_2}{18 N (1+N)^2}
        +\frac{2 \big(
                -72+193 N+121 N^2\big) S_3}{9 N (1+N)}
\nonumber\\ && 
        -4 S_2^2
        +\frac{22}{9} S_1^3
        -8 S_4
        +\Biggl(
                \frac{4 P_{61}}{(N-2) N^2 (1+N)^2 (3+N)}
                +\frac{8 (-3+4 N) S_1}{N (1+N)}
\nonumber\\ &&                
                +16 S_1^2
 -8 S_2
        \Biggr) S_{-2}
        +\Biggl(
                \frac{16}{1+N}
                -8 S_1
        \Biggr) S_{-3}
        -\frac{4 \big(
                -6+11 N+11 N^2\big) S_{2,1}}{3 N (1+N)}
        -24 S_{3,1}
\nonumber\\ &&        
        -20 S_{-4}
 -12 S_{-2}^2
        -\frac{16 (-1+2 N) S_{-2,1}}{N (1+N)}
        +8 S_{-2,2}
        +16 S_{-3,1}
        +24 S_{2,1,1}
\nonumber\\ &&
        +\frac{6 \big(
                6+N+9 N^2\big) \zeta_3}{N (1+N)}
\Biggr]
+ \textcolor{blue}{C_F T_F N_F} \Biggl[
        \frac{P_{235}}{54 N^3 (1+N)^3}
        +\Biggl(
                \frac{8}{3} S_2
                -\frac{2 P_{236}}{27 N^2 (1+N)^2}                
        \Biggr) S_1
        \nonumber \\ &&
        +\frac{2 \big(6-29 N-29 N^2\big)}{9 N (1+N)} S_1^2
        -\frac{8}{9} S_1^3
        -\frac{2 \big(6-85 N-85 N^2\big)}{9 N (1+N)} S_2
\nonumber \\ &&
        -\frac{88}{9} S_3
        +\frac{16}{3} S_{2,1}
\Biggr]
~,
\\
C_{F_2,q}^{\rm PS, (2)} &=& \textcolor{blue}{C_F T_F N_F} \Biggl[
        \frac{8 S_1 P_{228}}{(N-1) N^3 (1+N)^3 (2+N)^2}
        +\frac{4 P_{232}}{(N-1) N^4 (1+N)^4 (2+N)^3}
\nonumber\\ &&    
    +\frac{4 \big(
                2+N+N^2\big)^2 S_1^2}{(N-1) N^2 (1+N)^2 (2+N)}
        -\frac{4 \big(
                2+N+N^2\big)^2 S_2}{(N-1) N^2 (1+N)^2 (2+N)}
\nonumber\\ &&  
      +\frac{64 S_{-2}}{(N-1) N (1+N) (2+N)}
\Biggr] ~,
\\
C_{F_2,g}^{(2)} &=&
\textcolor{blue}{C_A T_F N_F} \Biggl[
        \frac{8 S_2 P_{220}}{(N-1) N (1+N)^2 (2+N)^2}
        -\frac{8 S_1^2 P_{221}}{(N-1) N (1+N)^2 (2+N)^2}
\nonumber\\ && 
        -\frac{4 P_{234}}{(N-1) N^4 (1+N)^4 (2+N)^4}
        +\Biggl(
                -\frac{4 P_{230}}{(N-1) N^3 (1+N)^3 (2+N)^3}
\nonumber\\ &&
                +
                \frac{20 \big(
                        2+N+N^2\big) S_2}{N (1+N) (2+N)}
        \Biggr) S_1
        -\frac{4 \big(
                2+N+N^2\big) S_1^3}{3 N (1+N) (2+N)}
        +\frac{8 \big(
                -2+5 N+5 N^2\big) S_3}{3 N (1+N) (2+N)}
\nonumber\\ &&
        +\Biggl(
                \frac{16 P_{38}}{(N-1) N (1+N)^2 (2+N)^2}
                +\frac{16 (N-1) S_1}{N (1+N)}
        \Biggr) S_{-2}
        -\frac{16 \big(
                4+N+N^2\big) S_{-3}}{N (1+N) (2+N)}
\nonumber\\ &&
        -\frac{16 \big(
                2+N+N^2\big) S_{2,1}}{N (1+N) (2+N)}
        +\frac{64 S_{-2,1}}{N (1+N) (2+N)}
        -\frac{24 (N-1) \zeta_3}{N (1+N)}
\Biggr]
\nonumber\\ &&
+\textcolor{blue}{C_F T_F N_F} \Biggl[
        \frac{2 S_2 P_{216}}{N^2 (1+N)^2 (2+N)}
        -\frac{2 P_{231}}{(N-2) N^4 (1+N)^4 (2+N) (3+N)}
\nonumber\\ &&
        -\frac{2 S_1^2 P_{217}}{N^2 (1+N)^2 (2+N)}
        +\Biggl(
                \frac{4 P_{226}}{N^3 (1+N)^3 (2+N)}
                +\frac{12 \big(
                        2+N+N^2\big) S_2}{N (1+N) (2+N)}
        \Biggr) S_1
\nonumber\\ &&    
        -\frac{20 \big(
                2+N+N^2\big) S_1^3}{3 N (1+N) (2+N)}
     -\frac{64 \big(
                -1+N+N^2\big) S_3}{3 N (1+N) (2+N)}
        +\Biggl(
                \frac{128 S_1}{N (1+N) (2+N)}
\nonumber\\ &&                 
+\frac{16 P_{59}}{(N-2) N^2 (1+N)^2 (2+N) (3+N)}
        \Biggr) S_{-2}
        +\frac{64 S_{-3}}{N (1+N) (2+N)}
\nonumber\\ &&    
     +\frac{16 \big(
                2+N+N^2\big) S_{2,1}}{N (1+N) (2+N)}
        -\frac{128 S_{-2,1}}{N (1+N) (2+N)}
        +
        \frac{48 (N-1) \zeta_3}{N (1+N)}
\Biggr] ~,
\\
C_{F_L,q}^{\rm NS, (2)}  &=& \textcolor{blue}{C_F} \Biggl\{
        \textcolor{blue}{T_F N_F} \Biggl[
                -\frac{8 \big(
                        -6+7 N+19 N^2\big)}{9 N (1+N)^2}
                -\frac{16 S_1}{3 (1+N)}
        \Biggr]
        +\textcolor{blue}{C_A} \Biggl[
                \frac{92 S_1}{3 (1+N)}
\nonumber\\ &&                
+\frac{2 P_{219}}{9 (N-2) N (1+N)^2 (3+N)}
                +\frac{16 S_3}{1+N}
                +\Biggl(
                        -\frac{32 P_{15}}{(N-2) N (1+N)^2 (3+N)}
\nonumber\\ &&          
               +\frac{32 S_1}{1+N}
                \Biggr) S_{-2}
                +\frac{16 S_{-3}}{1+N}
                -\frac{32 S_{-2,1}}{1+N}
                -\frac{48 \zeta_3}{1+N}
        \Biggr]
        \Biggr\}
+
\textcolor{blue}{C_F^2} \Biggl\{
        -\frac{4 \big(
                2+13 N+9 N^2\big) S_1}{N (1+N)^2}
\nonumber\\ &&
        -\frac{2 P_{227}}{(N-2) N^2 (1+N)^3 (3+N)}
        +\frac{8 S_1^2}{1+N}
        -\frac{8 S_2}{1+N}
        -\frac{32 S_3}{1+N}
        +\Biggl(
                -\frac{64 S_1}{1+N}
\nonumber\\ &&
                +\frac{64 P_{15}}{(N-2) N (1+N)^2 (3+N)}
        \Biggr) S_{-2}
        -\frac{32 S_{-3}}{1+N}
        +\frac{64 S_{-2,1}}{1+N}
        +\frac{96 \zeta_3}{1+N}
\Biggr\} ~,
\\
C_{F_L,q}^{\rm PS (2)} &=& \textcolor{blue}{C_F T_F N_F} \Biggl[
        -\frac{32 P_{222}}{(N-1) N^2 (1+N)^3 (2+N)^2}
        -\frac{32 \big(
                2+N+N^2\big) S_1}{(N-1) N (1+N)^2 (2+N)}
\Biggr] ~,
\\
C_{F_L,g}^{(2)} &=&
\textcolor{blue}{C_A T_F N_F} \Biggl[
        -\frac{32 P_{223}}{(N-1) N^2 (1+N)^3 (2+N)^3}
        +\frac{64 \big(
                -1-N-2 N^2+2 N^3\big) S_1}{(N-1) N (1+N)^2 (2+N)}
\nonumber\\ &&
        +\frac{32 S_1^2}{(1+N) (2+N)}
        -\frac{32 S_2}{(1+N) (2+N)}
        -\frac{64 S_{-2}}{(1+N) (2+N)}
\Biggr]
\nonumber\\ &&
+\textcolor{blue}{C_F T_F N_F} \Biggl[
        \frac{16 (N-1) P_{215}}{(N-2) N^2 (1+N)^3 (2+N) (3+N)}
        -\frac{16 \big(
                2+3 N+3 N^2\big) S_1}{N (1+N)^2 (2+N)}
\nonumber\\ &&
        +\frac{64 (N-1) S_{-2}}{(N-2) (1+N) (3+N)}
\Biggr] ~,
\end{eqnarray}

\noindent
with 
\begin{eqnarray}
P_{214} &=& -27 N^4-26 N^3+9 N^2+40 N+24~,
\\
P_{215} &=& 2 N^4+19 N^3+39 N^2+40 N+12~,
\\
P_{216} &=& 9 N^4+8 N^3+9 N^2+6 N-8~,
\\
P_{217} &=& 9 N^4+12 N^3+N^2-14 N-16~,
\\
P_{218} &=& 95 N^4+162 N^3+35 N^2-32 N-16~,
\\
P_{219} &=& 215 N^4+298 N^3-1597 N^2-888 N+396~,
\\
P_{220} &=& N^5-10 N^3-9 N^2-4 N-2~,
\\
P_{221} &=& N^5-2 N^4-6 N^3-3 N^2-12 N-2~,
\\
P_{222} &=& N^5+2 N^4+2 N^3-5 N^2-12 N-4~,
\\
P_{223} &=& 5 N^5+8 N^4-3 N^3-22 N^2-28 N-8~,
\\
P_{224} &=& 3155 N^5+11607 N^4+12279 N^3+3329 N^2+510 N+792~,
\\
P_{225} &=& -51 N^6-203 N^5-207 N^4-33 N^3-106 N^2-160 N-48~,
\\
P_{226} &=& N^6-7 N^5-3 N^4-5 N^3-30 N^2-40 N-16~,
\\
P_{227} &=& 17 N^6+39 N^5-157 N^4-299 N^3-64 N^2+104 N+24~,
\\
P_{228} &=& N^7-15 N^5-58 N^4-92 N^3-76 N^2-48 N-16~,
\\
P_{229} &=& -16395 N^8-47520 N^7+51416 N^6+162042 N^5+99843 N^4-7930 N^3
\nonumber\\ &&
-21432 N^2
+25848 N+23760~,
\\
P_{230} &=& 7 N^9+5 N^8-43 N^7+25 N^6+296 N^5+498 N^4+524 N^3+336 N^2
+144 N
\nonumber\\ &&
+32~,
\\
P_{231} &=& 2 N^{10}+18 N^9+98 N^8+98 N^7-425 N^6-1071 N^5-477 N^4
+651 N^3
\nonumber\\ &&
+886 N^2
+484 N+120~,
\\
P_{232} &=& 3 N^{10}+14 N^9+33 N^8+79 N^7+297 N^6+849 N^5+1373 
N^4+1312 
N^3
\nonumber\\ &&
+840 N^2
+368 N+80~,
\\
P_{233} &=& 331 N^{10}+1179 N^9-848 N^8-4754 N^7-2157 N^6+4247 N^5
+3474 N^4
\nonumber\\ &&
-2528 N^3
-4976 N^2-2704 N-480~,
\\
P_{234} &=& 4 N^{12}+34 N^{11}+100 N^{10}+116 N^9-81 N^8-637 N^7-1677 
N^6
\nonumber\\ &&
-3093 N^5
-3998 N^4
-3472 N^3-2064 N^2-816 N-160~,
\\
P_{235} &=& 360+648 N+140 N^2+31 N^3+1397 N^4+2517 N^5+1371 N^6~,
\\
P_{236} &=& 72+66 N+331 N^2+620 N^3+247 N^4.
\end{eqnarray}
\begin{eqnarray}
\label{eq:F3NS2}
C_{F_3,q}^{(2),\rm NS} &=& 
\textcolor{blue}{C_F} \Biggl\{
        \textcolor{blue}{T_F N_F} \Biggl[
                \frac{P_{241}}{54 N^3 (1+N)^3}
                +\Biggl(
                        -\frac{2 P_{239}}{27 N^2 (1+N)^2}
                        +\frac{8 S_2}{3}
                \Biggr) S_1
                -\frac{8}{9} S_1^3
\nonumber\\ && 
                -\frac{2 \big(
                        -6+29 N+29 N^2\big) S_1^2}{9 N (1+N)}
                +\frac{2 \big(
                        -6+85 N+85 N^2\big) S_2}{9 N (1+N)}
                -\frac{88}{9} S_3
                +\frac{16}{3} S_{2,1}
        \Biggr]
\nonumber\\ && 
        +\textcolor{blue}{C_A} \Biggl[
                \frac{P_{243}}{216 (N-1) N^4 (1+N)^4 (2+N)}
                +\Biggl(
                        \frac{P_{242}}{54 N^3 (1+N)^3}
                        +24 S_3
                        -16 S_{2,1}
\nonumber\\ && 
                        -16 S_{-2,1}
                        -\frac{2 \big(
                                6+11 N+11 N^2\big) S_2}{3 N (1+N)}
                        -48 \zeta_3
                \Biggr) S_1
                +\Biggl(
                        \frac{-66+367 N+367 N^2}{18 N (1+N)}
                        +4 S_2
                \Biggr) 
\nonumber\\ && \times
S_1^2
                +\frac{22}{9} S_1^3
                +\frac{\big(
                        66-929 N-2134 N^2-1067 N^3\big) S_2}{18 N (1+N)^2}
                -4 S_2^2
                -8 S_4
                -12 S_{-2}^2
\nonumber\\ && 
                +\frac{2 \big(
                        -36+121 N+121 N^2\big) S_3}{9 N (1+N)}
                +\Biggl(
                        \frac{4 P_{123}}{(N-1) N (1+N)^2 (2+N)}
                        -\frac{8 S_1}{N (1+N)}
\nonumber\\ &&               
                        +16 S_1^2
         -8 S_2
                \Biggr) S_{-2}
                +\Biggl(
                        \frac{8}{N (1+N)}
                        -8 S_1
                \Biggr) S_{-3}
                -20 S_{-4}
                +\frac{4 \big(
                        6-11 N-11 N^2\big) S_{2,1}}{3 N (1+N)}
\nonumber\\ &&                
 -24 S_{3,1}
                +8 S_{-2,2}
                +16 S_{-3,1}
                +24 S_{2,1,1}
                +\frac{6 (1+3 N) (2+3 N) \zeta_3}{N (1+N)}
        \Biggr]
\Biggr\}
\nonumber\\ && 
+\textcolor{blue}{C_F^2} \Biggl[
        \frac{S_2 P_{238}}{2 N^2 (1+N)^2}
        +\frac{P_{244}}{8 (N-1) N^4 (1+N)^4 (2+N)}
        +\Biggl(
                \frac{P_{240}}{2 N^3 (1+N)^3}
\nonumber\\ &&                
 -\frac{2 (-2+3 N) (5+3 N) S_2}{N (1+N)}
                -24 S_3
                +16 S_{2,1}
                +32 S_{-2,1}
                +48 \zeta_3
        \Biggr) S_1
        +\Biggl(
                -20 S_2
\nonumber\\ && 
+                \frac{P_{237}}{2 N^2 (1+N)^2}
        \Biggr) S_1^2
        +\frac{2 \big(
                -2+3 N+3 N^2\big) S_1^3}{N (1+N)}
        +2 S_1^4
        +6 S_2^2
        +12 S_4
\nonumber\\ && 
        -\frac{2 \big(
                -2+9 N+9 N^2\big) S_3}{N (1+N)}
        +\Biggl(
                -\frac{8 P_{123}}{(N-1) N (1+N)^2 (2+N)}
                +\frac{16 S_1}{N (1+N)}
\nonumber\\ &&                
 -32 S_1^2
                +16 S_2
        \Biggr) S_{-2}
        +24 S_{-2}^2
        +\Biggl(
                -\frac{16}{N (1+N)}
                +16 S_1
        \Biggr) S_{-3}
        +40 S_{-4}
\nonumber\\ &&      
  +\frac{4 \big(
                -2+3 N+3 N^2\big) S_{2,1}}{N (1+N)}
        +40 S_{3,1}
        -16 S_{-2,2}
        -32 S_{-3,1}
        -24 S_{2,1,1}
        -72 \zeta_3
\Biggr] ~,
\end{eqnarray}

\noindent
with the polynomials
\begin{eqnarray}
P_{237} &=& -27 N^4-42 N^3-15 N^2+32 N+24~,
\\
P_{238} &=& 95 N^4+178 N^3+59 N^2-24 N-16~,
\\
P_{239} &=& 247 N^4+548 N^3+223 N^2+30 N+72~,
\\
P_{240} &=& -51 N^6-131 N^5-115 N^4-49 N^3-54 N^2-72 N-16~,
\\
P_{241} &=& 1371 N^6+3429 N^5+2957 N^4+271 N^3-556 N^2+360 N+360~,
\\
P_{242} &=& 3155 N^6+9951 N^5+9867 N^4+3473 N^3+546 N^2-72 N-432~,
\\
P_{243} &=& -16395 N^{10}-74235 N^9-101500 N^8+5090 N^7+133973 N^6+105113 N^5
\nonumber\\ && 
+16970 N^4+6224 N^3+16200 N^2-3312 N-5184~,
\\
P_{244} &=& 331 N^{10}+1451 N^9+1492 N^8-1826 
N^7-4341 N^6-1673 N^5+246 N^4
\nonumber\\ &&
-928 N^3
-1232 N^2+112 N+224 
\end{eqnarray}
\begin{eqnarray}
C_{g_1,q}^{(2),\rm NS,L} &=& 
\textcolor{blue}{C_F} \Biggl\{
        \textcolor{blue}{T_F N_F} \Biggl[
                \frac{P_{251}}{54 N^3 (1+N)^3}
                +\Biggl(
                        -\frac{2 P_{239}}{27 N^2 (1+N)^2}
                        +\frac{8}{3} S_2
                \Biggr) S_1
                 -\frac{8}{9} S_1^3
\nonumber\\ && 
                -\frac{2 \big(
                        -6+29 N+29 N^2\big)}{9 N (1+N)} S_1^2
                +\frac{2 \big(
                        -6+85 N+85 N^2\big)}{9 N (1+N)} S_2
                -\frac{88}{9} S_3
                +\frac{16}{3} S_{2,1}
        \Biggr]
\nonumber\\ && 
        +\textcolor{blue}{C_A} \Biggl[
                \frac{P_{253}}{216 (N-1) N^4 (1+N)^4 (2+N)}
                +\Biggl(
                        -\frac{2 \big(
                                6+11 N+11 N^2\big)}{3 N (1+N)} S_2
\nonumber\\ && 
                        +\frac{P_{242}}{54 N^3 (1+N)^3}
                        +24 S_3
                        -16 S_{2,1}
                        -16 S_{-2,1}
                        -48 \zeta_3
                \Biggr) S_1
                +\frac{22}{9} S_1^3
                +\Biggl(
                        4 S_2
\nonumber\\ && 
                        + \frac{-66+367 N+367 N^2}{18 N (1+N)}
                \Biggr) S_1^2
                +\frac{\big(
                        66-929 N-2134 N^2-1067 N^3\big)}{18 N 
(1+N)^2} S_2
                -4 S_2^2
\nonumber\\ &&                
 +\frac{2 \big(
                        -36+121 N+121 N^2\big)}{9 N (1+N)} S_3
                -8 S_4
                +\Biggl(
                        \frac{4 P_{165}}{(N-1) N (1+N)^2 (2+N)}
\nonumber\\ && 
                        -\frac{8}{N (1+N)} S_1
                        +16 S_1^2
                        -8 S_2
                \Biggr) S_{-2}
                -12 S_{-2}^2
                +\Biggl(
                        \frac{8}{N (1+N)}
                        -8 S_1
                \Biggr) S_{-3}
\nonumber\\ &&                
 -20 S_{-4}
                -\frac{4 \big(
                        -6+11 N+11 N^2\big)}{3 N (1+N)} S_{2,1}
                -24 S_{3,1}
                +8 S_{-2,2}
                +16 S_{-3,1}
 +24 S_{2,1,1}
\nonumber\\ &&                
                +\frac{6 (1+3 N) (2+3 N)}{N (1+N)} \zeta_3
        \Biggr]
\Biggr\}
+\textcolor{blue}{C_F^2} \Biggl[
        \frac{P_{256}}{8 (N-1) N^4 (1+N)^4 (2+N)}
\nonumber\\ && 
        +\frac{P_{246}}{2 N^2 (1+N)^2} S_2
        +\Biggl(
                \frac{P_{248}}{2 N^3 (1+N)^3}
                -\frac{2 (-2+3 N) (5+3 N) S_2}{N (1+N)}
                -24 S_3
                +16 S_{2,1}
\nonumber\\ &&                
 +32 S_{-2,1}
                +48 \zeta_3
        \Biggr) S_1
        +\Biggl(
                \frac{P_{245}}{2 N^2 (1+N)^2}
                -20 S_2
        \Biggr) S_1^2
        +\frac{2 \big(
                -2+3 N+3 N^2\big)}{N (1+N)} S_1^3
\nonumber\\ && 
        +2 S_1^4
        +6 S_2^2
        -\frac{2 \big(
                -2+9 N+9 N^2\big) S_3}{N (1+N)}
        +12 S_4
        +\Biggl(
                 \frac{16}{N (1+N)} S_1
\nonumber\\ && 
                -\frac{8 P_{165}}{(N-1) N (1+N)^2 (2+N)}
                -32 S_1^2
                +16 S_2
        \Biggr) S_{-2}
        +24 S_{-2}^2
        +16 \Biggl(
                -\frac{1}{N (1+N)}
\nonumber\\ &&                
 +S_1
        \Biggr) S_{-3}
        +40 S_{-4}
        +\frac{4 \big(
                -2+3 N+3 N^2\big)}{N (1+N)} S_{2,1}
        +40 S_{3,1}
        -16 S_{-2,2}
        -32 S_{-3,1}
\nonumber\\ &&     
   -24 S_{2,1,1}
        -72 \zeta_3
\Biggr] ~,
\\
\label{eq:g1NS2}
 C_{g_1,q}^{(2),\rm NS,M} &=& C_{g_1,q}^{(2),\rm NS,L} 
- \Bigl( C_{g1,q}^{(1),\rm NS,L} - z_{qq}^{(1)} \Bigr) z_{qq}^{(1)} - z_{qq}^{(2),\rm NS}
 \nonumber\\ 
 &=& C_{g_1,q}^{(2),\rm NS,L} 
 + \textcolor{blue}{C_F} \Biggl\{
        - \textcolor{blue}{T_F N_F} \frac{16 \big(
                -3-N+5 N^2\big)}{9 N^2 (1+N)^2}
       +\textcolor{blue}{C_A} \Biggl[
                \frac{4 P_{247}}{9 N^3 (1+N)^3}
\nonumber\\ && 
                +\frac{16}{N (1+N)} S_{-2}
        \Biggr]
\Biggl\}
+\textcolor{blue}{C_F^2} \Biggl[
        -\frac{8 \big(
                11+16 N+11 N^2\big)}{N (1+N)^3}
        +\frac{8 (-4+3 N)}{N^2 (1+N)} S_1
\nonumber\\ &&        
 +\frac{16}{N (1+N)} S_1^2
        -\frac{32}{N (1+N)} S_2
        -\frac{32}{N (1+N)} 
S_{-2}
\Biggr] ~,
\\
C_{g_1,q}^{(2),\rm PS,L} &=& 
\textcolor{blue}{C_F T_F N_F} \Biggl[
        \frac{4 P_{252}
        }{(N-1) N^4 (1+N)^4 (2+N)}
        +
        \frac{8 (2+N) \big(
                2+N-N^2+2 N^3\big)}{N^3 (1+N)^3} S_1
\nonumber\\ &&  
      +\frac{4 (N-1) (2+N)}{N^2 (1+N)^2} S_1^2
        -\frac{4 (N-1) (2+N)}{N^2 (1+N)^2} S_2
        -\frac{64}{(N-1) N (1+N) (2+N)} 
\nonumber\\ &&  \times
S_{-2}
\Biggr] ~,
\\
\label{eq:g1PS2}
 C_{g_1,q}^{(2),\rm PS,M} &=& 
 C_{g_1,q}^{(2),\rm NS,L} -  z_{qq}^{(2),\rm PS}
 \nonumber\\ &=&
 C_{g_1,q}^{(2),\rm NS,L} + 
 \textcolor{blue}{C_F T_F N_F} \frac{8 (2+N) \big(-1-N+N^2\big)}{N^3 (1+N)^3} ~,
\\
C_{g_1,g}^{(2)} &=& 
\textcolor{blue}{C_A T_F N_F} 
\Biggl[
        -\frac{4 P_{255}}{(N-1) N^4 (1+N)^4 (2+N)^2}
        +\Biggl(
                -\frac{4 P_{250}}{N^3 (1+N)^3 (2+N)}
\nonumber\\ && 
                +\frac{20 (N-1)}{N (1+N)} S_2
        \Biggr) S_1
        -\frac{8 \big(
                3-N^2+N^3\big)}{N^2 (1+N)^2} S_1^2
        +\frac{8 \big(
                3-2 N+N^2+N^3\big)}{N^2 (1+N)^2} S_2
\nonumber\\ && 
        -\frac{4 (N-1)}{3 N (1+N)} S_1^3
        +\frac{8 \big(
                2+5 N+5 N^2\big)}{3 N (1+N) (2+N)} S_3
        +\Biggl(
                \frac{16 P_{163}}{(N-1) N (1+N)^2 (2+N)^2}
\nonumber\\ &&                
 +\frac{16 \big(
                        2+N+N^2\big)}{N (1+N) (2+N)} S_1
        \Biggr) S_{-2}
        -\frac{16 \big(
                -4+N+N^2\big)}{N (1+N) (2+N)} S_{-3}
        -\frac{16 (N-1)}{N (1+N)} S_{2,1}
\nonumber\\ &&        
 -\frac{64 }{N (1+N) (2+N)} S_{-2,1}
        -\frac{24 \big(
                2+N+N^2\big)}{N (1+N) (2+N)} \zeta_3
\Biggr]
\nonumber\\ && 
+\textcolor{blue}{C_F T_F N_F} \Biggl[
        -\frac{2 P_{254}}{(N-1) N^4 (1+N)^4 (2+N)^2}
        +\Biggl(
                \frac{4 P_{249}}{N^3 (1+N)^3 (2+N)}
\nonumber\\ &&                
 +\frac{12 (N-1)}{N (1+N)} S_2
        \Biggr) S_1
        -\frac{2 (N-1) \big(
                -8+3 N+9 N^2\big)}{N^2 (1+N)^2} S_1^2
        -\frac{20 (N-1)}{3 N (1+N)} S_1^3
\nonumber\\ &&   
      +\frac{2 \big(
                4-19 N-10 N^2+9 N^3\big)}{N^2 (1+N)^2} S_2
        -\frac{64 \big(
                1+N+N^2\big)}{3 N (1+N) (2+N)} S_3
\nonumber\\ && 
        +\Biggl(
                -\frac{128}{N (1+N) (2+N)} S_1
+                \frac{16 \big(
                        10+N+N^2\big)}{(N-1) (2+N)^2}
        \Biggr) S_{-2}
        -\frac{64}{N (1+N) (2+N)} S_{-3}
\nonumber\\ && 
        +\frac{16 (N-1)}{N (1+N)} S_{2,1}
        +\frac{128}{N (1+N) (2+N)} S_{-2,1}
        +\frac{48 \big(
                2+N+N^2\big)}{N (1+N) (2+N)} \zeta_3
\Biggr] ~,
\end{eqnarray}

\noindent
with the polynomials
\begin{eqnarray}
P_{245} &=& -27 N^4-42 N^3-47 N^2+24~,
\\
P_{246} &=& 95 N^4+178 N^3+123 N^2+40 N-16~,
\\
P_{247} &=& 103 N^4+140 N^3+58 N^2+21 N+36~,
\\
P_{239} &=& 247 N^4+548 N^3+223 N^2+30 N+72~,
\\
P_{248} &=& -51 N^6-131 N^5-163 N^4-81 N^3+26 N^2-8 N-16~,
\\
P_{249} &=& N^6+5 N^5+3 N^4-13 N^3-20 N^2+12 N+16~,
\\
P_{250} &=& 7 N^6+16 N^5-11 N^4-18 N^3+6 N^2-44 N-32~,
\\
P_{251} &=& 1371 N^6+3429 N^5+3437 N^4+655 N^3-940 N^2+72 N+360~,
\\
P_{242} &=& 3155 N^6+9951 N^5+9867 N^4+3473 N^3+546 N^2-72 N-432~,
\\
P_{252} &=& 6 N^8+20 N^7+2 N^6-36 N^5-43 N^4-62 N^3-47 N^2+12 N+20~,
\\
P_{253} &=& -16395 N^{10}-74235 N^9-111388 N^8-28126 N^7+111413 N^6
+125177 N^5
\nonumber\\ &&
+41930 N^4
+12464 N^3+23688 N^2+3600 N-5184~,
\\
P_{254} &=& 2 N^{10}+18 N^9+26 N^8-42 N^7-253 N^6-563 N^5-633 N^4
-181 N^3+126 N^2
\nonumber\\ &&
+4 N-40~,
\\
P_{255} &=& 4 N^{10}+18 N^9-2 N^8-74 N^7
+5 N^6+141 N^5+99 N^4+171 N^3+122 N^2
\nonumber\\ &&
-44 N-56~,
\\
P_{256} &=& 331 N^{10}+1451 N^9+2196 N^8+606 N^7-2293 
N^6-2697 N^5-2506 N^4
\nonumber\\ && 
-2336 N^3
-1232 N^2+112 N+224~,
\end{eqnarray}

\noindent
with the coefficients $z_{ij}^{(l)}$ given in \cite{Moch:2014sna,Blumlein:2021ryt}.
Up to two--loop order, the Wilson coefficients are given in different schemes \cite{Zijlstra:1993sh,
Vogt:2008yw,Blumlein:2019zux,Blumlein:2021xlc}. In particular the non--singlet Wilson coefficients
are also given in the $\overline{\sf MS}$ scheme \cite{Moch:2008fj,Behring:2015zaa}.
\section{The three--loop Wilson coefficients for the structure functions 
\boldmath  $F_2(x,Q^2)$ and $F_L(x,Q^2)$}
\label{sec:5}

\vspace*{1mm}
\noindent
Before we present the three--loop Wilson coefficients we would like to remark that the
unrenormalized three--loop forward Compton amplitudes depend on the complete 
anomalous 
dimensions up to two--loop order, cf. e.g. \cite{Blumlein:2022ndg}, and contain the following 
three--loop anomalous dimensions in their pole terms of $O(1/\ep)$: {$\gamma_{qq}^{(2), \rm NS},
\gamma_{qq}^{(2), \rm PS}, 
\Delta \gamma_{qq}^{(2), \rm NS}, \Delta \gamma_{qq}^{(2), \rm PS}
, \gamma_{qg}^{(2)}$ and $\Delta \gamma_{qg}^{(2)}$}. We confirm 
the previous results given in Refs.~\cite{Moch:2004pa,Vogt:2004mw,Blumlein:2021enk,
Ablinger:2014nga,
Ablinger:2017tan,Moch:2014sna,Behring:2019tus,Blumlein:2021ryt,HIDY,Gehrmann:2022euk}.

We now turn to the unpolarized three--loop Wilson coefficients for neutral--current deep--inelastic 
scattering for pure virtual photon exchange, $F_2(x,Q^2)$ and $F_L(x,Q^2)$. In the structure 
of the following expressions there are some evanescent poles at $N = 1,2$, which are, however, all
tractable. Therefore, the analytic continuation \cite{Blumlein:2009ta} to $N \in \mathbb{N}$ can be 
carried out directly.

The non--singlet and gluonic Wilson coefficients are split for convenience as
\begin{eqnarray}
\mathbb{C}_{i,q}^{(3)} &=& C_{i,q}^{\rm NS, (3)} + C_{i,q}^{d_{abc}, (3)} + C_{i,q}^{\rm PS, (3)}
\\
\mathbb{C}_{i,g}^{(3)} &=& C_{i,g}^{\rm NS, (3)} + C_{i,g}^{d_{abc}, (3)}.
\end{eqnarray}

The Wilson coefficients are given by

\begin{eqnarray}
&& C_{F_L,q}^{\rm NS, (3)} = 
\nonumber\\  &&
\textcolor{blue}{C_F^2} \Biggl\{
        \textcolor{blue}{T_F N_F}\Biggl[
                \frac{8 S_2 P_{303}}{3 N (1+N)^2 (2+N) (3+N)}
                -\frac{64 \zeta_3 P_{331}}{3 (N-2) (N-1) N (1+N) 
(2+N) (3+N)}
\nonumber\\ &&
                +\frac{2 P_{446}}{27 (N-2)^2 N^3 (1+N)^4 (3+N)}
                +\Biggl[
                        \frac{32 P_{425}
                        }{27 (N-2) (N-1) N^2 (1+N)^3 (2+N) (3+N)}
\nonumber\\ &&                       
 -
                        \frac{64 (4+3 N) \big(
                                -5+N^2\big) S_2}{3 (1+N)^2 (2+N) 
(3+N)}
                        +\frac{128 S_3}{3 (1+N)}
                        -\frac{256 S_{2,1}}{3 (1+N)}
                        -\frac{1024 S_{-2,1}}{3 (1+N)}
                        -\frac{1280 \zeta_3}{3 (1+N)}
                \Biggr] S_1
\nonumber\\ &&
                +\Biggl[
                        \frac{8 \big(
                                30+83 N+23 N^2\big)}{9 N (1+N)^2}
                        +\frac{64 S_2}{3 (1+N)}
                \Biggr] S_1^2
                +\frac{32 \big(
                        294+553 N+270 N^2+35 N^3\big) S_3}{9 (1+N)^2 
(2+N) (3+N)}
\nonumber\\ &&
                -\frac{128 S_1^3}{9 (1+N)}
                -\frac{128 S_2^2}{3 (1+N)}
                -\frac{1088 S_4}{3 (1+N)}
                +\Biggl[
                        \frac{512 S_1 P_{329}}{9 (N-2) (N-1) N 
(1+N) (2+N) (3+N)}
\nonumber\\ &&
                        -\frac{256 P_{436}}{9 (N-2)^2 (N-1) N^2 
(1+N)^3 (2+N) (3+N)}
                        +\frac{512 S_1^2}{3 (1+N)}
                        -\frac{512 S_2}{3 (1+N)}
                \Biggr] S_{-2}
                -\frac{256 S_{-2}^2}{3 (1+N)}
\nonumber\\ &&                
+\Biggl[
                        \frac{256 P_{343}}{9 (N-2) (N-1) N (1+N)^2 
(2+N) (3+N)}
                        -\frac{512 S_1}{3 (1+N)}
                \Biggr] S_{-3}
                -\frac{256 S_{-4}}{1+N}
\nonumber\\ &&
                +\frac{64 \big(
                        -10+25 N+38 N^2+11 N^3\big) S_{2,1}}{3 
(1+N)^2 (2+N) (3+N)}
                +\frac{128 S_{3,1}}{1+N}
                -\frac{512 \big(
                        -16-23 N+N^2+2 N^3\big) S_{-2,1}}{9 (N-1) 
(1+N)^2 (2+N)}
\nonumber\\ &&
                +\frac{512 S_{-3,1}}{3 (1+N)}
                +\frac{128 S_{2,1,1}}{1+N}
                +\frac{1024 S_{-2,1,1}
                }{3 (1+N)}
        \Biggr]
        +\textcolor{blue}{C_A} \Biggl[
                -
                \frac{64 S_{-2,2} P_{300}}{(N-2) N (1+N)^2 (3+N)}
\nonumber\\ &&            
    +\frac{128 S_{-2,1,1} P_{306}}{3 (N-2) N (1+N)^2 
(3+N)}
                -\frac{64 S_{-3,1} P_{310}}{3 (N-2) N (1+N)^2 (3+N)}
                -\frac{16 S_{2,1} P_{311}}{3 N (1+N)^2 (2+N) (3+N)}
\nonumber\\ &&              
  -\frac{64 S_{3,1} P_{344}}{(N-2) (N-1) N (1+N)^2 
(2+N) (3+N)}
                +\frac{32 S_4 P_{367}}{3 (N-2) (N-1) N (1+N)^2 
(2+N) (3+N)}
\nonumber\\ &&
                +\frac{16 \zeta_3 P_{407}}{3 (N-2) (N-1) N^2 
(1+N)^2 (2+N) (3+N)}
                +\frac{P_{434}}{54 (N-2)^2 N^3 (1+N)^4 (3+N)}
\nonumber\\ &&             
   +\frac{32 S_{-2,1} P_{408}}{9 (N-2) (N-1) N (1+N)^3 
(2+N) (3+N)}
\nonumber\\ &&
                -\frac{8 S_3 P_{410}}{9 (N-2) (N-1) N (1+N)^3 (2+N) 
(3+N)}
                +\Biggl(
	                        \frac{16 S_2 P_{294}}{3 N (1+N)^2 (2+N) 
(3+N)}
\nonumber\\ &&
                        -\frac{16 \zeta_3 P_{349}}{3 (N-2) (N-1) N 
(1+N)^2 (2+N) (3+N)}
                        +\frac{64 S_4}{1+N}
                        +\frac{512 S_{2,1}}{3 (1+N)}
                        -\frac{512 S_{3,1}}{1+N}
\nonumber\\ &&                    
                        +\frac{16 S_3 P_{360}}{3 (N-2) (N-1) N (1+N)^2 (2+N) (3+N)}
                        -\frac{32 S_{-2,1} P_{309}}{3 (N-2) N (1+N)^2 (3+N)} 
\nonumber \\ && 
                        +\frac{3712 S_{-2,2}}{1+N}
                        +\frac{3840 S_{-3,1}}{1+N}
                        -\frac{7168 S_{-2,1,1}}{1+N}
\nonumber\\ &&
                        -\frac{4 P_{464}}{27 (N-2)^2 (N-1) N^3 
(1+N)^4 (2+N) (3+N)}
                \Biggr) S_1
                +\Biggl(
                        -\frac{128 S_2}{3 (1+N)}
                        +\frac{160 S_3}{1+N}
                        +\frac{1728 S_{-2,1}}{1+N}
\nonumber\\ &&
                        +\frac{160 \zeta_3}{1+N}
                        -
                        \frac{2 P_{315}}{9 (N-2) N (1+N)^2 (3+N)}
                \Biggr) S_1^2
                +\frac{640 S_1^3}{9 (1+N)}
                +\Biggl(
                        -\frac{416 S_3}{1+N}
                        -\frac{5312 S_{-2,1}}{1+N}
                        +\frac{96 \zeta_3}{1+N}
\nonumber\\ &&
                        -\frac{2 P_{377}}{3 (N-2) N (1+N)^3 (2+N) 
(3+N)}
                \Biggr) S_2
                +\frac{304 S_2^2}{3 (1+N)}
                -\frac{64 (N-1) (2+N) S_5}{1+N}
\nonumber\\ &&           
     +\Biggl(
                        \frac{64 S_2 P_{299}}{3 (N-2) N (1+N)^2 
(3+N)}
                        +\frac{16 P_{461}}{9 (N-2)^2 (N-1) N^3 
(1+N)^4 (2+N) (3+N)}
\nonumber\\ &&
                        +\Biggl(
                                -\frac{32 P_{441}}{9 (N-2)^2 (N-1) 
N^2 (1+N)^3 (2+N) (3+N)}
                                -\frac{192 S_2}{1+N}
                        \Biggr) S_1
                        +\frac{64 S_1^3}{1+N}
\nonumber\\ &&
                        -\frac{64 \big(
                                -5+2 N+2 N^2\big) S_3}{1+N}
                        -\frac{32 \big(
                                -72-108 N+35 N^2+23 N^3\big) S_1^2}{3 
(N-2) N (1+N) (3+N)}
                        +\frac{5376 S_{2,1}}{1+N}
\nonumber\\ &&                        
-\frac{128 \big(
                                -3+N+N^2\big) S_{-2,1}}{1+N}
                        -\frac{64 \big(
                                -21+2 N+2 N^2\big) \zeta_3}{1+N}
                \Biggr) S_{-2}
\nonumber\\ &&
                +\Biggl(
                        \frac{16 P_{307}}{3 (N-2) N (1+N)^2 (3+N)}
                        +\frac{384 S_1}{1+N}
                \Biggr) S_{-2}^2
                +\Biggl(
                        \frac{16 S_1 P_{318}}{3 (N-2) N (1+N)^2 
(3+N)}
\nonumber\\ &&
                        -\frac{16 P_{427}
                        }{9 (N-2) (N-1) N^2 (1+N)^3 (2+N) (3+N)}
                        -
                        \frac{1120 S_1^2}{1+N}
                        +\frac{2656 S_2}{1+N}
                        +\frac{448 S_{-2}}{1+N}
                \Biggr) S_{-3}
\nonumber\\ &&
                +\Biggl(
                        \frac{32 P_{305}}{(N-2) N (1+N)^2 (3+N)}
                        -\frac{1600 S_1}{1+N}
                \Biggr) S_{-4}
                -\frac{64 \big(
                        -34+N+N^2\big) S_{-5}}{1+N}
                +\frac{384 S_{2,3}}{1+N}
\nonumber\\ &&
                +\frac{2688 S_{2,-3}}{1+N}
                -\frac{384 S_{4,1}}{1+N}
                +\frac{64 \big(
                        -65+2 N+2 N^2\big) S_{-2,3}}{1+N}
                +\frac{2496 S_{-4,1}}{1+N}
                -\frac{256 S_{2,1,1}}{1+N}
\nonumber\\ &&
                -\frac{5376 S_{2,1,-2}}{1+N}
                +\frac{768 S_{3,1,1}}{1+N}
                +\frac{128 \big(
                        -10+N+N^2\big) S_{-2,1,-2}}{1+N}
                -\frac{5376 S_{-2,2,1}}{1+N}
                -\frac{5376 S_{-3,1,1}}{1+N}
\nonumber\\ &&
                +\frac{10752 S_{-2,1,1,1}}{1+N}
                -\frac{160 \big(
                        -21+N+N^2\big) \zeta_5}{1+N}
        \Biggr]
\Biggr\}
\nonumber\\ &&
+\textcolor{blue}{C_F} \Biggl\{
        \textcolor{blue}{T_F^2 N_F^2} \Biggl[
                \frac{32 P_{313}}{81 N^2 (1+N)^3}
                +\frac{64 \big(
                        -6+7 N+19 N^2\big) S_1}{27 N (1+N)^2}
                +\frac{64 S_1^2}{9 (1+N)}
                -\frac{64 S_2}{9 (1+N)}
        \Biggr]
\nonumber\\ &&
        +\textcolor{blue}{C_A T_F N_F} \Biggl[
                \frac{64 S_{3,1} P_{274}}{(N-2) (N-1) (1+N) (2+N) 
(3+N)}
\nonumber\\ &&
                +\frac{32 S_4 P_{291}}{3 (N-2) (N-1) (1+N) (2+N) 
(3+N)}
                +\frac{64 S_3 P_{328}}{9 (N-2) (N-1) (1+N)^2 (2+N) 
(3+N)}
\nonumber\\ &&
                -
                \frac{16 P_{415}}{81 (N-2)^2 N^2 (1+N)^3 (3+N)}
                +\Biggl(
                        \frac{1024 \zeta_3 P_{272}}{3 (N-2) (N-1) 
(1+N) (2+N) (3+N)}
\nonumber\\ &&
                        -\frac{256 S_3 P_{273}}{3 (N-2) (N-1) (1+N) 
(2+N) (3+N)}
                        -\frac{16 P_{426}}{27 (N-2) (N-1) N^2 
(1+N)^3 (2+N) (3+N)}
\nonumber\\ &&
                        +\frac{64 (N-1) (3+2 N) S_2}{3 (1+N)^2 
(3+N)}
                        +\frac{128 S_{2,1}}{3 (1+N)}
                        +\frac{512 S_{-2,1}}{3 (1+N)}
                \Biggr) S_1
                +\Biggl(
                        -\frac{640}{9 (1+N)}
                        -\frac{32 S_2}{3 (1+N)}
                \Biggr) S_1^2
\nonumber\\ &&
                +\frac{64 \big(
                        21+28 N+N^2\big) S_2}{9 (1+N)^2 (3+N)}
                +\frac{64 S_2^2}{3 (1+N)}
                +\Biggl(
                        -\frac{256 S_1 P_{329}}{9 (N-2) (N-1) N 
(1+N) (2+N) (3+N)}
\nonumber\\ &&
                        +\frac{128 P_{436}}{9 (N-2)^2 (N-1) N^2 
(1+N)^3 (2+N) (3+N)}
                        -\frac{256 S_1^2}{3 (1+N)}
                        +\frac{256 S_2}{3 (1+N)}
                \Biggr) S_{-2}
                +\frac{128 S_{-2}^2}{3 (1+N)}
\nonumber\\ &&
                +\Biggl(
                        -\frac{128 P_{343}}{9 (N-2) (N-1) N (1+N)^2 
(2+N) (3+N)}
                        +\frac{256 S_1}{3 (1+N)}
                \Biggr) S_{-3}
                +\frac{128 S_{-4}}{1+N}
\nonumber\\ &&
                -\frac{128 (N-1) (3+2 N) S_{2,1}}{3 (1+N)^2 (3+N)}
                +\frac{256 \big(
                        -16-23 N+N^2+2 N^3\big) S_{-2,1}}{9 (N-1) 
(1+N)^2 (2+N)}
                -\frac{256 S_{-3,1}}{3 (1+N)}
                -\frac{64 S_{2,1,1}
                }{1+N}
\nonumber\\ &&
                -
                \frac{512 S_{-2,1,1}}{3 (1+N)}
                +\frac{128 \big(
                        36-45 N-22 N^2+2 N^3\big) \zeta_3}{3 (N-2) N 
(1+N) (3+N)}
        \Biggr]
\nonumber\\ &&
        +\textcolor{blue}{C_A^2} \Biggl[
                -\frac{1024 S_{-2,1,1} P_{279}}{3 (N-2) N (1+N)^2 
(3+N)}
                +\frac{64 S_{-2,2} P_{288}}{(N-2) N (1+N)^2 (3+N)}
\nonumber\\ &&
                +\frac{128 S_{-3,1} P_{297}}{3 (N-2) N (1+N)^2 
(3+N)}
                -\frac{16 S_{3,1} P_{342}}{(N-2) (N-1) N (1+N)^2 
(2+N) (3+N)}
\nonumber\\ &&
                -\frac{8 S_4 P_{365}}{3 (N-2) (N-1) N (1+N)^2 (2+N) 
(3+N)}
                +\frac{16 S_3 P_{368}}{9 (N-2) (N-1) N (1+N)^2 
(2+N) (3+N)}
\nonumber\\ &&
                -\frac{32 S_{-2,1} P_{373}}{9 (N-2) (N-1) N (1+N)^2 
(2+N) (3+N)}
                -\frac{8 \zeta_3 P_{374}}{3 (N-2) (N-1) N (1+N)^2 
(2+N) (3+N)}
\nonumber\\ &&
                +\frac{4 P_{420}}{81 (N-2)^2 N^2 (1+N)^3 (3+N)}
                +\Biggl(
                        \frac{512 S_{-2,1} P_{275}}{3 (N-2) N (1+N)^2 
(3+N)}
\nonumber\\ &&
                        +\frac{16 S_3 P_{352}}{3 (N-2) (N-1) N 
(1+N)^2 (2+N) (3+N)}
                        -\frac{32 \zeta_3 P_{355}}{3 (N-2) (N-1) N 
(1+N)^2 (2+N) (3+N)}
\nonumber\\ &&
                        +\frac{4 P_{456}}{27 (N-2) (N-1) N^3 
(1+N)^4 (2+N) (3+N)}
                        -\frac{176 (N-1) (3+2 N) S_2}{3 (1+N)^2 
(3+N)}
	                        -\frac{192 S_4}{1+N}
\nonumber\\ &&
                        -\frac{352 S_{2,1}}{3 (1+N)}
                        +\frac{384 S_{3,1}
                        }{1+N}
                        -
                        \frac{1024 S_{-2,2}}{1+N}
                        -\frac{1024 S_{-3,1}}{1+N}
                        +\frac{2048 S_{-2,1,1}}{1+N}
                \Biggr) S_1
                +\Biggl(
                        \frac{1276}{9 (1+N)}
\nonumber\\ &&                       
                        +\frac{88 S_2}{3 (1+N)}
 -\frac{96 S_3}{1+N}
                        -\frac{512 S_{-2,1}}{1+N}
                        -\frac{64 \zeta_3}{1+N}
                \Biggr) S_1^2
                +\Biggl(
                        \frac{44 \big(
                                -51-68 N+7 N^2\big)}{9 (1+N)^2 (3+N)}
                        +\frac{288 S_3}{1+N}
\nonumber\\ &&                        
+\frac{1536 S_{-2,1}}{1+N}
                \Biggr) S_2
                -\frac{176 S_2^2}{3 (1+N)}
                +\frac{16 \big(
                        4+N+N^2\big) S_5}{1+N}
                +\Biggl(
                        \frac{512 S_1^2}{3 (1+N)}
                        -\frac{704 S_2}{3 (1+N)}
\nonumber\\ &&                        
+\frac{32 S_1 P_{366}}{9 (N-2) (N-1) N 
(1+N)^2 (2+N) (3+N)}
                        +\frac{32 \big(
                                -4+N+N^2\big) S_3}{1+N}
\nonumber\\ &&
                        -\frac{16 P_{442}}{9 (N-2)^2 (N-1) N^2 
(1+N)^3 (2+N) (3+N)}
                        -\frac{1536 S_{2,1}}{1+N}
                        +\frac{32 \big(
                                -4+N+N^2\big) S_{-2,1}}{1+N}
\nonumber\\ &&                   
      +\frac{32 (N-3) (4+N) \zeta_3}{1+N}
                \Biggr) S_{-2}
                +\Biggl(
                        -\frac{16 P_{301}}{3 (N-2) N (1+N)^2 (3+N)}
                        -\frac{128 S_1}{1+N}
                \Biggr) S_{-2}^2
\nonumber\\ &&
                +\Biggl(
                        -\frac{128 S_1 P_{293}}{3 (N-2) N (1+N)^2 
(3+N)}
                        +\frac{16 P_{378}}{9 (N-2) (N-1) N (1+N)^2 
(2+N) (3+N)}
\nonumber\\ && 
                        +\frac{256 S_1^2}{1+N}
                        -\frac{768 S_2}{1+N}
                        -\frac{128 S_{-2}}{1+N}
                \Biggr) S_{-3}
                +\Biggl(
                        -
                        \frac{16 P_{302}}{(N-2) N (1+N)^2 (3+N)}
                        +\frac{384 S_1}{1+N}
                \Biggr) S_{-4}
\nonumber\\ &&
                +\frac{16 \big(
                        -40+N+N^2\big) S_{-5}}{1+N}
                +\frac{352 (N-1) (3+2 N) S_{2,1}}{3 (1+N)^2 (3+N)}
                -\frac{288 S_{2,3}}{1+N}
                -\frac{768 S_{2,-3}}{1+N}
                +\frac{288 S_{4,1}}{1+N}
\nonumber\\ &&
                -\frac{32 \big(
                        -38+N+N^2\big) S_{-2,3}}{1+N}
                -\frac{704 S_{-4,1}}{1+N}
                +\frac{176 S_{2,1,1}}{1+N}
                +\frac{1536 S_{2,1,-2}}{1+N}
                -\frac{576 S_{3,1,1}}{1+N}
\nonumber\\ &&
                -\frac{32 (N-3) (4+N) S_{-2,1,-2}}{1+N}
                +\frac{1536 S_{-2,2,1}}{1+N}
                +\frac{1536 S_{-3,1,1}}{1+N}
                -\frac{3072 S_{-2,1,1,1}}{1+N}
\nonumber\\ &&
                +\frac{40 \big(
                        -16+N+N^2\big) \zeta_5}{1+N}
        \Biggr]
\Biggr\}
\nonumber\\ &&
+\textcolor{blue}{C_F^3} \Biggl\{
        -\frac{256 S_{-2,1,1} P_{289}}{(N-2) N (1+N)^2 (3+N)}
        +\frac{128 S_{-2,2} P_{290}}{(N-2) N (1+N)^2 (3+N)}
\nonumber\\ &&   
     +\frac{128 S_{-3,1} P_{292}}{(N-2) N (1+N)^2 (3+N)}
        +\frac{32 S_{3,1} P_{357}}{(N-2) (N-1) N (1+N)^2 (2+N) 
(3+N)}
\nonumber\\ &&
        -\frac{16 S_4 P_{359}}{(N-2) (N-1) N (1+N)^2 (2+N) (3+N)}
        -\frac{64 S_{-2,1} P_{396}}{(N-2) (N-1) N (1+N)^3 (2+N) 
(3+N)}
\nonumber\\ &&
        -\frac{16 \zeta_3 P_{401}}{(N-2) (N-1) N^2 (1+N)^2 (2+N) 
(3+N)}
        +
        \frac{8 S_3 P_{403}}{(N-2) (N-1) N (1+N)^3 (2+N) (3+N)}
\nonumber\\ &&
        +\frac{P_{462}}{6 (N-2)^2 N^4 (1+N)^5 (3+N)}
        +\Biggl(
                -\frac{64 S_3 P_{345}}{(N-2) (N-1) N (1+N)^2 (2+N) 
(3+N)}
\nonumber\\ &&
               +\frac{64 S_{-2,1} P_{298}}{(N-2) N (1+N)^2 (3+N)}
                +\frac{32 \zeta_3 P_{354}}{(N-2) (N-1) N (1+N)^2 
(2+N) (3+N)}
\nonumber\\ &&
                +\frac{8 P_{460}}{(N-2)^2 (N-1) N^3 (1+N)^4 (2+N) 
(3+N)}
                +\frac{8 (1+3 N) (10+9 N) S_2}{N (1+N)^2}
                +\frac{640 S_4}{1+N}
                +\frac{64 S_{2,1}}{1+N}
\nonumber\\ &&
                -\frac{512 S_{3,1}}{1+N}
                -\frac{3328 S_{-2,2}}{1+N}
                -\frac{3584 S_{-3,1}}{1+N}
                +\frac{6144 S_{-2,1,1}}{1+N}
        \Biggr) S_1
        +\Biggl(
                -\frac{2 P_{330}}{(N-2) N^2 (1+N)^2 (3+N)}
\nonumber\\ &&            
    -\frac{80 S_2}{1+N}
                +\frac{64 S_3}{1+N}
                -\frac{1408 S_{-2,1}}{1+N}
                -\frac{64 \zeta_3}{1+N}
        \Biggr) S_1^2
        -\frac{8 \big(
                2+13 N+9 N^2\big) S_1^3}{N (1+N)^2}
        +\frac{8 S_1^4}{1+N}
\nonumber\\ &&
        +\Biggl(
                \frac{2 P_{346}}{(N-2) N^2 (1+N)^3 (3+N)}
                -\frac{320 S_3}{1+N}
                +\frac{4480 S_{-2,1}}{1+N}
                -\frac{192 \zeta_3}{1+N}
        \Biggr) S_2
        +\frac{24 S_2^2}{1+N}
\nonumber\\ &&
        +\frac{64 \big(
                -8+N+N^2\big) S_5}{1+N}
        +\Biggl(
                -
                \frac{32 P_{459}}{(N-2)^2 (N-1) N^3 (1+N)^4 (2+N) 
(3+N)}
\nonumber\\ &&
                -\frac{128 S_2 P_{278}}{(N-2) N (1+N)^2 (3+N)}
                +\Biggl(
                        \frac{64 P_{437}}{(N-2)^2 (N-1) N^2 (1+N)^3 
(2+N) (3+N)}
                        +\frac{384 S_2}{1+N}
                \Biggr) S_1
\nonumber\\ &&
                -\frac{64 \big(
                        24-28 N-N^2+3 N^3\big) S_1^2}{(N-2) N (1+N) 
(3+N)}
                -\frac{128 S_1^3}{1+N}
                +\frac{128 \big(
                        -1+N+N^2\big) S_3}{1+N}
                -\frac{4608 S_{2,1}}{1+N}
\nonumber\\ &&                
+\frac{128 (N-1) (2+N) S_{-2,1}}{1+N}
                +\frac{128 \big(
                        -9+N+N^2\big) \zeta_3}{1+N}
        \Biggr) S_{-2}
        +\Biggl(
                -\frac{256 S_1}{1+N}
\nonumber\\ &&                
-\frac{32 \big(
                        -8+N+N^2
                \big)
\big(-6+5 N+5 N^2\big)}{(N-2) N (1+N)^2 (3+N)}
        \Biggr) S_{-2}^2
        +\Biggl(
                -\frac{32 S_1 P_{304}}{(N-2) N (1+N)^2 (3+N)}
\nonumber\\ &&             
   +\frac{32 P_{424}}{(N-2) (N-1) N^2 (1+N)^3 (2+N) 
(3+N)}
                +\frac{1216 S_1^2}{1+N}
                -\frac{2240 S_2}{1+N}
                -\frac{384 S_{-2}}{1+N}
        \Biggr) S_{-3}
\nonumber\\ &&
        +\Biggl(
                -\frac{64 P_{295}}{(N-2) N (1+N)^2 (3+N)}
                +\frac{1664 S_1}{1+N}
        \Biggr) S_{-4}
        +\frac{64 \big(
                -28+N+N^2\big) S_{-5}}{1+N}
\nonumber\\ &&   
     +\frac{16 \big(
                -2+3 N+3 N^2\big) S_{2,1}}{N (1+N)^2}
        +\frac{384 S_{2,3}}{1+N}
        -\frac{2304 S_{2,-3}}{1+N}
        -\frac{384 S_{4,1}}{1+N}
        -\frac{128 \big(
                -27+N+N^2\big) S_{-2,3}
        }{1+N}
\nonumber\\ &&
        -\frac{2176 S_{-4,1}}{1+N}
        -\frac{96 S_{2,1,1}}{1+N}
        +\frac{4608 S_{2,1,-2}}{1+N}
        +\frac{768 S_{3,1,1}}{1+N}
        -\frac{128 \big(
                -8+N+N^2\big) S_{-2,1,-2}}{1+N}
\nonumber\\ &&
        +\frac{4608 S_{-2,2,1}}{1+N}
        +\frac{4608 S_{-3,1,1}}{1+N}
        -\frac{9216 S_{-2,1,1,1}}{1+N}
        +\frac{160 \big(
                -26+N+N^2\big) \zeta_5}{1+N}
\Biggr\}
\nonumber\\ &&
\end{eqnarray}


\noindent
with
\begin{eqnarray}
    P_{257}&=&-617 N^4-1108 N^3-281 N^2+222 N+148,\\ 
    P_{258}&=&-9 N^4-4 N^3+3 N^2+26 N+20,\\ 
    P_{259}&=&N^4-22 N^3-325 N^2+322 N-312,\\ 
    P_{260}&=&N^4-22 N^3-237 N^2+74 N-192,\\ 
    P_{261}&=&N^4-22 N^3-226 N^2-11 N-280,\\ 
    P_{262}&=&N^4-22 N^3-217 N^2+310 N-240,\\ 
    P_{263}&=&N^4-22 N^3-137 N^2+138 N-180,\\ 
    P_{264}&=&N^4-22 N^3-133 N^2-14 N-84,\\ 
    P_{265}&=&N^4-22 N^3-112 N^2+583 N-502,\\ 
    P_{266}&=&N^4-22 N^3-64 N^2+7 N-34,\\ 
    P_{267}&=&N^4-22 N^3-37 N^2+202 N-168,\\ 
    P_{268}&=&N^4-22 N^3-37 N^2+634 N-518,\\ 
    P_{269}&=&N^4-22 N^3-N^2+406 N-96,\\ 
    P_{270}&=&N^4-22 N^3+98 N^2+313 N-354,\\ 
    P_{271}&=&N^4-22 N^3+271 N^2+582 N-500,\\ 
    P_{272}&=&N^4+2 N^3-4 N^2-5 N+9,\\ 
    P_{273}&=&N^4+2 N^3-N^2-2 N+6,\\ 
    P_{274}&=&N^4+2 N^3+9 N^2+8 N-4,\\ 
    P_{275}&=&N^4+2 N^3+16 N^2+15 N-36,\\ 
    P_{276}&=&2 N^4-44 N^3-197 N^2-7 N-118,\\ 
    P_{277}&=&2 N^4-44 N^3+369 N^2+895 N-854,\\ 
    P_{278}&=&2 N^4+4 N^3-5 N^2-7 N-12,\\ 
    P_{279}&=&2 N^4+4 N^3+11 N^2+9 N-36,\\ 
    P_{280}&=&3 N^4-66 N^3-1501 N^2-1720 N+192,\\ 
    P_{281}&=&3 N^4-66 N^3-1387 N^2+1562 N-1188,\\ 
    P_{282}&=&3 N^4-66 N^3-221 N^2+2584 N-1776,\\ 
    P_{283}&=&3 N^4-66 N^3-166 N^2+2243 N-1665,\\ 
    P_{284}&=&5 N^4+10 N^3+N^2-4 N-4,\\ 
    P_{285}&=&6 N^4-132 N^3-2003 N^2-1577 N-792,\\ 
    P_{286}&=&6 N^4-132 N^3-1723 N^2+3311 N-2694,\\ 
    P_{287}&=&9 N^4-198 N^3-1123 N^2-52 N-576,\\ 
    P_{288}&=&9 N^4+18 N^3+11 N^2+2 N-96,\\ 
    P_{289}&=&9 N^4+18 N^3+43 N^2+34 N-144,\\ 
    P_{290}&=&10 N^4+20 N^3+39 N^2+29 N-150,\\ 
    P_{291}&=&11 N^4+22 N^3-125 N^2-136 N+180,\\ 
    P_{292}&=&11 N^4+22 N^3+35 N^2+24 N-156,\\ 
    P_{293}&=&13 N^4+26 N^3-23 N^2-36 N-72,\\ 
    P_{294}&=&15 N^4-64 N^3-366 N^2-343 N-18,\\ 
    P_{295}&=&15 N^4+30 N^3+8 N^2-7 N-138,\\ 
    P_{296}&=&18 N^4-396 N^3-1537 N^2-43 N-1296,\\ 
    P_{297}&=&19 N^4+38 N^3-11 N^2-30 N-144,\\ 
    P_{298}&=&23 N^4+42 N^3+55 N^2+60 N-276,\\ 
    P_{299}&=&28 N^4+56 N^3-125 
    N^2-153 N-36,\\ P_{300}&=&28 N^4+56 N^3+61 N^2+33 N-342,\\ P_{301}&=&31 N^4+62 
    N^3-203 N^2-234 N+144,\\ P_{302}&=&43 N^4+86 N^3-119 N^2-162 
    N-144,\\ P_{303}&=&43 N^4+118 N^3+19 N^2-104 N-60,\\ P_{304}&=&47 N^4+90 N^3+7 
    N^2-12 N-420,\\ P_{305}&=&58 N^4+116 N^3-111 N^2-169 N-282,\\ P_{306}&=&59 
    N^4+118 N^3+305 N^2+246 N-1008,\\ P_{307}&=&77 N^4+154 N^3-529 N^2-606 
    N+432,\\ P_{308}&=&85 N^4+86 N^3-47 N^2-136 N-84,\\ P_{309}&=&101 N^4+190 
    N^3+677 N^2+660 N-1980,\\ P_{310}&=&109 N^4+218 N^3+61 N^2-48 
    N-1044,\\ P_{311}&=&121 N^4+418 N^3+269 N^2-140 N-36,\\ P_{312}&=&147 N^4+426 
    N^3+80 N^2-211 N-18,\\ P_{313}&=&203 N^4+178 N^3-31 N^2-6 N+36,\\ P_{314}&=&235 
    N^4+596 N^3+319 N^2+66 N+72,\\ P_{315}&=&265 N^4+1478 N^3+745 N^2-6156 
    N-2844,\\ P_{316}&=&277 N^4+728 N^3+152 N^2-383 N-108,\\ P_{317}&=&291 N^4+368 
    N^3+21 N^2-380 N-272,\\ P_{318}&=&349 N^4+686 N^3-347 N^2-612 
    N-2412,\\ P_{319}&=&423 N^4+264 N^3+293 N^2+452 N+54,\\ P_{320}&=&536 N^4+1351 
    N^3+498 N^2-329 N-42,\\ P_{321}&=&536 N^4+1459 N^3+565 N^2-358 
    N-102,\\ P_{322}&=&562 N^4+1565 N^3+649 N^2-354 N-72,\\ P_{323}&=&575 N^4+1468 
    N^3+324 N^2-845 N-294,\\ P_{324}&=&683 N^4+1702 N^3+767 N^2+164 
    N+292,\\ P_{325}&=&1055 N^4+2236 N^3+1139 N^2+66 N+72,\\ P_{326}&=&1124 
    N^4+2320 N^3+823 N^2-451 N-93,\\ P_{327}&=&4513 N^4+11762 N^3+2455 
    N^2-4794 N-684,\\ P_{328}&=&2 N^5-9 N^4+113 N^3+273 N^2-61 N-174,\\ P_{329}&=&2 
    N^5+N^4-38 N^3-13 N^2+120 N-36,\\ P_{330}&=&17 N^5-62 N^4-413 N^3+218 
    N^2+504 N+144,\\ P_{331}&=&31 N^5-34 N^4-301 N^3+148 N^2+540 
    N-288,\\ P_{332}&=&83 N^5+404 N^4+643 N^3-50 N^2-57 N+126,\\ P_{333}&=&97 
    N^5+460 N^4+32 N^3-1669 N^2-1518 N+72,\\ P_{334}&=&131 N^5+308 N^4-347 
    N^3-3392 N^2-2004 N+2592,\\ P_{335}&=&328 N^5+1985 N^4+2899 N^3+927 
    N^2-231 N-180,\\ P_{336}&=&361 N^5+2020 N^4+899 N^3-7756 N^2-8724 
    N+1728,\\ P_{337}&=&1103 N^5+4825 N^4+6329 N^3+2733 N^2+66 
    N+360,\\ P_{338}&=&1687 N^5+7540 N^4+539 N^3-34066 N^2-30444 
    N+9576,\\ P_{339}&=&3153 N^5+3318 N^4-13819 N^3-2896 N^2-11820 
    N+9936,\\ P_{340}&=&8425 N^5+30483 N^4+30615 N^3+7625 N^2+2712 
    N+3212,\\ P_{341}&=&-279 N^6-1137 N^5-1215 N^4-63 N^3-714 N^2-1344 
    N-496,\\ P_{342}&=&13 N^6+39 N^5-33 N^4-131 N^3-380 N^2-308 
    N+48,\\ P_{343}&=&14 N^6+39 N^5-46 N^4-129 N^3+50 N^2+72 N+144,\\ P_{344}&=&17 
    N^6+51 N^5+11 N^4-63 N^3+284 N^2+324 N+48,\\ P_{345}&=&18 N^6+52 N^5-37 
    N^4-142 N^3+101 N^2+148 N+60,\\ P_{346}&=&21 N^6+15 N^5-383 N^4-463 
    N^3+378 N^2+608 N+96,\\ P_{347}&=&33 N^6+729 N^5-997 N^4-4225 N^3+5052 
    N^2+3240 N+6264,\\ P_{348}&=&34 N^6-450 N^5+657 N^4+3376 N^3-4743 N^2-3234 
    N-6192,\\ P_{349}&=&35 N^6+69 N^5+1763 N^4+3747 N^3+902 N^2-1548 
    N+3096,\\ P_{350}&=&36 N^6-390 N^5+1249 N^4+2912 N^3-7001 N^2-3342 
    N-7632,\\ P_{351}&=&37 N^6+169 N^5+149 N^4-749 N^3-1170 N^2+484 
    N-1032,\\ P_{352}&=&41 N^6+123 N^5-157 N^4-519 N^3-484 N^2-204 
    N+72,\\ P_{353}&=&49 N^6+210 N^5+17 N^4-848 N^3-807 N^2+29 
    N-318,\\ P_{354}&=&51 N^6+141 N^5+15 N^4-93 N^3+226 N^2+100 
    N+360,\\ P_{355}&=&55 N^6+165 N^5-395 N^4-1065 N^3+28 N^2+588 
    N-504,\\ P_{356}&=&61 N^6-50 N^5+490 N^4+842 N^3-3659 N^2-1536 
    N-3780,\\ P_{357}&=&65 N^6+195 N^5-93 N^4-511 N^3+284 N^2+572 
    N+288,\\ P_{358}&=&74 N^6+302 N^5-307 N^4-1512 N^3+137 N^2+826 
    N+840,\\ P_{359}&=&75 N^6+225 N^5-131 N^4-637 N^3+240 N^2+596 
    N+432,\\ P_{360}&=&77 N^6+219 N^5+161 N^4+69 N^3+1742 N^2+1548 
    N+216,\\ P_{361}&=&83 N^6-109 N^5-77 N^4-567 N^3-42 N^2+1096 
    N+576,\\ P_{362}&=&96 N^6+178 N^5-595 N^4-740 N^3-13 N^2+1274 
    N+1536,\\ P_{363}&=&100 N^6+313 N^5+235 N^4+31 N^3-79 N^2-60 
    N-20,\\ P_{364}&=&112 N^6+517 N^5+773 N^4+539 N^3+311 N^2+184 
    N+156,\\ P_{365}&=&115 N^6+345 N^5-671 N^4-1917 N^3+1756 N^2+2772 
    N-144,\\ P_{366}&=&122 N^6+300 N^5-871 N^4-1626 N^3+2009 N^2+1794 
    N+468,\\ P_{367}&=&134 N^6+402 N^5-373 N^4-1416 N^3+1121 N^2+1896 
    N+252,\\ P_{368}&=&146 N^6+603 N^5-733 N^4-2823 N^3+2234 N^2+3471 
    N+486,\\ P_{369}&=&158 N^6+178 N^5+85 N^4+28 N^3-4043 N^2-1126 
    N-3576,\\ P_{370}&=&182 N^6-1416 N^5+1896 N^4+9632 N^3-15849 N^2-8379 
    N-19494,\\ P_{371}&=&208 N^6+190 N^5+711 N^4+456 N^3-7635 N^2-2802 
    N-7488,\\ P_{372}&=&248 N^6+2595 N^5-1572 N^4-13771 N^3+6576 N^2+7356 
    N+12672,\\ P_{373}&=&257 N^6+705 N^5-913 N^4-2385 N^3+3410 N^2+3642 
    N-324,\\ P_{374}&=&385 N^6+99 N^5-3089 N^4-1239 N^3+9376 N^2+4644 
    N-2736,\\ P_{375}&=&388 N^6+6360 N^5-6891 N^4-36278 N^3+34155 N^2+24738 
    N+48240,\\ P_{376}&=&429 N^6+1849 N^5+285 N^4-7533 N^3-7626 N^2+716 
    N-3576,\\ P_{377}&=&449 N^6+705 N^5-2395 N^4-4669 N^3-198 N^2+3860 
    N+1896,\\ P_{378}&=&521 N^6+1497 N^5-1177 N^4-4233 N^3+1826 N^2+3114 
    N+2844,\\ P_{379}&=&676 N^6+5853 N^5-2718 N^4-32261 N^3+8853 N^2+19179 
    N+23166,\\ P_{380}&=&913 N^6-3282 N^5+8202 N^4+26842 N^3-64629 N^2-30582 
    N-73008,\\ P_{381}&=&1046 N^6+15756 N^5-11349 N^4-94906 N^3+60393 
    N^2+67464 N+102060,\\ P_{382}&=&1360 N^6+4524 N^5-4169 N^4-19314 N^3-7739 
    N^2+7914 N+5832,\\ P_{383}&=&1472 N^6+2121 N^5+2868 N^4-2785 N^3-43668 
    N^2-11448 N-39960,\\ P_{384}&=&1540 N^6+4053 N^5-150 N^4-13805 N^3-28359 
    N^2-3609 N-20682,\\ P_{385}&=&1540 N^6+9345 N^5-792 N^4-42161 N^3-13683 
    N^2+8649 N+13122,\\ P_{386}&=&1545 N^6+5807 N^5+5289 N^4+833 N^3+2254 
    N^2+3776 N+1232,\\ P_{387}&=&2882 N^6+12624 N^5-5307 N^4-59242 N^3-19839 
    N^2+22626 N+8424,\\ P_{388}&=&4357 N^6+20253 N^5+23997 N^4+10171 N^3+666 
    N^2-2700 N-1944,\\ P_{389}&=&4502 N^6+9708 N^5+465 N^4-27358 N^3-93105 
    N^2-17352 N-73548,\\ P_{390}&=&4816 N^6+5952 N^5+12135 N^4-1466 N^3-156051 
    N^2-48114 N-147312,\\ P_{391}&=&6457 N^6+43941 N^5-12141 N^4-206669 
    N^3-9264 N^2+63756 N+108864,\\ P_{392}&=&7241 N^6+50345 N^5+124855 
    N^4+122983 N^3+29388 N^2-3516 N+4032,\\ P_{393}&=&7531 N^6+26499 N^5+27861 
    N^4+8401 N^3+336 N^2+504 N-540,\\ P_{394}&=&9187 N^6+58477 N^5+134975 
    N^4+133571 N^3+52890 N^2+27036 N+24408,\\ P_{395}&=&-3221 N^7-26286 
    N^6-57930 N^5-47240 N^4-11865 N^3+942 N^2-72 N
    \nonumber\\ &&
    -1080,\\ P_{396}&=&23 N^7+88 
    N^6+42 N^5-144 N^4+271 N^3+824 N^2+408 N-72,\\ P_{397}&=&25 N^7-379 
    N^6-1119 N^5+2215 N^4+4787 N^3-4173 N^2-5436 N-4860,\\ P_{398}&=&79 
    N^7+169 N^6-407 N^5-1353 N^4-912 N^3+88 N^2+1536 N+144,\\ P_{399}&=&81 
    N^7-1463 N^6-5889 N^5-333 N^4+16148 N^3+10448 N^2+7072 
    N+3216,\\ P_{400}&=&83 N^7+487 N^6+582 N^5-1789 N^4-4016 N^3-339 N^2-180 
    N-1836,\\ P_{401}&=&115 N^7+309 N^6+141 N^5+79 N^4+1552 N^3+1180 N^2-48 
    N-288,\\ P_{402}&=&173 N^7+32 N^6-2263 N^5-2810 N^4+3582 N^3+4014 N^2+2120 
    N-120,\\ P_{403}&=&179 N^7+700 N^6+352 N^5-1242 N^4-407 N^3+1830 N^2+1572 
    N+216,\\ P_{404}&=&379 N^7+3227 N^6+2631 N^5-20843 N^4-22294 N^3+6252 
    N^2+21744 N
    \nonumber\\ &&
    +18576,\\ P_{405}&=&403 N^7+3797 N^6+6527 N^5-9957 N^4-29446 
    N^3-8996 N^2+4936 N+6240,\\ P_{406}&=&417 N^7+1362 N^6-523 N^5-4588 
    N^4-4220 N^3-2720 N^2+864 N+2880,\\ P_{407}&=&713 N^7+1029 N^6-2740 
    N^5-1095 N^4+11969 N^3+6480 N^2-2772 N-432,\\ P_{408}&=&721 N^7+2716 
    N^6-38 N^5-7892 N^4+4489 N^3+21520 N^2+10308 N-1296,\\ P_{409}&=&781 
    N^7-2755 N^6-22167 N^5-45617 N^4+12110 N^3+59256 N^2+100728 
    N
    \nonumber\\ &&
    +67824,\\ P_{410}&=&1669 N^7+7264 N^6+3778 N^5-16112 N^4-2843 N^3+29584 
    N^2+22020 N
    \nonumber\\ &&
    +3024,\\ P_{411}&=&1873 N^7+8797 N^6+5775 N^5-38361 N^4-70804 
    N^3-27520 N^2+20832 N
    \nonumber\\ &&
    -3888,\\ P_{412}&=&1964 N^7+11260 N^6+8946 N^5-44536 
    N^4-72725 N^3-15489 N^2+21528 N
    \nonumber\\ &&
    +12204,\\ P_{413}&=&2251 N^7+11006 N^6+6514 
    N^5-39436 N^4-64685 N^3-23230 N^2+1596 N
    \nonumber\\ &&
    -1656,\\ P_{414}&=&2284 N^7+12899 
    N^6+12631 N^5-41611 N^4-89702 N^3-32317 N^2+10896 N
    \nonumber\\ &&
    -2700,\\ P_{415}&=&2686 
    N^7-128 N^6-28605 N^5+15784 N^4+49639 N^3+2700 N^2-7740 
    N
    \nonumber\\ &&
    +4752,\\ P_{416}&=&3098 N^7+17416 N^6+16611 N^5-60352 N^4-122465 
    N^3-52368 N^2+12312 N
    \nonumber\\ &&
    -7128,\\ P_{417}&=&3691 N^7+18675 N^6+34038 N^5+26401 
    N^4+6405 N^3-618 N^2-36 N-540,\\ P_{418}&=&4591 N^7+32765 N^6+46752 
    N^5-90491 N^4-251779 N^3-89754 N^2+43020 N
    \nonumber\\ &&
    +10584,\\ P_{419}&=&4645 
    N^7+15287 N^6-15678 N^5-107039 N^4-63475 N^3+26208 N^2+51948 
    N
    \nonumber\\ &&
    +28080,\\ P_{420}&=&16828 N^7+802 N^6-194889 N^5+102862 N^4+378583 
    N^3+53370 N^2-56556 N
    \nonumber\\ &&
    +26136,\\ P_{421}&=&32287 N^7+193848 N^6+395214 
    N^5+329536 N^4+98223 N^3+5076 N^2+3996 N
    \nonumber\\ &&
    -324,\\ 
    P_{422}&=&-116957 N^8-894616 N^7-2631828 N^6-3627014 N^5-2222155 N^4-543882 N^3
    \nonumber\\ &&
    -286164 N^2-177768 N+23328,\\ 
    P_{423}&=&-28023 N^8-203826 N^7-566428 N^6-738922 N^5-430121 N^4-91484 N^3
    \nonumber\\ &&
    -69852 N^2-97776 N-34560,\\ 
    P_{424}&=&35 N^8+136 N^7+72 N^6-222 N^5+117 N^4+702 N^3+472 N^2+32 N+96,\\ 
    P_{425}&=&296 N^8+920 N^7-1687 N^6-6931 N^5-679 N^4+9827 N^3+5634 N^2-252 
    \nonumber\\ &&
    N-216,\\ 
    P_{426}&=&805 N^8+2632 N^7-2678 N^6-12656 N^5+661 N^4+16144 N^3+5856 N^2
    \nonumber\\ &&
    -1044 N+648,\\ 
    P_{427}&=&1357 N^8+5260 N^7+1288 N^6-12818 N^5-3761 N^4+16198 N^3+16164 N^2
    \nonumber\\ &&
    +5976 N+864,\\ 
    P_{428}&=&1511 N^8+9442 N^7+22584 N^6+26891 N^5+16961 N^4+5241 N^3+38 N^2-804 N
    \nonumber\\ &&
    -648,\\ 
    P_{429}&=&17529 N^8+60627 N^7-8987 N^6-211343 N^5-245062 N^4-225100 N^3+6936 N^2
    \nonumber\\ &&
    +155088 N+23328,\\ 
    P_{430}&=&28551 N^8+63048 N^7+57098 N^6+23004 N^5+3519 N^4+3044 N^3-9864 N^2
    \nonumber\\ &&
    -21888 N-9936,\\ 
    P_{431}&=&50689 N^8+231286 N^7+374706 N^6+252688 N^5+58645 N^4+3486 N^3
    \nonumber\\ &&
    +22752 N^2+19980 N+8100,\\ 
    P_{432}&=&-97249 N^9-674988 N^8-1131968 N^7+1296434 N^6+5370597 N^5+4078570 N^4
    \nonumber\\ &&
    -68076 N^3-537576 N^2+352224 N+114048,\\ 
    P_{433}&=&-71503 N^9-495343 N^8-1283176 N^7-1561498 N^6-898003 N^5-199315 N^4
    \nonumber\\ &&
    -12042 N^3-12852 N^2-4536 N-1620,\\ 
    P_{434}&=&-47295 N^9-50958 N^8+608888 N^7+213650 N^6-1767525 N^5-1823284 N^4
    \nonumber\\ &&
    +42212 N^3+300888 N^2-153216 N-85536,\\ 
    P_{435}&=&-5021 N^9-13792 N^8+14010 N^7+70516 N^6+68039 N^5+94644 N^4+83628 N^3
    \nonumber\\ &&
    -75096 N^2-108576 N-25056,\\ 
    P_{436}&=&10 N^9+14 N^8-108 N^7-168 N^6+213 N^5+438 N^4+29 N^3-824 N^2-180 N
    \nonumber\\ &&
    +288,\\ 
    P_{437}&=&14 N^9+20 N^8-133 N^7-173 N^6+407 N^5+397 N^4-384 N^3-964 N^2-192 N
    \nonumber\\ &&
    +288,\\
    P_{438}&=&25 N^9-121 N^8-775 N^7-546 N^6+1772 N^5-1113 N^4-5854 N^3-2060 N^2
    \nonumber\\ &&
    +1112 N+384,\\ 
    P_{439}&=&79 N^9+61 N^8-785 N^7-936 N^6+1418 N^5-2005 N^4-8840 
    N^3-7848 N^2
    \nonumber\\ &&
    -3104 N
    -1104,\\ 
    P_{440}&=&307 N^9+3168 N^8+6674 N^7-2144 N^6-18819 N^5-6472 N^4+12218 N^3-3564 N^2
    \nonumber\\ &&
    -15128 N-6288,\\ 
    P_{441}&=&370 N^9+536 N^8-4027 N^7-4267 N^6+14417 N^5+9647 N^4-14144 N^3
    \nonumber\\ &&
    -16788 N^2
    -3600 N+2592,\\ 
    P_{442}&=&430 N^9+728 N^8-3699 N^7-6279 N^6+7179 N^5+13731 N^4-616 N^3-21338 N^2
    \nonumber\\ &&
    -5472 N+6552,\\ 
    P_{443}&=&575 N^9+7672 N^8+19410 N^7-7588 N^6-68521 N^5-45276 N^4+23872 N^3
    \nonumber\\ &&
    +21648 N^2-8624 N-3840,\\ 
    P_{444}&=&1288 N^9+33279 N^8+109499 N^7-4538 N^6-391455 N^5-190195 N^4+311202 N^3
    \nonumber\\ &&
    +104832 N^2-74520 N-16848,\\ 
    P_{445}&=&1963 N^9+30012 N^8+88574 N^7-19280 N^6-343611 N^5-220246 N^4+153144 N^3
    \nonumber\\ &&
    +49212 N^2-44496 N-6480,\\ 
    P_{446}&=&2133 N^9+810 N^8-35584 N^7+4418 N^6+120351 N^5+97028 N^4-21724 N^3
    \nonumber\\ &&
    -15336 N^2+17856 N+7776,\\ 
    P_{447}&=&3761 N^9-14568 N^8-104870 N^7-23608 N^6+335763 N^5+109918 N^4
    \nonumber\\ &&
    -430176 N^3
    -423972 N^2-170640 N+9072,\\ 
    P_{448}&=&4258 
    N^9+34062 N^8+65735 N^7-85433 N^6-339093 N^5-148117 N^4+85182 N^3
    \nonumber\\ &&
    -99882 N^2-153360 N-48600,\\ 
    P_{449}&=&5894 N^9-12921 N^8-126065 N^7-48880 N^6+374049 N^5+55783 N^4
    \nonumber\\ &&
    -668856 N^3-635868 N^2-254448 N-20736,\\ 
    P_{450}&=&43949 N^9+268194 N^8+376282 N^7-637870 N^6-2123703 N^5-1675742 N^4
    \nonumber\\ &&
    -209052 N^3+96174 N^2-61992 N-8424,\\ 
    P_{451}&=&-5563 N^{10}-24261 N^9-16394 N^8+58642 N^7+172417 N^6+124231 N^5
    \nonumber\\ &&
    -76780 N^4-55652 N^3+153696 N^2+131616 N+34560,\\ 
    P_{452}&=&-4253 N^{10}-18719 N^9-3594 N^8+69026 N^7+89183 N^6+6165 N^5-37744 N^4
    \nonumber\\ &&
    +8424 N^3+50664 N^2+35696 N+8160,\\ 
    P_{453}&=&24 N^{10}-230 N^9-267 N^8+2755 N^7-219 
    N^6-13309 N^5+8190 N^4+29248 N^3
    \nonumber\\ &&
    -144 N^2
    -28400 N-17376,\\ 
    P_{454}&=&261 N^{10}+324 N^9-1405 N^8+651 N^7-1688 N^6-14839 N^5+15318 N^4+41614 
    N^3
    \nonumber\\ &&
    +29772 N^2-27720 N-28944,\\ 
    P_{455}&=&561 N^{10}+2115 N^9+794 N^8-3146 N^7-15755 N^6-9537 N^5+39368 N^4
    \nonumber\\ &&
    +15976 N^3
    -84440 N^2-83120 
    N-24864,\\ 
    P_{456}&=&6439 N^{10}+27179 N^9-8632 N^8-154834 N^7-149641 N^6+93575 N^5+183334 N^4
    \nonumber\\ &&
    +75768 N^3-2448 N^2-12636 N-9720,\\ 
    P_{457}&=&7868 N^{10}+57302 N^9+2431 N^8-508191 N^7-330541 N^6+1419481 N^5
    \nonumber\\ &&
    +541702 
    N^4
    -1330188 N^3+96696 N^2+1283040 N+663552,\\ P_{458}&=&-5729259 
    N^{11}-8768673 N^{10}+46945558 N^9+46575050 N^8-44721647 N^7
    \nonumber\\ &&
    -98312261 
    N^6-77570092 N^5-6099524 N^4+7097376 N^3-18265104 N^2
    \nonumber\\ &&
    +8239104 N
    +14427072,\\ 
    P_{459}&=&27 N^{11}+69 N^{10}-176 N^9-554 N^8+209 N^7+1231 
    N^6+396 N^5-1730 N^4
    \nonumber\\ &&
    -2104 N^3
    -792 N^2+352 N+192,\\ P_{460}&=&33 N^{11}+90 
    N^{10}-827 N^9-2100 N^8+2393 N^7+7786 N^6+2037 N^5-7676 N^4
    \nonumber\\ &&
    -8880 
    N^3
    -936 N^2+1712 N+1248,\\ P_{461}&=&1103 N^{11}+2937 N^{10}-7526 
    N^9-24942 N^8+3681 N^7+52899 N^6+29794 N^5
    \nonumber\\ &&
    -59478 N^4
    -72556 N^3-4968 
    N^2+16272 N+1728,\\ 
    P_{462}&=&1937 N^{11}+4659 N^{10}-17094 N^9-30326 
    N^8+45693 N^7+123927 N^6+59528 N^5
    \nonumber\\ &&
    -29940 N^4-13440 N^3+26832 
    N^2+17088 N+2880,\\ 
    P_{463}&=&6009 N^{11}+89967 N^{10}+289824 N^9-63490 
    N^8-1473715 N^7-1627513 N^6
    \nonumber\\ &&
    +315866 N^5
    +800300 N^4+227984 N^3+702960 N^2+787680 N+229824,\\ 
    P_{464}&=&8183 N^{11}+18327 N^{10}-109486 
    N^9-251134 N^8+289655 N^7+850715 N^6
    \nonumber\\ &&
    +154440 N^5
    -665500 N^4-552096 
    N^3-42048 N^2+84672 N+62208,\\ P_{465}&=&80453 N^{11}+688424 
    N^{10}+1757044 N^9+44376 N^8-6544585 N^7-10461842 N^6
    \nonumber\\ &&
    -6460412 
    N^5-1603590 N^4-41364 N^3+220968 N^2+31104 N-69984,\\ P_{466}&=&428649 
    N^{11}+601299 N^{10}-3536240 N^9-3241402 N^8+3843517 N^7+7383499 
    N^6
    \nonumber\\ &&
    +4538354 N^5+20668 N^4+528168 N^3+1514448 N^2-1263168 
    N-1311552,\\ P_{467}&=&463143 N^{11}+3002511 N^{10}+5392892 N^9-3891846 
    N^8-24947513 N^7
    \nonumber\\ &&
    -31706485 N^6
    -15500034 N^5-1653956 N^4-1227864 
    N^3-2638368 N^2
    \nonumber\\ &&
    -1159488 N
    -124416,\\ P_{468}&=&241 N^{12}+521 N^{11}-1523 
    N^{10}-3268 N^9+659 N^8-1815 N^7+2291 N^6+25482 N^5
    \nonumber\\ &&
    +45116 N^4+19656 
    N^3-27296 N^2-32512 N-9984,\\ P_{469}&=&10629 N^{12}+22185 N^{11}-70738 
    N^{10}-118858 N^9+91525 N^8-54183 N^7-58576 N^6
    \nonumber\\ &&
    +818608 N^5+1687828 
    N^4+888516 N^3-727992 N^2-580176 N-28512,\\ P_{470}&=&27765 N^{12}+58437 
    N^{11}-182597 N^{10}-325952 N^9+200843 N^8-157371 N^7
    \nonumber\\ &&
    -55295 
    N^6
    +2325230 N^5+4593788 N^4+2307744 N^3-2192976 N^2-2038176 
    N
    \nonumber\\ &&
    -326592,\\ P_{471}&=&-3069 N^{13}+31221 N^{12}+144399 N^{11}-301293 
    N^{10}-795373 N^9+250503 N^8
    \nonumber\\ &&
    +1886057 N^7+1623213 N^6-63486 N^5-894316 
    N^4-31760 N^3+890256 N^2
    \nonumber\\ &&
    +677088 N+157248,\\ P_{472}&=&494694 
    N^{13}+1075005 N^{12}-4137924 N^{11}-6752619 N^{10}+5087252 N^9
    \nonumber\\ &&
    +13650627 N^8
    +2421476 N^7-9000225 N^6-3343434 N^5+7931228 
    N^4
    \nonumber\\ &&
    +4200112 N^3
    -6598896 N^2
    -6774048 N-1729728,\\ P_{473}&=&599375 
    N^{13}+5709355 N^{12}+18035106 N^{11}+12196222 N^{10}-51833983 
    N^9
    \nonumber\\ &&
    -129886893 N^8-124003084 N^7-53668604 N^6-7785942 N^5+2544156 
    N^4
    \nonumber\\ &&
    +6838020 N^3+7118928 N^2+4129056 N+1049760,\\ P_{474}&=&-458487 
    N^{14}-3433746 N^{13}-4940481 N^{12}+20878856 N^{11}+64591263 
    N^{10}
    \nonumber\\ &&
    +10248134 N^9-139889479 N^8-158068804 N^7-33854512 N^6+17126664 
    N^5
    \nonumber\\ &&
    -5950960 N^4+25711872 N^3+46793088 N^2+27948672 
    N+5971968,\\ P_{475}&=&3003 N^{14}+22594 N^{13}+47885 N^{12}
    -148240 
    N^{11}-659235 N^{10}-247430 N^9
    \nonumber\\ &&
    +1482531 N^8+1874548 N^7+339992 
    N^6-373632 N^5-131264 N^4-1500992 N^3
    \nonumber\\ &&
    -2232704 N^2-1344000 
    N-301824,\\ P_{476}&=&-7255 N^{15}-25524 N^{14}+35986 N^{13}+183186 
    N^{12}+88824 N^{11}-295334 N^{10}
    \nonumber\\ &&
    -321342 N^9+60030 N^8+10607 
    N^7-441646 N^6-382164 N^5+194648 N^4
    \nonumber\\ &&
    +529968 N^3+385824 N^2+130752 
    N+17280.
    \end{eqnarray}

\noindent
Further non--singlet contributions are
\begin{eqnarray}
C_{F_2,q}^{d_{abc}, (3)} &=&
    \textcolor{blue}{\frac{d_{abc} d^{abc} N_F}{N_C}} \Biggl\{
            \frac{64 \big(
                    -3-N+N^2\big)}{(N-2) (3+N)}
            -\frac{128 (N-1) S_3 P_{479}}{(N-2) N (1+N) (2+N) (3+N)}
    \nonumber\\ &&  
          -\frac{256 S_4 P_{483}}{(N-2) N^2 (1+N)^2 (2+N) (3+N)}
            +\frac{512 S_{3,1} P_{483}}{(N-2) N^2 (1+N)^2 (2+N) (3+N)}
    \nonumber\\ &&   
         -\frac{256 S_{-2,1} P_{484}}{(N-2) N (1+N) (2+N) (3+N)}
            +\Biggl[
                    -\frac{128 \big(
                            -66+5 N-9 N^2+N^3\big)}{(N-2) N (1+N) (2+N) (3+N)}
    \nonumber\\ &&               
     -\frac{256 S_3 P_{483}}{(N-2) N^2 (1+N)^2 (2+N) (3+N)}
                    -\frac{1024 (N-3) S_{-2,1}}{(N-2) (1+N) (3+N)}
            \Biggr] S_1
    \nonumber\\ &&      
      -\Biggl[
                    \frac{256 S_1 P_{480}}{(N-2) N (1+N)^2 (2+N) (3+N)}
                    +\frac{64 P_{487}}{(N-2) N^2 (1+N)^2 (2+N) (3+N)}
            \Biggr] 
    \nonumber\\ &&  
    \times S_{-2}
            +\Biggl[
                    \frac{128 P_{484}}{(N-2) N (1+N) (2+N) (3+N)}
                    +\frac{512 (N-3) S_1}{(N-2) (1+N) (3+N)}
            \Biggr] S_{-3}
    \nonumber\\ &&     
          +\frac{64 (N-1) S_{-2}^2}{1+N}
       +\frac{64 \big(
                    -18+N+N^3\big) S_{-4}}{(N-2) (1+N) (3+N)}
            -\frac{1024 (N-3) S_{-2,2}}{(N-2) (1+N) (3+N)}
    \nonumber\\ &&
            -\frac{512 (N-1) S_{-2,3}}{N (1+N)}
            -\frac{1024 (N-3) S_{-3,1}}{(N-2) (1+N) (3+N)}
            +\frac{512 (N-1) S_{-4,1}}{N (1+N)}
    \nonumber\\ &&       
     +\frac{2048 (N-3) S_{-2,1,1}}{(N-2) (1+N) (3+N)}
            +\Biggl[
                    \frac{32 P_{485}}{3 (N-2) N (1+N) (2+N) (3+N)}
    \nonumber\\ &&               
     +\frac{1024 \big(
                            -12-4 N-3 N^2+N^3\big) S_1}{(N-2) N^2 (1+N)^2 (2+N) (3+N)}
                    +\frac{1024 (N-1) S_{-2}}{N (1+N)}
            \Biggr] \zeta_3
    \nonumber\\ &&
            -\frac{1280 (N-3) (N-2)}{3 N (1+N)} \zeta_5
    \Biggr\},
    \\
C_{F_L,q}^{d_{abc}, (3)} &=& \textcolor{blue}{\frac{d_{abc} d^{abc} N_F}{N_C}} \Biggl\{
            -\frac{128 N}{(N-2) (3+N)}
    -\frac{128 S_3 P_{478}}{(N-2) (N-1) (1+N) (2+N) (3+N)}
    \nonumber\\ &&        
            +\frac{512 S_{-2,1} P_{482}}{(N-2) (N-1) (1+N) (2+N) (3+N)}
            +\Biggl[
                    \frac{4096 S_{-2,1}}{(N-2) (1+N) (3+N)}
    \nonumber\\ &&
                    +\frac{1024 \big(
                            6+N+2 N^2+2 N^3+N^4\big) S_3}{(N-2) (N-1) N
    (1+N)^2 (2+N) (3+N)}
    \nonumber\\ &&
    + \frac{1536 \big( 
                            8+N+N^2\big)}{(N-2) (N-1) (1+N) (2+N)
    (3+N)}
            \Biggr] S_1
    \nonumber\\ &&
            +\frac{1024 \big(
                    6+N+2 N^2+2 N^3+N^4\big) S_4}{(N-2) (N-1) N (1+N)^2
    (2+N) (3+N)}
    \nonumber\\ &&     
       +\Biggl[
                    \frac{1536 S_1 P_{481}}{(N-2) (N-1) N (1+N)^2 (2+N)
    (3+N)}
    \nonumber\\ &&
                    +\frac{128 P_{486}}{(N-2) (N-1) N (1+N)^2 (2+N)
    (3+N)}
            \Biggr] S_{-2}
           -\frac{128 S_{-2}^2}{1+N}  
    \nonumber\\ && 
            +\Biggl[
                    -\frac{256 P_{482}}{(N-2) (N-1) (1+N) (2+N) (3+N)}
      -\frac{2048 S_1}{(N-2) (1+N) (3+N)}
            \Biggr] S_{-3}
    \nonumber\\ &&              
            -\frac{128 \big(
                    10+N+N^2\big) S_{-4}}{(N-2) (1+N) (3+N)}
       -\frac{2048 \big(
                    6+N+2 N^2+2 N^3+N^4\big) S_{3,1}}{(N-2) (N-1) N
    (1+N)^2 (2+N) (3+N)}
    \nonumber\\ &&     
            +\frac{4096 S_{-2,2}}{(N-2) (1+N) (3+N)} 
    -\frac{512 
    S_{-2,3}}{1+N}
            +\frac{4096 S_{-3,1}}{(N-2) (1+N) (3+N)}
            +\frac{512 S_{-4,1}}{1+N}
    \nonumber\\ && 
            -\frac{8192 S_{-2,1,1}}{(N-2) (1+N) (3+N)}
    +\Biggl[
                    -\frac{64 P_{477}}{(N-2) (N-1) (1+N) (2+N) (3+N)}
    \nonumber\\ &&              
                    +\frac{1024 S_{-2}}{1+N}
       -\frac{6144 \big(
                            2+N+N^2\big) S_1}{(N-2) (N-1) N (1+N)^2
    (2+N) (3+N)}
            \Biggr] \zeta_3
            +\frac{2560}{1+N} \zeta_5
    \Biggr\},
\end{eqnarray}


\noindent
with
\begin{eqnarray}   
    P_{477} &=& N^4+2 N^3-43 N^2-44 N+36,
    \\
    P_{478} &=& N^4+2 N^3+5 N^2+4 N+36,
    \\
    P_{479} &=& N^4+6 N^3+6 N^2-8 N-24,
    \\
    P_{480} &=& 2 N^4-4 N^3-17 N^2-23 N-6,
    \\
    P_{481} &=& 2 N^4+4 N^3+3 N^2+N+2,
    \\
    P_{482} &=& 4 N^4+8 N^3-31 N^2-35 N+18,
    \\
    P_{483} &=& N^5-5 N^3-12 N^2-8 N-24,
    \\
    P_{484} &=& 2 N^5-21 N^3+21 N^2+46 N-24,
    \\
    P_{485} &=& 7 N^5+22 N^4-25 N^3+128 N^2+60 N-432,
    \\
    P_{486} &=& 7 N^6+21 N^5-26 N^4-87 N^3-47 N^2-12,
    \\
    P_{487} &=& 5 N^7+11 N^6-31 N^5-49 N^4+28 N^3+68 N^2+40 N+48.
    \end{eqnarray}

\noindent
The pure singlet contributions are given by
\begin{eqnarray}
&& C_{F_2,q}^{\rm PS, (3)} = 
    \nonumber\\ 
    &&\textcolor{blue}{C_F} \Biggl\{ \textcolor{blue}{T_F^2 N_F^2}
    \Biggl[
                    -\frac{512 S_{-2} P_{495}}{9 (N-2) (N-1) N^2 (1+N) 
    (2+N)^2 (3+N)}
    \nonumber\\ &&
                  +  \frac{128 \zeta_3 P_{492}}{9 (N-1) N^2 (1+N)^2 (2+N)}
                    +\frac{16 S_1^2 P_{516}}{27 (N-1) N^3 (1+N)^3 
    (2+N)^2}
    \nonumber\\ &&
                    -\frac{32 P_{547}}{243 (N-2) (N-1) N^5 (1+N)^5 
    (2+N)^4 (3+N)}
    \nonumber\\ &&
                    +\frac{16 S_2 P_{522}}{27 (N-1) N^3 (1+N)^3 (2+N)^2}
                    +\Biggl(
                            -\frac{32 P_{537}}{81 (N-1) N^4 (1+N)^4 
    (2+N)^3}
    \nonumber\\ &&
                            -\frac{32 \big(
                                    2+N+N^2\big)^2 S_2}{9 (N-1) N^2 
    (1+N)^2 (2+N)}
                    \Biggr) S_1
                    +\frac{64 \big(
                            2+N+N^2\big)^2 S_1^3}{27 (N-1) N^2 (1+N)^2 
    (2+N)}
    \nonumber\\ &&
                    -\frac{160 \big(
                            2+N+N^2\big)^2 S_3}{27 (N-1) N^2 (1+N)^2 
    (2+N)}
                    +\frac{1024 S_{-3}}{3 (N-1) N (1+N) (2+N)}
            \Biggr]
    \nonumber\\ &&
            +\textcolor{blue}{C_A T_F N_F} \Biggl[
                    -\frac{48 \big(
                            2+N+N^2\big)^2}{(N-1) N^2 (1+N)^2 
    (2+N)} \zeta_4
                    -\frac{16 S_{-2,2} P_{488}}{(N-1) N^2 (1+N)^2 (2+N)}
    \nonumber\\ &&              
      +\frac{64 S_{-3,1} P_{493}}{3 (N-1) N^2 (1+N)^2 
    (2+N)}
                    +\frac{64 S_{-2,1,1} P_{494}}{3 (N-1) N^2 (1+N)^2 
    (2+N)}
    \nonumber\\ &&
                    -\frac{64 S_{3,1} P_{497}}{(N-2) N^2 (1+N)^2 (2+N) 
    (3+N)}
                    -\frac{8 S_{-4} P_{501}}{3 (N-1) N^2 (1+N)^2 (2+N)}
    \nonumber\\ &&           
         -\frac{4 S_4 P_{508}}{3 (N-2) (N-1) N^2 (1+N)^2 
    (2+N) (3+N)}
                    -
                    \frac{8 \big(
                            2+N+N^2\big) S_1^3 P_{510}}{27 (N-1)^2 N^3 
    (1+N)^3 (2+N)^2}
    \nonumber\\ &&
                    -\frac{16 S_{2,1} P_{519}}{3 (N-1) N^3 (1+N)^3 
    (2+N)^2}
                    -\frac{16 S_{-2,1} P_{520}}{3 (N-1) N^3 (1+N)^3 
    (2+N)^2}
    \nonumber\\ &&
                    +\frac{8 S_3 P_{538}}{27 (N-2) (N-1)^2 N^3 (1+N)^3 
    (2+N)^2 (3+N)}
                    -\frac{4 S_2 P_{542}}{27 (N-1)^2 N^4 (1+N)^4 (2+N)^3}
    \nonumber\\ &&              
      +\frac{8 P_{550}}{243 (N-2) (N-1)^2 N^6 (1+N)^6 
    (2+N)^5 (3+N)}
                    +\Biggl(
                            -\frac{16 S_{-2,1} P_{499}}{3 (N-1) N^2 
    (1+N)^2 (2+N)}
    \nonumber\\ &&
                            +\frac{8 S_3 P_{513}}{9 (N-2) (N-1) N^2 
    (1+N)^2 (2+N) (3+N)}
                            +\frac{8 S_2 P_{526}}{9 (N-1)^2 N^3 (1+N)^3 
    (2+N)^2}
    \nonumber\\ &&
                            +\frac{4 P_{548}}{81 (N-2) (N-1)^2 N^5 
    (1+N)^5 (2+N)^4 (3+N)}
                    \Biggr) S_1
    \nonumber\\ &&                
    +\Biggl(
                            \frac{4 P_{541}}{27 (N-1)^2 N^4 (1+N)^4 
    (2+N)^3}
                            -\frac{80 \big(
                                    2+N+N^2\big)^2 S_2}{3 (N-1) N^2 
    (1+N)^2 (2+N)}
                    \Biggr) S_1^2
    \nonumber\\ &&                
    +\frac{28 \big(
                            2+N+N^2\big)^2 S_1^4}{9 (N-1) N^2 (1+N)^2 
    (2+N)}
                    +\frac{16 \big(
                            2+N+N^2
                    \big)
    \big(5+2 N+2 N^2\big) S_2^2}{3 (N-1) N^2 (1+N)^2 (2+N)}
    \nonumber\\ &&                 
    +\Biggl(
                            -\frac{16 S_1 P_{531}}{3 (N-2) (N-1)^2 N^3 
    (1+N)^3 (2+N)^2 (3+N)}
    \nonumber\\ &&
                            -\frac{8 P_{543}}{9 (N-2) (N-1)^2 N^4 
    (1+N)^4 (2+N)^3 (3+N)}
    \nonumber\\ &&                        
    -
                            \frac{16 \big(
                                    -22+N+N^2
                            \big)
    \big(2+N+N^2\big) S_1^2}{3 (N-1) N^2 (1+N)^2 (2+N)}
                            +\frac{32 \big(
                                    -8+N+N^2
                            \big)
    \big(2+N+N^2\big) S_2}{3 (N-1) N^2 (1+N)^2 (2+N)}
                    \Biggr) S_{-2}
    \nonumber\\ &&                
    -\frac{16 \big(
                            2+N+N^2
                    \big)
    \big(2+3 N+3 N^2\big) S_{-2}^2}{3 (N-1) N^2 (1+N)^2 (2+N)}
                    +\Biggl(
                            \frac{8 S_1 P_{503}}{3 (N-1) N^2 (1+N)^2 
    (2+N)}
    \nonumber\\ &&
                            +\frac{8 P_{527}}{3 (N-1)^2 N^3 (1+N)^3 
    (2+N)^2}
                    \Biggr) S_{-3}
                    +\frac{32 \big(
                            2+N+N^2\big)^2 S_{2,1,1}}{3 (N-1) N^2 
    (1+N)^2 (2+N)}
    \nonumber\\ &&
                    +\Biggl(
                            -\frac{16 S_1 P_{509}}{3 (N-2) (N-1) N^2 
    (1+N)^2 (2+N) (3+N)}
    \nonumber\\ &&
                            -\frac{16 P_{534}}{9 (N-2) (N-1)^2 N^3 
    (1+N)^3 (2+N)^2 (3+N)}
                    \Biggr) \zeta_3
            \Biggr]
    \Biggr\}
    \nonumber\\ &&
    +\textcolor{blue}{C_F^2 T_F N_F} \Biggl\{
            \frac{48 \big(
                    2+N+N^2\big)^2}{(N-1) N^2 (1+N)^2 (2+N)} \zeta_4
            -\frac{256 S_{-2,1} P_{496}}{(N-1) N^3 (1+N)^3 (2+N)^2}
    \nonumber\\ &&
            -\frac{64 S_{3,1} P_{506}}{(N-2) (N-1) N^2 (1+N)^2 (2+N) 
    (3+N)}
    \nonumber\\ &&
            +\frac{8 S_4 P_{511}}{3 (N-2) (N-1) N^2 (1+N)^2 (2+N) (3+N)}
    \nonumber\\ &&       
     +\frac{64 S_{2,1} P_{518}}{3 (N-1) N^3 (1+N)^3 (2+N)^2}
            +\frac{16 S_1^3 P_{521}}{9 (N-1) N^3 (1+N)^3 (2+N)^2}
    \nonumber\\ &&        
    -\frac{8 S_3 P_{530}
            }{9 (N-2) (N-1) N^3 (1+N)^3 (2+N)^2 (3+N)}
            -
            \frac{4 S_2 P_{532}}{3 (N-1) N^4 (1+N)^4 (2+N)^3}
    \nonumber\\ &&       
     -\frac{4 P_{549}}{3 (N-2)^2 (N-1) N^6 (1+N)^6 (2+N)^4 (3+N)}
    \nonumber\\ &&        
    +\Biggl(
                    \frac{16 S_3 P_{515}}{9 (N-2) (N-1) N^2 (1+N)^2 
    (2+N) (3+N)}
                    -\frac{8 S_2 P_{523}}{3 (N-1) N^3 (1+N)^3 (2+N)^2}
    \nonumber\\ &&              
      +\frac{8 P_{546}}{3 (N-2) (N-1) N^5 (1+N)^5 (2+N)^4 
    (3+N)}
                    -\frac{256 S_{-2,1}}{N^2 (1+N)^2}
            \Biggr) S_1
    \nonumber\\ &&        
    +\Biggl(
                    \frac{4 P_{535}}{3 (N-1) N^4 (1+N)^4 (2+N)^3}
                    -\frac{112 \big(
                            2+N+N^2\big)^2 S_2}{3 (N-1) N^2 (1+N)^2 
    (2+N)}
            \Biggr) S_1^2
    \nonumber\\ &&
            +\frac{44 \big(
                    2+N+N^2\big)^2 S_1^4}{9 (N-1) N^2 (1+N)^2 (2+N)}
            +\frac{68 \big(
                    2+N+N^2\big)^2 S_2^2}{3 (N-1) N^2 (1+N)^2 (2+N)}
    \nonumber\\ &&        
    +\Biggl(
                    -\frac{32 S_1 P_{524}}{(N-2) (N-1) N^3 (1+N)^2 
    (2+N)^2 (3+N)}
    \nonumber\\ &&
                    +\frac{32 P_{539}}{(N-2)^2 (N-1) N^4 (1+N)^4 (2+N)^2 
    (3+N)}
                    -\frac{128 \big(
                            4+N+N^2\big) S_1^2}{(N-1) N^2 (1+N)^2 (2+N)}
    \nonumber\\ &&             
       +\frac{128 \big(
                            4+N+N^2\big) S_2}{(N-1) N^2 (1+N)^2 (2+N)}
            \Biggr) S_{-2}
            +\Biggl(
                    \frac{128 \big(
                            6+N+N^2\big) S_1}{(N-1) N^2 (1+N)^2 (2+N)}
    \nonumber\\ &&
                    +\frac{64 P_{528}}{(N-2) (N-1) N^3 (1+N)^3 (2+N)^2 
    (3+N)}
            \Biggr) S_{-3}
            +\frac{128 \big(
                    6+5 N+5 N^2\big) S_{-4}
            }{(N-1) N^2 (1+N)^2 (2+N)}
    \nonumber\\ &&        
    -
            \frac{256 \big(
                    2+3 N+3 N^2\big) S_{-2,2}}{(N-1) N^2 (1+N)^2 (2+N)}
            -\frac{1024 \big(
                    1+N+N^2\big) S_{-3,1}}{(N-1) N^2 (1+N)^2 (2+N)}
    \nonumber\\ &&        
    -\frac{32 \big(
                    2+N+N^2\big)^2 S_{2,1,1}}{3 (N-1) N^2 (1+N)^2 (2+N)}
            +\frac{1024 S_{-2,1,1}}{(N-1) N (1+N) (2+N)}
    \nonumber\\ &&        
    +\Biggl(
                    -\frac{128 S_1 P_{507}}{3 (N-2) (N-1) N^2 (1+N)^2 
    (2+N) (3+N)}
    \nonumber\\ &&
                    +\frac{32 P_{529}}{3 (N-2) (N-1) N^3 (1+N)^3 (2+N)^2 
    (3+N)}
            \Biggr) \zeta_3 \Biggr\} 
    \end{eqnarray} 
and
\begin{eqnarray} 
&&  C_{F_L,q}^{\rm PS, (3)} = 
    \nonumber\\ 
    && \textcolor{blue}{C_F} \Biggl\{
            \textcolor{blue}{T_F^2 N_F^2} \Biggl[
                    \frac{128 P_{536}}{27 (N-2) (N-1) N^3 (1+N)^4 
    (2+N)^3 (3+N)}
    \nonumber\\ &&
                    +\frac{128 \big(
                            16+27 N+13 N^2+8 N^3\big) S_1}{9 (N-1) N 
    (1+N)^3 (2+N)}
                    -\frac{64 \big(
                            2+N+N^2\big) S_1^2}{3 (N-1) N (1+N)^2 (2+N)}
    \nonumber\\ &&              
      -\frac{64 \big(
                            2+N+N^2\big) S_2}{3 (N-1) N (1+N)^2 (2+N)}
                    -\frac{2048 S_{-2}}{3 (N-2) (N-1) (1+N) (2+N) 
    (3+N)}
            \Biggr]
    \nonumber\\ &&
            +\textcolor{blue}{C_A T_F N_F} \Biggl[
                    -\frac{32 S_3 P_{502}}{3 (N-2) (N-1) N (1+N)^2 (2+N) 
    (3+N)}
                    -\frac{16 S_1^2 P_{512}}{3 (N-1)^2 N^2 (1+N)^3 
    (2+N)^2}
    \nonumber\\ &&
                    +\frac{16 S_2 P_{514}}{3 (N-1)^2 N^2 (1+N)^3 (2+N)^2}
                    -\frac{32 P_{545}}{27 (N-2) (N-1)^2 N^4 (1+N)^5 
    (2+N)^4 (3+N)}
    \nonumber\\ &&
                    +\Biggl(
                            -
                            \frac{32 P_{540}}{9 (N-2) (N-1)^2 N^3 
    (1+N)^4 (2+N)^3 (3+N)}
                            +\frac{160 \big(
                                    2+N+N^2\big) S_2}{(N-1) N (1+N)^2 
    (2+N)}
    \nonumber\\ &&
                            +\frac{256 \big(
                                    2+N+N^2\big) S_3}{(N-2) (N-1) (1+N) 
    (2+N) (3+N)}
                    \Biggr) S_1
                    -\frac{160 \big(
                            2+N+N^2\big) S_1^3}{3 (N-1) N (1+N)^2 (2+N)}
    \nonumber\\ &&
                    +\frac{256 \big(
                            2+N+N^2\big) S_4}{(N-2) (N-1) (1+N) (2+N) 
    (3+N)}
                    +\Biggl(
                            \frac{128 S_1 P_{490}}{(N-2) (N-1) N (1+N)^2 
    (2+N) (3+N)}
    \nonumber\\ &&
                            +\frac{64 P_{525}}{3 (N-2) (N-1)^2 N^2 
    (1+N)^3 (2+N)^2 (3+N)}
                    \Biggr) S_{-2}
                    -\frac{576 \big(
                            2+N+N^2\big) S_{-3}}{(N-1) N (1+N)^2 (2+N)}
    \nonumber\\ &&
                    -\frac{512 \big(
                            2+N+N^2\big) S_{3,1}}{(N-2) (N-1) (1+N) 
    (2+N) (3+N)}
                    +\frac{384 \big(
                            2+N+N^2\big) S_{-2,1}}{(N-1) N (1+N)^2 
    (2+N)}
    \nonumber\\ &&
                    +\Biggl(
                            \frac{192 \big(
                                    18+5 N+5 N^2\big)}{(N-2) N (1+N)^2 
    (3+N)}
                            -\frac{512 \big(
                                    2+N+N^2\big) S_1}{(N-2) (N-1) (1+N) 
    (2+N) (3+N)}
                    \Biggr) \zeta_3
        \Biggr] 
    \Biggr\}
    \nonumber\\ &&
    +\textcolor{blue}{C_F^2 T_F N_F} \Biggl\{
            \frac{256 S_{-3} P_{491}}{(N-2) (N-1) N (1+N)^2 (2+N) (3+N)}
     +\frac{16 S_2 P_{504}}{(N-1) N^2 (1+N)^3 (2+N)^2}
    \nonumber\\ &&
            +\frac{64 S_3 P_{500}}{3 (N-2) (N-1) N (1+N)^2 (2+N) (3+N)}
            -\frac{16 S_1^2 P_{505}}{(N-1) N^2 (1+N)^3 (2+N)^2}
    \nonumber\\ &&
            +\frac{32 P_{544}}{(N-2)^2 (N-1) N^4 (1+N)^5 (2+N)^3 (3+N)}
            +\Biggl(
                    \frac{96 \big(
                            2+N+N^2\big) S_2}{(N-1) N (1+N)^2 (2+N)}
    \nonumber\\ &&
                    -\frac{32 P_{533}}{(N-2) (N-1) N^3 (1+N)^4 (2+N)^3 
    (3+N)}
                    -\frac{512 \big(
                            2+N+N^2\big) S_3}{(N-2) (N-1) (1+N) (2+N) 
    (3+N)}
            \Biggr) 
    \nonumber\\ && \times
    S_1
            -\frac{32 \big(
                    2+N+N^2\big) S_1^3}{3 (N-1) N (1+N)^2 (2+N)}
            -\frac{512 \big(
                    2+N+N^2\big) S_4}{(N-2) (N-1) (1+N) (2+N) (3+N)}
    \nonumber\\ &&
            +  \Biggl(
                    -\frac{128 S_1 P_{489}}{(N-2) (N-1) N (1+N)^2 (2+N) 
    (3+N)}
    \nonumber\\ &&
                    +\frac{256 P_{517}}{(N-2)^2 (N-1) N^2 (1+N)^3 (2+N) 
    (3+N)}
            \Biggr) S_{-2}
     \nonumber\\ &&
            +\frac{1024 \big(
                    2+N+N^2\big) S_{3,1}}{(N-2) (N-1) (1+N) (2+N) 
    (3+N)}
    -\frac{1024 S_{-2,1}}{(N-1) N (1+N)^2 (2+N)}
    \nonumber\\ &&
            +\Biggl(
                    -\frac{128 P_{498}}{(N-2) (N-1) N (1+N)^2 (2+N) 
    (3+N)}
    \nonumber\\ &&
                    +\frac{1024 \big(
                            2+N+N^2\big) S_1}{(N-2) (N-1) (1+N) (2+N) 
    (3+N)}
            \Biggr) \zeta_3
    \Biggr\}
\end{eqnarray}


\noindent
with
\begin{eqnarray}
    P_{488} &=& N^4+2 N^3-71 N^2-72 N-20,
    \\
    P_{489} &=& N^4+2 N^3-15 N^2-16 N+44,
    \\
    P_{490}&=&N^4+2 N^3-13 N^2-14 N+16,
    \\
    P_{491}&=&N^4+2 N^3+3 N^2+2 N-16,
    \\
    P_{492}&=&N^4+2 N^3+41 N^2+40 N+4,
    \\
    P_{493}&=&N^4+2 N^3+65 N^2+64 N+28,
    \\
    P_{494}&=&5 N^4+10 N^3-83 N^2-88 N-4,
    \\
    P_{495}&=&5 N^4+15 N^3-38 N^2-36 N+72,
    \\
    P_{496}&=&7 N^4+16 N^3+13 N^2+12 N+4,
    \\
    P_{497}&=&8 N^4+13 N^3+15 N^2-20 N-60,
    \\
    P_{498}&=&11 N^4+22 N^3+39 N^2+28 N-84,
    \\
    P_{499}&=&25 N^4+50 N^3-87 N^2-112 N+60,
    \\
    P_{500}&=&41 N^4+82 N^3+117 N^2+76 N-204,
    \\
    P_{501}&=&47 N^4+94 N^3+487 N^2+440 N+308,
    \\
    P_{502}&=&49 N^4+98 N^3+105 N^2+56 N-372,
    \\
    P_{503}&=&63 N^4+126 N^3-N^2-64 N+4,
    \\
    P_{504}&=&N^5-6 N^4-19 N^3-52 N^2-76 N-32,
    \\
    P_{505}&=&11 N^5+30 N^4+47 N^3+36 N^2-36 N-32,
    \\
    P_{506}&=&N^6-13 N^5-7 N^4-3 N^3+34 N^2+44 N-120,
    \\
    P_{507}&=&4 N^6-12 N^5-11 N^4-18 N^3-43 N^2-20 N+84,
    \\
    P_{508}&=&5 N^6-177 N^5+65 N^4+297 N^3+190 N^2+140 N-3720,
    \\
    P_{509}&=&7 N^6+117 N^5+67 N^4+3 N^3-382 N^2-428 N-600,
    \\
    P_{510}&=&7 N^6+246 N^5+558 N^4+532 N^3+377 N^2-340 N-228,
    \\
    P_{511}&=&19 N^6-135 N^5+67 N^4+231 N^3-178 N^2-188 N-3336,
    \\
    P_{512}&=&19 N^6+9 N^5-111 N^4-329 N^3-364 N^2+32 N+24,
    \\
    P_{513}&=&37 N^6+399 N^5+301 N^4+129 N^3-2326 N^2-2516 N+264,
    \\
    P_{514}&=&41 N^6+87 N^5+3 N^4-175 N^3-284 N^2-80 N-24,
    \\
    P_{515}&=&43 N^6-159 N^5-173 N^4-273 N^3+278 N^2+580 N-1896,
    \\
    P_{516}&=&N^7-37 N^6-248 N^5-799 N^4-1183 N^3-970 N^2-580 N-168,
    \\
    P_{517}&=&N^7+2 N^6-10 N^5+37 N^3-2 N^2+4 N-40,
    \\
    P_{518}&=&N^7+3 N^6+2 N^5+33 N^4+73 N^3+52 N^2+36 N+16,
    \\
    P_{519}&=&3 N^7+7 N^6+7 N^5+125 N^4+274 N^3+248 N^2+200 N+80,
    \\
    P_{520}&=&15 N^7-26 N^6-187 N^5-464 N^4-734 N^3-644 N^2-504 N-176,
    \\
    P_{521}&=&23 N^7+94 N^6+147 N^5-28 N^4-288 N^3-348 N^2-384 N-176,
    \\
    P_{522}&=&49 N^7+185 N^6+340 N^5+287 N^4+65 N^3+62 N^2-196 N-168,
    \\
    P_{523}&=&51 N^7+175 N^6+139 N^5-435 N^4-1022 N^3-1060 N^2-952 N-384,
    \\
    P_{524}&=&N^8+9 N^7-9 N^6-165 N^5-148 N^4+408 N^3+464 N^2+432 N+576,
    \\
    P_{525}&=&6 N^8-6 N^7-17 N^6+141 N^5+61 N^4+297 N^3+58 N^2-900 N-216,
    \\
    P_{526}&=&11 N^8+221 N^7+754 N^6+1826 N^5+2301 N^4+897 N^3-150 N^2
    -1180 N
    \nonumber\\ &&
    -648,
    \\
    P_{527}&=&37 N^8-119 N^7-883 N^6-2063 N^5-986 N^4+1994 N^3+2252 N^2
    +1576 N
    \nonumber\\ &&
    +496,
    \\
    P_{528}&=&N^9+10 N^8+24 N^7+16 N^6-35 N^5-214 N^4-246 N^3+100 N^2
    +168 N
    \nonumber\\ &&
    +144,
    \\
    P_{529}&=&9 N^9-54 N^8-154 N^7-140 N^6-1159 N^5-1550 N^4+1192 N^3+2896 N^2
    \nonumber\\ &&
    +3664 N+2208,
    \\
    P_{530}&=&37 N^9-744 N^8-2698 N^7-1652 N^6+1309 N^5+2820 N^4+4416 N^3
    \nonumber\\ &&
    +560 N^2+560 N-768,
    \\
    P_{531}&=&3 N^{10}-15 N^9-121 N^8+106 N^7+993 N^6+693 N^5-1643 N^4-2376 N^3
    \nonumber\\ &&
    -2088 N^2-688 N+528,
    \\
    P_{532}&=&6 N^{10}+53 N^9+192 N^8-30 N^7+160 N^6+4697 N^5+12778 N^4+17152 N^3
    \nonumber\\ &&
    +14848 N^2+7920 N+1888,
    \\
    P_{533}&=&28 N^{10}+205 N^9+641 N^8+1139 N^7+1035 N^6+92 N^5-748 N^4-1656 N^3
    \nonumber \\ &&
    -2944 N^2-2464 N-768,
    \\
    P_{534}&=&67 N^{10}-142 N^9-27 N^8+1572 N^7-5543 N^6-12210 N^5+5535 N^4+16444 N^3
    \nonumber\\ &&
    +20992 N^2+3696 N-5040,
    \\
    P_{535}&=&114 N^{10}+681 N^9+1398 N^8-264 N^7-4306 N^6-2417 N^5+7466 N^4+14832 N^3
    \nonumber\\ &&
    +15008 N^2+9040 N+2400,
    \\
    P_{536}&=&127 N^{10}+842 N^9+1753 N^8-236 N^7-8468 N^6-17624 N^5-14130 N^4
    \nonumber\\ &&
    -2296 N^3
    +288 N^2-1440 N-864,
    \\
    P_{537}&=&176 N^{10}+973 N^9+1824 N^8-948 N^7-10192 N^6-19173 N^5-20424 N^4
    \nonumber\\ &&
    -16036 N^3
    -7816 N^2-1248 N+288,
    \\
    P_{538}&=&220 N^{10}-673 N^9-3885 N^8-2682 N^7+1450 N^6+10779 N^5+19215 N^4
    \nonumber\\ &&
    -3632 N^3
    +712 N^2-2928 N-2448,
    \\
    P_{539}&=&9 N^{11}+23 N^{10}-100 N^9-182 N^8+333 N^7+519 N^6+446 N^5+520 N^4-
    \nonumber\\ &&
    1040 N^3
    -2288 N^2-2784 N-1152,
    \\
    P_{540}&=&64 N^{11}+343 N^{10}-190 N^9-3834 N^8-7845 N^7-4410 N^6+6491 N^5+4865 N^4
    \nonumber\\ &&
    -9500 N^3-11760 N^2-9216 N-3024,
    \\
    P_{541}&=&82 N^{11}-107 N^{10}+1405 N^9+19278 N^8+66823 N^7+123752 N^6+143766 N^5
    \nonumber\\ &&
    +112177 N^4+57120 N^3+5000 N^2-7584 N -1584,
    \\
    P_{542}&=&824 N^{11}+3749 N^{10}+6953 N^9+11052 N^8+22103 N^7+50950 N^6+83556 N^5
    \nonumber\\ &&
    +75245 N^4+44472 N^3+22840 N^2+7584 N+720,
    \\
    P_{543}&=&39 N^{13}+66 N^{12}-1159 N^{11}-6196 N^{10}-3405 N^9+38262 N^8+50307 N^7
    \nonumber\\ &&
    -84768 N^6-167366 N^5+10252 N^4+198720 N^3+200928 N^2+92064 N
    \nonumber\\ &&
    +17856,
    \\
    P_{544}&=&43 N^{13}+253 N^{12}+286 N^{11}-985 N^{10}-2923 N^9-1277 N^8+5318 N^7
    \nonumber\\ &&
    +9049 N^6+4744 N^5+1348 N^4+5408 N^3+8624 N^2+4736 N+960,
    \\
    P_{545}&=&1844 N^{14}+16184 N^{13}+50096 N^{12}+39487 N^{11}-152593 N^{10}-478032 N^9
    \nonumber\\ &&
    -483740 N^8+106345 N^7+667861 N^6 +593144 N^5+330028 N^4+334800 N^3
    \nonumber\\ &&
    +331056 N^2+152064 N+25920,
    \\
    P_{546}&=&78 N^{15}+542 N^{14}+845 N^{13}-3182 N^{12}-15821 N^{11}-29382 N^{10}
    -37143 N^9
    \nonumber\\ &&
    -60252 N^8
    -74387 N^7+58234 N^6+375052 N^5+647480 N^4+655920 N^3+
    \nonumber\\ &&
    432480 N^2
    +173632 N+31872,
    \\
    P_{547}&=&1733 N^{15}+14466 N^{14}+39452 N^{13}+4062 N^{12}-191762 N^{11}-320533 N^{10}
    \nonumber\\ &&
    +130420 N^9+775488 N^8+348241 N^7-756255 N^6-802132 N^5+107380 N^4
    \nonumber\\ &&
    +801072 N^3
    +878256 N^2+502848 N+119232,
    \\
    P_{548}&=&4373 N^{16}+39739 N^{15}+116771 N^{14}-491 N^{13}-811481 N^{12}-1974485 N^{11}
    \nonumber\\ &&
    -1526011 N^{10}+1448375 N^9+6113428 N^8+13620182 N^7+22903928 N^6
    \nonumber\\ &&
    +27507368 N^5
    +24009536 N^4+15346080 N^3+7254144 N^2+2401920 N
    \nonumber\\ &&
    +414720,
    \\
    P_{549}&=&424 N^{18}+3206 N^{17}+5830 N^{16}-13976 N^{15}-63725 N^{14}-46169 N^{13}
    \nonumber\\ &&
    +105366 N^{12}+112942 N^{11}-193953 N^{10}-59123 N^9+1048298 N^8+1685392 N^7
    \nonumber\\ &&
    +242464 N^6
    -2245408 N^5-3487424 N^4-2893056 N^3-1520896 N^2-
    \nonumber\\ &&
    480000 N
    -69120,
    \\
    P_{550}&=&24754 N^{19}+255605 N^{18}+933618 N^{17}+783522 N^{16}-4107236 N^{15}
    \nonumber\\ &&
    -12029850 N^{14}-4276384 N^{13}+29789918 N^{12}+46452820 N^{11}-4027635 N^{10}
    \nonumber\\ &&
    -75545302 N^9-98368740 N^8-114975366 N^7-183461620 N^6-250266040 N^5
    \nonumber\\ &&
    -242010000 N^4
    -166111776 N^3-78065856 N^2-22156416 N-2799360.
    \end{eqnarray}

\noindent
Finally, the gluonic contributions read



\noindent
with
\begin{eqnarray}
P_{551} & =& -249 N^4-346 N^3-65 N^2+336 N+452,\\ 
P_{552} & =& -117 N^4+14 N^3-109 N^2-256 N+244,\\ 
P_{553} & =& 2 N^4+N^3+6 N^2+34 N+21,\\ 
P_{554} & =& 3 N^4+6 N^3+7 N^2+4 N+4,\\ 
P_{555} & =& 17 N^5+14 N^4-45 N^3-74 N^2-68 N-24,\\ 
P_{556} & =& 19 N^5+12 N^4-48 N^3-29 N^2-112 N-58,\\ 
P_{557} & =& 19 N^5+40 N^4+37 N^3+44 N^2-32 N-36,\\ 
P_{558} & =& 44 N^5+149 N^4+221 N^3+160 N^2-10 N-36,\\ 
P_{559} & =& 53 N^5+33 N^4+12 N^3+74 N^2-68 N+112,\\ 
P_{560} & =& 177 N^5+500 N^4+317 N^3-134 N^2-692 N-712,\\ 
P_{561} & =& -305 N^6-531 N^5+1305 N^4+2359 N^3+2600 N^2+1676 N+1248,\\ 
P_{562} & =& -293 N^6+201 N^5+1473 N^4+595 N^3+1628 N^2+380 N-240,\\ 
P_{563} & =& 3 N^6+9 N^5-5 N^4-25 N^3-14 N^2-16,\\ 
P_{564} & =& 5 N^6-21 N^5-147 N^4-181 N^3+100 N^2+220 N+120,\\ 
P_{565} & =& 6 N^6+24 N^5-37 N^4-22 N^3-93 N^2+210 N+144,\\ 
P_{566} & =& 6 N^6+36 N^5-65 N^4-146 N^3+81 N^2+228 N+468,\\ 
P_{567} & =& 6 N^6+48 N^5-31 N^4-154 N^3-51 N^2-66 N+720,\\ 
P_{568} & =& 8 N^6+15 N^5-14 N^4-17 N^3+60 N^2+44 N-24,\\ 
P_{569} & =& 11 N^6-3 N^5-132 N^4-193 N^3+187 N^2+298 N+120,\\ 
P_{570} & =& 12 N^6+60 N^5-139 N^4-238 N^3+159 N^2+618 N+720,\\ 
P_{571} & =& 12 N^6+84 N^5-121 N^4-346 N^3+165 N^2+294 N+1152,\\ 
P_{572} & =& 15 N^6+83 N^5+81 N^4-315 N^3-482 N^2-370 N-12,\\ 
P_{573} & =& 19 N^6+27 N^5+12 N^4+52 N^3+41 N^2+47 N+18,\\ 
P_{574} & =& 21 N^6+105 N^5-9 N^4-425 N^3-368 N^2+84 N+432,\\ 
P_{575} & =& 23 N^6+13 N^5-3 N^4+299 N^3+484 N^2-72 N+24,\\ 
P_{576} & =& 35 N^6-3 N^5-177 N^4-65 N^3+298 N^2+328 N+160,\\ 
P_{577} & =& 42 N^6+81 N^5-61 N^4-53 N^3+383 N^2+228 N-92,\\ 
P_{578} & =& 67 N^6+102 N^5-12 N^4-35 N^3-304 N^2-106 N+72,\\ 
P_{579} & =& 127 N^6+375 N^5+165 N^4+737 N^3+1668 N^2+1960 N+872,\\ 
P_{580} & =& 206 N^6+522 N^5+777 N^4-556 N^3-3197 N^2-884 N-180,\\ 
P_{581} & =& 349 N^6+1029 N^5+588 N^4-1499 N^3-3295 N^2-616 N+420,\\ 
P_{582} & =& 368 N^6+1404 N^5+1419 N^4-190 N^3-3353 N^2-2636 N-324,\\ 
P_{583} & =& 497 N^6-9 N^5-1449 N^4-1027 N^3-4304 N^2-1388 N-96,\\ 
P_{584} & =& 3 N^7-260 N^6-461 N^5+1640 N^4+2358 N^3-1832 N^2-5040 N-7056,\\ 
P_{585} & =& 12 N^7+74 N^6-315 N^5-1024 N^4+829 N^3+4244 N^2+2996 N+5568,\\ 
P_{586} & =& 15 N^7+155 N^6+292 N^5-489 N^4-1527 N^3-594 N^2+916 N+1128,\\ 
P_{587} & =& 16 N^7+36 N^6-89 N^5-342 N^4-395 N^3-246 N^2-124 N-24,\\ 
P_{588} & =& 23 N^7-83 N^6-933 N^5-1405 N^4+670 N^3+2664 N^2+2616 N+1072,\\ 
P_{589} & =& 85 N^7+701 N^6+305 N^5-1809 N^4-1686 N^3+412 N^2+8520 N+27360,\\ 
P_{590} & =& 93 N^7+801 N^6-2057 N^5-9261 N^4+4880 N^3+31656 N^2+23760 N+57456,\\ 
P_{591} & =& 147 N^7+483 N^6-1091 N^5-3551 N^4+844 N^3+9776 N^2+7552 N+6384,\\ 
P_{592} & =& 213 N^7+985 N^6-2533 N^5-10357 N^4+2972 N^3+33504 N^2+27296 N
\nonumber\\ &&
+33936,\\ 
P_{593} & =& 3 N^8+24 N^7-16 N^6-204 N^5+71 N^4+588 N^3-58 N^2-336 N-648,\\ 
P_{594} & =& 3 N^8+24 N^7+34 N^6-84 N^5-269 N^4-12 N^3+592 N^2+144 N-1008,\\ 
P_{595} & =& 9 N^8+512 N^7+186 N^6-4406 N^5-1547 N^4+9986 N^3+9116 N^2+1936 N
\nonumber\\ &&
-13488,\\ 
P_{596} & =& 9 N^8+512 N^7+244 N^6-4234 N^5-2009 N^4+8834 N^3+9964 N^2+3096 N
\nonumber\\ &&
-11808,\\ 
P_{597} & =& 9 N^8+512 N^7+302 N^6-4062 N^5-2471 N^4+7682 N^3+10812 N^2+4256 N
\nonumber\\ &&
-10128,\\ 
P_{598} & =& 20 N^8-34 N^7-295 N^6+167 N^5+885 N^4+243 N^3-726 N^2-1452 N+936,\\ 
P_{599} & =& 27 N^8+174 N^7+76 N^6-902 N^5-931 N^4+1008 N^3+1380 N^2+32 N
\nonumber\\ &&
-1632,\\ 
P_{600} & =& 30 N^8+493 N^7+213 N^6-3539 N^5-1915 N^4+5886 N^3+6880 N^2
+3280 N
\nonumber\\ &&
-6720,\\ 
P_{601} & =& 33 N^8+194 N^7-666 N^6-1212 N^5+2993 N^4+5026 N^3-440 N^2+6888 N
\nonumber\\ &&
-15120,\\ 
P_{602} & =& 39 N^8+4 N^7+188 N^6+886 N^5-895 N^4-2378 N^3-6316 N^2-5352 N
\nonumber\\ &&
+4608,\\ 
P_{603} & =& 39 N^8+200 N^7-1006 N^5-1299 N^4+166 N^3+3348 N^2+1816 N+1344,\\ 
P_{604} & =& 60 N^8-96 N^7-793 N^6+835 N^5+2809 N^4-603 N^3-4488 N^2-4252 N
\nonumber\\ &&
+2688,\\ 
P_{605} & =& 66 N^8-106 N^7-877 N^6+933 N^5+3225 N^4-427 N^3-5330 N^2-5284 
\nonumber\\ &&
N+2040,\\ 
P_{606} & =& 69 N^8+176 N^7-576 N^6-1906 N^5-521 N^4+6258 N^3+7340 N^2
\nonumber\\ &&
-4888 N-1344,\\ 
P_{607} & =& 72 N^8-116 N^7-961 N^6+1031 N^5+3641 N^4-251 N^3-6172 N^2
\nonumber\\ &&
-6316 N +1392,\\ 
P_{608} & =& 124 N^8+481 N^7+609 N^6+181 N^5-759 N^4-1226 N^3-730 N^2
\nonumber\\ &&
-192 N-216,\\ 
P_{609} & =& 143 N^8-454 N^7-2349 N^6+3611 N^5+12288 N^4+4331 N^3-27470 N^2
\nonumber\\ &&
-32364 N-7272,\\ 
P_{610} & =& 176 N^8-10 N^7-2127 N^6-153 N^5+5809 N^4+3419 N^3-2982 N^2
\nonumber\\ &&
-8524 N+1320,\\ 
P_{611} & =& 200 N^8+1179 N^7+952 N^6-2962 N^5-4454 N^4-1081 N^3+530 N^2
\nonumber\\ &&
+236 N +216,\\ 
P_{612} & =& 205 N^8+646 N^7+7902 N^6+6508 N^5-42471 N^4-77522 N^3+3764 N^2
\nonumber\\ &&
-16824 N+137376,\\ 
P_{613} & =& 233 N^8+392 N^7-3288 N^6-4054 N^5+3891 N^4 +9254 N^3+28540 N^2
\nonumber\\ &&
+22920 N-10656,\\ 
P_{614} & =& 263 N^8+602 N^7-4467 N^6-6067 N^5+9090 N^4+17633 N^3+33166 N^2
\nonumber\\ &&
+27756 N-35352,\\ 
P_{615} & =& 274 N^8+654 N^7-544 N^6-1793 N^5-1288 N^4-1853 N^3-782 N^2
\nonumber\\ &&
-68 N+216,\\ 
P_{616} & =& 434 N^8+1964 N^7-5853 N^6-14263 N^5+25449 N^4+54611 N^3
\nonumber\\ &&
-15746 N^2+19140 N-129672,\\ 
P_{617} & =& 485 N^8+1172 N^7-6414 N^6-8680 N^5+23145 N^4+38828 N^3-5048 N^2
\nonumber\\ &&
-6192 N-96048,\\ 
P_{618} & =& 641 N^8+1412 N^7-8862 N^6-11728 N^5+24861 N^4+49412 N^3
+9640 N^2
\nonumber\\ &&
+8784 N-51120,\\ 
P_{619} & =& -205 N^9-474 N^8+744 N^7+3298 N^6+5881 N^5+10528 N^4+17388 N^3
\nonumber\\ &&
+20760 N^2+13184 N+3232,\\ 
P_{620} & =& 13 N^9+88 N^8+136 N^7-122 N^6-177 N^5-738 N^4-2500 N^3-2060 N^2
\nonumber\\ &&
-1984 N-1776,\\ 
P_{621} & =& 35 N^9-40 N^8-760 N^7-28 N^6+3417 N^5+2604 N^4-1580 N^3-2032 N^2
\nonumber\\ &&
+928 N+768,\\ 
P_{622} & =& 71 N^9+84 N^8-660 N^7+36 N^6+3113 N^5+1232 N^4-4772 N^3-8240 N^2
\nonumber\\ &&
-3968 N-2112,\\ 
P_{623} & =& 93 N^9+1002 N^8+5144 N^7+12790 N^6+13999 N^5+2456 N^4-9140 N^3
\nonumber\\ &&
-11528 N^2-6880 N-1632,\\ 
P_{624} & =& 114 N^9+1125 N^8+2886 N^7-2796 N^6-16498 N^5-15241 N^4+830 N^3
\nonumber\\ &&
+6052 N^2+6248 N+4992,\\ 
P_{625} & =& 140 N^9+1139 N^8+2790 N^7+1892 N^6-1836 N^5-4153 N^4-3614 N^3
\nonumber\\ &&
-1902 N^2-720 N-648,\\ 
P_{626} & =& 293 N^9+2208 N^8+754 N^7-7064 N^6+1245 N^5+38896 N^4+54988 N^3
\nonumber\\ &&
+40360 N^2+512 N-6816,\\ 
P_{627} & =& 388 N^9+2233 N^8+4500 N^7+4324 N^6+2124 N^5-3167 N^4-8164 N^3
\nonumber\\ &&
-6414 N^2-2088 N-648,\\ 
P_{628} & =& 1067 N^9+4674 N^8+6232 N^7+2504 N^6-2267 N^5-3250 N^4+6992 N^3
\nonumber\\ &&
+17600 N^2+13680 N+3744,\\ 
P_{629} & =& 1721 N^9+10404 N^8+25600 N^7+22820 N^6-13643 N^5-11908 N^4
\nonumber\\ &&
+45842 N^3+52772 N^2+23496 N+4464,\\ 
P_{630} & =& 30 N^{10}+177 N^9-26 N^8-932 N^7-534 N^6+589 N^5-282 N^4
-314 N^3
\nonumber\\ &&
+2852 N^2+2520 N+144,\\ 
P_{631} & =& 191 N^{10}+663 N^9-914 N^8-3481 N^7+3013 N^6+7856 N^5-9352 N^4
\nonumber\\ &&
-14488 N^3+5280 N^2+6048 N-5184,\\ 
P_{632} & =& 383 N^{10}+2370 N^9+5056 N^8+4994 N^7+5029 N^6+5516 N^5+2420 N^4
\nonumber\\ &&
+5360 N^3+13992 N^2+13344 N+3744,\\ 
P_{633} & =& 764 N^{10}+436 N^9-4817 N^8+1183 N^7+16949 N^6+12811 N^5
\nonumber\\ &&
+9684 N^4+4650 N^3+2836 N^2-2160 N-864,\\ 
P_{634} & =& 1427 N^{10}+2932 N^9-2816 N^8+1429 N^7+18422 N^6+15883 N^5
\nonumber\\ &&
+15795 N^4+6756 N^3-6620 N^2-12456 N-4464,\\ 
P_{635} & =& 1471 N^{10}+17463 N^9+74204 N^8+138364 N^7+86603 N^6-47561 N^5
\nonumber\\ &&
-29066 N^4+91534 N^3+81060 N^2+9720 N-5616,\\ 
P_{636} & =& 19 N^{11}+134 N^{10}-120 N^9-1514 N^8+237 N^7+5580 N^6-1912 N^5
\nonumber\\ &&
-7208 N^4+4848 N^3+6464 N^2-1920 N-17280,\\ 
P_{637} & =& 42 N^{11}-21 N^{10}-902 N^9-1743 N^8-123 N^7+3687 N^6+5601 N^5
\nonumber\\ &&
+3531 N^4+6320 N^3+14040 N^2+13488 N+5040,\\ 
P_{638} & =& 90 N^{11}+399 N^{10}-448 N^9+2160 N^8+17304 N^7+9873 N^6-29826 N^5
\nonumber\\ &&
-27432 N^4-1208 N^3-7968 N^2-36096 N-16704,\\ 
P_{639} & =& 240 N^{11}+633 N^{10}-1246 N^9+5868 N^8+31038 N^7-5913 N^6
-104064 N^5
\nonumber\\ &&
-90228 N^4+28888 N^3+118368 N^2+64800 N+27648,\\ 
P_{640} & =& 339 N^{11}+4164 N^{10}+17783 N^9-2526 N^8-133535 N^7-158392 N^6
\nonumber\\ &&
+173869 N^5+322418 N^4-27256 N^3-251632 N^2-232368 N-113760,\\ 
P_{641} & =& 1565 N^{11}+11465 N^{10}+30742 N^9+35334 N^8+10185 N^7-2331 N^6
\nonumber\\ &&
+34756 N^5+92380 N^4+111968 N^3+73104 N^2+41472 N+15552,\\ 
P_{642} & =& 4724 N^{11}+32993 N^{10}+49126 N^9-60426 N^8-161406 N^7-29073 N^6
\nonumber\\ &&
+68536 N^5+87850 N^4+173588 N^3-14280 N^2+92016 N+95040,\\ 
P_{643} & =& 39 N^{12}+1656 N^{11}+7771 N^{10}-9028 N^9-78031 N^8-36088 N^7
+227761 N^6
\nonumber\\ &&
+194300 N^5-257892 N^4-310104 N^3-149632 N^2+33696 N+71424,\\ 
P_{644} & =& 63 N^{12}-381 N^{11}-2219 N^{10}+1918 N^9+14631 N^8-6291 N^7
-35967 N^6
\nonumber\\ &&
+62346 N^5+83132 N^4-143680 N^3-160656 N^2-10464 N-6336,\\ 
P_{645} & =& 153 N^{12}+49 N^{11}-1696 N^{10}+126 N^9+5857 N^8-3075 N^7
-10674 N^6
\nonumber\\ &&
+19412 N^5+39320 N^4-14912 N^3-43712 N^2-41344 N-19200,\\ 
P_{646} & =& 226 N^{12}+579 N^{11}-1701 N^{10}-2617 N^9+8109 N^8+1923 N^7
\nonumber\\ &&
-33361 N^6-11565 N^5+57183 N^4+42280 N^3-7200 N^2-6768 N-5616,\\ 
P_{647} & =& 430 N^{12}+2883 N^{11}+2713 N^{10}-8201 N^9-3396 N^8+3522 N^7
-44182 N^6
\nonumber\\ &&
-83352 N^5-43465 N^4+53360 N^3+113496 N^2+99072 N+31536,\\ 
P_{648} & =& 651 N^{12}+4796 N^{11}+12675 N^{10}+13520 N^9+6837 N^8
+19740 N^7+57489 N^6
\nonumber\\ &&
+94024 N^5+124188 N^4+127120 N^3+85008 N^2+46656 N
\nonumber\\ &&
+15552,\\ 
P_{649} & =& 1973 N^{12}+8607 N^{11}-2581 N^{10}-29530 N^9-4464 N^8-5301 N^7
-29117 N^6
\nonumber\\ &&
+129132 N^5+193261 N^4-30332 N^3+31872 N^2-54864 N-63504,\\ 
P_{650} & =& 5399 N^{12}+21717 N^{11}-8182 N^{10}-141466 N^9-174693 N^8
+104127 N^7
\nonumber\\ &&
+435784 N^6+484686 N^5+327700 N^4+206368 N^3+101088 N^2
+31968 N
\nonumber\\ &&
+5184,\\ 
P_{651} & =& 254 N^{13}+807 N^{12}-2683 N^{11}-10266 N^{10}-3135 N^9
+52217 N^8+158159 N^7
\nonumber\\ &&
+189724 N^6+56845 N^5-102970 N^4-172072 N^3-143952 N^2
-66544 N
\nonumber\\ &&
-12576,\\ 
P_{652} & =& 5459 N^{13}+43255 N^{12}+82772 N^{11}-53316 N^{10}-306641 N^9
-409197 N^8
\nonumber\\ &&
-673052 N^7-1415068 N^6-1869746 N^5-1266354 N^4-356752 N^3
+99792 N^2
\nonumber\\ &&
+139104 N+38880,\\ 
P_{653} & =& 13789 N^{13}+50057 N^{12}-25496 N^{11}-233268 N^{10}-154315 N^9
-125559 N^8
\nonumber\\ &&
-1154968 N^7-2581772 N^6-2738026 N^5-1636170 N^4-717392 N^3
\nonumber\\ &&
-119664 N^2
+52704 N+7776,\\ 
P_{654} & =& -12043 N^{14}-178942 N^{13}-923869 N^{12}-1714868 N^{11}
+1262555 N^{10}
\nonumber\\ &&
+9673170 N^9
+13425517 N^8+4916752 N^7-5024960 N^6
-9988720 N^5
\nonumber\\ &&
-15826400 N^4
-20182656 N^3-16420608 N^2-7236864 N-1306368,\\ 
P_{655} & =& 205 N^{14}-1070 N^{13}-6997 N^{12}+9444 N^{11}+57515 N^{10}
-20678 N^9
\nonumber\\ &&
-188539 N^8+10016 N^7+350968 N^6-25824 N^5-496784 N^4+19520 N^3
\nonumber\\ &&
+470400 N^2+418560 N+156672,\\ 
P_{656} & =& 9429 N^{14}+70894 N^{13}+159471 N^{12}-47508 N^{11}
-712173 N^{10}-1057954 N^9
\nonumber\\ &&
-675639 N^8-551176 N^7-224640 N^6+1703952 N^5
+3636768 N^4
\nonumber\\ &&
+4242688 N^3+3770112 N^2+1951488 N+380160,\\ 
P_{657} & =& 617 N^{15}+1213 N^{14}+3647 N^{13}-5135 N^{12}-62711 N^{11}
-59557 N^{10}
\nonumber\\ &&
+245733 N^9+472003 N^8-3502 N^7-620780 N^6-330504 N^5
+457568 N^4
\nonumber\\ &&
+783264 N^3+552896 N^2+210048 N+34560,\\ 
P_{658} & =& 3805 N^{15}+25467 N^{14}+31782 N^{13}-102029 N^{12}
-229612 N^{11}-7783 N^{10}
\nonumber\\ &&
-46240 N^9-603459 N^8+610299 N^7+3183968 N^6
+1748734 N^5-3156692 N^4
\nonumber\\ &&
-5185536 N^3-3830112 N^2-1561248 N-212544,\\ 
P_{659} & =& 1538 N^{16}+6349 N^{15}-4604 N^{14}-4419 N^{13}+247590 N^{12}
+595697 N^{11}
\nonumber\\ &&
-155918 N^{10}-1529965 N^9-660880 N^8+802290 N^7-3242454 N^6
\nonumber\\ &&
-12852560 N^5
-19050600 N^4-16010208 N^3-9510048 N^2-4292352 N
\nonumber\\ &&
-1026432,\\ 
P_{660} & =& 4920 N^{17}+59829 N^{16}+268187 N^{15}+460995 N^{14}
-275823 N^{13}-2426947 N^{12}
\nonumber\\ &&
-3950033 N^{11}-2689715 N^{10}-806789 N^9
-436062 N^8+2361354 N^7
\nonumber\\ &&
+9732268 N^6
+15162440 N^5+13427504 N^4+7410144 N^3
+2426688 N^2
\nonumber\\ &&
+328320 N-20736,\\ 
P_{661} & =& 32968 N^{18}+200687 N^{17}-76724 N^{16}-2085743 N^{15}
+226312 N^{14}
\nonumber\\ &&
+17762935 N^{13}
+21073412 N^{12}-46005765 N^{11}-129344416 N^{10}
-84661090 N^9
\nonumber\\ &&
+89668640 N^8
+206886704 N^7+116843424 N^6-124546688 N^5
-362045568 N^4
\nonumber\\ &&
-451275264 N^3
-326343168 N^2-127858176 N-20404224,\\ 
P_{662} & =& 34763 N^{18}+253932 N^{17}+235627 N^{16}-2269948 N^{15}
-5497382 N^{14}
\nonumber\\ &&
+4841882 N^{13}
+30639918 N^{12}+32530946 N^{11}
-9999541 N^{10}
-49961642 N^9
\nonumber\\ &&
-76271245 N^8
-149423714 N^7-245858156 N^6
-261281936 N^5
\nonumber\\ &&
-186926496 N^4
-92761632 N^3
-31078080 N^2-7713792 N-1306368,\\ 
P_{663} & =& -10701 N^{20}-117180 N^{19}-383698 N^{18}+362182 N^{17}
+4985840 N^{16}
\nonumber\\ &&
+7991006 N^{15}
-14095594 N^{14}-52618222 N^{13}
-4462979 N^{12}+154482966 N^{11}
\nonumber\\ &&
+143027116 N^{10}-208752656 N^9-456973712 N^8
-131024384 N^7+630587904 N^6
\nonumber\\ &&
+1328877824 N^5+1500652800 N^4+1142401536 N^3
+629462016 N^2
\nonumber\\ &&
+238215168 N+44789760,\\ 
P_{664} & =& 295317 N^{20}+3132540 N^{19}+8925794 N^{18}-15503126 N^{17}
-128158720 N^{16}
\nonumber\\ &&
-159520558 N^{15}+378838970 N^{14}+1195612190 N^{13}+526578043 N^{12}
\nonumber\\ &&
-1946889990 N^{11}-3106593740 N^{10}-1641770960 N^9-862068320 N^8
\nonumber\\ &&
-1429854848 N^7
+580862976 N^6+5174659328 N^5+7525373184 N^4
\nonumber\\ &&
+5839105536 N^3+2748681216 N^2
+747823104 N+89579520,\\ 
P_{665} & =& 23379 N^{21}+334890 N^{20}+1845809 N^{19}+4416076 N^{18}
+910717 N^{17}
\nonumber\\ &&
-19493138 N^{16}-43527868 N^{15}-31459862 N^{14}+7207567 N^{13}
-2971600 N^{12}
\nonumber\\ &&
-47214885 N^{11}+27754946 N^{10}+294517297 N^9+665343056 N^8
+1059128256 N^7
\nonumber\\ &&
+1391632944 N^6+1471240544 N^5+1188614208 N^4
+711979776 N^3
\nonumber\\ &&
+304328448 N^2+83960064 N+11197440
\end{eqnarray}

\noindent
and
\begin{eqnarray}
C_{F_L,g}^{(3)} &=& \textcolor{blue}{C_F} \Biggl\{
        \textcolor{blue}{T_F^2 N_F^2} \Biggl[
                \frac{64 S_1^2 P_{685}}{3 (N-1) N^2 (1+N)^3 (2+N)^2}
                +\frac{64 S_2 P_{698}}{3 (N-1) N^2 (1+N)^3 (2+N)^2 (3+N)}
\nonumber\\ &&               
  -\frac{32 P_{721}}{9 (N-2)^2 (N-1) N^4 (1+N)^5 (2+N)^4 (3+N)}
                +\Biggl(
                        -\frac{128 (5+3 N) S_2}{3 (1+N) (2+N) (3+N)}
\nonumber\\ && 
+                        \frac{32 P_{716}}{9 (N-2) (N-1) N^3 (1+N)^4 (2+N)^3 (3+N)}
                \Biggr) S_1
\nonumber\\ &&                
 -\frac{128 (5+3 N) S_3}{3 (1+N) (2+N) (3+N)}
                +\Biggl(
                        -\frac{128 P_{699}}{9 (N-2)^2 (N-1) N (1+N)^2 (2+N)^2 
(3+N)}
\nonumber\\ && 
                        -\frac{256 (N-1) S_1}{3 (N-2) (1+N) (3+N)}
                \Biggr) S_{-2}
                +\frac{512 (N-1) S_{-3}}{3 (N-2) (1+N) (3+N)}
\nonumber\\ &&               
 +\frac{256 (5+3 N) S_{2,1}}{3 (1+N) (2+N) (3+N)}
                -\frac{256 \big(
                        -2-3 N+N^2\big) \zeta_3}{(N-2) (1+N) (2+N) (3+N)}
\Biggr]
        +\textcolor{blue}{C_A T_F N_F} \Biggl[
\nonumber\\ &&
                -\frac{128 S_{-4} P_{667}}{(N-2) N (1+N)^2 (2+N) (3+N)}
                +\frac{512 S_{-3,1} P_{669}}{(N-2) N (1+N)^2 (2+N) (3+N)}
\nonumber\\ &&          
       +\frac{128 S_{-2,2} P_{672}}{(N-2) N (1+N)^2 (2+N) (3+N)}
                -\frac{256 S_{3,1} P_{683}}{(N-2) (N-1) N (1+N)^2 (2+N)^2 (3+N)}
\nonumber\\ &&                
 +\frac{64 S_4 P_{688}}{(N-2) (N-1) N (1+N)^2 (2+N)^2 (3+N)}
\nonumber\\ &&         
       +\frac{256 S_{-2,1} P_{697}}{(N-2) (N-1) N (1+N)^3 (2+N)^2 (3+N)}
\nonumber\\ &&                
 -
                \frac{32 S_2 P_{703}}{3 (N-2) (N-1) N (1+N)^3 (2+N)^2 (3+N)}
\nonumber\\ &&          
      -\frac{32 S_3 P_{704}}{3 (N-2) (N-1) N (1+N)^3 (2+N)^2 (3+N)}
\nonumber\\ &&                
 -\frac{64 \zeta_3 P_{705}}{(N-2) (N-1) N^2 (1+N)^2 (2+N)^2 (3+N)}
\nonumber\\ &&               
 -\frac{32 S_1^2 P_{706}}{3 (N-2) (N-1) N (1+N)^3 (2+N)^2 (3+N)}
\nonumber\\ &&              
  -\frac{8 P_{722}}{9 (N-2)^2 (N-1) N^4 (1+N)^5 (2+N)^4 (3+N)}
\nonumber\\ &&               
 +\Biggl(
                        -\frac{64 S_3 P_{678}}{(N-2) (N-1) N (1+N)^2 (2+N)^2 (3+N)}
\nonumber\\ &&                   
                        +\frac{64 \zeta_3 P_{692}}{(N-2) (N-1) N (1+N)^2 (2+N)^2 
(3+N)}
\nonumber\\ && 
                        -\frac{8 P_{719}}{9 (N-2)^2 (N-1) N^3 (1+N)^4 (2+N)^3 
(3+N)}
\nonumber\\ && 
      +\frac{128 S_{-2,1} P_{673}}{(N-2) N (1+N)^2 (2+N) (3+N)}
                        +\frac{16 \big(
                                90+293 N+380 N^2+117 N^3\big) S_2}{3 N (1+N)^2 (2+N) 
(3+N)}
                \Biggr) S_1
\nonumber\\ &&              
                -\frac{16 \big(
                        2+13 N+13 N^2\big) S_1^3}{3 N (1+N)^2 (2+N)}
                +\Biggl(
                        -\frac{256 S_1^2 P_{666}}{(N-2) N (1+N)^2 (2+N) (3+N)}
\nonumber\\ &&                
         +\frac{256 S_1 P_{707}}{3 (N-2)^2 (N-1) N^2 (1+N)^2 (2+N)^2 
(3+N)}
\nonumber\\ && 
                        +\frac{32 P_{717}}{9 (N-2)^2 (N-1) N^3 (1+N)^4 (2+N)^2 
(3+N)}
                        +\frac{3072 \zeta_3}{(1+N) (2+N)}
\nonumber\\ && 
                        -\frac{256 \big(
                                6+N+N^2\big) S_2}{(N-2) N (1+N)^2 (2+N) (3+N)}
                \Biggr) S_{-2}
                +
                \frac{64 \big(
                        -54+13 N+13 N^2\big) S_{-2}^2}{(N-2) (1+N) (2+N) (3+N)}
\nonumber\\ &&                
 +\Biggl(
                        -\frac{64 P_{709}}{3 (N-2) (N-1) N^2 (1+N)^3 (2+N)^2 (3+N)}
\nonumber\\ && 
                        -\frac{64 S_1 P_{674}}{(N-2) N (1+N)^2 (2+N) (3+N)}
                \Biggr) S_{-3}
                -\frac{128 \big(
                        9+35 N+50 N^2+18 N^3\big) S_{2,1}}{3 N (1+N)^2 (2+N) (3+N)}
\nonumber\\ &&                
 -\frac{1536 S_{-2,3}}{(1+N) (2+N)}
                +\frac{1536 S_{-4,1}}{(1+N) (2+N)}
                -\frac{4608 (N-1) S_{-2,1,1}}{(N-2) (1+N) (3+N)}
\nonumber\\ &&                
 +\frac{7680 \zeta_5}{(1+N) (2+N)}
        \Biggr]
\Biggr\}
\nonumber\\ &&
+\textcolor{blue}{C_A T_F^2 N_F^2} \Biggl\{
        \frac{128 S_{-2} P_{691}}{9 (N-2) (N-1) N (1+N)^2 (2+N)^2 (3+N)}
\nonumber\\ &&    
     -\frac{32 P_{718}}{27 (N-2) (N-1) N^3 (1+N)^4 (2+N)^4 (3+N)}
- \frac{64 \zeta_3}{(N+1)(N+2)}
\nonumber\\ && 
        +\Biggl(
                -\frac{64 P_{677}}{27 (N-1) N (1+N)^3 (2+N)^2}
                +\frac{128 S_2}{(1+N) (2+N)}
                +\frac{256 S_{-2}}{3 (1+N) (2+N)}
        \Biggr) S_1
\nonumber\\ &&   
      -\frac{128 \big(
                -3-8 N-6 N^2+11 N^3\big) S_1^2}{9 (N-1) N (1+N)^2 (2+N)}
        +\frac{128 \big(
                -6-19 N+11 N^2+5 N^3\big) S_2}{9 (N-1) N (1+N) (2+N)^2}
\nonumber\\ && 
        -\frac{128 S_1^3}{9 (1+N) (2+N)}
        +\frac{512 S_3}{9 (1+N) (2+N)}
        -\frac{512 S_{-3}}{3 (1+N) (2+N)}
        -\frac{512 S_{2,1}}{3 (1+N) (2+N)}
\Biggr\}
\nonumber\\ &&
+
\textcolor{blue}{C_F^2 T_F N_F} \Biggl\{
        \frac{4096 S_{3,1} P_{668}}{(N-2) (N-1) N (1+N)^2 (2+N)^2 (3+N)}
        +\frac{8 S_1^2 P_{676}}{N^2 (1+N)^3 (2+N)}
\nonumber\\ &&       
  -\frac{128 S_{-4} P_{670}}{(N-2) N (1+N)^2 (2+N) (3+N)}
        -\frac{128 S_4 P_{687}}{(N-2) (N-1) N (1+N)^2 (2+N)^2 (3+N)}
\nonumber\\ && 
        -\frac{8 S_2 P_{671}}{N^2 (1+N)^3 (2+N)}
        +\frac{64 S_3 P_{700}}{3 (N-2) (N-1) N (1+N)^3 (2+N)^2 (3+N)}
\nonumber\\ &&       
  -\frac{128 S_{-2,1} P_{701}}{(N-2) (N-1) N (1+N)^3 (2+N)^2 (3+N)}
\nonumber\\ &&      
   +\frac{128 \zeta_3 P_{702}}{(N-2) (N-1) N^2 (1+N)^2 (2+N)^2 (3+N)}
\nonumber\\ &&   
     -\frac{8 P_{714}}{(N-2)^2 N^4 (1+N)^5 (2+N) (3+N)}
\nonumber\\ &&
        +\Biggl(
                \frac{8 P_{712}}{(N-2) (N-1) N^3 (1+N)^4 (2+N)^2 (3+N)}
\nonumber\\ &&                
 +\frac{128 S_3 P_{679}}{(N-2) (N-1) N (1+N)^2 (2+N)^2 (3+N)}
\nonumber\\ &&               
         -\frac{128 \zeta_3 P_{690}}{(N-2) (N-1) N (1+N)^2 (2+N)^2 (3+N)}
 +\frac{48 \big(
                        2+N+N^2\big) S_2}{N (1+N)^2 (2+N)}
\nonumber\\ &&              
                -\frac{768 (N-1) \big(
                        2+N+N^2\big) S_{-2,1}}{(N-2) N (1+N)^2 (3+N)}
        \Biggr) S_1
        -\frac{80 \big(
                2+N+N^2\big) S_1^3}{3 N (1+N)^2 (2+N)}
\nonumber\\ && 
        +\Biggl(
                \frac{128 S_1 P_{696}}{(N-2) (N-1) N^2 (1+N)^2 (2+N)^2 (3+N)}
                +
                \frac{256 \big(
                        2+N+N^2\big) S_1^2}{N (1+N)^2 (2+N)}
\nonumber\\ &&                
 -\frac{32 P_{715}}{(N-2)^2 (N-1) N^3 (1+N)^4 (2+N)^2 (3+N)}
                -\frac{256 \big(
                        2+N+N^2\big) S_2}{N (1+N)^2 (2+N)}
\nonumber\\ &&                
 -\frac{3072 \zeta_3}{(1+N) (2+N)}
        \Biggr) S_{-2}
        -\frac{512 \big(
                -4+N+N^2\big) S_{-2}^2}{(N-2) (1+N) (2+N) (3+N)}
\nonumber\\ &&  
      +\Biggl(
                \frac{64 P_{708}}{(N-2) (N-1) N^2 (1+N)^3 (2+N)^2 (3+N)}
\nonumber\\ &&               
  -\frac{128 \big(
                        -18+N+N^2
                \big)
\big(2+N+N^2\big) S_1}{(N-2) N (1+N)^2 (2+N) (3+N)}
        \Biggr) S_{-3}
        +\frac{64 \big(
                2+N+N^2\big) S_{2,1}}{N (1+N)^2 (2+N)}
\nonumber\\ && 
       -\frac{256 \big(
                2+N+N^2
        \big)
\big(6+N+N^2\big) S_{-2,2}}{(N-2) N (1+N)^2 (2+N) (3+N)}
        -\frac{3072 \big(
                2+N+N^2\big) S_{-3,1}}{(N-2) N (1+N)^2 (2+N) (3+N)}
\nonumber\\ && 
        +\frac{1536 S_{-2,3}}{(1+N) (2+N)}
        -\frac{1536 S_{-4,1}}{(1+N) (2+N)}
        +\frac{1024 \big(
                2+N+N^2\big) S_{-2,1,1}}{(N-2) (1+N) (2+N) (3+N)}
\nonumber\\ &&   
     -\frac{7680 \zeta_5}{(1+N) (2+N)}
\Biggr\}
\nonumber\\ &&
+ \textcolor{blue}{C_A^2 T_F N_F} \Biggl\{
        \frac{64 S_4 P_{680}}{(N-2) (N-1) N (1+N)^2 (2+N)^2 (3+N)}
\nonumber\\ &&
        +\frac{256 S_{3,1} P_{682}}{(N-2) (N-1) N (1+N)^2 (2+N)^2 (3+N)}
\nonumber\\ &&
        +\frac{128 S_{-2,1} P_{689}}{(N-2) (N-1) N (1+N)^2 (2+N)^2 (3+N)}
\nonumber\\ &&
        +\frac{16 \zeta_3 P_{693}
        }{(N-2) (N-1) N (1+N)^2 (2+N)^2 (3+N)}
\nonumber\\ &&    
     -
        \frac{32 S_3 P_{694}}{9 (N-2) (N-1) N (1+N)^2 (2+N)^2 (3+N)}
        -\frac{32 S_2 P_{710}}{9 (N-1)^2 N^2 (1+N)^3 (2+N)^3}
\nonumber\\ &&    
     +\frac{16 P_{723}}{27 (N-2) (N-1)^2 N^4 (1+N)^5 (2+N)^5 (3+N)}
\nonumber\\ && 
        +\Biggl[
                -\frac{32 S_2 P_{675}}{(N-1) N (1+N)^2 (2+N)^2}
                +\frac{128 S_3 P_{681}}{(N-2) (N-1) N (1+N)^2 (2+N)^2 (3+N)}
\nonumber\\ &&                
 -\frac{128 \zeta_3 P_{686}}{(N-2) (N-1) N (1+N)^2 (2+N)^2 (3+N)}
                -\frac{512 \big(
                        -14+3 N+3 N^2\big) S_{-2,1}}{(N-2) (1+N) (2+N) (3+N)}
\nonumber\\ &&                 
+\frac{32 P_{720}}{27 (N-2) (N-1)^2 N^3 (1+N)^4 (2+N)^4 (3+N)}
        \Biggr] S_1
        +\Biggl[
                -\frac{256 S_2}{(1+N) (2+N)}
\nonumber\\ && 
+                \frac{32 P_{711}}{9 (N-1)^2 N^2 (1+N)^3 (2+N)^3}
        \Biggr] S_1^2
        +\frac{32 \big(
                -36-47 N-72 N^2+83 N^3\big) S_1^3}{9 (N-1) N (1+N)^2 (2+N)}
\nonumber\\ &&   
      +\frac{32 S_1^4}{(1+N) (2+N)}
        +\frac{32 S_2^2}{(1+N) (2+N)}
        +\Biggl[
                -\frac{128 S_1^2}{(1+N) (2+N)}
\nonumber\\ &&
                -\frac{64 S_1 P_{695}}{3 (N-2) (N-1) N (1+N)^2 (2+N)^2 (3+N)}
                +\frac{256 S_2}{(1+N) (2+N)}
\nonumber\\ &&
                -\frac{32 P_{713}}{9 (N-2) (N-1)^2 N^2 (1+N)^3 (2+N)^3 (3+N)}
                -\frac{768 \zeta_3}{(1+N) (2+N)}
        \Biggr] S_{-2}
\nonumber\\ &&       
 -\frac{32 \big(
                -14+5 N+5 N^2\big) S_{-2}
        ^2}{(N-2) (1+N) (2+N) (3+N)}
        +\Biggl[
                \frac{512 \big(
                        -16+3 N+3 N^2\big) S_1}{(N-2) (1+N) (2+N) (3+N)}
\nonumber\\ &&          
       -\frac{64 P_{684}}{3 (N-2) (N-1) N (1+N)^2 (2+N)^2 (3+N)}
        \Biggr] S_{-3}
        +\frac{1408 S_{2,1}}{3 (1+N) (2+N)}
\nonumber\\ && 
        +\frac{32 \big(
                -122+23 N+23 N^2\big) S_{-4}}{(N-2) (1+N) (2+N) (3+N)}
        -\frac{128 \big(
                -62+13 N+13 N^2\big) S_{-2,2}}{(N-2) (1+N) (2+N) (3+N)}
\nonumber\\ &&   
      +\frac{384 S_{-2,3}}{(1+N) (2+N)}
        -\frac{256 \big(
                -34+7 N+7 N^2\big) S_{-3,1}}{(N-2) (1+N) (2+N) (3+N)}
        -\frac{384 S_{-4,1}}{(1+N) (2+N)}
\nonumber\\ &&   
      +\frac{1024 \big(
                -14+3 N+3 N^2\big) S_{-2,1,1}}{(N-2) (1+N) (2+N) (3+N)}
        -\frac{1920 \zeta_5}{(1+N) (2+N)}
\Biggr\}.
\end{eqnarray}


\noindent
with
\begin{eqnarray}
P_{666} &=& N^4+2 N^3-6 N^2-7 N-6,\\ 
P_{667} &=& N^4+2 N^3+16 N^2+15 N+18,\\ 
P_{668} &=& 2 N^4+4 N^3+N^2-N+12,\\ 
P_{669} &=& 4 N^4+8 N^3-3 N^2-7 N+6,\\ 
P_{670} &=& 5 N^4+10 N^3-27 N^2-32 N-36,\\ 
P_{671} &=& 15 N^4+48 N^3+71 N^2+58 N+8,\\ 
P_{672} &=& 17 N^4+34 N^3-15 N^2-32 N+12,\\ 
P_{673} &=& 21 N^4+42 N^3-31 N^2-52 N-12,\\ 
P_{674} &=& 25 N^4+50 N^3-27 N^2-52 N+36,\\ 
P_{675} &=& 35 N^4+46 N^3-79 N^2-66 N-32,\\ 
P_{676} &=& 35 N^4+84 N^3+99 N^2+78 N+16,\\ 
P_{677} &=& 302 N^5+597 N^4-286 N^3-1017 N^2-844 N-120,\\ 
P_{678} &=& N^6+3 N^5-87 N^4-179 N^3-46 N^2+44 N-312,\\ 
P_{679} &=& N^6+3 N^5-35 N^4-75 N^3-26 N^2+12 N-168,\\ 
P_{680} &=& N^6+3 N^5-33 N^4-71 N^3-4 N^2+32 N-72,\\ 
P_{681} &=& N^6+3 N^5-19 N^4-43 N^3+22 N-36,\\ 
P_{682} &=& N^6+3 N^5+9 N^4+13 N^3+8 N^2+2 N+36,\\ 
P_{683} &=& N^6+3 N^5+37 N^4+69 N^3+22 N^2-12 N+168,\\ 
P_{684} &=& 2 N^6-6 N^5+143 N^4+396 N^3-403 N^2-720 N-612,\\ 
P_{685} &=& 2 N^6+6 N^5+7 N^4+4 N^3+9 N^2+8 N+12,\\ 
P_{686} &=& 3 N^6+9 N^5-35 N^4-85 N^3-4 N^2+40 N-72,\\ 
P_{687} &=& 3 N^6+9 N^5+23 N^4+31 N^3-14 N^2-28 N+264,\\ 
P_{688} &=& 5 N^6+15 N^5+65 N^4+105 N^3+14 N^2-36 N+408,\\ 
P_{689} &=& 8 N^6+20 N^5-13 N^4-26 N^3-9 N^2-32 N-60,\\ 
P_{690} &=& 11 N^6+33 N^5-85 N^4-225 N^3-130 N^2-12 N-168,\\ 
P_{691} &=& 16 N^6+48 N^5-83 N^4-246 N^3+13 N^2+144 N-36,\\ 
P_{692} &=& 17 N^6+51 N^5-211 N^4-507 N^3-154 N^2+108 N-456,\\ 
P_{693} &=& 33 N^6+99 N^5-421 N^4-1007 N^3+68 N^2+588 N-1152,\\ 
P_{694} &=& 35 N^6+33 N^5-976 N^4-1407 N^3+131 N^2+132 N-3132,\\ 
P_{695} &=& 59 N^6+129 N^5-427 N^4-669 N^3+794 N^2+678 N-612,\\ 
P_{696} &=& N^7-N^6-11 N^5+9 N^4-66 N^3-68 N^2+168 N-96,\\ 
P_{697} &=& 2 N^7+7 N^6-48 N^5-160 N^4-120 N^3+29 N^2+138 N+72,\\ 
P_{698} &=& 6 N^7+24 N^6+19 N^5-37 N^4-73 N^3-59 N^2-36 N-36,\\ 
P_{699} &=& 13 N^7+13 N^6-103 N^5-113 N^4+74 N^3+172 N^2+520 N-864,\\ 
P_{700} &=& 17 N^7+56 N^6+216 N^5+554 N^4+891 N^3+854 N^2+748 N+120,\\ 
P_{701} &=& 21 N^7+80 N^6-74 N^5-468 N^4-343 N^3+164 N^2+316 N+48,\\ 
P_{702} &=& 23 N^7+75 N^6-152 N^5-473 N^4-109 N^3+136 N^2-156 N+144,\\ 
P_{703} &=& 27 N^7+42 N^6-68 N^5-51 N^4+109 N^3+15 N^2-554 N-384,\\ 
P_{704} &=& 31 N^7+156 N^6+722 N^5+1524 N^4+1595 N^3+960 N^2+1828 N+1248,\\ 
P_{705} &=& 32 N^7+146 N^6-269 N^5-972 N^4-149 N^3+416 N^2-612 N+144,\\ 
P_{706} &=& 77 N^7+278 N^6-260 N^5-1501 N^4-725 N^3+1145 N^2+1466 N+384,\\ 
P_{707} &=& 14 N^8+11 N^7-137 N^6-85 N^5+514 N^4-109 N^3-880 N^2+828 N-144,\\ 
P_{708} &=& 29 N^8+112 N^7-42 N^6-484 N^5-463 N^4-12 N^3+300 N^2+112 N+192,\\ 
P_{709} &=& 49 N^8+190 N^7-244 N^6-1346 N^5-1213 N^4+4 N^3+1096 N^2+696 N+288,\\ 
P_{710} &=& 67 N^8+334 N^7-73 N^6-1247 N^5-406 N^4+1645 N^3+1708 N^2-12 N-72,\\ 
P_{711} &=& 358 N^8+1078 N^7-397 N^6-2729 N^5-529 N^4+2743 N^3+2944 N^2-12 N
\nonumber\\ &&
-216,\\ 
P_{712} &=& 91 N^{10}+417 N^9-128 N^8-2686 N^7-6937 N^6-9783 N^5-4234 N^4+2324 N^3
\nonumber\\ &&
+424 N^2-1440 N-576,\\ 
P_{713} &=& 281 N^{10}+1405 N^9-105 N^8-7986 N^7-4629 N^6+13197 N^5+7549 N^4
\nonumber\\ &&
-15904 N^3-2448 N^2+16848 N+2160,\\ 
P_{714} &=& 18 N^{11}+68 N^{10}+179 N^9-304 N^8-1767 N^7-2263 N^6-230 N^5+1039 N^4
\nonumber\\ &&
+76 N^3-1424 N^2-1040 N-240,\\ 
P_{715} &=& 59 N^{11}+161 N^{10}-392 N^9-1310 N^8+31 N^7+2625 N^6+1478 N^5-1220 N^4
\nonumber\\ &&
-152 N^3-1088 N^2+64 N+768,\\ 
P_{716} &=& 93 N^{11}+543 N^{10}+650 N^9-1782 N^8-5427 N^7-4905 N^6-900 N^5+4272 N^4
\nonumber\\ &&
+7024 N^3+5616 N^2+7488 N+3456,\\ 
P_{717} &=& 581 N^{11}+1653 N^{10}-3824 N^9-13086 N^8+1281 N^7+26757 N^6+12850 N^5
\nonumber\\ && 
-15036 N^4-808 N^3-5184 N^2-6336 N+3456,\\ 
P_{718} &=& 51 N^{12}+423 N^{11}-314 N^{10}-9178 N^9-21183 N^8+4305 N^7+82314 N^6
\nonumber\\ &&
+134826 N^5+92180 N^4+11968 N^3-18048 N^2-11232 N-1728,\\ 
P_{719} &=& 1347 N^{12}+5385 N^{11}-13244 N^{10}-70178 N^9-33309 N^8+167445 N^7
\nonumber\\ &&
+305262 N^6+206148 N^5-13096 N^4-53200 N^3+70560 N^2+100800 N
\nonumber\\ && 
+31104,\\ 
P_{720} &=& 1820 N^{13}+12757 N^{12}+15253 N^{11}-73492 N^{10}-200208 N^9
-38667 N^8
\nonumber\\ &&
+354250 N^7+351875 N^6-74335 N^5-203477 N^4+26664 N^3+91944 N^2
\nonumber\\ &&
+58320 N
+19440,\\ 
P_{721} &=& 12 N^{15}+134 N^{14}+501 N^{13}+453 N^{12}-1684 N^{11}-5780 N^{10}
-10781 N^9
\nonumber\\ &&
-18749 N^8
-8256 N^7+28470 N^6+10840 N^5-52984 N^4-74944 N^3
\nonumber\\ &&
-83808 N^2
-58752 N
-17280,\\ 
P_{722} &=& 96 N^{15}+266 N^{14}-5889 N^{13}-31881 N^{12}-27580 N^{11}+147580 N^{10}
+402667 N^9
\nonumber\\ &&
+282187 N^8-247206 N^7-501336 N^6-287480 N^5-222400 N^4-437536 N^3
\nonumber\\ &&
-461952 N^2-206208 N-34560,\\ 
P_{723} &=& 150 N^{16}+1614 N^{15}-1820 N^{14}-61808 N^{13}-210369 N^{12}
-122317 N^{11}
\nonumber\\ &&
+723938 N^{10}+1759218 N^9+1253237 N^8-923191 N^7-2589744 N^6
\nonumber\\ &&
-2736284 N^5-2333312 N^4-1942608 N^3-1354752 N^2-565056 N-103680.
\end{eqnarray}

\noindent
The $d_{abc}$ contribution read
\begin{eqnarray}
&& C_{F_2,g}^{d_{abc},\rm (3)} =
\textcolor{blue}{\frac{d_{abc}d^{abc} N_F^2}{N_A}} \Biggl\{
\nonumber\\ &&
        -\frac{256 S_{-2,1} P_{725}}{45 N (1+N) (2+N)^2}
        +\frac{128 S_3 P_{726}}{(N-2) N (1+N) (2+N)^2 (3+N)}
\nonumber\\ &&
        -\frac{256 S_4 P_{734}}{3 (N-2) N^2 (1+N)^2 (2+N)^2 (3+N)}
        +\frac{512 S_{3,1} P_{734}}{3 (N-2) N^2 (1+N)^2 (2+N)^2 
(3+N)}
\nonumber\\ &&
        +\frac{128 (N-1) S_5 P_{724}}{45 N (1+N) (2+N)}
        +\frac{32 P_{742}}{225 (N-2) (N-1)^3 N^3 (1+N)^5 (2+N)^5 
(3+N)^3}
\nonumber\\ &&
        +\Biggl(
                -\frac{64 P_{740}}{225 (N-2) (N-1)^2 N^2 (1+N)^4 
(2+N)^4 (3+N)^3}
                -\frac{1024 (N-4) (N-1) S_{3,1}}{N (1+N) (2+N)}
\nonumber\\ &&
                -\frac{128 S_3 P_{733}}{(N-2) N^2 (1+N)^2 (2+N)^2 
(3+N)}
                +\frac{512 (N-4) (N-1) S_4}{N (1+N) (2+N)}
\nonumber\\ &&          
      +\frac{512 \big(
                        8-3 N+3 N^2\big) S_{-2,1}}{3 N (1+N) (2+N)}
        \Biggr) S_1
        +\Biggl(
                \frac{128 \big(
                        19-30 N+3 N^2\big)}{3 N (1+N) (2+N)}
                +\frac{256 (N-4) (N-1) S_3}{N (1+N) (2+N)}
        \Biggr) 
\nonumber\\ && \times
S_1^2
        -\frac{768 (N-4) (N-1) S_2 S_3}{N (1+N) (2+N)}
        +\Biggl(
                \frac{256 S_1 P_{736}}{45 (N-2) N^2 (1+N)^2 (2+N)^2 
(3+N)}
\nonumber\\ &&
                +\frac{64 P_{738}}{45 (N-2) N^2 (1+N)^3 (2+N)^3 
(3+N)}
                -\frac{256 \big(
                        8-3 N+3 N^2\big) S_1^2}{3 N (1+N) (2+N)}
\nonumber\\ &&                
+
                \frac{256}{45} (N-1) \big(
                        21-2 N+N^2\big) S_3
                +\frac{256}{45} (N-1) \big(
                        21-2 N+N^2\big) S_{-2,1}
        \Biggr) S_{-2}
\nonumber\\ &&
        -\frac{128 (N-1) S_{-2}^2}{(1+N) (2+N)}
        +\Biggl(
                \frac{128 P_{725}}{45 N (1+N) (2+N)^2}
                -\frac{256 \big(
                        8-3 N+3 N^2\big) S_1}{3 N (1+N) (2+N)}
        \Biggr) S_{-3}
\nonumber\\ &&
        -\frac{256 \big(
                4-3 N+3 N^2\big) S_{-4}}{3 N (1+N) (2+N)}
        +\frac{128}{45} (N-1) \big(
                21-2 N+N^2\big) S_{-5}
\nonumber\\ &&   
        +\frac{768 (N-4) (N-1) S_{2,3}}{N (1+N) (2+N)}
     -\frac{768 (N-4) (N-1) S_{4,1}}{N (1+N) (2+N)}
        +\frac{256 \big(
                8-3 N+3 N^2\big) S_{-2,2}}{3 N (1+N) (2+N)}
\nonumber\\ &&   
        -\frac{256}{45} (N-1) \big(
                21-2 N+N^2\big) S_{-2,3}
        +\frac{256 \big(
                8-3 N+3 N^2\big) S_{-3,1}}{3 N (1+N) (2+N)}
\nonumber\\ && 
        +\frac{1536 (N-4) (N-1) S_{3,1,1}}{N (1+N) (2+N)}
-\frac{512 \big(
                8-3 N+3 N^2\big) S_{-2,1,1}}{3 N (1+N) (2+N)}
\nonumber\\ &&        
        -\frac{256}{45} (N-1) \big(
                21-2 N+N^2\big) S_{-2,1,-2}
+\Biggl(
                -\frac{64 P_{732}}{45 (N-2) N (1+N) (2+N)^2 (3+N)}
\nonumber\\ &&
                +\frac{512 S_1 P_{735}}{3 (N-2) N^2 (1+N)^2 (2+N)^2 
(3+N)}
                -\frac{512 (N-4) (N-1) S_1^2}{N (1+N) (2+N)}
\nonumber\\ &&           
     +\frac{256}{45} (N-1) \big(
                        21-2 N+N^2\big) S_{-2}
        \Biggr) \zeta_3
        +\frac{64 (N-1) P_{724}}{9 N (1+N) (2+N)} \zeta_5
\Biggr\},
\\
&&  C_{F_L,g}^{d_{abc},\rm (3)} =
\textcolor{blue}{\frac{d_{abc} d^{abc} N_F^2}{N_A}} \Biggl\{
\nonumber\\ &&
        \frac{256 S_4 P_{729}}{(N-2) (N-1) N (1+N)^2 (2+N)^2 (3+N)}
        -\frac{512 S_{3,1} P_{729}}{(N-2) (N-1) N (1+N)^2 (2+N)^2 
(3+N)}
\nonumber\\ &&
        -\frac{256 P_{741}}{225 (N-2) (N-1)^3 N^3 (1+N)^5 (2+N)^5 
(3+N)^3}
        +\Biggl(
                \frac{6144 S_{3,1}}{(1+N) (2+N)}
                -\frac{3072 S_4}{(1+N) (2+N)}
\nonumber\\ &&                
+\frac{512 S_3 P_{728}}{(N-2) (N-1) N (1+N)^2 (2+N)^2 
(3+N)}
                -\frac{1024 S_{-2,1}}{(1+N) (2+N)}
\nonumber\\ &&
                +\frac{256 P_{739}}{225 (N-2) (N-1)^2 N^2 (1+N)^4 
(2+N)^4 (3+N)^3}
    \Biggr) S_1
        +\Biggl(
                -\frac{1408}{(1+N) (2+N)}
\nonumber\\ &&               
 -\frac{1536 S_3}{(1+N) (2+N)}
        \Biggr) S_1^2
        -\frac{1536 \big(
                1+N+N^2
        \big)
\big(-10+3 N+3 N^2\big) S_3}{(N-2) (N-1) (1+N) (2+N)^2 (3+N)}
        +\frac{4608 S_2 S_3}{(1+N) (2+N)}
\nonumber\\ &&        
-\frac{128 (N-4) (5+N) \big(
                18+N+N^2\big) S_5}{15 (1+N) (2+N)}
        +\Biggl(
                 \frac{512 S_1^2}{(1+N) (2+N)}
                -\frac{256}{15} (N-1) N S_3
\nonumber\\ &&              
  -\frac{256 S_1 P_{731}}{15 (N-2) (N-1) N (1+N)^2 
(2+N)^2 (3+N)}
                -\frac{256}{15} (N-1) N S_{-2,1}
\nonumber\\ &&             
   -\frac{128 P_{737}}{15 (N-2) (N-1) N (1+N)^3 (2+N)^3 
(3+N)}
        \Biggr) S_{-2}
        +\frac{256 S_{-2}^2}{(1+N) (2+N)}
\nonumber\\ &&        
+\Biggl(
                -\frac{128 \big(
                        10+N+N^2
                \big)
\big(30+N+N^2\big)}{15 (N-1) (1+N) (2+N)^2}
                +\frac{512 S_1}{(1+N) (2+N)}
        \Biggr) S_{-3}
        +\frac{512 S_{-4}}{(1+N) (2+N)}
\nonumber\\ &&     
   -\frac{128}{15} (N-1) N S_{-5}
        -\frac{4608 S_{2,3}}{(1+N) (2+N)}
        +\frac{256 \big(
                10+N+N^2
        \big)
\big(30+N+N^2\big) S_{-2,1}}{15 (N-1) (1+N) (2+N)^2}
\nonumber\\ &&
        +\frac{4608 S_{4,1}}{(1+N) (2+N)}
        -\frac{512 S_{-2,2}}{(1+N) (2+N)}
        +\frac{256}{15} (N-1) N S_{-2,3}
        -\frac{512 S_{-3,1}}{(1+N) (2+N)}
\nonumber\\ &&    
    -\frac{9216 S_{3,1,1}}{(1+N) (2+N)}
        +\frac{1024 S_{-2,1,1}}{(1+N) (2+N)}
        +\frac{256}{15} (N-1) N S_{-2,1,-2}
        +\Biggl(
                 \frac{3072 S_1^2}{(1+N) (2+N)}
\nonumber\\ &&           
                -\frac{256}{15} (N-1) N S_{-2}
                +\frac{128 P_{727}}{15 (N-2) (N-1) (1+N) (2+N)^2 (3+N)}
\nonumber\\ &&                
-\frac{512 S_1 P_{730}}{(N-2) (N-1) N (1+N)^2 (2+N)^2 
(3+N)}
        \Biggr) \zeta_3
        -\frac{64 (N-4) (5+N) \big(
                18+N+N^2\big)}{3 (1+N) (2+N)} \zeta_5
\Biggr\},
\nonumber\\
\end{eqnarray}


\noindent
with the polynomials
\begin{eqnarray}
P_{724}&=&N^5+N^4+17 N^3+59 N^2-138 N+720,
\\
P_{725}&=&N^5+N^4+59 N^3+59 N^2+225 N-450,
\\
P_{726}&=&5 N^5-16 N^4-58 N^3+25 N^2+32 N+60,
\\
P_{727}&=&N^6+3 N^5+1162 N^4+2319 N^3-1661 N^2-2820 N-4860,
\\
P_{728}&=&11 N^6+33 N^5-15 N^4-85 N^3-50 N^2-2 N+36,
\\
P_{729}&=&21 N^6+63 N^5-25 N^4-155 N^3-104 N^2-16 N+72,
\\
P_{730}&=&23 N^6+69 N^5-35 N^4-185 N^3-96 N^2+8 N+72,
\\
P_{731}&=&45 N^6+135 N^5+215 N^4+205 N^3-1340 N^2-1420 N+624,
\\
P_{732}&=&2 N^7+4 N^6+873 N^5-3466 N^4-11255 N^3+6582 N^2+7020 N+16200,
\\
P_{733}&=&5 N^7-19 N^6-105 N^5+23 N^4+240 N^3+104 N^2-8 N-96,
\\
P_{734}&=&6 N^7-33 N^6-154 N^5+41 N^4+358 N^3+202 N^2+36 N-144,
\\
P_{735}&=&9 N^7-24 N^6-161 N^5+28 N^4+362 N^3+110 N^2-60 N-144,
\\
P_{736}&=&45 N^7+90 N^6+50 N^5-130 N^4-1895 N^3-740 N^2+984 N-1944,
\\
P_{737}&=&N^{10}+7 N^9-58 N^8-438 N^7-1017 N^6-987 N^5-130 N^4+674 N^3
\nonumber\\ &&
+3940 N^2+9960 N+6480,\\
P_{738}&=&2 N^{11}+12 N^{10}-315 N^9-2240 N^8-4508 N^7+2140 N^6
+22169 N^5
\nonumber\\ &&
+36208 N^4+30892 N^3+10680 N^2-23040 N-21600,
\\
P_{739}&=&1065 N^{14}+14910 N^{13}+91950 N^{12}+334830 N^{11}
+786240 N^{10}+1070370 N^9
\nonumber\\ &&
+148290 N^8-2319270 N^7-3721785 N^6
-1334280 N^5+2039760 N^4
\nonumber\\ &&
+2233440 N^3+654480 N^2,
\\
P_{740}&=&220 N^{15}-960 N^{14}-28415 N^{13}-151530 N^{12}
-369410 N^{11}-556890 N^{10}
\nonumber\\ &&
-716080 N^9-97590 N^8+2725310 N^7
+4299810 N^6-627745 N^5-5315880 N^4
\nonumber\\ &&
-2040120 N^3+1823040 N^2
+1056240 N,
\\
P_{741}&=&405 N^{18}+6480 N^{17}+43335 N^{16}+152280 N^{15}
+272970 N^{14}+103680 N^{13}
\nonumber\\ &&
-541890 N^{12}-1017360 N^{11}
-393255 N^{10}+810000 N^9+1042875 N^8
\nonumber\\ &&
+217080 N^7-366120 N^6
	-272160 N^5-58320 N^4,
\\
P_{742}&=&555 N^{19}+6195 N^{18}+17955 N^{17}-54135 N^{16}-471780 N^{15}
-1092330 N^{14}
\nonumber\\ &&
-398730 N^{13}
+2590050 N^{12}+4158675 N^{11}-126225 N^{10}
-5426505 N^9
\nonumber\\ &&
-3556395 N^8
+1998870 N^7+2874360 N^6+341280 N^5-641520 N^4
\nonumber\\ &&
-220320 N^3.
\end{eqnarray}

\noindent
Despite the fact that the above equations contain terms in which the power of $N$ is larger in the 
numerator than the denominator, the whole expressions behave maximally $\propto \ln(N)/N$ for large
values of $N$ for the Wilson coefficient {of} $F_2$ and $\propto 
\ln(N)/N^2$ for that of $F_L$.

\section{The three--loop Wilson coefficients for the structure function 
\boldmath $xF_3(x,Q^2)$}
\label{sec:6}

\vspace*{1mm}
\noindent

The Wilson coefficient for the structure function $xF_3(x,Q^2)$ has already received the 
renormalizations of the coupling and those of the axial-vector coupling, as well the collinear
singularities have been removed, to arrive at $\mathbb{C}_{\hat{F}_3,q}^{(3)}$, which still 
requires the {finite} renormalization to restore the Ward identities performed by
\begin{eqnarray}
\mathbb{C}_{F_3,q}^{(3)} &=& Z_5 \mathbb{C}_{\hat{F}_3,q}^{(3)},
\end{eqnarray}
with, \cite{Moch:2008fj},
\begin{eqnarray}
Z_5 &=& 1 + a_s \textcolor{blue}{C_F} \Biggl[
        - 4
        - 10 \ep
        + \ep^2 \big(
                -22
                +2 \zeta_2
        \big)
\Biggr]
+ a_s^2 \Biggl[
        -\frac{107}{9} \textcolor{blue}{C_A C_F} 
        +22 \textcolor{blue}{C_F^2}
        +\frac{4}{9} \textcolor{blue}{C_F T_F N_F}
\nonumber\\ &&        
+ \ep \Biggl(
                \frac{662}{27} \textcolor{blue}{C_F T_F N_F} 
                +\textcolor{blue}{C_F^2} \big(
                        132
                        -48 \zeta_3
                \big)
                +\textcolor{blue}{C_A C_F} \Biggl(
                        -\frac{7229}{54}
                        +48 \zeta_3
                \Biggr)
        \Biggr)
\Biggr]
\nonumber\\ &&
+ a_s^3 \Biggl[
        \frac{208}{81} \textcolor{blue}{C_F T_F^2 N_F^2}
        +\textcolor{blue}{C_A C_F^2} \Biggl(
                \frac{5834}{27}
                -160 \zeta_3
        \Biggr)
        +2 \textcolor{blue}{C_F^2 T_F N_F}  \Biggl(
                -\frac{62}{27}
                -\frac{32 \zeta_3}{3}
        \Biggr)
\nonumber\\ &&      
   +2 \textcolor{blue}{C_A C_F T_F N_F} \Biggl(
                \frac{356}{81}
                +\frac{32 \zeta_3}{3}
        \Biggr)
        +\textcolor{blue}{C_A^2 C_F} \Biggl(
                -\frac{2147}{27}
                +56 \zeta_3
        \Biggr)
        +\textcolor{blue}{C_F^3} \Biggl(
                -\frac{370}{3}
                +96 \zeta_3
        \Biggr)
\Biggr].
\nonumber\\
\end{eqnarray}
One obtains
{
\begin{eqnarray}
\mathbb{C}_{F_3,q}^{(3)} &=& C_{F_3,q}^{\rm NS, (3)} + C_{F_3,q}^{d_{abc}, (3)}
\end{eqnarray}
}
with the contributions

\vspace*{1mm} 
\noindent 



\noindent
with the polynomials
\begin{eqnarray}
P_{743}&=&-617 N^4-1180 N^3-485 N^2+90 N+52,\\ P_{744}&=&-9 N^4-12 N^3-9 N^2+22 
N+20,\\ P_{745}&=&N^4+2 N^3-293 N^2-294 N+356,\\ P_{746}&=&N^4+2 N^3-120 N^2-121 
N+258,\\ P_{747}&=&N^4+2 N^3+15 N^2+14 N+60,\\ P_{748}&=&N^4+2 N^3+15 N^2+14 
N+194,\\ P_{749}&=&N^4+2 N^3+24 N^2+23 N+68,\\ P_{750}&=&N^4+2 N^3+42 N^2+41 
N+10,\\ P_{751}&=&N^4+2 N^3+90 N^2+89 N+166,\\ P_{752}&=&N^4+2 N^3+111 N^2+110 
N+36,\\ P_{753}&=&N^4+2 N^3+115 N^2+114 N+54,\\ P_{754}&=&N^4+2 N^3+195 N^2+194 
N-156,\\ P_{755}&=&N^4+2 N^3+204 N^2+203 N+40,\\ P_{756}&=&N^4+2 N^3+215 N^2+214 
N+48,\\ P_{757}&=&N^4+2 N^3+303 N^2+302 N-144,\\ P_{758}&=&2 N^4+4 N^3-413 
N^2-415 N+614,\\ P_{759}&=&2 N^4+4 N^3+153 N^2+151 N+46,\\ P_{760}&=&3 N^4+6 
N^3+100 N^2+97 N+495,\\ P_{761}&=&3 N^4+6 N^3+155 N^2+152 N+408,\\ P_{762}&=&3 
N^4+6 N^3+1321 N^2+1318 N-252,\\ P_{763}&=&3 N^4+6 N^3+1435 N^2+1432 
N-480,\\ P_{764}&=&5 N^4+10 N^3+9 N^2+4 N+4,\\ P_{765}&=&6 N^4+12 N^3+1591 
N^2+1585 N+246,\\ P_{766}&=&6 N^4+12 N^3+1871 N^2+1865 N-216,\\ P_{767}&=&9 
N^4+18 N^3+925 N^2+916 N+144,\\ P_{768}&=&18 N^4+36 N^3+1141 N^2+1123 
N+756,\\ P_{769}&=&29 N^4+51 N^3+16 N^2-20 N-6,\\ P_{770}&=&83 N^4+166 N^3+239 
N^2+192 N+63,\\ P_{771}&=&85 N^4+134 N^3+57 N^2-80 N-52,\\ P_{772}&=&97 N^4+194 
N^3-121 N^2-218 N-6,\\ P_{773}&=&131 N^4+262 N^3-23 N^2-154 
N+216,\\ P_{774}&=&147 N^4+294 N^3+38 N^2-121 N+6,\\ P_{775}&=&235 N^4+524 
N^3+211 N^2+30 N+72,\\ P_{776}&=&277 N^4+536 N^3+200 N^2-143 
N-36,\\ P_{777}&=&291 N^4+488 N^3+217 N^2-304 N-256,\\ P_{778}&=&361 N^4+722 
N^3-433 N^2-794 N-72,\\ P_{779}&=&423 N^4+792 N^3+413 N^2+44 
N-90,\\ P_{780}&=&536 N^4+1075 N^3+601 N^2+62 N+162,\\ P_{781}&=&536 N^4+1087 
N^3+414 N^2-149 N+6,\\ P_{782}&=&575 N^4+1084 N^3+420 N^2-365 
N-150,\\ P_{783}&=&683 N^4+1510 N^3+479 N^2+68 N+292,\\ P_{784}&=&1055 N^4+2164 
N^3+1031 N^2+30 N+72,\\ P_{785}&=&1108 N^4+2261 N^3+1099 N^2-126 
N+288,\\ P_{786}&=&1124 N^4+2224 N^3+847 N^2-331 N-57,\\ P_{787}&=&1687 N^4+3374 
N^3-1657 N^2-3344 N+372,\\ P_{788}&=&4513 N^4+9026 N^3+2815 N^2-1698 
N+612,\\ P_{789}&=&112 N^5+313 N^4+188 N^3-73 N^2+8 N+52,\\ P_{790}&=&328 
N^5+1513 N^4+2150 N^3+959 N^2-102 N-120,\\ P_{791}&=&1103 N^5+4010 N^4+4309 
N^3+1096 N^2-60 N+240,\\ P_{792}&=&7241 N^5+26102 N^4+28297 N^3+8428 
N^2-540 N-384,\\ P_{793}&=&9187 N^5+33508 N^4+33155 N^3+2138 N^2+2844 
N+8136,\\ P_{794}&=&-279 N^6-705 N^5-663 N^4-159 N^3-402 N^2-816 
N-304,\\ P_{795}&=&25 N^6+75 N^5-161 N^4-303 N^3+901 N^2+705 N+54,\\ P_{796}&=&33 
N^6+9 N^5-669 N^4-1433 N^3-1180 N^2+368 N-8,\\ P_{797}&=&34 N^6+114 N^5+695 
N^4+1472 N^3+1051 N^2-430 N-632,\\ P_{798}&=&36 N^6+138 N^5+969 N^4+2040 
N^3+1355 N^2-682 N-400,\\ P_{799}&=&37 N^6+49 N^5+161 N^4+491 N^3+242 
N^2-188 N-216,\\ P_{800}&=&49 N^6+148 N^5+110 N^4-4 N^3-24 N^2-61 
N-74,\\ P_{801}&=&61 N^6+190 N^5+530 N^4+926 N^3+617 N^2-320 
N-276,\\ P_{802}&=&74 N^6+206 N^5+5 N^4-328 N^3-203 N^2+94 N+8,\\ P_{803}&=&79 
N^6+215 N^5+73 N^4-143 N^3-52 N^2+36 N-96,\\ P_{804}&=&81 N^6+103 N^5-651 
N^4-1335 N^3-570 N^2+748 N+568,\\ P_{805}&=&83 N^6-293 N^5+15 N^4-23 
N^3+170 N^2+1040 N+704,\\ P_{806}&=&83 N^6+249 N^5+164 N^4-33 N^3+14 N^2-63 
N-90,\\ P_{807}&=&96 N^6+250 N^5+49 N^4-300 N^3-333 N^2+78 N+16,\\ P_{808}&=&100 
N^6+349 N^5+301 N^4+31 N^3+79 N^2+128 N+76,\\ P_{809}&=&158 N^6+454 N^5+651 
N^4+780 N^3+471 N^2-290 N-496,\\ P_{810}&=&173 N^6+466 N^5-273 N^4-1264 
N^3-472 N^2+462 N+380,\\ P_{811}&=&182 N^6+528 N^5+2678 N^4+5724 N^3+4385 
N^2-1383 N-1746,\\ P_{812}&=&208 N^6+622 N^5+1151 N^4+1664 N^3+1113 N^2-598 
N-704,\\ P_{813}&=&248 N^6+771 N^5-1216 N^4-4599 N^3-3616 N^2+828 
N+672,\\ P_{814}&=&379 N^6+1065 N^5-3293 N^4-9309 N^3-5402 N^2+2304 
N+3240,\\ P_{815}&=&388 N^6+1128 N^5-5339 N^4-15318 N^3-11297 N^2+3702 
N+2544,\\ P_{816}&=&403 N^6+1191 N^5+633 N^4-599 N^3-440 N^2-116 
N-16,\\ P_{817}&=&417 N^6+900 N^5-253 N^4-1298 N^3-458 N^2+20 
N-192,\\ P_{818}&=&429 N^6+1233 N^5+1041 N^4+459 N^3+50 N^2-676 
N-808,\\ P_{819}&=&676 N^6+2109 N^5-2858 N^4-10617 N^3-5747 N^2+2811 
N+3258,\\ P_{820}&=&781 N^6+1227 N^5-5909 N^4-14895 N^3-12548 N^2+5496 
N+8568,\\ P_{821}&=&913 N^6+2766 N^5+10126 N^4+19782 N^3+14323 N^2-5646 
N-5976,\\ P_{822}&=&1046 N^6+3192 N^5-11971 N^4-34482 N^3-20953 N^2+9492 
N+12204,\\ P_{823}&=&1360 N^6+3972 N^5+1355 N^4-3874 N^3-2955 N^2+310 
N-600,\\ P_{824}&=&1472 N^6+4425 N^5+7064 N^4+9099 N^3+6692 N^2-3048 
N-4968,\\ P_{825}&=&1540 N^6+4629 N^5+4822 N^4+3087 N^3+1945 N^2-1785 
N-3870,\\ P_{826}&=&1540 N^6+4701 N^5+1444 N^4-5937 N^3-7331 N^2-1671 
N-522,\\ P_{827}&=&1545 N^6+4511 N^5+3537 N^4+737 N^3+1222 N^2+2384 
N+784,\\ P_{828}&=&1873 N^6+5331 N^5+2385 N^4-3299 N^3-2470 N^2+44 
N-1368,\\ P_{829}&=&1964 N^6+5856 N^5+68 N^4-10116 N^3-6055 N^2+1461 
N+1314,\\ P_{830}&=&2251 N^6+6552 N^5+1826 N^4-6688 N^3-3669 N^2+292 
N-132,\\ P_{831}&=&2284 N^6+6951 N^5+3143 N^4-5233 N^3-2970 N^2-53 
N-450,\\ P_{832}&=&2882 N^6+8592 N^5+2333 N^4-10050 N^3-5899 N^2+1926 
N-1080,\\ P_{833}&=&3098 N^6+9240 N^5+3107 N^4-8898 N^3-5665 N^2-18 
N-1188,\\ P_{834}&=&3153 N^6+7047 N^5-1489 N^4-9635 N^3-5120 N^2-148 
N-720,\\ P_{835}&=&4357 N^6+16149 N^5+16977 N^4+9091 N^3+3798 N^2-1404 
N-1944,\\ P_{836}&=&4502 N^6+13344 N^5+15503 N^4+13194 N^3+8129 N^2-6180 
N-12204,\\ P_{837}&=&4591 N^6+13935 N^5+6934 N^4-8871 N^3-5873 N^2-1464 
N-612,\\ P_{838}&=&4645 N^6+13341 N^5-4124 N^4-31599 N^3-16541 N^2+8430 
N+8568,\\ P_{839}&=&4816 N^6+14448 N^5+24487 N^4+33174 N^3+23401 N^2-11478 
N-16272,\\ P_{840}&=&6457 N^6+18885 N^5-245 N^4-36645 N^3-39944 N^2-3372 
N-2160,\\ P_{841}&=&7531 N^6+23619 N^5+23253 N^4+7825 N^3+2064 N^2+1080 
N-252,\\ P_{842}&=&8425 N^6+26643 N^5+24471 N^4+6857 N^3+2520 N^2+1484 
N-864,\\ P_{843}&=&-116957 N^7-556705 N^6-889137 N^5-480947 N^4-48370 
N^3-93780 N^2
\nonumber\\ &&
-76536 N+2592,\\ 
P_{844}&=&-28023 N^7-129045 N^6-212881 N^5-149635 N^4-36908 N^3-4628 N^2
\nonumber\\ &&
-17376 N-9216,\\ 
P_{845}&=&-3221 N^7-19897 N^6-43683 N^5-43139 N^4-15736 N^3+3828 N^2+3720 N
\nonumber\\ &&
+720,\\ 
P_{846}&=&1511 N^7+6361 N^6+8811 N^5+3500 N^4-61 N^3+1842 N^2+1772 N+552,
\\ 
P_{847}&=&3691 N^7+16568 N^6+25248 N^5+13189 N^4+236 N^3+1188 N^2+1860 N
\nonumber\\ &&
+360,
\\ 
P_{848}&=&32287 N^7+161273 N^6+288291 N^5+220675 N^4+64886 N^3+4416 N^2
\nonumber\\ &&
+1188 N-792,\\ 
P_{849}&=&-97249 N^8-360322 N^7-305212 N^6+226586 N^5+292481 N^4-76528 N^3
\nonumber\\ &&
-50508 N^2+100320 N+63072,\\ 
P_{850}&=&-5021 N^8-15158 N^7-6690 N^6+17488 N^5+14927 N^4-3058 N^3+1392 N^2
\nonumber\\ &&
+6920 N+4560,\\ 
P_{851}&=&25 N^8+105 N^7+101 N^6+52 N^5+1332 N^4+2547 N^3+890 N^2-932 N
\nonumber\\ &&
-664,
\\ 
P_{852}&=&79 N^8+323 N^7+315 N^6+22 N^5+1546 N^4+3031 N^3+1088 N^2-1076 N
\nonumber\\ &&
-720,\\ 
P_{853}&=&307 N^8+1174 N^7+1142 N^6-608 N^5-2451 N^4-2370 N^3-250 N^2
\nonumber\\ &&
+888 N+824,\\ 
P_{854}&=&575 N^8+2226 N^7+1974 N^6-1440 N^5+15 N^4+5006 N^3+1172 N^2-3320 N
\nonumber\\ &&
-1984,\\ 
P_{855}&=&1288 N^8+4747 N^7+10549 N^6+7126 N^5-99269 N^4-195599 N^3-68574 N^2
\nonumber\\ &&
+70380 N+51624,\\ 
P_{856}&=&1963 N^8+7582 N^7+13276 N^6+8530 N^5-63305 N^4-126830 N^3-44544 N^2
\nonumber\\ &&
+45216 N+33696,\\ 
P_{857}&=&3761 N^8+15422 N^7+6284 N^6-22930 N^5+78905 N^4+182846 N^3+61872 N^2
\nonumber\\ &&
-51408 N-36288,\\ 
P_{858}&=&4258 N^8+16762 N^7+18841 N^6-1469 N^5-28625 N^4-39305 N^3-12672 N^2
\nonumber\\ &&
+15102 N+12636,\\ 
P_{859}&=&5894 N^8+24143 N^7+14789 N^6-22336 N^5+120647 N^4+264683 N^3
\nonumber\\ &&
+91248 N^2
-80460 N-55728,\\ 
P_{860}&=&17529 N^8+56958 N^7+34840 N^6-52470 N^5-72773 N^4-21696 N^3-4220 N^2
\nonumber\\ &&
-10440 N-7344,\\ 
P_{861}&=&28551 N^8+92280 N^7+118370 N^6+46548 N^5-11961 N^4+10748 N^3
\nonumber\\ &&
+12600 N^2-14112 N-9936,\\ 
P_{862}&=&43949 N^8+169784 N^7+155918 N^6-99922 N^5-174643 N^4-20848 N^3
\nonumber\\ &&
+15276 N^2-15534 N-7884,\\ 
P_{863}&=&50689 N^8+208318 N^7+323478 N^6+236812 N^5+86401 N^4+16554 N^3
\nonumber\\ &&
-2304 N^2-2772 N-540,\\ 
P_{864}&=&-71503 N^9-425688 N^8-980826 N^7-1101180 N^6-615171 N^5
\nonumber\\ &&
-147456 N^4
-4292 N^3+4428 N^2+2940 N+72,\\ 
P_{865}&=&24 N^9+82 N^8-N^7-166 N^6+590 N^5+262 N^4-1321 N^3-1098 N^2
\nonumber\\ &&
+52 N+424,\\ 
P_{866}&=&261 N^9+783 N^8-127 N^7-1760 N^6-493 N^5-1289 N^4-1879 N^3
\nonumber\\ &&
-758 N^2+114 N-36,\\ 
P_{867}&=&7868 N^9+22794 N^8+4185 N^7-36138 N^6-70242 N^5-28434 N^4
\nonumber\\ &&
+88933 N^3+85194 N^2-6336 N-36720,\\ 
P_{868}&=&-5563 N^{10}-22141 N^9-36762 N^8-34822 N^7+5681 N^6+48719 N^5
\nonumber\\ &&
+42980 N^4+28996 N^3+19056 N^2-7936 
N-10560,\\ 
P_{869}&=&-4253 N^{10}-19223 N^9-23706 N^8+10042 N^7+39207 N^6+16757 N^5
\nonumber\\ &&
-6072 N^4+1800 N^3+6168 N^2-528 N-1760,\\ 
P_{870}&=&561 N^{10}+2523 N^9+4538 N^8+4510 N^7+1005 N^6-5025 N^5-8992 N^4
\nonumber\\ &&
-7960 N^3-8936 N^2-1968 N+1312,\\ 
P_{871}&=&6009 N^{10}+21117 N^9+38586 N^8+68306 N^7+61705 N^6+28925 N^5
\nonumber\\ &&
-2468 N^4-31900 N^3+38584 N^2+73824 N+29088,\\ 
P_{872}&=&80453 N^{10}+440578 N^9+771262 N^8+284116 N^7-538565 N^6-656852 N^5
\nonumber\\ &&
-328702 N^4-44730 N^3+83592 N^2+19008 
N-16848,\\ 
P_{873}&=&463143 N^{10}+2173209 N^9+3165914 N^8+374654 N^7-3058933 N^6
\nonumber\\ &&
-2659411 N^5-640204 N^4-29428 N^3-117840 N^2+26496 N+53568,\\ 
P_{874}&=&245 N^{11}+944 N^{10}+786 N^9-975 N^8-1959 N^7-1998 N^6-1756 N^5
\nonumber\\ &&
-695 N^4+748 N^3+228 N^2-112 N-64,\\ 
P_{875}&=&10683 N^{11}+42732 N^{10}+25931 N^9-78468 N^8-83703 N^7-19116 N^6
\nonumber\\ &&
-61407 N^5-68436 N^4-10052 N^3+14856 N^2-468 N-648,\\ 
P_{876}&=&27981 N^{11}+110952 N^{10}+73084 N^9-183261 N^8-220299 N^7
\nonumber\\ &&
-92178 N^6-170226 N^5-155637 N^4+92 N^3+35868 N^2-3960 N-3024,\\ 
P_{877}&=&599375 N^{12}+3815193 N^{11}+8947106 N^{10}+8383052 N^9-1037899 N^8
\nonumber\\ &&
-9404623 N^7-8442036 N^6-2234074 N^5+1430550 N^4+814536 N^3
\nonumber\\ &&
-384156 N^2-321408 N-50544,\\ 
P_{878}&=&-458487 N^{13}-2189691 N^{12}-3476151 N^{11}-706147 N^{10}
\nonumber\\ &&
+5181991 N^9+6607155 N^8-353013 N^7-6891201 N^6-4645260 N^5
\nonumber\\ &&
+669484 N^4
+2725928 N^3+920640 N^2-844704 N-521856,\\ 
P_{879}&=&-3069 N^{13}+2556 N^{12}+79368 N^{11}+221328 N^{10}+43770 N^9-547648 N^8
\nonumber\\ &&
-435404 N^7-15640 N^6-194549 N^5-563692 N^4-68324 N^3+170136 N^2
\nonumber\\ &&
+10848 N-26784,\\ 
P_{880}&=&3003 N^{13}+13255 N^{12}+27059 N^{11}+32543 N^{10}+397 N^9-54447 N^8
-26095 N^7
\nonumber\\ &&
+141989 N^6+192852 N^5+62644 N^4-92576 N^3-56336 N^2+29952 N
\nonumber\\ &&
+20672,\\ 
P_{881}&=&428649 N^{13}+1845381 N^{12}+2026189 N^{11}-1935731 N^{10}-5520669 N^9
\nonumber\\ &&
-1634037 N^8+3712287 N^7+2628591 N^6+266216 N^5+1213892 N^4
\nonumber\\ &&
+461088 N^3
-542160 N^2
-82944 N+119232,\\ 
P_{882}&=&-5729259 N^{15}-30802914 N^{14}-53071856 N^{13}-478888 N^{12}
\nonumber\\ &&
+104531506 N^{11}+96490716 N^{10}-31073520 N^9-90914040 N^8-63564367 N^7
\nonumber\\ &&
-54366650 N^6-31332904 N^5+18604224 N^4+17762832 N^3-2637792 N^2
\nonumber\\ &&
-4240512 N
-559872,\\ 
P_{883}&=&-7255 N^{15}-40527 N^{14}-54850 N^{13}+77690 N^{12}+228204 N^{11}
+45864 N^{10}
\nonumber\\ &&
-240778 N^9-240422 N^8-203989 N^7-270809 N^6-72500 N^5+144732 
\nonumber\\ &&
N^4+110144 N^3-17904 N^2-37248 N-10176,\\ 
P_{884}&=&494694 N^{15}+2631933 N^{14}+3703662 N^{13}-3856542 N^{12}
-13210842 N^{11}
\nonumber\\ &&
-3078616 N^{10}+14028102 N^9+13282506 N^8+9865380 N^7
+14654523 N^6
\nonumber\\ &&
+7299204 N^5-6176300 N^4-4977096 N^3+807456 N^2+1390176 N
\nonumber\\ &&
+300672.
\end{eqnarray}

\noindent
and 
\begin{eqnarray}
C_{F_3}^{d_{abc}} &=& \textcolor{blue}{\frac{d_{abc} d^{abc} N_F}{N_c}} 
\Biggl\{
-\frac{64 S_{-2,1} P_{887}}{3 (N-1) N^3 (1+N)^3 (2+N)}
-\frac{32 S_3 P_{888}}{3 (N-1) N^3 (1+N)^3 (2+N)}
\nonumber\\ &&
+\frac{64 S_{2,1} P_{885}}{3 N^3 (1+N)^3}
+\frac{16 S_2 P_{890}}{3 (N-1) N^4 (1+N)^4 (2+N)}
-\frac{80 S_1^2 P_{891}}{3 (N-1) N^4 (1+N)^4 (2+N)}
\nonumber\\ &&
+\frac{64 P_{894}}{3 (N-1)^2 N^6 (1+N)^6 (2+N)^2}
-\Biggl[
        \frac{16 P_{895}}{3 (N-1)^2 N^5 (1+N)^5 (2+N)^2}
\nonumber\\ &&
        +\frac{32 \big(
                2+N+N^2
        \big)
\big(18+5 N+5 N^2\big) S_3}{3 (N-1) N^2 (1+N)^2 (2+N)}
        +\frac{64 \big(
                2+N+N^2
        \big)
\big(17 N^2 + 17 N - 10\big) S_{-2,1}}{3 (N-1) N^2 (1+N)^2 (2+N)}
\nonumber\\ &&
        +\frac{32 S_2 P_{885}}{3 N^3 (1+N)^3}
\Biggr] S_1
-\frac{16 \big(
        2+N+N^2\big) S_2^2}{3 N^2 (1+N)^2}
+\frac{16 \big(
        2+N+N^2
\big)
\big(-14+3 N+3 N^2\big) S_4}{(N-1) N^2 (1+N)^2 (2+N)}
\nonumber\\ &&
+\Biggl[
        -\frac{64 S_1 P_{892}}{3 (N-1)^2 N^3 (1+N)^3 (2+N)^2}
        -\frac{32 P_{893}}{3 (N-1)^2 N^4 (1+N)^4 (2+N)^2}
\nonumber\\ &&        
-\frac{320 \big(
                2+N+N^2\big)^2 S_1^2}{3 (N-1) N^2 (1+N)^2 (2+N)}
        +\frac{256 \big(
                2+N+N^2\big) S_2}{3 (N-1) N^2 (1+N)^2 (2+N)}
\Biggr] S_{-2}
\nonumber\\ &&
-\frac{64 \big(
        2+N+N^2\big) S_{-2}^2}{3 N^2 (1+N)^2}
+\Biggl[
        \frac{32 \big(
                2+N+N^2
        \big)
\big(18+35 N+35 N^2\big) S_1}{3 (N-1) N^2 (1+N)^2 (2+N)}
\nonumber\\ &&
+        \frac{32 P_{889}}{3 (N-1) N^3 (1+N)^3 (2+N)}
\Biggr] S_{-3}
+\frac{32 \big(
        2+N+N^2
\big)
\big(6+N+N^2\big) S_{-4}}{(N-1) N^2 (1+N)^2 (2+N)}
\nonumber\\ &&
+\frac{256 \big(
        2+N+N^2\big) S_{3,1}}{(N-1) N^2 (1+N)^2 (2+N)}
-\frac{64 \big(
        2+N+N^2
\big)
\big(2+3 N+3 N^2\big) S_{-2,2}}{(N-1) N^2 (1+N)^2 (2+N)}
\nonumber\\ &&
-\frac{128 \big(
        2+N+N^2\big)^2 S_{-3,1}}{(N-1) N^2 (1+N)^2 (2+N)}
+\frac{512 \big(
        2+N+N^2\big) S_{-2,1,1}}{(N-1) N (1+N) (2+N)}
\nonumber\\ &&
+\Biggl[
        \frac{64 P_{886}}{(N-1) N^3 (1+N)^3 (2+N)}
        +\frac{256 \big(
                2+N+N^2
        \big)
\big(4+N+N^2\big) S_1}{(N-1) N^2 (1+N)^2 (2+N)}
\Biggr] \zeta_3
\Biggr\},
\end{eqnarray}


\noindent
with the polynomials
\begin{eqnarray}
P_{885} &=& N^4+2 N^3-5 N^2-6 N-4,
\\
P_{886} &=& 3 N^6+9 N^5-9 N^4-33 N^3-74 N^2-56 N-32,
\\
P_{887} &=& 7 N^6+12 N^5-73 N^4-170 N^3-104 N^2+56 N+80,
\\
P_{888} &=& 11 N^6+30 N^5-26 N^4-106 N^3-113 N^2-24 N+4,
\\
P_{889} &=& 13 N^6+28 N^5-139 N^4-326 N^3-288 N^2-40 N+48,
\\
P_{890} &=& N^8+6 N^7+20 N^6+28 N^5+32 N^4+44 N^3+73 N^2+40 N+12,
\\
P_{891} &=& 3 N^8+12 N^7+16 N^6+6 N^5+30 N^4+64 N^3+73 N^2+40 N+12,
\\
P_{892} &=& 8 N^8+24 N^7-51 N^6-281 N^5-459 N^4-383 N^3-190 N^2+76 N+104,
\\
P_{893} &=&9 N^{10}+37 N^9+10 N^8+2 N^7+491 N^6+1291 N^5+1566 N^4+926 N^3
\nonumber\\ &&
+300 N^2
-8 N-16,
\\
P_{894} &=& 4 N^{12}+21 N^{11}+20 N^{10}-158 N^9-683 N^8-1644 N^7-2947 N^6
\nonumber\\ &&
-3115 N^5
-1578 N^4+92 N^3+484 N^2+240 N+48,
\\
P_{895} &=& 43 N^{12}+232 N^{11}+319 N^{10}-336 N^9-1485 N^8-2782 N^7-5229 N^6
\nonumber\\ &&
-7186 N^5
-5296 N^4-304 N^3+1896 N^2+1344 N+352.
\end{eqnarray}
\section{The three--loop Wilson coefficients for the structure function \boldmath $g_1(x,Q^2)$}
\label{sec:7}

\vspace*{1mm}
\noindent
The non--singlet Wilson coefficient for the polarized structure function {$g_1(x,Q^2)$} has the 
representation
\begin{eqnarray}
\Delta \mathbb{C}_{g_1,q}^{(3)} &=& 
\Delta C_{g_1,q}^{\rm NS, (3)} + \Delta C_{g_1,q}^{d_{abc}, (3)}
+ \Delta C_{g_1,q}^{\rm PS, (3)},
\\
\Delta \mathbb{C}_{g_1,g}^{(3)} &=& 
\Delta C_{g_1,g}^{(3)},
\end{eqnarray}
with 
\begin{eqnarray}
\Delta {C}_{g_1,q}^{\rm NS, (3), M} &=& 
{C}_{F_3,q}^{\rm NS, (3)}
\end{eqnarray}
in the $\overline{\sf MS}$ scheme similar to the observation at one-- and two--loop order
in Eqs.~(\ref{eq:F3NS1}, \ref{eq:g1NS1}) and (\ref{eq:F3NS2}, \ref{eq:g1NS2}).
$\Delta {C}_{g_1,q}^{\rm NS, (3), L}$ is given by
\begin{eqnarray}
\Delta {C}_{g_1,q}^{\rm NS, (3), M}(N,a_s) &=& Z_5^{-1}(N,a_s) \, \Delta {C}_{g_1,q}^{\rm NS, (3), L}(N,a_s)
\end{eqnarray}
The $Z$-factor $Z_5^{-1}(N,a_s)$ providing the finite renormalization is calculated in 
Appendix~\ref{sec:A}.

The function $\Delta C_{g_1,q}^{d_{abc}, (3)}$ is given by
\begin{eqnarray}
    \Delta C_{g_1,q}^{d_{abc}, (3)} &=&
    \textcolor{blue}{\frac{d_{abc} d^{abc}}{N_C} N_F}  \Biggl\{
            \frac{64 P_{896}}{(N-1) N (1+N) (2+N)}
            -\frac{512 P_{897}}{(N-1) N (1+N) (2+N)} S_{-2,1}
    \nonumber\\ &&     
        -\frac{256 P_{898}}{(N-1) N^2 (1+N)^2 (2+N)} S_4
            +\frac{512 P_{898}}{(N-1) N^2 (1+N)^2 (2+N)} S_{3,1}
    \nonumber\\ &&     
        +\frac{32 P_{900}}{3 (N-1) N (1+N) (2+N)} \zeta_3
            +\Biggl(
                    -\frac{128 \big(
                            -16+5 N+5 N^2\big)}{(N-1) N (1+N) (2+N)}
    \nonumber\\ && 
                    -\frac{256 P_{898}}{(N-1) N^2 (1+N)^2 (2+N)} S_3
     -\frac{1024}{(N-1) (2+N)} S_{-2,1}
    \nonumber\\ &&                
     -\frac{1024 \big(
                            2+N+N^2\big)}{(N-1) N^2 (1+N)^2 (2+N)} \zeta_3
            \Biggr) S_1
            -\frac{128 \big(
                    2-N+2 N^3+N^4\big)}{(N-1) N (1+N) (2+N)} S_3
    \nonumber\\ &&   
          +\Biggl(
                    -\frac{64 P_{899}}{(N-1) N (1+N) (2+N)}
                    -\frac{256 \big(
                            -1+2 N+2 N^2\big)}{(N-1) N (1+N) (2+N)} S_1
    \nonumber\\ &&                
     +\frac{1024}{N (1+N)} \zeta_3
            \Biggr) S_{-2}
            +64 S_{-2}^2
            +\Biggl(
                    \frac{256 P_{897}}{(N-1) N (1+N) (2+N)}
    \nonumber\\ &&                
     +\frac{512}{(N-1) (2+N)} S_1
            \Biggr) S_{-3}
            +\frac{64 \big(
                    6+N+N^2\big)}{(N-1) (2+N)} S_{-4}
            -\frac{1024}{(N-1) (2+N)} S_{-2,2}
    \nonumber\\ &&    
         -\frac{512}{N (1+N)} S_{-2,3}
            -\frac{1024}{(N-1) (2+N)} S_{-3,1}
            +\frac{512}{N (1+N)} S_{-4,1}
    \nonumber\\ &&
            +\frac{2048}{(N-1) (2+N)} S_{-2,1,1}
            -\frac{1280 (N-2) (3+N)}{3 N (1+N)} \zeta_5
    \Biggr\},
\end{eqnarray}

\noindent
with
\begin{eqnarray}
    P_{896} &=& N^4+2 N^3-7 N^2-8 N+30,
    \\
    P_{897} &=& N^4+2 N^3-4 N^2-5 N+2,
    \\
    P_{898} &=& N^4+2 N^3-N^2-2 N-4,
    \\ 
    P_{899} &=& 5 N^4+10 N^3-19 N^2-24 N+10,
    \\
    P_{900} &=& 7 N^4+14 N^3-19 N^2-26 N+216.
\end{eqnarray}

\noindent
The Wilson coefficient $\Delta C_{g_1,q}^{d_{abc}, (3)}$ obeys
\begin{eqnarray}
\label{eq:BjSR}
\Delta C_{g_1,q}^{d_{abc}, (3)}(N=1) = 0
\end{eqnarray}
and does not contribute to the Bjorken sum rule \cite{Larin:1991tj,Baikov:2010je}. Four-index $d_{abcd}$ 
structures do, however, contribute in four--loop order. Let us also present this term in $z$--space. 
One obtains
\begin{eqnarray}
\Delta C_{g_1,q}^{d_{abc}, (3),\delta(1-z)}(z) &=& \textcolor{blue}{\frac{d_{abc}d^{abc} N_F}{N_C}}
\Biggl[
        64
        +160 \zeta_2
        + \frac{224}{3} \zeta_3
        - \frac{32}{5} \zeta_2^2
        - \frac{1280}{3} \zeta_5
\Biggr] \delta(1-z)
\\
\Delta C_{g_1,q}^{d_{abc}, (3),+}(z) &=& 
0
\\
\Delta C_{g_1,q}^{d_{abc}, (3),reg}(z) &=& \textcolor{blue}{\frac{d_{abc}d^{abc} N_F}{N_C}} \Biggl\{
- 768 (1 - z)
+\frac{1}{1+z} \Biggl[
        \frac{1}{1-z} \Biggl(
                32 \big(
                        2+z+3 z^2-3 z^3+z^4\big) \HA_0^2
\nonumber\\ &&             
    -\frac{64}{3} z^4 \HA_0^3
                +\frac{128}{3} \big(
                        -6-11 z+3 z^2+11 z^3+6 z^4\big) \HA_0 \zeta_2
        \Biggr)
        +64 \big(
                -3
\nonumber\\ && 
+3 z+11 z^2\big) \HA_0
        -128 \big(
                1+6 z+z^2\big) \HA_{0,1}
        +64 \big(
                -5+5 z+3 z^2
\nonumber\\ && 
+z^3\big) \zeta_2
\Biggr]
+(1-z) \Biggl[
        -896 \HA_1
        +\frac{64}{3} \HA_0^2 \HA_1
        +\frac{64 \big(
                1+7 z+z^2\big)}{3z} \HA_0^2 \HA_1^2
\nonumber\\ &&    
     -\frac{128}{3} \HA_0 \HA_{0,1}
        -\frac{256 \big(
                1+7 z+z^2\big)}{3z} \HA_0 \HA_1 \HA_{0,1}
        +\frac{128 \big(
                1+7 z+z^2\big)}{3z} \HA_{0,1}^2
\nonumber\\ && 
        +\frac{256 \big(
                1+7 z+z^2\big)}{3z} \HA_0 \HA_{0,1,1}
        -\frac{256 \big(
                1+7 z+z^2\big)}{3z} \HA_{0,0,1,1}
        +\frac{512}{3} z \HA_{0,-1,0,1}
\nonumber\\ &&     
    +\frac{256 \big(
                1+7 z+z^2\big)}{3z} \HA_0 \HA_1 \zeta_2
        -\frac{256 \big(
                1+7 z+z^2\big)}{3z} \HA_{0,1} \zeta_2
        -\frac{512}{3} z \HA_{0,-1} \zeta_2
\nonumber\\ &&       
  -\frac{512 \big(
                1+4 z+z^2\big)}{3z} \HA_1 \zeta_3
\Biggr]
+(1+z) \Biggl[
        \frac{(1-z)^2}{z} \big(
                64 \HA_{-1}^2 \HA_0
                -128 \HA_{-1} \HA_{0,-1}     
\nonumber\\ &&            
                +128 \HA_{0,-1,-1}
        \big)
        +\frac{512(1-z)}{3z} \big(
                  \HA_{0,-1,0,1}
                - \HA_{0,-1} \zeta_2
        \big)
        +\frac{64 \big(
                1-8 z+z^2\big)}{z} 
\nonumber\\ &&  \times 
\HA_{-1} \HA_0
        -\frac{64 \big(
                3+4 z+3 z^2\big)}{3z} \HA_{-1} \HA_0^2
        +\frac{128 \big(
                1-z+z^2\big)}{3z} \HA_{-1}^2 \HA_0^2
\nonumber\\ && 
        +256 \HA_{-1} \HA_0^2 \HA_{0,1}
        -\frac{128 \big(
                3+14 z+3 z^2\big)}{3z} \HA_{-1} \HA_{0,1}
        +\frac{512 \big(
                1-z+z^2\big)}{3z} 
\nonumber\\ &&  \times
\HA_{-1}^2 \HA_{0,1}
 -256 \HA_0 \HA_{0,1}^2
        -\frac{64 \big(
                1-8 z+z^2\big)}{z} \HA_{0,-1}
        -\frac{512 \big(
                1-z+z^2\big)}{3z} 
\nonumber\\ && 
\HA_{-1} \HA_0 H_{0,-1}
        +\frac{128}{3} (7+3 z) \HA_{0,0,1}
        -1024 \HA_{-1} \HA_0 \HA_{0,0,1}
        -\frac{512 \big(
                1-z+z^2\big)}{3z} 
\nonumber\\ &&  \times
\HA_{-1} \HA_{0,0,1}
        +\frac{512 \big(
                1-z+z^2\big)}{3z} \HA_{-1} \HA_{0,0,-1}
        +\frac{128 \big(
                3+14 z+3 z^2\big)}{3 z} \HA_{0,1,-1}
\nonumber\\ && 
        +256  (\HA_{0,1,1} - \HA_{0,1,-1}) \HA_0^2
        -\frac{1024 \big(
                1-z+z^2\big)}{3z} \HA_{-1} \HA_{0,1,-1}
        -256  \HA_{0,-1,1} \HA_0^2
\nonumber\\ &&    
     +\frac{128 \big(
                3+14 z+3 z^2\big)}{3z} \HA_{0,-1,1}
        -\frac{1024 \big(
                1-z+z^2\big)}{3z} \HA_{-1} \HA_{0,-1,1}
\nonumber\\ && 
        +\frac{512 \big(
                1-z+z^2\big)}{3z} \HA_0 \HA_{0,-1,-1}
        +1024 \HA_{-1} \HA_{0,0,0,1}
        +
        \frac{512 \big(
                1-z+z^2\big)}{3 z} 
\nonumber\\ &&  \times
\HA_{0,0,1,-1}
        +1024 \HA_0 \HA_{0,0,1,-1}
        +\frac{512 \big(
                1-z+z^2\big)}{3z} \HA_{0,0,-1,1}
        +1024 \HA_0 \HA_{0,0,-1,1}
\nonumber\\ && 
        -\frac{512 \big(
                1-z+z^2\big)}{3z} \HA_{0,0,-1,-1}
        +\frac{1024 \big(
                1-z+z^2\big)}{3z} \HA_{0,1,-1,-1}
        +512 \HA_0 
\nonumber\\ &&  \times
\HA_{0,-1,0,1}
        +\frac{1024 \big(
                1-z+z^2\big)}{3z} \HA_{0,-1,1,-1}
        +\frac{1024 \big(
                1-z+z^2\big)}{3z} \HA_{0,-1,-1,1}
\nonumber\\ &&     
    +1536 \HA_{0,0,0,1,1}
        -1024 \HA_{0,0,0,1,-1}
        -1024 \HA_{0,0,0,-1,1}
        +1024 \HA_{0,0,1,0,1}
\nonumber\\ &&    
     -512  \HA_{0,0,-1,0,1}
        +\frac{512 \big(
                1-z+z^2\big)}{3z} \HA_{-1} \HA_0 \zeta_2
        +256 \HA_{-1} \HA_0^2 \zeta_2
\nonumber\\ &&   
      +\frac{64 \big(
                9+22 z+9 z^2\big)}{3z} \HA_{-1} \zeta_2
        -\frac{512 \big(
                1-z+z^2\big)}{3z} \HA_{-1}^2 \zeta_2
        +512 \HA_0 \HA_{0,1} \zeta_2
\nonumber\\ &&    
     -512 \HA_0 \HA_{0,-1} \zeta_2
        -1024 \HA_{0,0,1} \zeta_2
        +512 \HA_{0,0,-1} \zeta_2
        -\frac{2048}{5} \HA_{-1} \zeta_2^2
\nonumber\\ &&   
      -512 \HA_{-1} \HA_0 \zeta_3
        +\frac{256 \big(
                1-z+z^2\big)}{z} \HA_{-1} \zeta_3
        -512 \HA_{0,1} \zeta_3
        +512 \HA_{0,-1} \zeta_3
\Biggr]
\nonumber\\ && 
-\frac{128}{3} z (6+z) \HA_0^2 \HA_{0,1}
-\frac{128 \big(
        -3-4 z-6 z^2+10 z^3\big)}{3 (1-z) z} \HA_0 \HA_{0,-1}
\nonumber\\ && 
-\frac{256}{3} z^2 \HA_0^2 \HA_{0,-1}
+\frac{512}{3} z (6+z) \HA_0 
\HA_{0,0,1}
+\frac{128 \big(
        -3-z-13 z^2+3 z^3\big)}{3z} 
\nonumber\\ &&  \times
\HA_{0,0,-1}
+\frac{1024}{3} z^2 \HA_0 \HA_{0,0,-1}
-1024 z \HA_{0,0,0,1}
-512 z^2 \HA_{0,0,0,-1}
\nonumber\\ && 
-128 z (2+z) \HA_0^2 \zeta_2
+\frac{64 (1-z) (1+z)^2}{z} \HA_1 \zeta_2
+\frac{512}{5} z (4+z) \zeta_2^2
\nonumber\\ && 
+\Biggl(
        -\frac{64}{3} \big(
                79-25 z+15 z^2\big)
        +256 \zeta_2
\Biggr) \zeta_3
-\frac{256}{3} (-6+z) z \HA_0 \zeta_3
+2560 \zeta_5
\nonumber \\ &&
-\left(\frac{256 \HA_{0,0,-1}
-192 \zeta_3}{1-z} \right)
\Biggr\}.
\nonumber \\ 
\end{eqnarray}
Illustrating three--loop heavy flavor effects on the polarized non--singlet structure function $g_1^{\rm NS}(x,Q^2)$ 
in Ref.~\cite{Behring:2015zaa} it was assumed that there is no contribution by $\Delta C_{g_1,q}^{d_{abc}, (3)}$, 
despite this function is only known now. This is correct for the three massless quark flavors dealt with there, 
but changes if four or five massless flavors are considered at very high virtualities or energies. One may generally 
assume some suppression due to (\ref{eq:BjSR}), but there is a finite contribution.  
{Note that $\Delta C_{g_1,q}^{d_{abc}, (3)}$ is independent of the 
scheme employed for the treatment of $\gamma_5$, since its contribution 
is finite.}

The other Wilson coefficients are calculated in the Larin scheme.
The Wilson coefficient $\Delta C_{g_1,q}^{\rm PS, (3),L}$ reads
\begin{eqnarray}
\lefteqn{\Delta C_{g_1,q}^{\rm PS, (3), L} =} 
\nonumber\\ &&
        \textcolor{blue}{C_F} \Biggl\{
                \textcolor{blue}{T_F^2 N_F^2} 
\Biggl[
                        \frac{128 \zeta_3 P_{902}}{9 (N-1) N^2 (1+N)^2 
(2+N)}
                        -\frac{32 P_{936}}{243 (N-1)^2 N^5 (1+N)^5 
(2+N)}
\nonumber\\ &&
                        +\Biggl(
                                -\frac{32 (2+N) P_{919}}{81 N^4 
(1+N)^4}
                                -\frac{32 (N-1) (2+N) S_2}{9 N^2 
(1+N)^2}
                        \Biggr) S_1
                        +\frac{64 (N-1) (2+N) S_1^3}{27 N^2 (1+N)^2}
\nonumber\\ &&
                        +\frac{16 (2+N) \big(
                                21+17 N-6 N^2+16 N^3\big) S_1^2}{27 
N^3 (1+N)^3}
                        +\frac{16 (2+N) \big(
                                21+5 N-24 N^2+46 N^3\big)}{27 N^3 
(1+N)^3}
\nonumber\\ && \times S_2
                        -\frac{160 (N-1) (2+N) S_3}{27 N^2 (1+N)^2}
                        +\frac{512 \big(
                                -5-12 N+5 N^2\big) S_{-2}}{9 (N-1)^2 
N (1+N)^2 (2+N)}
\nonumber\\ &&
                        -\frac{1024 S_{-3}}{3 (N-1) N (1+N) (2+N)}
                \Biggr]
\nonumber\\ &&               
  +\textcolor{blue}{C_A T_F N_F}  \Biggl[
                        \frac{64 S_{-3,1} P_{901}}{3 (N-1) N^2 (1+N)^2 
(2+N)}
                        -\frac{20 S_4 P_{903}}{3 (N-1) N^2 (1+N)^2 
(2+N)}
\nonumber\\ && 
                        -\frac{16 S_{-2,2} P_{905}}{(N-1) N^2 (1+N)^2 
(2+N)}
                        +\frac{16 S_{2,1} P_{906}}{3 N^3 (1+N)^3}
                        +\frac{64 S_{-2,1,1} P_{907}}{3 (N-1) N^2 
(1+N)^2 (2+N)}
\nonumber\\ && 
                        +\frac{8 S_1^3 P_{910}}{27 N^3 (1+N)^3}
                        -\frac{8 S_{-4} P_{915}}{3 (N-1) N^2 (1+N)^2 
(2+N)}
                        -\frac{16 S_{-2,1} P_{923}}{3 (N-1) N^3 
(1+N)^3 (2+N)}
\nonumber\\ && 
                        +\frac{8 S_3 P_{927}}{27 (N-1) N^3 (1+N)^3 
(2+N)}
                        +\frac{8 P_{938}}{243 (N-1)^2 N^6 (1+N)^6 
(2+N)}
\nonumber\\ &&                
         -\Biggl(
                                \frac{8 S_2 P_{913}}{9 N^3 (1+N)^3}
                                +\frac{16 S_{-2,1} P_{911}}{3 (N-1) 
N^2 (1+N)^2 (2+N)}
                                -\frac{8 S_3 P_{912}}{9 (N-1) N^2 
(1+N)^2 (2+N)}
\nonumber\\ && 
                                -\frac{8 P_{934}}{81 (N-1) N^5 
(1+N)^5 (2+N)}
                        \Biggr) S_1
                        +\Biggl(
                                \frac{4 P_{926}}{27 N^4 (1+N)^4}
                                -\frac{80 (N-1) (2+N) S_2}{3 N^2 
(1+N)^2}
                        \Biggr) 
\nonumber\\ && \times
S_1^2
                        +\frac{28 (N-1) (2+N) S_1^4}{9 N^2 (1+N)^2}
                        -\frac{16 \big(
                                5-2 N-2 N^2\big) S_2^2}{3 N^2 
(1+N)^2}
                        -\Biggl(
                                \frac{16 \big(
                                        22+N+N^2\big) S_1^2}{3 N^2 
(1+N)^2}
\nonumber\\ && 
                                +\frac{16 S_1 P_{922}}{3 (N-1) N^3 
(1+N)^3 (2+N)}
                                +\frac{16 P_{933}}{9 (N-1)^2 N^4 
(1+N)^4 (2+N)}
\nonumber\\ && 
                                -\frac{32 \big(
                                        8+N+N^2\big) S_2}{3 N^2 
(1+N)^2}
                        \Biggr) S_{-2}
                        -\frac{16 \big(
                                -2+3 N+3 N^2\big) S_{-2}^2}{3 N^2 
(1+N)^2}                        -
                        \frac{4 S_2 P_{928}}{27 N^4 (1+N)^4}
\nonumber\\ && 
                        +\Biggl(
                                \frac{8 P_{925}}{3 (N-1) N^3 (1+N)^3 
(2+N)}
+                                \frac{8 S_1 P_{916}}{3 (N-1) N^2 
(1+N)^2 (2+N)}
                        \Biggr) S_{-3}
\nonumber\\ &&                
         +\frac{320 \big(
                                2+N+N^2\big) S_{3,1}}{(N-1) N^2 
(1+N)^2 (2+N)}
                        +\frac{32 (N-1) (2+N) S_{2,1,1}}{3 N^2 
(1+N)^2}
                        -\frac{96 (N-1) (2+N) \zeta_2^2}{5 N^2 
(1+N)^2}
\nonumber\\ && 
                        +\Biggl(
                                -
                                \frac{16 S_1 P_{908}}{3 (N-1) N^2 
(1+N)^2 (2+N)}
                                -\frac{16 P_{924}}{9 (N-1) N^3 
(1+N)^3 (2+N)}
                        \Biggr) \zeta_3
                \Biggr]
        \Biggr\}
\nonumber\\ && 
        +\textcolor{blue}{C_F^2 T_F N_F} 
\Biggl\{
                -\frac{64 S_{3,1} P_{904}}{(N-1) N^2 (1+N)^2 (2+N)}
                +\frac{8 S_4 P_{909}}{3 (N-1) N^2 (1+N)^2 (2+N)}
\nonumber\\ &&            
    -\frac{8 S_3 P_{920}}{9 (N-1) N^3 (1+N)^3 (2+N)}
                -\frac{4 S_2 P_{931}}{3 (N-1) N^4 (1+N)^4 (2+N)}
\nonumber\\ &&              
  -\frac{8 P_{937}}{3 (N-1)^2 N^6 (1+N)^6 (2+N)}
                +\Biggl(
                        \frac{16 S_3 P_{914}}{9 (N-1) N^2 (1+N)^2 
(2+N)}
\nonumber\\ &&
                        -\frac{8 S_2 P_{917}}{3 N^3 (1+N)^3}
 +\frac{8 P_{935}}{3 (N-1)^2 N^5 (1+N)^5 
(2+N)}
\nonumber\\ &&                        
                        +\frac{256 \big(
                                2+N+N^2\big) S_{-2,1}}{(N-1) N^2 
(1+N)^2 (2+N)}
                \Biggr) S_1
\nonumber\\ &&               
 +\Biggl(
                        \frac{4 P_{932}}{3 (N-1) N^4 (1+N)^4 (2+N)}
                        -\frac{112 (N-1) (2+N) S_2}{3 N^2 (1+N)^2}
                \Biggr) S_1^2
\nonumber\\ && 
                +\frac{16 (2+N) \big(
                        22+2 N-15 N^2+27 N^3\big) S_1^3}{9 N^3 
(1+N)^3}
                +\frac{68 (N-1) (2+N) S_2^2}{3 N^2 (1+N)^2}
\nonumber\\ &&                 
+\Biggl(
                        -\frac{32 P_{929}}{(N-1)^2 N^3 (1+N)^3 (2+N)}
                        -\frac{32 S_1 P_{930}}{(N-1)^2 N^3 (1+N)^3 
(2+N)}
\nonumber\\ && 
                        +\frac{128 \big(
                                -4+N+N^2\big) S_1^2}{(N-1) N^2 
(1+N)^2 (2+N)}
                        -
                        \frac{128 \big(
                                -4+N+N^2\big) S_2}{(N-1) N^2 (1+N)^2 
(2+N)}
                \Biggr) S_{-2}
\nonumber\\ &&                
 +\Biggl(
                        \frac{64 P_{918}}{(N-1) N^3 (1+N)^2 (2+N)}
                        -\frac{128 (N-2) (3+N) S_1}{(N-1) N^2 
(1+N)^2 (2+N)}
                \Biggr) S_{-3}
\nonumber\\ &&                
 -\frac{128 \big(
                        -6+5 N+5 N^2\big) S_{-4}}{(N-1) N^2 (1+N)^2 
(2+N)}
                -\frac{64 \big(
                        4+3 N-3 N^2+2 N^3\big) S_{2,1}}{3 N^3 
(1+N)^3}
\nonumber\\ && 
                -\frac{256 \big(
                        6-N-2 N^2+N^3\big) S_{-2,1}}{(N-1) N^3 
(1+N)^2 (2+N)}
                +\frac{256 \big(
                        -2+3 N+3 N^2\big) S_{-2,2}}{(N-1) N^2 
(1+N)^2 (2+N)}
\nonumber\\ && 
                +\frac{1024 \big(
                        -1+N+N^2\big) S_{-3,1}}{(N-1) N^2 (1+N)^2 
(2+N)}
                -\frac{32 (N-1) (2+N) S_{2,1,1}}{3 N^2 (1+N)^2}
\nonumber\\ &&          
      +\frac{96 (N-1) (2+N) \zeta_2^2}{5 N^2 (1+N)^2}
 -\frac{1024 S_{-2,1,1}}{(N-1) N (1+N) (2+N)}
                +\frac{44 (N-1) (2+N) S_1^4}{9 N^2 (1+N)^2}
\nonumber\\ &&               
                +\Biggl(
                        \frac{32 P_{921}}{3 (N-1) N^3 (1+N)^3 (2+N)}
         -\frac{128 \big(
                                7+4 N+4 N^2\big) S_1}{3 N^2 (1+N)^2}
                \Biggr) \zeta_3
\Biggr\},
\end{eqnarray}

\noindent
with the polynomials
\begin{eqnarray}
P_{901} &=& N^4+2 N^3-63 N^2-64 N+28,\\ 
P_{902} &=& N^4+2 N^3-39 N^2-40 N+4,\\ 
P_{903} &=& N^4+2 N^3-15 N^2-16 N+124,\\ 
P_{904} &=& N^4+2 N^3+5 N^2+4 N+20,\\ 
P_{905} &=& N^4+2 N^3+73 N^2+72 N-20,\\ 
P_{906} &=& N^4+10 N^3-5 N^2+18 N+12,\\ 
P_{907} &=& 5 N^4+10 N^3+93 N^2+88 N-4,\\ 
P_{908} &=& 7 N^4+14 N^3-81 N^2-88 N+100,\\ 
P_{909} &=& 19 N^4+38 N^3-105 N^2-124 N+556,\\ 
P_{910} &=& 23 N^4+N^3-107 N^2+167 N+222,\\ 
P_{911} &=& 25 N^4+50 N^3+137 N^2+112 N+60,\\ 
P_{912} &=& 37 N^4+74 N^3-375 N^2-412 N-44,\\ 
P_{913} &=& 37 N^4+77 N^3-97 N^2+151 N+246,\\ 
P_{914} &=& 43 N^4+86 N^3+87 N^2+44 N+316,\\ 
P_{915} &=& 47 N^4+94 N^3-393 N^2-440 N+308,\\ 
P_{916} &=& 63 N^4+126 N^3+127 N^2+64 N+4,\\ 
P_{917} &=& 65 N^4+86 N^3-49 N^2+66 N+96,\\ 
P_{918} &=& N^5+2 N^4+7 N^3-6 N+12,\\ 
P_{919} &=& 161 N^5+23 N^4+55 N^3+55 N^2-12 N-18,\\ 
P_{920} &=& 3 N^6+27 N^5+157 N^4+461 N^3+1212 N^2+380 N+64,\\ 
P_{921} &=& 3 N^6+33 N^5+71 N^4+115 N^3+294 N^2+4 N-88,\\ 
P_{922} &=& 10 N^6+24 N^5+7 N^4-30 N^3-135 N^2-64 N+92,\\ 
P_{923} &=& 21 N^6+34 N^5-191 N^4-248 N^3-40 N^2-184 N-160,\\ 
P_{924} &=& 58 N^6+237 N^5-811 N^4-1741 N^3+873 N^2+652 N-852,\\ 
P_{925} &=& 63 N^6+138 N^5+23 N^4+228 N^3+308 N^2-456 N-240,\\ 
P_{926} &=& 79 N^6+78 N^5-1526 N^4-1632 N^3-794 N^2-3351 N-1710,\\ 
P_{927} &=& 199 N^6+588 N^5+20 N^4-208 N^3+3177 N^2+1612 N-1068,\\ 
P_{928} &=& 953 N^6+2298 N^5-664 N^4-2490 N^3+980 N^2-1455 N-1422,\\ 
P_{929} &=& N^7+N^6-10 N^5-14 N^4-35 N^3+9 N^2+20 N-4,\\ 
P_{930} &=& N^7+2 N^6+20 N^5+10 N^4-25 N^3-92 N^2-12 N+32,\\ 
P_{931} &=& 56 N^8+205 N^7+154 N^6+40 N^5-84 N^4-977 N^3-942 N^2+308
N+472,\\ 
P_{932} &=& 156 N^8+501 N^7+164 N^6-374 N^5-242 N^4-1139 N^3-958 N^2
+524 N
\nonumber\\ &&
+600,\\ 
P_{933} &=& 45 N^9+57 N^8+386 N^7+232 N^6-1896 N^5-2008 N^4+937 N^3
+63 N^2
\nonumber\\ &&
-228 N-84,\\ 
P_{934} &=& 2701 N^{10}+11348 N^9+6606 N^8-13281 N^7+8301 N^6+20247 N^5
+980 N^4
\nonumber\\ &&
+32572 N^3
+10626 N^2-17460N-10800,\\ 
P_{935} &=& 96 N^{11}+302 N^{10}-111 N^9-890 N^8-689 N^7+706 N^6
+2477 N^5+2544 N^4
\nonumber\\ &&
-497 N^3
-1854 N^2+324 N+664,\\ 
P_{936} &=& 2042 N^{11}+6095 N^{10}-2575 N^9-15042 N^8-2412 N^7
+15972 N^6+22057 N^5
\nonumber\\ &&
+18358 N^4
+2452 N^3-8427 N^2+468 N+2484,\\ 
P_{937} &=& 260 N^{13}+1033 N^{12}+579 N^{11}-2050 N^{10}-2768 N^9
-582 N^8+146 N^7-680 N^6
\nonumber\\ &&
+752N^5
+1237 N^4-765 N^3-950 N^2+356 N+360,\\ 
P_{938} &=& 30520 N^{13}+123245 N^{12}+50588 N^{11}-274955 N^{10}
-241017 N^9+129972 N^8
\nonumber\\ &&
+331898 N^7+404113 N^6+59461 N^5-137557 N^4
-25818 N^3+116766 N^2
\nonumber\\ &&
+1080 N
-29160.
\end{eqnarray}

Finally, the Wilson coefficient $\Delta C_{g_1,g}^{(3)}$ is given by
\begin{eqnarray}
\Delta C_{g_1,g}^{(3)} &=& 
\Delta C_{g_1,g}^{(3),a} +
\Delta C_{g_1,g}^{(3),d_{abc}},
\end{eqnarray}
with 



\noindent
and
\begin{eqnarray}
P_{939} &=& -305 N^4-514 N^3-19 N^2+142 N-1416,
\\
P_{940} &=& -117 N^4+14 N^3+791 N^2+676 N-116,
\\
P_{941} &=& 2 N^4+8 N^3-6 N^2-35 N-20,
\\
P_{942} &=& 3 N^4+15 N^3+14 N^2+2 N+8,
\\
P_{943} &=& 11 N^4-14 N^3+148 N^2+191 N-48,
\\
P_{944} &=& 17 N^4-20 N^3+51 N^2+112 N-36,
\\
P_{945} &=& 44 N^4+61 N^3+55 N^2+74 N-6,
\\
P_{946} &=& 124 N^4+233 N^3+49 N^2-60 N+54,
\\
P_{947} &=& 177 N^4+146 N^3-731 N^2-492 N+324,
\\
P_{948} &=& 349 N^4+680 N^3-550 N^2-245 N+1038,
\\
P_{949} &=& 368 N^4+1036 N^3-1145 N^2-1741 N+954,
\\
P_{950} &=& 497 N^4+646 N^3-785 N^2+482 N+1848,
\\
P_{951} &=& 6 N^5+12 N^4+65 N^3+138 N^2+225 N+130,
\\
P_{952} &=& 12 N^5+36 N^4+191 N^3+318 N^2+369 N+226,
\\
P_{953} &=& 16 N^5-47 N^3+19 N^2-4 N+12,
\\
P_{954} &=& 19 N^5+76 N^4+45 N^3-52 N^2+20 N+36,
\\
P_{955} &=& 42 N^5+15 N^4+23 N^3-141 N^2-153 N+70,
\\
P_{956} &=& 53 N^5+158 N^4+9 N^3-119 N^2+34 N-108,
\\
P_{957} &=& 67 N^5+277 N^4+216 N^3-145 N^2+44 N+180,
\\
P_{958} &=& 127 N^5+N^4-421 N^3+267 N^2-114 N-436,
\\
P_{959} &=& 3 N^6-14 N^5-551 N^4-838 N^3-644 N^2-572 N-264,
\\
P_{960} &=& 3 N^6+9 N^5-N^4+7 N^3+82 N^2+20 N+24,
\\
P_{961} &=& 3 N^6+9 N^5+49 N^4+77 N^3-28 N^2-50 N+84,
\\
P_{962} &=& 6 N^6+24 N^5+113 N^4+158 N^3+159 N^2+112 N+4,
\\
P_{963} &=& 6 N^6+36 N^5+147 N^4+158 N^3+87 N^2+78 N+64,
\\
P_{964} &=& 8 N^6+27 N^5+18 N^4-25 N^3-80 N^2-44 N+24,
\\
P_{965} &=& 9 N^6+71 N^5+981 N^4+1675 N^3+1936 N^2+1224 N-136,
\\
P_{966} &=& 9 N^6+95 N^5+955 N^4+1477 N^3+2132 N^2+1620 N-528,
\\
P_{967} &=& 9 N^6+119 N^5+929 N^4+1279 N^3+2328 N^2+2016 N-920,
\\
P_{968} &=& 12 N^6+50 N^5+565 N^4+958 N^3+433 N^2+86 N+488,
\\
P_{969} &=& 12 N^6+60 N^5+261 N^4+314 N^3+267 N^2+222 N+16,
\\
P_{970} &=& 15 N^6+23 N^5-101 N^4-131 N^3-32 N^2-82 N+20,
\\
P_{971} &=& 15 N^6+47 N^5+107 N^4+139 N^3+108 N^2+24 N-8,
\\
P_{972} &=& 19 N^6+63 N^5-28 N^4-154 N^3-63 N^2-35 N-18,
\\
P_{973} &=& 21 N^6+57 N^5+29 N^4+3 N^3+126 N^2+76 N-24,
\\
P_{974} &=& 23 N^6-201 N^5-955 N^4-583 N^3+828 N^2-48 N-536,
\\
P_{975} &=& 23 N^6+61 N^5+93 N^4-117 N^3-692 N^2-256 N+600,
\\
P_{976} &=& 27 N^6+87 N^5+179 N^4+213 N^3+94 N^2-40 N+16,
\\
P_{977} &=& 30 N^6+163 N^5+788 N^4+865 N^3+1322 N^2+1408 N-544,
\\
P_{978} &=& 33 N^6+113 N^5+815 N^4+1319 N^3+1140 N^2+708 N+1632,
\\
P_{979} &=& 35 N^6+93 N^5-17 N^4-97 N^3-238 N^2-320 N+256,
\\
P_{980} &=& 39 N^6+13 N^5-439 N^4-421 N^3-2156 N^2-2412 N+1344,
\\
P_{981} &=& 39 N^6+161 N^5+45 N^4-345 N^3+600 N^2+908 N-832,
\\
P_{982} &=& 60 N^6+126 N^5-491 N^4-724 N^3-271 N^2-584 N-420,
\\
P_{983} &=& 60 N^6+132 N^5-367 N^4-652 N^3-711 N^2-782 N+16,
\\
P_{984} &=& 66 N^6+140 N^5-337 N^4-550 N^3-835 N^2-1024 N+236,
\\
P_{985} &=& 69 N^6+203 N^5-153 N^4+21 N^3+1536 N^2+308 N-1408,
\\
P_{986} &=& 72 N^6+148 N^5-307 N^4-448 N^3-959 N^2-1266 N+456,
\\
P_{987} &=& 85 N^6+243 N^5+443 N^4+425 N^3+2592 N^2+2644 N+3072,
\\
P_{988} &=& 93 N^6+39 N^5+3465 N^4+7681 N^3+5118 N^2+924 N+3416,
\\
P_{989} &=& 143 N^6+231 N^5+298 N^4+1495 N^3-4263 N^2-6796 N+3708,
\\
P_{990} &=& 147 N^6+453 N^5+811 N^4+755 N^3+606 N^2+500 N+760,
\\
P_{991} &=& 176 N^6+390 N^5-829 N^4-1288 N^3-429 N^2-1208 N-268,
\\
P_{992} &=& 205 N^6+225 N^5-7345 N^4-15529 N^3-14712 N^2-4748 N-9936,
\\
P_{993} &=& 213 N^6+703 N^5+3205 N^4+4849 N^3+3122 N^2+1340 N+2120,
\\
P_{994} &=& 233 N^6+591 N^5-713 N^4-683 N^3+7632 N^2+5372 N-4656,
\\
P_{995} &=& 263 N^6+771 N^5-848 N^4-1349 N^3+9819 N^2+7532 N-1932,
\\
P_{996} &=& 274 N^6+624 N^5-278 N^4-615 N^3+625 N^2+108 N-540,
\\
P_{997} &=& 434 N^6+810 N^5+5245 N^4+10912 N^3+3351 N^2-2956 N+16764,
\\
P_{998} &=& 485 N^6+975 N^5+2077 N^4+4777 N^3+666 N^2-3820 N+13848,
\\
P_{999} &=& 641 N^6+1995 N^5+3601 N^4+5437 N^3+11178 N^2+6116 N-1320,
\\
P_{1000} &=& 764 N^6+1896 N^5-592 N^4-1797 N^3+413 N^2-3762 N-3600,
\\
P_{1001} &=& 1721 N^6+3837 N^5+29 N^4+909 N^3+4838 N^2-5394 N-5508,
\\
P_{1002} &=& 30 N^7+51 N^6-5 N^5-173 N^4-63 N^3-22 N^2-118 N+12,
\\
P_{1003} &=& 140 N^7+643 N^6+902 N^5+N^4-676 N^3+94 N^2+264 N-216,
\\
P_{1004} &=& 200 N^7+1357 N^6+2729 N^5+1069 N^4-481 N^3+1174 N^2-324 N-1080,
\\
P_{1005} &=& 293 N^7+513 N^6+2089 N^5+4287 N^4+11362 N^3+9848 N^2+3496 N-784,
\\
P_{1006} &=& 383 N^7+1702 N^6+689 N^5-4466 N^4-4516 N^3-176 N^2-312 N-336,
\\
P_{1007} &=& 388 N^7+1781 N^6+1660 N^5-2725 N^4-5366 N^3-2458 N^2-192 N-216,
\\
P_{1008} &=& 1067 N^7+5191 N^6+5663 N^5-2933 N^4-2176 N^3+4672 N^2-288 N-1008,
\\
P_{1009} &=& 1427 N^7+6712 N^6+6878 N^5-5021 N^4-4822 N^3+280 N^2-12132 N-8856,
\\
P_{1010} &=& 1471 N^7+7367 N^6+12949 N^5+9107 N^4+6508 N^3+4334 N^2-12144 N
\nonumber\\ &&
-10440,
\\
P_{1011} &=& -205 N^8-904 N^7+88 N^6+3218 N^5+2309 N^4+254 N^3-944 N^2-2584 N
\nonumber\\ &&
-1424,
\\
P_{1012} &=& 13 N^8+50 N^7+104 N^6+130 N^5+127 N^4+136 N^3+100 N^2+36 N-24,
\\
P_{1013} &=& 35 N^8+142 N^7+16 N^6-424 N^5+341 N^4+1418 N^3+704 N^2+24 N+48,
\\
P_{1014} &=& 39 N^8+390 N^7-2192 N^6-4350 N^5+10577 N^4+8384 N^3-28776 N^2-9800 N
\nonumber\\ &&
+13632,
\\
P_{1015} &=& 42 N^8+141 N^7+284 N^6+351 N^5-1843 N^4-2925 N^3-730 N^2-1272 N
\nonumber\\ &&
-960,
\\
P_{1016} &=& 71 N^8+290 N^7+108 N^6-624 N^5+437 N^4+1838 N^3+384 N^2
\nonumber\\ &&
-248 N+48,
\\
P_{1017} &=& 90 N^8+417 N^7-155 N^6-1779 N^5+7189 N^4+14142 N^3+6376 N^2+4776 N
\nonumber\\ &&
+3504,
\\
P_{1018} &=& 93 N^8+1048 N^7+2200 N^6+1622 N^5+1891 N^4+890 N^3-936 N^2+376 N
\nonumber\\ &&
+624,
\\
P_{1019} &=& 114 N^8+345 N^7-222 N^6-1074 N^5-1960 N^4-3335 N^3-2280 N^2-68 N
\nonumber\\ &&
+416,
\\
P_{1020} &=& 153 N^8+725 N^7+380 N^6-2494 N^5-3431 N^4-399 N^3-510 N^2-1672 N
\nonumber\\ &&
-1008,
\\
P_{1021} &=& 191 N^8+684 N^7-217 N^6-1949 N^5-1246 N^4+2305 N^3+4728 N^2+1120 N
\nonumber\\ &&
-432,
\\
P_{1022} &=& 240 N^8+1089 N^7-917 N^6-6465 N^5+6181 N^4+19980 N^3+8596 N^2+3840 N
\nonumber\\ &&
+3168,
\\
P_{1023} &=& 339 N^8+843 N^7-2092 N^6-2274 N^5+1147 N^4-14645 N^3-26082 N^2+116 N
\nonumber\\ &&
+6360,
\\
P_{1024} &=& 430 N^8+1951 N^7-340 N^6-7385 N^5+2891 N^4+11059 N^3+970 N^2+8136 N
\nonumber\\ &&
+3888,
\\
P_{1025} &=& 1565 N^8+5906 N^7+3476 N^6-7570 N^5-4705 N^4+152 N^3-8328 N^2+3888,
\\
P_{1026} &=& 1973 N^8+5921 N^7+1402 N^6-1417 N^5+14062 N^4+21596 N^3+41243 N^2
\nonumber\\ &&
+14832 N
-14076,
\\
P_{1027} &=& 4724 N^8+15473 N^7+3469 N^6-22813 N^5+48721 N^4+140300 N^3+112538 N^2
\nonumber\\ &&
+15108 N-6480,
\\
P_{1028} &=& 19 N^9+63 N^8+192 N^7+162 N^6-513 N^5-513 N^4+230 N^3-720 N^2
\nonumber\\ &&
-216 N+144,
\\
P_{1029} &=& 651 N^9+3056 N^8+1492 N^7-12986 N^6-24959 N^5-15422 N^4-6056 N^3
\nonumber\\ &&
-5680 N^2
+1296 N+2592,
\\
P_{1030} &=& -12043 N^{10}-52199 N^9-68592 N^8-22926 N^7+115557 N^6+458949 N^5
\nonumber\\ &&
+428806 N^4
-302176 
N^3-337584 N^2+149040 N+140832,\\P_{1031} &=& 63 N^{10}+291 N^9+581 N^8+1646 N^7+2335 N^6+4289 N^5
+13213 N^4+5150 N^3
\nonumber\\ &&
-12616 N^2+552 N+5232,
\\
P_{1032} &=& 226 N^{10}+1061 N^9+860 N^8-1831 N^7-2444 N^6+2045 N^5+8496 N^4+6381 N^3
\nonumber\\ &&
-4438 N^2
-2268 N+2280,
\\
P_{1033} &=& 5399 N^{10}+21865 N^9+4191 N^8-66642 N^7-36909 N^6+68193 N^5+32287 N^4
\nonumber\\ &&
+4016 N^3
+45360 N^2+3888 N-19440,\\P_{1034} &=& 5459 N^{10}+35950 N^9+52398 N^8-48294 N^7-52359 N^6
+143592 N^5+37432 N^4
\nonumber\\ &&
-26596 N^3+134946 N^2-70632 
N-87480,
\\
P_{1035} &=& 9429 N^{10}+33109 N^9-26912 N^8-199686 N^7-184155 N^6+5745 N^5+39702 N^4
\nonumber\\ &&
+97328 N^3
+107792 N^2-12816 N-35424,
\\
P_{1036} &=& 13789 N^{10}+60242 N^9+31062 N^8-128262 N^7-48825 N^6
+174492 N^5
\nonumber\\ && 
+110408 N^4
+144844 N^3+66474 N^2-183816 N-115992,
\\
P_{1037} &=& 205 N^{12}+1122 N^{11}-2485 N^{10}-21156 N^9-27453 N^8+17874 N^7+42893 N^6
\nonumber\\ &&
-5568 N^5-25832 N^4+2784 N^3+10512 N^2+576 N-384,
\\
P_{1038} &=& 254 N^{12}+1391 N^{11}+413 N^{10}-8542 N^9-8007 N^8+22639 N^7+38167 N^6
\nonumber\\ &&
+10124 N^5
-9715 N^4-4980 N^3+2736 N^2+3344 N+1328,
\\
P_{1039} &=& 1538 N^{12}+8703 N^{11}+20491 N^{10}+20280 N^9-17340 N^8-20277 N^7
\nonumber\\ &&
+105025 N^6
+77580 N^5-83254 N^4+66642 N^3+131004 N^2-18792 N
\nonumber\\ &&
-42768,
\\
P_{1040} &=& 3805 N^{12}+23880 N^{11}+31020 N^{10}-63109 N^9-153501 N^8-33090 N^7
+131168 N^6
\nonumber\\ &&
+144597 N^5-48240 N^4-172070 N^3+49356 N^2+44280 N+10800,
\\
P_{1041} &=& 4920 N^{13}+22305 N^{12}+218 N^{11}-95044 N^{10}-22165 N^9+174177 N^8
+27618 N^7
\nonumber\\ &&
-156150 N^6-99101 N^5-98972 N^4-72698 N^3+67716 N^2+20376 N-22032,
\\
P_{1042} &=& 617 N^{14}+4175 N^{13}+14349 N^{12}+30747 N^{11}+2959 N^{10}-140311 N^9
-226729 N^8
\nonumber\\ &&
-109247 N^7-131196 N^6-283508 N^5-248768 N^4-18848 N^3+45408 N^2
\nonumber\\ &&
+6528 N
-2944,
\\
P_{1043} &=& 32968 N^{14}+214291 N^{13}+230166 N^{12}-1212607 N^{11}-2507302 N^{10}
+1955271 N^9
\nonumber\\ &&
+5076722 N^8-7562245 N^7-17306682 N^6-3182438 N^5+7389328 N^4+339888 N^3
\nonumber\\ &&
-2572992 N^2+515808 N+673920,
\\
P_{1044} &=& 34763 N^{14}+211157 N^{13}+229596 N^{12}-789770 N^{11}-1434236 N^{10}
+1014978 N^9
\nonumber\\ &&
+2087956 N^8-1947938 N^7-3974499 N^6-2470711 N^5+66548 N^4+2887812 N^3
\nonumber\\ &&
+1290168 N^2-982368 N-702432,
\\
P_{1045} &=& -10701 N^{15}-42606 N^{14}-112294 N^{13}-235574 N^{12}-177548 N^{11}
-133306 N^{10}
\nonumber\\ &&
-388110 N^9+1317318 N^8+3159337 N^7+1521824 N^6+1661012 N^5+2875000 N^4
\nonumber\\ &&
-65712 N^3-1912320 N^2+39744 N+466560,
\\
P_{1046} &=& 23379 N^{16}+176637 N^{15}+336908 N^{14}-418342 N^{13}-1616541 N^{12}
+367147 N^{11}
\nonumber\\ &&
+3663481 N^{10}+1241253 N^9-2594354 N^8-1895204 N^7+1153767 N^6
\nonumber\\ &&
+3040589 N^5-340384 N^4-2307312 N^3-1311840 N^2+714096 N+513216,
\\
P_{1047} &=& 295317 N^{16}+2034216 N^{15}+4013522 N^{14}-2161486 N^{13}-8694264 N^{12}
\nonumber\\ &&
+30856978 N^{11}+88306930 N^{10}+17465238 N^9-110878589 N^8-41117522 N^7
\nonumber\\ && +85727340 N^6
+17006960 N^5-58178704 N^4-9306528 N^3+13384512 N^2
\nonumber\\ &&
-1838592 N
-3027456.
\end{eqnarray}
\begin{eqnarray}
\Delta C_{g_1,g}^{(3),d_{abc}} &=&
\textcolor{blue}{\frac{d_{abc} d^{abc} N_F^2 }{N_A}}
\Biggl\{
        -\frac{128 (N-2) (3+N)P_{1048}}{45 N (1+N) (2+N)} S_5
        -\frac{64 (N-2) (3+N)P_{1048}}{9 N (1+N) (2+N)} \zeta_5
\nonumber\\ && 
        +\frac{128 P_{1049}}{(N-1) N (1+N) (2+N)^2} S_3
        +\frac{32 P_{1052}}{45 (N-1) N (1+N) (2+N)^2}
\nonumber\\ && 
        -\frac{256 P_{1053}}{45 N^2 (1+N)^2 (2+N)} S_{-2,1}
        -\frac{512  P_{1055}}{3 (N-1) N^2 (1+N)^2 (2+N)^2} S_4
\nonumber\\ && 
        +\frac{1024 P_{1055}}{3 (N-1) N^2 (1+N)^2 (2+N)^2} S_{3,1}
        -\frac{64P_{1058}}{45 (N-1) N^2 (1+N)^2 (2+N)^2} \zeta_3
\nonumber\\ &&
        +\Biggl[
                -\frac{128 \big(
                        -4+N+N^2\big)P_{1050}}{(N-1) N^2 (1+N)^2 
(2+N)^2} S_3
                -\frac{64 P_{1051}}{45 (N-1) N (1+N) (2+N)^2}
\nonumber\\ && 
                +\frac{512 P_{1056}}{3 (N-1) N^2 (1+N)^2 (2+N)^2} 
\zeta_3
                +\frac{512 \big(
                        -4+N+N^2\big)}{N (1+N) (2+N)} S_4
\nonumber\\ &&               
 -\frac{1024 \big(
                        -4+N+N^2\big) S_{3,1}}{N (1+N) (2+N)}
                +\frac{512 \big(
                        -8+3 N+3 N^2\big)}{3 N (1+N) (2+N)} 
S_{-2,1}
        \Biggr] S_1
\nonumber\\ && 
        +\Biggl[
                \frac{128 \big(
                        -19+3 N+3 N^2\big)}{3 N (1+N) (2+N)}
                +\frac{256 \big(
                        -4+N+N^2\big)}{N (1+N) (2+N)} S_3
                -\frac{512 \big(
                        -4+N+N^2\big)}{N (1+N) (2+N)} 
\nonumber\\ &&  \times \zeta_3 
        \Biggr] S_1^2
        -\frac{768 \big(
                -4+N+N^2\big)}{N (1+N) (2+N)} S_2 S_3
        +\Biggl[
                -
                \frac{64 P_{1054}}{45 (N-1) N (1+N) (2+N)^2}
\nonumber\\ &&                
 +\frac{256 P_{1057}}{45 (N-1) N^2 (1+N)^2 (2+N)^2} 
S_1
                -\frac{256 \big(
                        -8+3 N+3 N^2\big)}{3 N (1+N) (2+N)} S_1^2
\nonumber\\ &&                
 -\frac{256}{45} (N-1) \big(
                        -21+N+N^2\big) [S_3 + S_{-2,1} + \zeta_3]
        \Biggr] S_{-2}
        -\frac{128}{2+N} S_{-2}^2
\nonumber\\ &&        
 +\Biggl[
                \frac{128 P_{1053}}{45 N^2 (1+N)^2 (2+N)}
                -\frac{256 \big(
                        -8+3 N+3 N^2\big)}{3 N (1+N) (2+N)} S_1
        \Biggr] S_{-3}
\nonumber\\ &&   
      -\frac{256 \big(
                -4+3 N+3 N^2\big)}{3 N (1+N) (2+N)} S_{-4}
        -\frac{128}{45} (N-1) \big(
                -21+N+N^2\big) S_{-5}
\nonumber\\ &&    
     +\frac{768 \big(
                -4+N+N^2\big)}{N (1+N) (2+N)} [S_{2,3} - S_{4,1}]
        +\frac{256 \big(
                -8+3 N+3 N^2\big)}{3 N (1+N) (2+N)} S_{-2,2}
\nonumber\\ &&  
       +\frac{256}{45} (N-1) \big(
                -21+N+N^2\big) [S_{-2,3} + S_{-2,1,-2}]
        +\frac{256 \big(
                -8+3 N+3 N^2\big)}{3 N (1+N) (2+N)} S_{-3,1}
\nonumber\\ &&       
  +\frac{1536 \big(
                -4+N+N^2\big)}{N (1+N) (2+N)} S_{3,1,1}
        -\frac{512 \big(
                -8+3 N+3 N^2\big)}{3 N (1+N) (2+N)} 
S_{-2,1,1}
\Biggr\}
\end{eqnarray}

\noindent
and
\begin{eqnarray}
P_{1048} &=& N^4+2 N^3-16 N^2-17 N+120,
\\
P_{1049} &=& 5 N^4+10 N^3-16 N^2-21 N+10,
\\
P_{1050} &=& 5 N^4+10 N^3+3 N^2-2 N-4,
\\
P_{1051} &=& 46 N^4+92 N^3+1049 N^2+1003 N-3342,
\\
P_{1052} &=& 65 N^4+130 N^3-593 N^2-658 N+5232,
\\
P_{1053} &=& N^6+3 N^5+24 N^4+43 N^3-142 N^2-163 N-156,
\\
P_{1054} &=& 2 N^6+6 N^5+315 N^4+620 N^3-1391 N^2-1700 N+60,
\\
P_{1055} &=& 3 N^6+9 N^5-4 N^4-23 N^3-14 N^2-N+12,
\\
P_{1056} &=& 9 N^6+27 N^5-13 N^4-71 N^3-26 N^2+14 N+24,
\\
P_{1057} &=& 45 N^6+135 N^5-140 N^4-505 N^3+35 N^2+310 N+156,
\\
P_{1058} &=& 2 N^8+8 N^7+815 N^6+2417 N^5-1509 N^4-7037 N^3-1780 N^2
\nonumber\\ &&
+2140 
N+624.
\end{eqnarray}

\noindent
The above Wilson coefficients complete the massless parts of the polarized single and two mass heavy flavor Wilson 
coefficients still missing in Refs.~\cite{Behring:2015zaa,Ablinger:2019etw,Ablinger:2019gpu,
Ablinger:2020snj,Blumlein:2021xlc}.

\section{The small- and \boldmath large $x$ expansions of the Wilson coefficients}
\label{sec:8}

\vspace*{1mm}
\noindent
We will represent the small $x$ and large $x$ behaviour of the 
different Wilson coefficients first in Mellin $N$ space, where the former corresponds to the
expansion around $N=0$ and $N=1$, while the latter corresponds to the expansion in the limit 
$N \rightarrow \infty$. Finally, we also consider the large $N_F$ limit.

The non--singlet Wilson coefficients receive their leading small $x$ contributions from the expansion
around $N=0$, retaining the pole terms. The singlet ones also receive contributions from the pole 
terms at $N=1$. This is due to the Mellin transforms
\begin{eqnarray}
\frac{1}{(N-1)^k} &=& \frac{(-1)^{k-1}}{(k-1)!} 
\Mvec\left[\frac{\ln^{k-1}(x)}{x}\right](N)
\\
\frac{1}{N^k}     &=& \frac{(-1)^{k-1}}{(k-1)!} \Mvec[\ln^{k-1}(x)](N).
\end{eqnarray}
In the large $x$ region we will retain all terms up to $\propto 1/N$ in the asymptotic expansion.
Here, terms of the kind
\begin{eqnarray}
L_a^l \equiv (\ln(N) + \gamma_E)^l
\end{eqnarray}
contribute, which we will replace by
\begin{eqnarray}
\ln(N) + \gamma_E = S_1(N) - \frac{1}{2 N} + O\left(\frac{1}{N^2}\right),
\end{eqnarray}
to derive in a more direct way the inverse Mellin transforms. One obtains asymptotically up to terms of
$O(L_a^k/N^2),~~k \geq 0$, and smaller
\begin{eqnarray}
\Mvec^{-1}[S_1(N)](x) &\simeq& -\frac{1}{1-x} + 1,
\\
\Mvec^{-1}[S_1^2(N)](x) &\simeq& \frac{2 \ln(1-x)}{1-x} - 2 \ln(1-x) - \zeta_2 \delta(1-x) + 1,
\\
\Mvec^{-1}[S_1^3(N)](x) &\simeq&
-\frac{3 \ln^2(1-x)}{1-x}
+\frac{3 \zeta_2}{1-x}
+3 \ln^2(1-x)
-3 \ln(1-x) - 2 \zeta_3 \delta(1-x) 
\nonumber\\ &&
- 3 \zeta_2,
\\
\Mvec^{-1}[S_1^4(N)](x) &\simeq&
\frac{4 \ln^3(1-x)}{1-x}
-\frac{12 \zeta_2 \ln(1-x)}{1-x}
+\frac{8 \zeta_3}{1-x}
-4 \ln^3(1-x)
+6 \ln^2(1-x)
\nonumber\\ &&
+12 \zeta_2 \ln(1-x) + \frac{3}{5} \zeta_2^2 \delta(1-x) - 6 \zeta_2 - 8 \zeta_3,
\\
\Mvec^{-1}[S_1^5(N)](x) &\simeq&
-\frac{5 \ln^4(1-x)}{1-x}
+\frac{30 \zeta_2 \ln^2(1-x)}{1-x}
-\frac{40 \zeta_3 \ln(1-x)}{1-x}
-\frac{3 \zeta_2^2}{1-x}
\nonumber\\ &&
+5 \ln^4(1-x)
-10 \ln^3(1-x)
-30 \zeta_2 \ln^2(1-x)
+(30 \zeta_2 + 40 \zeta_3)
\nonumber\\ &&
\times \ln(1-x) + (20 \zeta_2 \zeta_3 - 24 \zeta_5) \delta(1-x) + 3 \zeta_2^2 - 20 \zeta_3,
\\
\Mvec^{-1}[S_1^6(N)](x) &\simeq&
\frac{6 \ln^5(1-x)}{1-x}
-\frac{60 \zeta_2 \ln^3(1-x)}{1-x}
+\frac{120 \zeta_3 \ln^2(1-x)}{1-x}
+\frac{18 \zeta_2^2 \ln(1-x)}{1-x}
\nonumber\\ &&
{ -(120 \zeta_2 \zeta_3 - 144 \zeta_5)} \frac{1}{1-x}
-6 \ln^5(1-x)
+15 \ln^4(1-x)
\nonumber\\ &&
+60 \zeta_2 \ln^3(1-x)
-(90 \zeta_2 + 120 \zeta_3) \ln^2(1-x)
+(120 \zeta_3 - 18  \zeta_2^2) \ln(1-x)
\nonumber\\  &&
+ \left(40 \zeta_3^2 - \frac{45}{7} \zeta_2^3\right) \delta(1-x) + 9 \zeta_2^2 + 120 \zeta_2 
\zeta_3 -144 \zeta_5,
\end{eqnarray}
and
\begin{eqnarray}
\Mvec^{-1}\left[\frac{S_1(N)}{N}\right](x) &\simeq& 
- \ln(1-x),
\\
\Mvec^{-1}\left[\frac{S_1^2(N)}{N}\right](x) &\simeq& 
 \ln^2(1-x) - \zeta_2,
\\
\Mvec^{-1}\left[\frac{S_1^3(N)}{N}\right](x) &\simeq& 
-\ln^3(1-x)
+3 \zeta_2 \ln(1-x) - 2 \zeta_3,
\\
\Mvec^{-1}\left[\frac{S_1^4(N)}{N}\right](x) &\simeq& 
\ln^4(1-x)
-6 \zeta_2 \ln^2(1-x)
+8 \zeta_3 \ln(1-x)
+ \frac{3}{5} \zeta_2^2,
\\
\Mvec^{-1}\left[\frac{S_1^5(N)}{N}\right](x) &\simeq&
-\ln^5(1-x)
+10 \zeta_2 \ln^3(1-x)
-20 \zeta_3 \ln^2(1-x)
-3 \zeta_2^2 \ln(1-x)
\nonumber\\ &&
{ +} 20 \zeta_2 \zeta_3 - 24 \zeta_5,
\\
\Mvec^{-1}\left[\frac{S_1^6(N)}{N}\right](x) &\simeq&
\ln^6(1-x)
-15 \zeta_2 \ln^4(1-x)
+40 \zeta_3 \ln^3(1-x)
+9 \zeta_2^2 \ln^2(1-x)
\nonumber\\ &&
+(144 \zeta_5 -120 \zeta_2 \zeta_3) \ln(1-x)
- \frac{45}{7} \zeta_2^3 + 40 \zeta_3^2.
\end{eqnarray}
We will now present the asymptotic behaviour of the Wilson coefficients in the small $x$ and large $x$ 
region. For the pure singlet and gluonic Wilson coefficients of the structure 
function $g_1(x,Q^2)$ we refer to the Larin scheme.\footnote{To one- and two-loop order it is known, that
the gluonic Wilson coefficients are the same in the Larin and {\sf M}-scheme \cite{Matiounine:1998re}.}
\subsection{The \boldmath small $x$ limit}
\label{sec:81}

\vspace*{1mm}
\noindent
{In this section we will show the small $x$ expansion of the Wilson coefficients calculated above.} 
The non--singlet unpolarized Wilson coefficients have leading singularities at $N = 0$, $\propto \ln^k(x)$ 
and the unpolarized pure singlet and gluonic Wilson coefficients have their leading singularity at $N = 1$, 
i.e. terms as $\ln(x)/x$ and $1/x$. Here we also list the logarithmic contributions up to 
the constant terms. One obtains 



\noindent
Note, that the expansion has to be performed in $z$--space, since the Mellin space expression 
is either applicable for even or odd moments. 
A small $x$ expansion of the non--singlet Wilson coefficient of the structure function $g_1(x,Q^2)$ has been 
considered in \cite{Kiyo:1996si} based on \cite{Bartels:1995iu}, however, 
in a different not fully specified scheme.
\subsection{The \boldmath large $x$ limit}
\label{sec:82}

\vspace*{1mm}
\noindent
In the large $x$ limit one obtains the following leading term behaviour for the Wilson 
coefficients calculated in the present paper. Here we dropped the terms $O(L_a^k/N^2),~~k \geq 0$.  

and
\begin{eqnarray}
\Delta C_{g_1,q}^{(1),\rm NS,L} &\simeq&  
\textcolor{blue}{C_F} \Biggl[
         2 S_1^2
        +3 S_1
        -9
        -2 \zeta_2
        +\frac{5}{N}
\Biggr],
\\
\Delta C_{g_1,q}^{(2),\rm NS,L} &\simeq& 
\textcolor{blue}{C_F} \Biggl\{
\textcolor{blue}{T_F N_F} \Biggl[
                -\frac{58}{9} S_1^2
                -\frac{8}{9} S_1^3
                +\Biggl(
                        -\frac{494}{27}
                        +\frac{8 \zeta_2}{3}
                        -\frac{12}{N}
                \Biggr) S_1
                +\frac{457}{18}
                +\frac{170}{9} \zeta_2
                +\frac{8}{9} \zeta_3
\nonumber\\ &&
                -\frac{332}{9 N}
        \Biggr]
        +\textcolor{blue}{C_A} \Biggl[
                 \frac{22}{9} S_1^3
                +\Biggl(
                        \frac{367}{18}
                        -4 \zeta_2
                \Biggr) S_1^2
                +\Biggl(
                        \frac{3155}{54}
                        +\frac{23}{N}
                        -\frac{22 \zeta_2}{3}
                        -40 \zeta_3
                \Biggr) S_1
\nonumber\\ &&         
                -\frac{5465}{72}
                -\frac{1139}{18} \zeta_2
       +\frac{464}{9} \zeta_3
                +\frac{51}{5} \zeta_2^2
                +\frac{772}{9 N}
                -\frac{8}{N} \zeta_2
        \Biggr]
\Biggr\}
+\textcolor{blue}{C_F^2} \Biggl[
         2 S_1^4
        +6 S_1^3
\nonumber\\ &&               
        +\Biggl(
                -\frac{27}{2}
                +\frac{22}{N}
                -4 \zeta_2
        \Biggr) S_1^2
        +\Biggl(
                -\frac{51}{2}
 +\frac{25}{N}
                -18 \zeta_2
                +24 \zeta_3
        \Biggr) S_1
        +\frac{331}{8}
        +\frac{111}{2} \zeta_2
\nonumber\\ && 
        -66 \zeta_3
        +\frac{4}{5} \zeta_2^2
       -\frac{61}{N}
        +\frac{2 \zeta_2}{N}
\Biggr],
\\
\\
\Delta C_{g_1,q}^{(2),\rm PS,L} &\simeq&  0,
\\
\Delta C_{g_1,q}^{(3),\rm PS,L} &\simeq&  0,
\\
\Delta C_{g_1,g}^{(1)} &\simeq&  -\frac{4}{N} \textcolor{blue}{T_F N_F} \big(
        1 + S_1 \big),
\\
\Delta C_{g_1,g}^{(2)} &\simeq& 
\frac{1}{N} \Biggl\{
        \textcolor{blue}{C_A T_F N_F}  \Biggl[
                -\frac{4}{3} S_1^3
                -8 S_1^2
                +\Biggl(
                        -28
                        +12 \zeta_2
                \Biggr) S_1
                -16
                -\frac{92}{3} \zeta_3
        \Biggr]
\nonumber\\ &&
        +\textcolor{blue}{C_F T_F N_F} \Biggl[
                -\frac{20}{3} S_1^3
                -18 S_1^2
                +\Biggl(
                        4
                        +12 \zeta_2
                \Biggr) S_1
                -4
                +10 \zeta_2
                +\frac{176}{3} \zeta_3
        \Biggr]
\Biggr\},
\\
\Delta C_{g_1,g}^{(3)} &\simeq& 
\frac{1}{N} \textcolor{blue}{T_F N_F} \Biggl\{
	        -\Biggl[
                \frac{4}{3} \textcolor{blue}{C_A^2}
                +\frac{20}{3} \textcolor{blue}{C_F^2}
        \Biggr] S_1^5
        +\Biggl[
                -\frac{293}{27} \textcolor{blue}{C_A^2}
                -\frac{83}{3} \textcolor{blue}{C_F^2}
                +\frac{28}{27} \textcolor{blue}{C_A T_F N_F} 
\nonumber\\ && 
                +\textcolor{blue}{C_F} \Biggl(
                        -\frac{412}{27} \textcolor{blue}{C_A}
                        +\frac{68}{27} \textcolor{blue}{T_F N_F}
                \Biggr)
        \Biggr] S_1^4
        +\Biggl[
                \frac{608}{81} \textcolor{blue}{C_A T_F N_F} 
                +\textcolor{blue}{C_F} \Biggl(
                        \frac{1984}{81} \textcolor{blue}{T_F N_F} 
                        +\textcolor{blue}{C_A} \Biggl(
\nonumber\\ &&                
                 -\frac{6884}{81}
                                +\frac{272}{9} \zeta_2
                        \Biggr)
                \Biggr)
                +\textcolor{blue}{C_A^2} \Biggl(
                        -\frac{6112}{81}
                        +\frac{152}{9} \zeta_2
                \Biggr)
                +\textcolor{blue}{C_F^2} \Biggl(
                        -\frac{254}{9}
                        +\frac{152}{9} \zeta_2
                \Biggr)
        \Biggr] S_1^3
\nonumber\\ && 
        +\Biggl[
                \textcolor{blue}{C_F} \Biggl(
                        \textcolor{blue}{T_F N_F} \Biggl(
                                \frac{6260}{81}
                                -\frac{56}{9} \zeta_2
                        \Biggr)
                        +\textcolor{blue}{C_A} \Biggl(
                                -\frac{12043}{81}
                                +\frac{568}{9} \zeta_2
                                +\frac{104}{3} \zeta_3
                        \Biggr)
                \Biggr)
                +\textcolor{blue}{C_A T_F N_F}
\nonumber\\ &&  \times
\Biggl(
                        \frac{4384}{81}
                        -\frac{104}{9} \zeta_2
                \Biggr)
                +\textcolor{blue}{C_F^2} \Biggl(
                        -\frac{205}{3}
                        +118 \zeta_2
                        +\frac{32}{3} \zeta_3
                \Biggr)
                +\textcolor{blue}{C_A^2} \Biggl(
                        -\frac{27578}{81}
                        +\frac{718}{9} \zeta_2
\nonumber\\ &&                
         +\frac{272}{3} \zeta_3
                \Biggr)
        \Biggr] S_1^2
        +\Biggl[
                \textcolor{blue}{C_F} \Biggl(
                        \textcolor{blue}{C_A} \Biggl(
                                \frac{65936}{243}
                                +\frac{7396}{27} \zeta_2
                                -\frac{10796}{27} \zeta_3
                                +\frac{976}{15} \zeta_2^2
                        \Biggr)
\nonumber\\ &&                
         +\textcolor{blue}{T_F N_F} \Biggl(
                                -\frac{12304}{243}
                                -\frac{1952}{27} \zeta_2
                                -\frac{224}{27} \zeta_3
                        \Biggr)
                \Biggr)
                +\textcolor{blue}{C_A T_F N_F} 
\Biggl(
                        \frac{43192}{243}
                        -\frac{1376}{27} \zeta_2
\nonumber\\ &&                
         +\frac{32}{27} \zeta_3
                \Biggr)
                +\textcolor{blue}{C_A^2} \Biggl(
                        -\frac{139052}{243}
                        +\frac{5992}{27} \zeta_2
                        +\frac{3368}{27} \zeta_3
                        -\frac{136}{3} \zeta_2^2
                \Biggr)
                +\textcolor{blue}{C_F^2} \Biggl(
                        -\frac{508}{3}
                        -54 \zeta_2
\nonumber\\ &&                
         +\frac{2072}{3} \zeta_3
                        -\frac{2732}{15} \zeta_2^2
                \Biggr)
        \Biggr] S_1
+        \textcolor{blue}{C_F} \Biggl\{
                \textcolor{blue}{C_A} \Biggl[
                        \frac{32813}{108}
                        +\frac{1321}{3} \zeta_2
                        -\frac{11908}{81} \zeta_3
                        -\frac{1172}{9} \zeta_2^2
\nonumber\\ &&                
                        +\frac{352}{9} \zeta_2 \zeta_3
                        +\frac{1496}{3} \zeta_5
                \Biggr]
 +\textcolor{blue}{T_F N_F} \Biggl(
                        -\frac{1189}{27}
                        -\frac{188}{3} \zeta_2
                        -\frac{16672}{81} \zeta_3
                        +\frac{1028}{45} \zeta_2^2
                \Biggr)
                \Biggr\}
\nonumber\\ && 
        +\textcolor{blue}{C_F^2} \Biggl[
                \frac{617}{12}
                -173 \zeta_2
                +\frac{5564}{9} \zeta_3
                +\frac{193}{5} \zeta_2^2
                -\frac{1040}{9} \zeta_2 \zeta_3
                -\frac{2384}{3} \zeta_5
        \Biggr]
        +\textcolor{blue}{C_A^2} \Biggl[
                -\frac{31172}{81}
\nonumber\\ &&          
                -\frac{478}{9} \zeta_2
 -\frac{10988}{81} \zeta_3
                +\frac{163}{5} \zeta_2^2
                -\frac{344}{9} \zeta_2 \zeta_3
                -16 \zeta_5
        \Biggr]
        +\textcolor{blue}{C_A T_F N_F} \Biggl[
                \frac{13120}{81}
                -\frac{16}{9} \zeta_2
\nonumber\\ &&               
                +\frac{17824}{81} \zeta_3
  -\frac{4}{5} \zeta_2^2
        \Biggr]
\Biggr\} ,
\\
\Delta C_{g_1,g}^{(3),d_{abc}} &\simeq& 
\frac{1}{N} \textcolor{blue}{\frac{d_{abc} d^{abc} N_F^2}{N_A}} 
\Biggl\{
        \Biggl[
                128
                +128 \zeta_2
                -256 \zeta_3
        \Biggr] S_1^2
+        48
        +\Biggl[
                -64
\nonumber\\ && 
                -128 \zeta_2
                +768 \zeta_3
                -\frac{1536}{5} \zeta_2^2
        \Biggr] S_1
        +256 \zeta_2
        -512 \zeta_3
        +\frac{1728}{5} \zeta_2^2
        -640 \zeta_5
\Biggr\}.
\end{eqnarray}

\subsection{The \boldmath large $N_F$ limit}
\label{sec:83}

\vspace*{1mm}

\noindent
There are also some predictions on the large $N_F$ behaviour of deep--inelastic Wilson coefficients.
The $O(C_F T_F^2 N_F^2)$ contribution to the flavor non--singlet Wilson coefficient of the structure function
$F_2(x,Q^2)$ agrees with Eq.~(15a) of \cite{Mankiewicz:1997gz}, which can be brought into a more compact form. 

For the structure function $F_L^{\rm NS}(N,a_s)$ the generating functional of the Wilson coefficients
reads \cite{Gracey:1995aj}.
\begin{eqnarray}
G_L(g) = \textcolor{blue}{C_F} \sum_{k=0}^\infty \frac{1}{k!}\frac{\partial^k}{\partial x^k} \left[
\frac{ g e^{5/3x} \Gamma(x+N)}{(1-x)(2-x)(1+x+n) \Gamma(1+x) \Gamma(N)}\right]_{x= (4/3) T_F N_F g}.
\end{eqnarray}
The $O(a_s^k)$ Wilson coefficient is given by its $k$th expansion coefficient in the variable 
$g$, $f_L^{(k)}(g=a_s)$ to be multiplied by $(-1)^{k+1}$. This prediction agrees with the corresponding 
contributions in our direct calculation to three loop order and implies a modification of 
\cite{Gracey:1995aj}. 
\section{Conclusions}
\label{sec:9}

\vspace*{1mm}
\noindent
We have calculated the unpolarized massless three--loop Wilson coefficients of the 
structure functions $F_2(x,Q^2)$ and $F_L(x,Q)$ for pure photon exchange and for the 
charged current structure function $xF_3(x,Q^2)$ to three--loop order. In the polarized 
case, we calculated the Wilson coefficients to the structure function $g_1(x,Q^2)$ to 
three--loop order in the Larin scheme and for the non--singlet case also in the 
$\overline{\sf MS}$ scheme. We applied the forward Compton amplitude for the respective 
scattering cross section and followed the other technical calculation steps having been 
described in our previous calculations of the three--loop anomalous dimensions in 
Refs.~\cite{Blumlein:2021enk,Blumlein:2021ryt}. By using the method of arbitrarily high
Mellin moments \cite{Blumlein:2017dxp} we have obtained a sufficiently large basis to 
compute the Wilson coefficients without any reference to further structural assumptions.
In parallel, we also used the method of differential equations \cite{Ablinger:2018zwz} since we face
only first order factorizable problems, which, even further, relate to harmonic polylogarithms only.  
This method works for any basis of master integrals. The Wilson 
coefficients depend on 60 harmonic sums, weighted by rational functions in the
Mellin variable $N$, after applying the algebraic relations, and on 31 harmonic sums by also
applying the structural relations. We confirm former results on $F_2, F_L$ and $xF_3$ 
at three--loop order in Refs.~\cite{Vermaseren:2005qc,Moch:2008fj} for the first time 
and all earlier results at one-- and two--loop order, also for the structure function 
$g_1(x,Q^2)$. The three--loop results for the structure function $g_1(x,Q^2)$ are new.
Concerning the polarized case we would like to remark that one may perfectly work in the 
Larin scheme, expressing the Wilson coefficients, the anomalous dimensions and the parton 
densities at the starting scale in this scheme, representing the polarized deep--inelastic 
structure functions or any other observables. For convenience, we also provide the expansions
of the Wilson coefficients in the small $x$ and large $x$ region. The latter information may 
be of relevance studying the high energy Regge limit and the soft region, respectively.
We also checked some predictions in the large $N_F$ limit.
The present results are of importance for the application in experimental and phenomenological 
analyzes of precision deep--inelastic data and in particular for precision measurements of the
strong coupling constant.

\vspace*{1cm}
\vspace{5mm}\noindent
{\bf Acknowledgments.}~~We would like to thank A.~Behring and S.~Moch for discussions. This project 
has received funding from the European Union's Horizon 2020 research and innovation programme under 
the Marie Sklodowska--Curie grant agreement No. 764850, SAGEX and from the Austrian Science Fund (FWF) 
grants SFB F50 (F5009-N15) and P33530.

\appendix
\section{\boldmath $Z_5^{\rm NS}(N)$}
\label{sec:A}

\vspace*{1mm} \noindent
In the following we calculate the function $Z_5^{\rm NS}(N)$ needed to transform the 
non--singlet Wilson coefficient of the structure function $g_1$ from the Larin into the 
$\overline{\sf MS}$ scheme to three--loop order. To two--loop order it has been 
calculated in \cite{Matiounine:1998re} in unrenormalized form.\footnote{Note that Eq.~(A11) in 
Ref.~\cite{Matiounine:1998re} contains typographical errors.} In \cite{Blumlein:2022ndg} we provided the 
expansion coefficients needed to two loop order at the required depth in $\ep$ for the 
present calculation in the unrenormalized case. 

The practical approach consists in deriving $Z_5^{\rm NS}(N)$ in renormalized form, since it provides a finite 
renormalization. It is given by
\begin{eqnarray}
Z_5^{\rm NS}(N) &=& \frac{A_{qq}^{\rm NS, phys, ac,ren}}{A_{qq}^{\rm NS, phys, Larin,ren}}
= 1 
+ {a}_s z_5^{(1),\rm NS}
+ {a}_s^2 z_5^{(2),\rm NS}
+ {a}_s^3 z_5^{(3),\rm NS} + O(\hat{a}_s^4),
\end{eqnarray}
where the OMEs are calculated for anticommuting $\gamma_5$ and in the Larin scheme. 
The gauge parameter $\xi$ is defined in \cite{Blumlein:2021enk}. Note that in the Larin scheme one
has to add the physical and EOM expansion coefficients projected in Ref.~\cite{Blumlein:2022ndg}, although
the two projections are orthogonal.

For odd integers $N$ it is given by 
\begin{eqnarray}
z_5^{(1),\rm NS,odd} &=& - \textcolor{blue}{C_F} \frac{8}{N (1+N)}
\\
z_5^{(2),\rm NS,odd} &=& 
        \textcolor{blue}{C_F T_F N_F} \frac{16\big(
                -3-N+5 N^2\big)}{9 N^2 (1+N)^2}
        +\textcolor{blue}{C_F^2} \Biggl[
                \frac{16 (1+2 N)}{N^2 (1+N)^2} S_1    
                +\frac{16}{N (1+N)} S_2          
\nonumber\\ && 
                + \frac{8 Q_2}{N^3 (1+N)^3}
                +\frac{32}{N (1+N)} S_{-2}
        \Biggr]
        -\textcolor{blue}{C_A C_F}  \Biggl[
                \frac{4 Q_5}{9 N^3 (1+N)^3}
                +\frac{16}{N (1+N)} S_{-2}
        \Biggr],
        \nonumber \\ &&
\\
z_5^{(3),\rm NS,odd} &=& 
        \textcolor{blue}{C_F T_F^2 N_F^2} \frac{128 Q_1}{81 N^3 (1+N)^3}
        +\textcolor{blue}{C_A} \Biggl\{
                \textcolor{blue}{C_F T_F N_F} \Biggl[
                        -\frac{32 Q_3}{9 N^3 (1+N)^3} S_1
                        +\frac{16 Q_{13}}{81 N^4 (1+N)^4}
\nonumber\\ &&                     
    -\frac{64}{3 N (1+N)} S_3
                        +\frac{128 \big(
                                -3+4 N+10 N^2\big)}{27 N^2 (1+N)^2} S_{-2}
                        -\frac{128}{9 N (1+N)} S_{-3}
\nonumber\\ &&                
         -\frac{256}{9 N (1+N)} S_{-2,1}
                        +\frac{128}{3 N (1+N)} \zeta_3
                \Biggr]
                +\textcolor{blue}{C_F^2} \Biggl[
                        \frac{64 Q_4}{27 N^3 (1+N)^3} S_2
\nonumber\\ &&                
         +\frac{8 Q_{15}}{27 (N-1) N^5 (1+N)^5 (2+N)}
                        +\Biggl(
                                -\frac{16 Q_{11}}{27 N^4 (1+N)^4}
                                -\frac{512}{3 N (1+N)} S_3
\nonumber\\ &&                
                 -\frac{3584}{3 N (1+N)} S_{-2,1}
                        \Biggr) S_1
                        -\frac{32 \big(
                                -30-7 N+5 N^2\big)}{9 N^2 (1+N)^2} S_3
                        -\frac{320}{3 N (1+N)} S_4
\nonumber\\ &&                
         +\Biggl(
                                \frac{32 Q_{12}}{27 (N-1) N^3 (1+N)^3 (2+N)}
                                +\frac{64 \big(
                                        -10+19 N+11 N^2\big) S_1}{3 N^2 (1+N)^2}
\nonumber\\ &&                           
      +\frac{256}{3 N (1+N)} S_2
                        \Biggr) S_{-2}
                        -\frac{512}{3 N (1+N)} S_{-2}^2
                        +\Biggl(
                                -\frac{64 \big(
                                        78-N+5 N^2\big)}{9 N^2 (1+N)^2}
\nonumber\\ &&                
                 +\frac{768}{N (1+N)} S_1
                        \Biggr) S_{-3}
                        +\frac{1472 
                        }{3 N (1+N)} S_{-4}
                        -\frac{256 \big(
                                -33+16 N+13 N^2\big)}{9 N^2 (1+N)^2} S_{-2,1}
\nonumber\\ && 
                        +
                        \frac{1024}{3 N (1+N)} S_{3,1}
                        -\frac{3712}{3 N (1+N)} S_{-2,2}
                        -\frac{1280}{N (1+N)} S_{-3,1}
                        +\frac{7168}{3 N (1+N)} S_{-2,1,1}
\nonumber\\ &&                
         -\frac{96 \big(
                                -2+5 N+5 N^2\big)}{N^2 (1+N)^2} \zeta_3
                \Biggr]
        \Biggr\}
        +\textcolor{blue}{C_F^2 T_F N_F} \Biggl[
                \frac{8 Q_7}{27 N^4 (1+N)^4}
                +\frac{256 \big(
                        3+N-5 N^2\big)}{27 N^2 (1+N)^2} 
\nonumber\\ && \times S_2
                +\frac{128 \big(
                        12+17 N-14 N^3+3 N^4\big)}{27 N^3 (1+N)^3} S_1
                +\frac{512}{9 N (1+N)} S_3
                +\frac{256}{9 N (1+N)} S_{-3}
\nonumber\\ && 
                -\frac{256 \big(
                        -3+4 N+10 N^2\big)}{27 N^2 (1+N)^2} S_{-2}
                +\frac{512}{9 N (1+N)} S_{-2,1}
                -\frac{128}{3 N (1+N)} \zeta_3
        \Biggr]
\nonumber\\ && 
        +\textcolor{blue}{C_A^2 C_F} \Biggl[
                -\frac{4 Q_{16}}{81 (N-1) N^5 (1+N)^5 (2+N)}
                +\Biggl(
                        \frac{16 Q_9}{9 N^4 (1+N)^4}
                        +\frac{256}{3 N (1+N)} S_3
\nonumber\\ &&                
         +\frac{1024}{3 N (1+N)} S_{-2,1}
                \Biggr) S_1
                -\frac{16 \big(
                        8+N+N^2\big)}{3 N^2 (1+N)^2} S_3
                +\frac{256}{3 N (1+N)} S_4
\nonumber\\ &&                
 +\Biggl(
                        -\frac{64 Q_{10}}{27 (N-1) N^3 (1+N)^3 (2+N)}
                        -\frac{256 \big(
                                -1+N+N^2\big)}{3 N^2 (1+N)^2} S_1
                \Biggr) S_{-2}
\nonumber\\ &&               
                +\frac{64 }{N (1+N)} S_{-2}^2 
                +\Biggl(
                        \frac{64 \big(
                                21+N+N^2\big)}{9 N^2 (1+N)^2}
                        -\frac{512}{3 N (1+N)} S_1
                \Biggr) S_{-3}
                -\frac{320}{3 N (1+N)} S_{-4}
\nonumber\\ &&                
 -\frac{512}{3 N (1+N)} S_{3,1}
                +\frac{128 \big(
                        -21+10 N+10 N^2\big)}{9 N^2 (1+N)^2} S_{-2,1}
                +\frac{1024}{3 N (1+N)} S_{-2,2}
\nonumber\\ &&                
                +\frac{1024}{3 N (1+N)} S_{-3,1}
 -\frac{2048}{3 N (1+N)} S_{-2,1,1}
                +\frac{32 \big(
                        -2+5 N+5 N^2\big)}{N^2 (1+N)^2} \zeta_3
        \Biggr]
\nonumber\\ && 
        +\textcolor{blue}{C_F^3} \Biggl[
                -\frac{8 Q_{14}}{3 (N-1) N^5 (1+N)^5 (2+N)}
                +\Biggl(
                        \frac{16 Q_8}{3 N^4 (1+N)^4}
                        -\frac{256 (1+2 N)}{3 N^2 (1+N)^2} S_2
\nonumber\\ &&                
         +\frac{1024}{N (1+N)} S_{-2,1}
                \Biggr) S_1
                -\frac{128 \big(
                        1+3 N+3 N^2\big)}{3 N^3 (1+N)^3} S_1^2
                -\frac{32 \big(
                        3+11 N-4 N^2+4 N^3\big)}{3 N^2 (1+N)^3} S_2
\nonumber\\ &&                
 -\frac{128}{3 N (1+N)} S_2^2
                -\frac{64 \big(
                        2+5 N+N^2\big) S_3}{3 N^2 (1+N)^2}
                -\frac{128}{3 N (1+N)} S_4
\nonumber\\ &&                
 +\Biggl(
                        -\frac{64 Q_6}{3 (N-1) N^3 (1+N)^3 (2+N)}
                        -\frac{128 \big(
                                -2+11 N+3 N^2\big)}{3 N^2 (1+N)^2} S_1
\nonumber\\ &&                
         -\frac{512}{3 N (1+N)} S_2
                \Biggr) S_{-2}
                +\frac{256}{3 N (1+N)} S_{-2}^2
                +\Biggl(
                        \frac{128 \big(
                                12-N+N^2\big)}{3 N^2 (1+N)^2}
\nonumber\\ &&                
         -\frac{2560}{3 N (1+N)} S_1
                \Biggr) S_{-3}
                -\frac{1664}{3 N (1+N)} S_{-4}
                +\frac{512 \big(
                        -4+2 N+N^2\big)}{3 N^2 (1+N)^2} S_{-2,1}
\nonumber\\ && 
                -\frac{512}{3 N (1+N)} S_{3,1}
                +\frac{3328}{3 N (1+N)} S_{-2,2}
                +\frac{3584}{3 N (1+N)} S_{-3,1}
                -\frac{2048}{N (1+N)} S_{-2,1,1}
\nonumber\\ &&               
 +\frac{64 \big(
                        -2+5 N+5 N^2\big)}{N^2 (1+N)^2} \zeta_3
        \Biggr]
\end{eqnarray}
with
\begin{eqnarray}
Q_1&=&N^4+12 N^3+7 N^2-4 N-3,
\\
Q_2&=&2 N^4+N^3+8 N^2+5 N+2,
\\
Q_3&=&3 N^4+6 N^3+5 N^2+2 N+2,
\\
Q_4&=&85 N^4+104 N^3+13 N^2-6 N+18,
\\
Q_5&=&103 N^4+140 N^3+58 N^2+21 N+36,
\\
Q_6&=&N^6-3 N^5+9 N^3-33 N^2-6 N+8,
\\
Q_7&=&17 N^6+207 N^5-685 N^4-691 N^3-312 N^2+80 N+48,
\\
Q_8&=&22 N^6+50 N^5+41 N^4-132 N^3-153 N^2-104 N-40,
\\
Q_9&=&24 N^6+72 N^5+44 N^4-32 N^3-35 N^2-7 N-12,
\\
Q_{10}&=&85 N^6+222 N^5-38 N^4-336 N^3-92 N^2+69 N+36,
\\
Q_{11}&=&165 N^6-185 N^5-1034 N^4-1285 N^3-895 N^2-726 N-396,
\\
Q_{12}&=&349 N^6+861 N^5-152 N^4-1263 N^3-665 N^2+222 N+216,
\\
Q_{13}&=&485 N^6+643 N^5+253 N^4+85 N^3+326 N^2-96 N-144,
\\
Q_{14}&=&24 N^{10}+99 N^9+259 N^8+308 N^7-186 N^6-853 N^5-1153 N^4-82 N^3
\nonumber\\ &&
+344 N^2
+328 N+144,
\\
Q_{15}&=&845 N^{10}+3292 N^9+7545 N^8+11366 N^7-121 N^6-19168 N^5-19017 N^4
\nonumber\\ &&
-2522 N^3+5420 N^2+4008 N+1440,
\\
Q_{16}&=&6087 N^{10}+24679 N^9+32532 N^8+11838 N^7-18471 N^6-38727 N^5
\nonumber\\ &&
-37968 N^4
-12190 N^3+11772 N^2+8352 N+1728.
\end{eqnarray}
To derive the above relations one has to apply the relations \cite{Blumlein:2009cf}
\begin{eqnarray}
S_{n_1,...,n_p}\left(\frac{N}{2}\right) = 2^{n_1+...+n_p - p} \sum_\pm S_{\pm n_1, ..., \pm n_p} (N).
\end{eqnarray}

{\footnotesize

}
\end{document}